\input amstexl
% LAMSTEX.TEX   VERSION 2.01
% COPYRIGHT (C) 1989, 1990, 1991 BY THE TEXPLORATORS CORPORATION
%  3701 W. ALABAMA, SUITE 450-273, HOUSTON, TX 77027
% ALL RIGHTS RESERVED

% ABSOLUTELY NO CHANGES SHOULD BE MADE TO THIS FILE;
% CHANGES SHOULD BE MADE ONLY IN STYLE FILES.

\catcode`\@=11
\ifx\amstexloaded@\relax\else
 \errmessage{AmS-TeX must be loaded before LamS-TeX}\fi
\ifx\laxread@\undefined\else\catcode`\@=\active \fi
\def\err@#1{\errmessage{LamS-TeX error: #1}}
\def^^L{\par}
\let\+\tabalign
\def\newcount{\alloc@0\count\countdef\insc@unt}
\def\newdimen{\alloc@1\dimen\dimendef\insc@unt}
\def\newskip{\alloc@2\skip\skipdef\insc@unt}
\def\newmuskip{\alloc@3\muskip\muskipdef\@cclvi}
\def\newbox{\alloc@4\box\chardef\insc@unt}
\let\newtoks\relax
\def\newhelp#1#2{\newtoks#1#1\expandafter{\csname#2\endcsname}}
\def\newtoks{\alloc@5\toks\toksdef\@cclvi}
\def\newread{\alloc@6\read\chardef\sixt@@n}
\def\newwrite{\alloc@7\write\chardef\sixt@@n}
\def\newfam{\alloc@8\fam\chardef\sixt@@n}
\def\newlanguage{\alloc@9\language\chardef\@cclvi}
\def\newinsert#1{\global\advance\insc@unt by\m@ne
  \ch@ck0\insc@unt\count
  \ch@ck1\insc@unt\dimen
  \ch@ck2\insc@unt\skip
  \ch@ck4\insc@unt\box
  \allocationnumber=\insc@unt
  \global\chardef#1=\allocationnumber
  \wlog{\string#1=\string\insert\the\allocationnumber}}
\def\newif#1{\count@\escapechar \escapechar\m@ne
  \expandafter\expandafter\expandafter
   \edef\@if#1{true}{\let\noexpand#1=\noexpand\iftrue}%
  \expandafter\expandafter\expandafter
   \edef\@if#1{false}{\let\noexpand#1=\noexpand\iffalse}%
  \@if#1{false}\escapechar\count@}

\def\Err@#1{\errhelp\defaulthelp@\err@{#1}}
{\catcode`\@=\active
 \edef\next{\gdef\noexpand@{\futurelet\noexpand\next
  \csname at\string@\endcsname}}
 \next
}
\def\at@{\ifcat\noexpand\next a\let\next@\at@@\else
 \ifcat\noexpand\next0\let\next@\at@@\else
 \ifcat\noexpand\next\relax\let\next@\at@@\else
 \let\next@\at@@@\fi\fi\fi\next@}
\def\at@@@{\errhelp\athelp@\err@{Invalid use of @}}
\def\at@@#1{\expandafter
 \ifx\csname\string#1@at\endcsname\relax\let\next@\at@@@\else
 \DN@{\csname\string#1@at\endcsname}\fi\next@}
\def\atdef@#1{\expandafter\def\csname\string#1@at\endcsname}
\newif\iftest@
\def\tagin@#1{\tagin@false
 \DN@##1\tag##2##3\next@{\test@true\ifx\tagin@##2\test@false\fi}%
 \next@#1\tag\tagin@\next@\tagin@false\iftest@\tagin@true\fi}
\let\lkerns@\relax
\def\nolinebreak{\RIfM@\mathmodeerr@\nolinebreak\else
 \ifhmode\saveskip@\lastskip\unskip
 \nobreak\ifdim\saveskip@>\z@\hskip\saveskip@\fi\lkerns@
 \else\vmodeerr@\nolinebreak\fi\fi}
\def\allowlinebreak{\RIfM@\mathmodeerr@\allowlinebreak\else
 \ifhmode\saveskip@\lastskip\unskip
 \allowbreak\ifdim\saveskip@>\z@\hskip\saveskip@\fi\lkerns@
 \else\vmodeerr@\allowlinebreak\fi\fi}
\def\linebreak{\RIfM@\mathmodeerr@\linebreak\else
 \ifhmode\unskip\unkern\break\lkerns@
 \else\vmodeerr@\linebreak\fi\fi}
\let\nkerns@\relax
\def\newline{\RIfM@\mathmodeerr@\newline\else
 \ifhmode\unskip\unkern\null\hfill\break\nkerns@
 \else\vmodeerr@\newline\fi\fi}%
\def\newbox@{\alloc@@4\box\chardef\insc@unt}
\def\newcount@{\alloc@@0\count\countdef\insc@unt}
\def\accentedsymbol#1#2{\expandafter\newbox@\csname\exstring@#1@box\endcsname
 \setbox\csname\exstring@#1@box\endcsname\hbox{$\m@th#2$}%
 \define#1{\copy\csname\exstring@#1@box\endcsname{}}}
\def\rightadd@#1\to#2{\toks@{\\#1}\toks@@\expandafter{#2}\xdef#2{\the\toks@@
 \the\toks@}\toks@{}\toks@@{}}
\def\fontlist@{\\\tenrm\\\sevenrm\\\fiverm\\\teni\\\seveni\\\fivei
 \\\tensy\\\sevensy\\\fivesy\\\tenex\\\tenbf\\\sevenbf\\\fivebf
 \\\tensl\\\tenit}
\def\font@#1=#2 {\rightadd@#1\to\fontlist@\font#1=#2 }
\def\ismember@#1#2{\global\let\Next@ F\let\next@= #2%
 {\def\\##1{\let\nextii@##1\ifx\nextii@\next@\global\let\Next@ T\fi}#1}%
 \test@false\ifx\Next@ T\test@true\fi\let\next@\relax}
\def\FNSS@#1{\let\FNSS@@#1\FN@\FNSS@@@}
\def\FNSS@@@{\ifx\next\space@\def\FNSS@@@@. {\FN@\FNSS@@@}\else
 \def\FNSS@@@@.{\FNSS@@}\fi\FNSS@@@@.}
\atdef@"{\unskip
 \DN@{\ifx\next`\DN@`{\FN@\nextii@}%
  \else\ifx\next\lq\DN@\lq{\FN@\nextii@}%
  \else\DN@####1{\FN@\nextiii@}\fi\fi
  \next@}%
 \DNii@{\ifx\next`\DN@`{\sldl@``}%
  \else\ifx\next\lq\DN@\lq{\sldl@``}%
  \else\DN@{\dlsl@`}\fi\fi\next@}%
 \def\nextiii@{\ifx\next'\DN@'{\srdr@''}%
  \else\ifx\next\rq\DN@\rq{\srdr@''}%
  \else\DN@{\drsr@'}\fi\fi\next@}%
 \FNSS@\next@}
\def\root{%
  \DN@{\ifx\next\uproot\let\next@\nextii@\else
   \ifx\next\leftroot\let\next@\nextiii@\else
   \let\next@\plainroot@\fi\fi\next@}%
  \DNii@\uproot##1{\uproot@##1\relax\FNSS@\nextiv@}%
  \def\nextiv@{\ifx\next\leftroot\let\next@\nextv@\else
   \let\next@\plainroot@\fi\next@}%
  \def\nextv@\leftroot##1{\leftroot@##1\relax\plainroot@}%
  \def\nextiii@\leftroot##1{\leftroot@##1\relax\FNSS@\nextvi@}%
  \def\nextvi@{\ifx\next\uproot\let\next@\nextvii@\else
   \let\next@\plainroot@\fi\next@}%
  \def\nextvii@\uproot##1{\uproot@##1\relax\plainroot@}%
  \bgroup\uproot@\z@\leftroot@\z@
 \FNSS@\next@}
\def\loop#1\repeat{\def\iterate{#1\relax\expandafter\iterate\fi}%
 \iterate\let\iterate\relax}
\def\gloop@#1\repeat{\gdef\iterate@{#1\relax\expandafter\iterate@\fi}%
 \iterate@\global\let\iterate@\relax}
\def\printoptions{\W@{Do you want S(yntax check),
  G(alleys) or P(ages)?^^JType S, G or P, follow by <return>: }\loop
 \read\m@ne to\ans@
 \edef\next@{\def\noexpand\Ans@{\ans@}}\uppercase\expandafter{\next@}%
 \ifx\Ans@\S@\test@true\syntax\else
 \ifx\Ans@\G@\test@true\galleys\else
 \ifx\Ans@\P@\test@true\else
 \test@false\fi\fi\fi
 \iftest@\else\W@{Type S, G or P, follow by <return>: }%
 \repeat}
\expandafter\let\csname A@;\endcsname;
\expandafter\let\csname A@:\endcsname:
\expandafter\let\csname A@?\endcsname?
\expandafter\let\csname A@!\endcsname!
\def\APdef#1{\def\next@{\expandafter\let\csname A@\string#1\endcsname#1}%
 \afterassignment\next@\def#1}
\let\fextra@\,
\def\tdots@{\unskip
 \DN@{$\m@th\mathinner{\ldotp\ldotp\ldotp}\,
   \ifx\next,\,$\else\ifx\next.\,$\else
   \ifx\next;\,$\else
   \expandafter\ifx\csname A@\string;\endcsname\next\fextra@$\else
   \ifx\next:\,$\else
   \expandafter\ifx\csname A@\string:\endcsname\next\fextra@$\else
   \ifx\next?\,$\else
   \expandafter\ifx\csname A@\string?\endcsname\next\fextra@$\else
   \ifx\next!\,$\else
   \expandafter\ifx\csname A@\string!\endcsname\next\fextra@$\else
   $ \fi\fi\fi\fi\fi\fi\fi\fi\fi\fi}%
 \ \FN@\next@}
\def\extrap@#1{%
 \ifx\next,\DN@{#1\,}\else
 \ifx\next;\DN@{#1\,}\else
 \expandafter\ifx\csname A@\string;\endcsname\next\DN@{#1\fextra@}\else
 \ifx\next.\DN@{#1\,}\else\extra@
 \ifextra@\DN@{#1\,}\else
 \let\next@#1\fi\fi\fi\fi\fi\next@}
\def\dotsc{\DN@{\ifx\next;\plainldots@\,\else
 \expandafter\ifx\csname A@\string;\endcsname\next\plainldots@\fextra@\else
 \ifx\next.\plainldots@\,\else\extra@\plainldots@
 \ifextra@\,\fi\fi\fi\fi}%
 \FN@\next@}
\def\keybin@{\keybin@true
 \ifx\next+\else\ifx\next=\else\ifx\next<\else\ifx\next>\else\ifx\next-\else
 \ifx\next*\else\ifx\next:\else
 \expandafter\ifx\csname A@\string;\endcsname\next\else
 \keybin@false\fi\fi\fi\fi\fi\fi\fi\fi}
\def\boldkey#1{\ifcat\noexpand#1A%
  \ifcmmibloaded@{\fam\cmmibfam#1}\else
   \Err@{First bold symbol font not loaded}\fi
 \else
 \let\next=#1%
 \ifx#1!\mathchar"5\bffam@21 \else
 \expandafter\ifx\csname A@\string!\endcsname\next\mathchar"5\bffam@21 \else
 \ifx#1(\mathchar"4\bffam@28 \else\ifx#1)\mathchar"5\bffam@29 \else
 \ifx#1+\mathchar"2\bffam@2B \else\ifx#1:\mathchar"3\bffam@3A \else
 \expandafter\ifx\csname A@\string:\endcsname\next\mathchar"3\bffam@3A \else
 \ifx#1;\mathchar"6\bffam@3B \else
 \expandafter\ifx\csname A@\string;\endcsname\next\mathchar"6\bffam@3B \else
 \ifx#1=\mathchar"3\bffam@3D \else
 \ifx#1?\mathchar"5\bffam@3F \else
 \expandafter\ifx\csname A@\string?\endcsname\next\mathchar"5\bffam@3F \else
 \ifx#1[\mathchar"4\bffam@5B \else
 \ifx#1]\mathchar"5\bffam@5D \else
 \ifx#1,\mathchari@63B \else
 \ifx#1-\mathcharii@200 \else
 \ifx#1.\mathchari@03A \else
 \ifx#1/\mathchari@03D \else
 \ifx#1<\mathchari@33C \else
 \ifx#1>\mathchari@33E \else
 \ifx#1*\mathcharii@203 \else
 \ifx#1|\mathcharii@06A \else
 \ifx#10\bold0\else\ifx#11\bold1\else\ifx#12\bold2\else\ifx#13\bold3\else
 \ifx#14\bold4\else\ifx#15\bold5\else\ifx#16\bold6\else\ifx#17\bold7\else
 \ifx#18\bold8\else\ifx#19\bold9\else
  \Err@{\noexpand\boldkey can't be used with #1}%
 \fi\fi\fi\fi\fi\fi\fi\fi\fi\fi\fi\fi\fi\fi\fi
 \fi\fi\fi\fi\fi\fi\fi\fi\fi\fi\fi\fi\fi\fi\fi\fi\fi\fi}
\def\arabic#1{#1}
\def\alph#1{\count@#1\relax\advance\count@96 \ifnum\count@>122
 \Err@{\noexpand\alph invalid for numbers > 26}\else\char\count@\fi}
\def\Alph#1{\count@#1\relax\advance\count@64 \ifnum\count@>90
 \Err@{\noexpand\Alph invalid for numbers > 26}\else\char\count@\fi}

\def\Roman#1{\uppercase\expandafter{\romannumeral#1}}
\def\fnsymbol#1{\count@#1\relax
 \count@@\count@
 \advance\count@\m@ne\divide\count@7
 \count@@@\count@\advance\count@@@\@ne
 \multiply\count@7 \advance\count@@-\count@
 \count@\count@@@
 {\loop
  \ifcase\count@@\or*\or\dag\or\ddag\or\P\or\S\or\text{$\|$}\or\#\fi
  \advance\count@\m@ne\ifnum\count@>\z@\repeat}}
\def\cardnine@#1{\ifcase#1\or one\or two\or three\or four\or five\or
 six\or seven\or eight\or nine\fi}
\let\alloc@\alloc@@
\newcount\ten@
\ten@10
\def\cardinal#1{\count@#1\relax
 \ifnum\count@>99 \number\count@
 \else
  \ifnum\count@=\z@ zero%
  \else
   \ifnum\count@<\ten@\cardnine@\count@
   \else
    \ifnum\count@<20
     \advance\count@-\ten@
     \ifcase\count@ ten\or eleven\or twelve\or thirteen\or fourteen\or
      fifteen\or sixteen\or seventeen\or eighteen\or nineteen\fi
    \else
     \count@@\count@\count@@@\count@@
     \divide\count@\ten@\multiply\count@\ten@
     \advance\count@@@-\count@\divide\count@\ten@
     \ifcase\count@\or\or twenty\or thirty\or forty\or fifty\or sixty\or
      seventy\or eighty\or ninety\fi
     \ifnum\count@@@=\z@\else-\cardnine@\count@@@\fi
    \fi
   \fi
  \fi
 \fi}
\def\ordnine@#1{\ifcase#1\or first\or second\or third\or fourth\or fifth\or
 sixth\or seventh\or eighth\or ninth\fi}
\newcount\count@@@@
\def\ordsuffix@{\count@@@@\count@
 \divide\count@\ten@
 \count@@@\count@\count@@\count@
 \divide\count@@\ten@\multiply\count@@\ten@
 \advance\count@@@-\count@@
 \ifnum\count@@@=\@ne th%
 \else
  \count@@@\count@@@@
  \count@@\count@@@@
  \divide\count@@\ten@\multiply\count@@\ten@
  \advance\count@@@-\count@@
  \ifcase\count@@@ th\or st\or nd\or rd\else th\fi
 \fi}
\def\nordinal#1{\count@#1\relax\number\count@\ordsuffix@}
\def\spordinal#1{\count@#1\relax\number\count@$^{\text{\ordsuffix@}}$}
\def\ordinal#1{\count@#1\relax
 \ifnum\count@>99 \number\count@\ordsuffix@
 \else
   \ifnum\count@=\z@ zeroth%
  \else
    \ifnum\count@<\ten@\ordnine@\count@
    \else
     \ifnum\count@<20 \advance\count@-\ten@
      \ifcase\count@ tenth\or eleventh\or twelfth\or thirteenth\or
       fourteenth\or fifteenth\or sixteenth\or seventeenth\or eighteenth\or
       nineteenth\fi
     \else
      \count@@\count@
      \divide\count@\ten@\multiply\count@\ten@
      \count@@@\count@@\advance\count@@@-\count@
      \divide\count@\ten@
      \ifcase\count@\or\or twent\or thirt\or fort\or fift\or sixt\or sevent\or
       eight\or ninet\fi
      \ifnum\count@@@=\z@ ieth\else y-\ordnine@\count@@@\fi
     \fi
    \fi
  \fi
 \fi}
\font@\tensmc=cmcsc10
\textonlyfont@\smc\tensmc
\newtoks\noexpandtoks@
\noexpandtoks@{\let\arabic\relax\let\alph\relax\let\Alph\relax
 \let\Roman\relax\let\fnsymbol\relax\let\rm\relax
 \let\it\relax\let\bf\relax\let\sl\relax\let\smc\relax
 \let\/\relax\let\null\relax}
\def\noexpands@{\the\noexpandtoks@}
\def\Nonexpanding#1{\global\noexpandtoks@
 \expandafter{\the\noexpandtoks@\let#1\relax}}
\def\prevanish@{\saveskip@\z@\ifhmode\saveskip@\lastskip\unskip\fi}
\def\postvanish@{\ifdim\saveskip@>\z@\hskip\saveskip@\fi\FN@\postvanish@@}
\def\postvanish@@{\DN@.{}%
 \ifx\next\space@\ifdim\saveskip@>\z@\DN@. {}\fi\fi\next@.}
\def\invisible#1{\prevanish@\ignorespaces#1\unskip\postvanish@}
\def\vanishlist@{\\\invisible}
\let\noindent@\noindent
\def\noindent{\par\noindent@\FN@\pretendspace@}
\def\pretendspace@{\ismember@\vanishlist@\next
 \iftest@\nobreak\hskip-\p@\hskip\p@\fi}

\newtoks\everypartoks@
\def\noindent@@{\par\everypartoks@\expandafter{\the\everypar}\everypar{}%
 \noindent@\everypar\expandafter{\the\everypartoks@}}
\def\page{\Err@{\noexpand\page has no meaning by itself}}
\let\page@C\pageno
\let\page@P\empty
\let\page@Q\empty
\def\page@S#1{#1\/}
\def\page@F{\rm}
\def\page@N{\arabic}   % cannot be \let
\newif\ifindexing@
\def\indexfile{\ifindexing@\else
 \alloc@@7\write\chardef\sixt@@n\ndx@
 \immediate\openout\ndx@=\jobname.ndx
 \global\indexing@true\fi}
\global\advance\insc@unt\m@ne
\ch@ck0\insc@unt\count
\ch@ck1\insc@unt\dimen
\ch@ck2\insc@unt\skip
\ch@ck4\insc@unt\box
\allocationnumber\insc@unt
\global\chardef\margin@\allocationnumber
\dimen\margin@\maxdimen
\count\margin@\z@
\skip\margin@\z@
\newif\ifindexproofing@
\def\indexproofing{\indexproofing@true}
\def\noindexproofing{\indexproofing@false}
\def\unmacro@#1:#2->#3\unmacro@{\def\macpar@{#2}\def\macdef@{#3}}
\def\starparts@#1{\def\stari@{#1}\def\starii@{#1}\let\stariii@\empty
 \test@false
 \DN@##1*##2##3\next@{\ifx\starparts@##2\test@false\else\test@true\fi}%
 \next@#1*\starparts@\next@
 \iftest@\DN@{\starparts@@#1\starparts@@}\else\let\next@\relax\fi\next@}
\def\starparts@@#1*#2\starparts@@{\def\starii@{#1}\def\stariii@{*#2}}
\def\windex@{\ifindexing@
 \expandafter\unmacro@\meaning\stari@\unmacro@
 \edef\macdef@{\string"\macdef@\string"}%
 \edef\next@{\write\ndx@{\macdef@}}\next@
 \write\ndx@{{\number\pageno}{\page@N}{\page@P}{\page@Q}}%
 \fi
 \ifindexproofing@
  \ifx\stariii@\empty\else
   \expandafter\unmacro@\meaning\stariii@\unmacro@\fi
  \insert\margin@{\hbox{\rm\vrule\height9\p@\depth2\p@\width\z@\starii@
  \ifx\stariii@\empty\else\tt\macdef@\fi}}\fi}
\catcode`\"=\active
\def"{\FN@\quote@}
\def\quote@{\ifx\next"\expandafter\quote@@\else\expandafter\quote@@@\fi}
\def\quote@@@#1"{\starparts@{#1}\starii@\windex@}
\def\quote@@"#1"{\prevanish@\starparts@{#1}\windex@\FN@\quote@@@@}
\def\quote@@@@{\ifx\next"\DN@"{\postvanish@}\else
 \let\next@\postvanish@\fi\next@}
\rightadd@"\to\vanishlist@
\def\idefine#1{\DN@{#1}\DNii@{\noexpand#1}%
 \afterassignment\idefine@\def\nextiii@}
\def\idefine@{\ifindexing@
 \expandafter\let\next@\nextiii@
 \expandafter\unmacro@\meaning\nextiii@\unmacro@
 \immediate\write\ndx@{\noexpand\define\nextii@\macpar@{\macdef@}}\fi}
\def\iabbrev*#1#2{\ifindexing@\toks@{#2}%
 \immediate\write\ndx@{\noexpand\abbrev*\noexpand#1{\the\toks@}}\fi}
\newread\laxread@
\newwrite\laxwrite@
\let\fnpages@\empty
\def\Finit@#1#2\Finit@{\let\nextii@#1\def\nextiii@{#2}}
\catcode`\~=11
\def\getparts@ @#1~#2~#3~#4~#5~#6{\def\nextiv@{#1}%
 \def\nextiii@{#2~#3~#4~#5~}\count@#6\relax}
\newif\ifdocument@
\def\document{\ifdocument@\else\global\document@true
 \let\fontlist@\empty
 \immediate\openin\laxread@=\jobname.lax\relax
 {\endlinechar\m@ne\noexpands@\catcode`\@=11 \catcode`\~=11
  \loop\ifeof\laxread@\else
   \read\laxread@ to\next@
   \ifx\next@\empty
   \else
    \expandafter\Finit@\next@\Finit@
    \if\nextii@ F%
     \expandafter\rightadd@\nextiii@\to\fnpages@
    \else
     \expandafter\getparts@\next@
     \edef\next@{\gdef\csname\nextiv@ @L\endcsname{\nextiii@\number\count@}}%
     \next@
    \fi
   \fi
  \repeat}%
 \immediate\closein\laxread@
 \immediate\openout\laxwrite@=\jobname.lax\relax\fi}
\let\thelabel@\relax
\def\thelabels@{\thelabel@ ~\thelabel@@ ~\thelabel@@@ ~\thelabel@@@@ ~}
\def\label#1{\prevanish@
 \ifx\thelabel@\relax
  \Err@{There's nothing here to be labelled}%
 \else
  {\noexpands@
  \expandafter\ifx\csname#1@L\endcsname\relax
   \expandafter\xdef\csname#1@L\endcsname{\thelabels@0}%
   \immediate\write\laxwrite@{@#1~\thelabels@1}%
  \else
   \edef\next@{@~\csname#1@L\endcsname}%
    \expandafter\getparts@\next@
    \ifodd\count@
    \expandafter\xdef\csname#1@L\endcsname{\thelabels@0}%
    \immediate\write\laxwrite@{@#1~\thelabels@1}%
   \else
    \Err@{Label #1 already used}%
   \fi
  \fi
  }%
 \fi
 \postvanish@}
\rightadd@\label\to\vanishlist@
\def\thepages@{\page@N{\number\page@C}~%
 \page@S{\page@P\page@N{\number\page@C}\page@Q}~%
 \number\page@C ~\page@P\page@N{\number\page@C}\page@Q ~}
\def\pagelabel#1{\prevanish@
 \expandafter\ifx\csname#1@L\endcsname\relax
  {\noexpands@
  \expandafter\xdef\csname#1@L\endcsname{\thepages@2}}%
  \write\laxwrite@{@#1~\thepages@3}%
 \else
  {\noexpands@
  \edef\next@{@~\csname#1@L\endcsname}%
  \expandafter\getparts@\next@
  \ifodd\count@
   \ifnum\count@=\@ne
    \expandafter\xdef\csname#1@L\endcsname{\thelabels@2}%
   \fi
   \write\laxwrite@{@#1~\thepages@3}%
  \else
   \Err@{Label #1 already used}%
  \fi
  }%
 \fi
 \postvanish@}
\rightadd@\pagelabel\to\vanishlist@
\newif\ifreferr@
\referr@true
\def\RefErrors{\global\referr@true}
\def\RefWarnings{\global\referr@false}
\setbox\z@\hbox{\global\count@=`^^30}
\ifnum\count@=48 \let\versionthree@\relax\fi
\def\nolabel@#1#2#3{\expandafter\ifx\csname#2@L\endcsname\relax
 \ifreferr@\Err@{No \noexpand\label found for #2}\else
 \W@{Warning: No \noexpand\label found for #2.}%
 \ifx\versionthree@\relax\W@{l.\number\inputlineno\space ... \string#1{#2}}\fi
 \fi#3\else}
\def\csL@#1{{\noexpands@\xdef\Next@{\csname#1@L\endcsname}}}
\def\ref#1{\nolabel@\ref{#1}\relax
 \DNii@##1~##2\nextii@{##1}%
 \csL@{#1}\expandafter\nextii@\Next@\nextii@\fi}
\def\Ref#1{\nolabel@\Ref{#1}\relax
 \DNii@##1~##2~##3\nextii@{##2}%
 \csL@{#1}\expandafter\nextii@\Next@\nextii@\fi}
\def\nref#1{\nolabel@\nref{#1}\relax
 \DNii@##1~##2~##3~##4\nextii@{##3}%
 \csL@{#1}\expandafter\nextii@\Next@\nextii@\fi}
\def\pref#1{\nolabel@\pref{#1}\relax
 \DNii@##1~##2~##3~##4~##5\nextii@{##4}%
 \csL@{#1}\expandafter\nextii@\Next@\nextii@\fi}
\let\pref@\pref
\def\Evaluatenref#1{\nolabel@\Evaluatenref{#1}{\gdef\Nref{-10000 }}%
 \DNii@##1~##2~##3~##4\nextii@{\DNii@{##3}}%
 \csL@{#1}\expandafter\nextii@\Next@\nextii@
 \xdef\Nref{\nextii@}\fi}
\def\Evaluatepref#1{\nolabel@\Evaluatepref{#1}{\global\let\Pref\empty}%
 \DNii@##1~##2~##3~##4~##5\nextii@{\DNii@{##4}}%
 \csL@{#1}\expandafter\nextii@\Next@\nextii@
 \xdef\Pref{\nextii@}\fi}
\def\readlax#1{\immediate\openin\laxread@=#1.lax\relax
 \ifeof\laxread@\W@{}\W@{File #1.lax not found.}\W@{}\fi
 {\endlinechar\m@ne\noexpands@\catcode`\@=11 \catcode`\~=11
  \loop\ifeof\laxread@\else
   \read\laxread@ to\nextv@
   \ifx\nextv@\empty
   \else
    \expandafter\Finit@\nextv@\Finit@
    \ifx\nextii@ F%
    \else
     \expandafter\getparts@\nextv@
     \expandafter\ifx\csname\nextiv@ @L\endcsname\relax
      \edef\next@{\gdef\csname\nextiv@ @L\endcsname
       {\nextiii@\ifnum\count@=\@ne0\else2\fi}}%
      \next@
     \else
      \Err@{Label \nextiv@\space in #1.lax already used}%
     \fi
    \fi
   \fi
  \repeat}%
 \immediate\closein\laxread@}
\catcode`\~=\active

\def\FNSSP@{\FNSS@\pretendspace@}
\everydisplay{\csname displaymath \endcsname}
\expandafter\def\csname displaymath \endcsname#1$${#1$$\FNSSP@}
\def\locallabel@{\let\thelabel@\Thelabel@\let\thelabel@@\Thelabel@@
 \let\thelabel@@@\Thelabel@@@\let\thelabel@@@@\Thelabel@@@@}
\newcount\tag@C
\tag@C\z@
\let\tag@P\empty
\let\tag@Q\empty
\def\tag@S#1{{\rm(}{#1\/}{\rm)}}
\let\tag@N\arabic
\def\tag@F{\rm}
\def\maketag@{\FN@\maketag@@}
\def\maketag@@{\ifx\next\relax\DN@\relax{\FN@\maketag@@}\else
 \ifx\next"\let\next@\maketag@@@\else
 \let\next@\maketag@@@@\fi\fi\next@}
\def\xdefThelabel@#1{\xdef\Thelabel@{#1{\Thelabel@@@}}}
\def\xdefThelabel@@#1{\xdef\Thelabel@@{#1{\Thelabel@@@@}}}
\def\maketag@@@@#1\maketag@{\global\advance\tag@C\@ne
 {\noexpands@
  \xdef\Thelabel@@@{\number\tag@C}%
  \xdefThelabel@\tag@N
  \xdef\Thelabel@@@@{\ifmathtags@$\tag@P\Thelabel@\tag@Q$\else
   \tag@P\Thelabel@\tag@Q\fi}%
  \xdefThelabel@@\tag@S
  }%
 \locallabel@
 \hbox{\tag@F\thelabel@@}%
 #1}
\def\Qlabel@#1{{\noexpands@\xdef\Thelabel@@{#1}%
 \let\style\empty\xdef\Thelabel@@@@{#1}%
 \let\pre\empty\let\post\empty\xdef\Thelabel@{#1}%
 \let\numstyle\empty\xdef\Thelabel@@@{#1}}}
\def\maketag@@@"#1"#2\maketag@{%
 {\let\pre\tag@P\let\post\tag@Q\let\style\tag@S\let\numstyle\tag@N
  \hbox{\tag@F#1}%
  \noexpands@
  \Qlabel@{#1}%
  }%
 \locallabel@
 #2}
\def\align@{\inalign@true\inany@true
 \vspace@\allowdisplaybreak@\displaybreak@\intertext@
 \def\tag{\global\tag@true\ifnum\and@=\z@
  \DN@{&\omit\global\rwidth@\z@&\relax}\else
  \DN@{&\relax}\fi\next@}%
 \iftagsleft@\DN@{\csname align \endcsname}\else
  \DN@{\csname align \space\endcsname}\fi\next@}
\def\noset@{\def\Offset##1##2{\prevanish@\postvanish@}%
 \def\Reset##1##2{\prevanish@\postvanish@}}
\def\measure@#1\endalign{\global\lwidth@\z@\global\rwidth@\z@
 \global\maxlwidth@\z@\global\maxrwidth@\z@
 \global\and@\z@
 \setbox\z@\vbox
  {\noset@\everycr{\noalign{\global\tag@false\global\and@\z@}}\Let@
  \halign{\setboxz@h{$\m@th\displaystyle{\@lign##}$}%
   \global\lwidth@\wdz@
   \ifdim\lwidth@>\maxlwidth@\global\maxlwidth@\lwidth@\fi
   \global\advance\and@\@ne
   &\setboxz@h{$\m@th\displaystyle{{}\@lign##}$}\global\rwidth@\wdz@
   \ifdim\rwidth@>\maxrwidth@\global\maxrwidth@\rwidth@\fi
   \global\advance\and@\@ne
   &\Tag@\eat@{##}\crcr#1\crcr}}%
 \totwidth@\maxlwidth@\advance\totwidth@\maxrwidth@}
\def\prepost@{\global\let\tag@P@\tag@P\global\let\tag@Q@\tag@Q}
\def\reprepost@{\let\tag@P\tag@P@\let\tag@Q\tag@Q@}
\expandafter\def\csname align \space\endcsname#1\endalign
 {\measure@#1\endalign\global\and@\z@
 \ifingather@\everycr{\noalign{\global\and@\z@}}\else\displ@y@\fi
 \Let@\tabskip\centering@
 \halign to\displaywidth
  {\hfil\strut@\setboxz@h{$\m@th\displaystyle{\@lign##\prepost@}$}%
  \boxz@\global\advance\and@\@ne
  \tabskip\z@skip
  &\setboxz@h{$\m@th\displaystyle{{}\@lign##\prepost@}$}%
  \global\rwidth@\wdz@\boxz@\hfil\global\advance\and@\@ne
  \tabskip\centering@
  &\setboxz@h{\@lign\strut@\reprepost@\maketag@##\maketag@}%
  \dimen@\displaywidth\advance\dimen@-\totwidth@
  \divide\dimen@\tw@\advance\dimen@\maxrwidth@\advance\dimen@-\rwidth@
  \ifdim\dimen@<\tw@\wdz@\llap{\vtop{\normalbaselines\null\boxz@}}%
  \else\llap{\boxz@}\fi
  \tabskip\z@skip
  \crcr#1\crcr
  \black@\totwidth@}}
\expandafter\def\csname align \endcsname#1\endalign{\measure@#1\endalign
 \global\and@\z@
 \ifdim\totwidth@>\displaywidth\let\displaywidth@\totwidth@\else
  \let\displaywidth@\displaywidth\fi
 \ifingather@\everycr{\noalign{\global\and@\z@}}\else\displ@y@\fi
 \Let@\tabskip\centering@\halign to\displaywidth
  {\hfil\strut@\setboxz@h{$\m@th\displaystyle{\@lign##\prepost@}$}%
  \global\lwidth@\wdz@\global\lineht@\ht\z@
  \boxz@\global\advance\and@\@ne
  \tabskip\z@skip&\setboxz@h{$\m@th\displaystyle{{}\@lign##\prepost@}$}%
  \ifdim\ht\z@>\lineht@\global\lineht@\ht\z@\fi
  \boxz@\hfil\global\advance\and@\@ne
  \tabskip\centering@&\kern-\displaywidth@
  \setboxz@h{\@lign\strut@\reprepost@\maketag@##\maketag@}%
  \dimen@\displaywidth\advance\dimen@-\totwidth@
  \divide\dimen@\tw@\advance\dimen@\maxlwidth@\advance\dimen@-\lwidth@
  \ifdim\dimen@<\tw@\wdz@
   \rlap{\vbox{\normalbaselines\boxz@\vbox to\lineht@{}}}\else
   \rlap{\boxz@}\fi
  \tabskip\displaywidth@\crcr#1\crcr\black@\totwidth@}}
\def\attag@#1{\let\Maketag@\maketag@\let\TAG@\Tag@
 \let\Prepost@\prepost@\let\Reprepost@\reprepost@
 \let\Tag@\relax\let\maketag@\relax
 \let\prepost@\relax\let\reprepost@\relax
 \ifmeasuring@
  \def\llap@##1{\setboxz@h{##1}\hbox to\tw@\wdz@{}}%
  \def\rlap@##1{\setboxz@h{##1}\hbox to\tw@\wdz@{}}%
 \else\let\llap@\llap\let\rlap@\rlap\fi
 \toks@{\hfil\strut@
  $\m@th\displaystyle{\@lign\the\hashtoks@\prepost@}$%
  \tabskip\z@skip\global\advance\and@\@ne&
  $\m@th\displaystyle{{}\@lign\the\hashtoks@\prepost@}$\hfil
  \ifxat@\tabskip\centering@\fi\global\advance\and@\@ne}%
 \iftagsleft@
  \toks@@{\tabskip\centering@&\Tag@\kern-\displaywidth
   \rlap@{\@lign\reprepost@\maketag@\the\hashtoks@\maketag@}%
   \global\advance\and@\@ne\tabskip\displaywidth}\else
  \toks@@{\tabskip\centering@&\Tag@\llap@{\@lign\reprepost@\maketag@
   \the\hashtoks@\maketag@}\global\advance\and@\@ne\tabskip\z@skip}\fi
 \atcount@#1\relax\advance\atcount@\m@ne
 \loop\ifnum\atcount@>\z@
  \toks@\expandafter{\the\toks@&\hfil$\m@th\displaystyle{\@lign
  \the\hashtoks@\prepost@}$\global\advance\and@\@ne
  \tabskip\z@skip
  &$\m@th\displaystyle{{}\@lign\the\hashtoks@\prepost@}$\hfil\ifxat@
  \tabskip\centering@\fi\global\advance\and@\@ne}\advance\atcount@\m@ne
 \repeat
 \edef\preamble@{\the\toks@\the\toks@@}%
 \edef\preamble@@{\preamble@}%
 \let\maketag@\Maketag@\let\Tag@\TAG@
 \let\prepost@\Prepost@\let\reprepost@\Reprepost@}
\def\unlabel@{\def\label##1{\prevanish@\postvanish@}%
 \def\pagelabel##1{\prevanish@\postvanish@}}
\newcount\tag@CC
\expandafter\def\csname alignat \endcsname#1#2\endalignat
 {\inany@true\xat@false
 \def\tag{\global\tag@true
  \count@#1\relax\multiply\count@\tw@\advance\count@\m@ne
  \gdef\tag@{&}%
  \loop\ifnum\count@>\and@\xdef\tag@{&\omit\tag@}%
  \advance\count@\m@ne\repeat
  \tag@\relax}%
 \vspace@\allowdisplaybreak@\displaybreak@\intertext@
 \displ@y@\measuring@true\tag@CC\tag@C
 \setbox\savealignat@\hbox{\noset@\unlabel@$\m@th\displaystyle\Let@
  \attag@{#1}\vbox{\halign{\span\preamble@@\crcr#2\crcr}}$}%
 \measuring@false
 \Let@\attag@{#1}\tag@C\tag@CC
 \tabskip\centering@\halign to\displaywidth
  {\span\preamble@@\crcr#2\crcr\black@{\wd\savealignat@}}}
\expandafter\def\csname xalignat \endcsname#1#2\endxalignat
 {\inany@true\xat@true
 \def\tag{\global\tag@true
  \count@#1\relax\multiply\count@\tw@\advance\count@\m@ne
  \gdef\tag@{&}%
  \loop\ifnum\count@>\and@\xdef\tag@{&\omit\tag@}%
  \advance\count@\m@ne\repeat
  \tag@\relax}%
 \vspace@\allowdisplaybreak@\displaybreak@\intertext@
 \displ@y@\measuring@true\tag@CC\tag@C
 \setbox\savealignat@\hbox{\noset@\unlabel@$\m@th\displaystyle\Let@
  \attag@{#1}\vbox{\halign{\span\preamble@@\crcr#2\crcr}}$}%
 \measuring@false\Let@\attag@{#1}\tag@C\tag@CC
 \tabskip\centering@\halign to\displaywidth
 {\span\preamble@@\crcr#2\crcr\black@{\wd\savealignat@}}}
\def\gather{\RIfMIfI@\DN@{\onlydmatherr@\gather}\else
 \ingather@true\inany@true\def\tag{&\relax}%
 \vspace@\allowdisplaybreak@\displaybreak@\intertext@
 \displ@y\Let@
 \iftagsleft@\DN@{\csname gather \endcsname}\else
  \DN@{\csname gather \space\endcsname}\fi\fi
 \else\DN@{\onlydmatherr@\gather}\fi\next@}
\def\exstring@{\expandafter\eat@\string}
\def\newcounter#1{\define#1{}%
 \edef\next@{\def\noexpand#1{\futurelet\noexpand\next
  \csname\exstring@#1@Z\endcsname}}\next@
 \edef\next@{\def\csname\exstring@#1@Z\endcsname
  {\global\advance\csname\exstring@#1@C\endcsname\@ne
  {\csname\exstring@#1@F\endcsname\csname\exstring@#1@S\endcsname
   {\csname\exstring@#1@P\endcsname\csname\exstring@#1@N\endcsname
   {\noexpand\number\csname\exstring@#1@C\endcsname}%
   \csname\exstring@#1@Q\endcsname}}%
  \noexpand\ifx\noexpand\next\noexpand\label
   \def\noexpand\next@\noexpand\label########1{{\noexpand\noexpands@
    \xdef\noexpand\Thelabel@{\csname\exstring@#1@N\endcsname
     {\noexpand\number\csname\exstring@#1@C\endcsname}}%
    \xdef\noexpand\Thelabel@@@{\noexpand\number
     \csname\exstring@#1@C\endcsname}%
    \xdef\noexpand\Thelabel@@{\csname\exstring@#1@S\endcsname
     {\csname\exstring@#1@P\endcsname
     \csname\exstring@#1@N\endcsname
     {\noexpand\number\csname\exstring@#1@C\endcsname}%
     \csname\exstring@#1@Q\endcsname}}%
    \xdef\noexpand\Thelabel@@@@{\csname\exstring@#1@P\endcsname
     \csname\exstring@#1@N\endcsname
     {\noexpand\number\csname\exstring@#1@C\endcsname}%
     \csname\exstring@#1@Q\endcsname}}%
    {\noexpand\locallabel@\noexpand\label{########1}}}%
   \noexpand\else\let\noexpand\next@\relax\noexpand\fi\noexpand\next@}}\next@
 \expandafter\newcount@\csname\exstring@#1@C\endcsname
 \expandafter\let\csname\exstring@#1@N\endcsname\arabic
 \expandafter\def\csname\exstring@#1@S\endcsname##1{##1\/}%
 \expandafter\let\csname\exstring@#1@P\endcsname\empty
 \expandafter\let\csname\exstring@#1@Q\endcsname\empty
 \expandafter\def\csname\exstring@#1@F\endcsname{\rm}%
 }
\def\HASH@#1#2{\ifnum#2=\z@\else
 \edef\next@{\toks@{\the\toks@\the\hashtoks@#2}%
 \toks@@{\the\toks@@{\the\hashtoks@#2}}}\next@\expandafter\HASH@\fi}
\def\HASH@@{\toks@{}\toks@@{}\expandafter\HASH@\macpar@00}
\def\usecounter#1#2{\expandafter\ifx\csname\exstring@#1@Z\endcsname
 \relax\Err@{\noexpand#1not created with \string\newcounter}\fi
 \expandafter\let\csname\exstring@#1@@Z\endcsname\relax
 \expandafter\let\csname\exstring@#1@@Z@\endcsname\relax
 \expandafter\let\csname\exstring@#1@@Z@@\endcsname\relax
 \edef\next@{\def\noexpand#2{\futurelet\noexpand\next
  \csname\exstring@#1@@Z\endcsname}}\next@
 \edef\next@{\def\csname\exstring@#1@@Z\endcsname{\noexpand\ifx
  \noexpand\next\noexpand\label\def\noexpand\next@\noexpand\label
   ########1{\csname\exstring@#1@@Z@\endcsname
   {\noexpand#1\noexpand\label{########1}}}%
   \noexpand\else\noexpand\ifx\noexpand\next
   \noexpand"\def\noexpand\next@\noexpand"########1\noexpand"%
   {\csname\exstring@#1@@Z@\endcsname{{\expandafter\noexpand
   \csname\exstring@#1@F\endcsname
   \let\noexpand\pre\expandafter\noexpand\csname\exstring@#1@P\endcsname
   \let\noexpand\post\expandafter\noexpand\csname\exstring@#1@Q\endcsname
   \let\noexpand\style\expandafter\noexpand\csname\exstring@#1@S\endcsname
   \let\noexpand\numstyle\expandafter\noexpand\csname\exstring@#1@N\endcsname
   ########1}}}\noexpand\else
   \def\noexpand\next@{\csname\exstring@#1@@Z@\endcsname{\noexpand#1}}%
   \noexpand\fi\noexpand\fi\noexpand\next@}}\next@
 \def\next@{\expandafter\expandafter\expandafter\unmacro@\expandafter
  \meaning\csname\exstring@#1@@Z@@\endcsname\unmacro@
  \HASH@@
  \edef\next@{\def\csname\exstring@#1@@Z@\endcsname\the\toks@{%
   \expandafter\noexpand\csname\exstring@#1@@Z@@\endcsname\the\toks@@
   \noexpand\FNSSP@}}\next@}%
 \afterassignment\next@
 \expandafter\def\csname\exstring@#1@@Z@@\endcsname}
\def\listbi@{\penalty50 \medskip}
\def\listbii@{\penalty100 \smallskip}
\let\listbiii@\relax
\let\listbiv@\relax
\let\listbv@\relax
\def\listmi@{\advance\leftskip30\p@\relax}
\let\listmii@\listmi@
\let\listmiii@\listmi@
\let\listmiv@\listmi@
\let\listmv@\listmi@
\def\itemi@#1{\noindent@@\llap{#1\hskip5\p@}}
\let\itemii@\itemi@
\let\itemiii@\itemi@
\let\itemiv@\itemi@
\let\itemv@\itemi@
\def\liste@{\penalty-50 \medskip}
\def\listei@{\penalty-100 \smallskip}
\let\listeii@\relax
\let\listeiii@\relax
\let\listeiv@\relax
\expandafter\newcount\csname list@C1\endcsname
\csname list@C1\endcsname\z@
\expandafter\newcount\csname list@C2\endcsname
\csname list@C2\endcsname\z@
\expandafter\newcount\csname list@C3\endcsname
\csname list@C3\endcsname\z@
\expandafter\newcount\csname list@C4\endcsname
\csname list@C4\endcsname\z@
\expandafter\newcount\csname list@C5\endcsname
\csname list@C5\endcsname\z@
\expandafter\let\csname list@P1\endcsname\empty
\expandafter\let\csname list@P2\endcsname\empty
\expandafter\let\csname list@P3\endcsname\empty
\expandafter\let\csname list@P4\endcsname\empty
\expandafter\let\csname list@P5\endcsname\empty
\expandafter\let\csname list@Q1\endcsname\empty
\expandafter\let\csname list@Q2\endcsname\empty
\expandafter\let\csname list@Q3\endcsname\empty
\expandafter\let\csname list@Q4\endcsname\empty
\expandafter\let\csname list@Q5\endcsname\empty
\expandafter\def\csname list@S1\endcsname#1{{\rm(}{#1\/}{\rm)}}
\expandafter\def\csname list@S2\endcsname#1{{\rm(}{#1\/}{\rm)}}
\expandafter\def\csname list@S3\endcsname#1{{\rm(}{#1\/}{\rm)}}
\expandafter\def\csname list@S4\endcsname#1{{\rm(}{#1\/}{\rm)}}
\expandafter\def\csname list@S5\endcsname#1{{\rm(}{#1\/}{\rm)}}
\expandafter\let\csname list@N1\endcsname\arabic
\expandafter\let\csname list@N2\endcsname\arabic
\expandafter\let\csname list@N3\endcsname\arabic
\expandafter\let\csname list@N4\endcsname\arabic
\expandafter\let\csname list@N5\endcsname\arabic
\expandafter\def\csname list@F1\endcsname{\rm}
\expandafter\def\csname list@F2\endcsname{\rm}
\expandafter\def\csname list@F3\endcsname{\rm}
\expandafter\def\csname list@F4\endcsname{\rm}
\expandafter\def\csname list@F5\endcsname{\rm}
\newcount\listlevel@
\listlevel@\z@
\def\list@@C{\csname list@C\number\listlevel@\endcsname}
\def\list@@P{\csname list@P\number\listlevel@\endcsname}
\def\list@@Q{\csname list@Q\number\listlevel@\endcsname}
\def\list@@S{\csname list@S\number\listlevel@\endcsname}
\def\list@@N{\csname list@N\number\listlevel@\endcsname}
\def\list@@F{\csname list@F\number\listlevel@\endcsname}
\newif\iffirstitemi@
\newif\iffirstitemii@
\newif\iffirstitemiii@
\newif\iffirstitemiv@
\newif\iffirstitemv@
\def\Firstitem@true{\csname firstitem\romannumeral\listlevel@
 @true\endcsname}
\def\Firstitem@false{\csname firstitem\romannumeral\listlevel@
 @false\endcsname}
\def\Listm@{\csname listm\romannumeral\listlevel@ @\endcsname}
\def\Item@{\csname item\romannumeral\listlevel@ @\endcsname}
\def\Liste@{\csname liste\romannumeral\listlevel@ @\endcsname}
\newif\iflistcontinue@
\def\keepitem{\listcontinue@true}
\newcount\list@C@
\def\list{%
 \iflistcontinue@\csname list@C1\endcsname\csname list@C@\endcsname\fi
 \global\csname list@C2\endcsname\z@
 \global\csname list@C3\endcsname\z@
 \global\csname list@C4\endcsname\z@
 \global\csname list@C5\endcsname\z@
 \begingroup
 \firstitemi@true
 \listlevel@\@ne
 \def\item{\FN@\item@}%
 \FN@\list@}
\Invalid@\runinitem
\def\list@{\ifx\next\par
 \DN@\par{\FN@\list@}\else
 \ifx\next\runinitem
  \DN@\runinitem{\FN@\runinitem@}\else
  \DN@{\par\dimen@\parskip\parskip\dimen@}\fi\fi\next@}
\newif\ifoutlevel@
\newif\ifrunin@
\def\item@{%
 \ifoutlevel@\Liste@\outlevel@false\fi
 \ifrunin@\runin@false\par
  \dimen@\parskip\parskip\dimen@
  \Listm@\fi
 \iffirstitemi@\listbi@\listmi@\firstitemi@false\else\par\fi
 \iffirstitemii@\listbii@\listmii@\firstitemii@false\else\par\fi
 \iffirstitemiii@\listbiii@\listmiii@\firstitemiii@false\else\par\fi
 \iffirstitemiv@\listbiv@\listmiv@\firstitemiv@false\else\par\fi
 \iffirstitemv@\listbv@\listmv@\firstitemv@false\else\par\fi
 \DN@"##1"{{\let\pre\list@@P\let\post\list@@Q
  \let\style\list@@S\let\numstyle\list@@N
  \vskip-\parskip
  \Item@{\list@@F##1}%
  \noexpands@
  \Qlabel@{##1}}%
  \locallabel@
  \FNSSP@}%
 \DNii@{\global\advance\list@@C\@ne
  {\noexpands@
   \xdef\Thelabel@@@{\number\list@@C}%
   \xdefThelabel@\list@@N
   \xdef\Thelabel@@@@{\list@@P\Thelabel@\list@@Q}%
   \xdefThelabel@@\list@@S
  }%
  \locallabel@
  \vskip-\parskip
  \Item@{\list@@F\thelabel@@}%
  \FN@\pretendspace@}%
 \ifx\next"\expandafter\next@\else\expandafter\nextii@\fi}
\def\runinitem@{%
  \runin@true
  \Firstitem@false
  \DN@"##1"{{\let\pre\list@@P\let\post\list@@Q
   \let\style\list@@S\let\numstyle\list@@N
   \unskip\space{\list@@F##1} %
   \noexpands@
   \Qlabel@{##1}}%
   \locallabel@
   \ignorespaces}%
  \DNii@{\global\advance\list@@C\@ne
   {\noexpands@
    \xdef\Thelabel@@@{\number\list@@C}%
    \xdefThelabel@\list@@N
    \xdef\Thelabel@@@@{\list@@P\Thelabel@\list@@Q}%
    \xdefThelabel@@\list@@S
   }%
   \locallabel@
   \unskip\space{\list@@F\thelabel@@} }%
  \ifx\next"\expandafter\next@\else\expandafter\nextii@\fi}
\def\inlevel{\ifnum\listlevel@=5
 \DN@{\Err@{Already 5 levels down}}\else
 \DN@{\begingroup\advance\listlevel@\@ne
 \Firstitem@true\FN@\inlevel@}\fi\next@}
\def\inlevel@{\ifx\next\par
 \DN@\par{\FN@\inlevel@}\else
 \ifx\next\runinitem
  \DN@\runinitem{\FN@\runinitem@}\else
  \let\next@\relax\fi\fi\next@}
\def\outlevel{\ifnum\listlevel@=\@ne
 \Err@{At top level}\else
 \par\global\list@@C\z@\endgroup\outlevel@true\fi}
\def\endlist{%
 \expandafter\global\csname list@C@\endcsname\csname list@C1\endcsname
 \par
 \global\toks\@ne{}\count@\listlevel@
 {\loop
  \ifnum\count@>\z@\global\toks\@ne\expandafter{\the\toks\@ne\endgroup}%
  \advance\count@\m@ne
  \repeat}%
 \the\toks\@ne
 \liste@
 \listcontinue@false\global\csname list@C1\endcsname\z@
 \vskip-\parskip
 \noindent@@
 \FN@\pretendspace@}
\newif\iffirstdescribe@
\def\describe{\par
 \begingroup\firstdescribe@true
 \def\item##1{%
  \iffirstdescribe@\penalty50 \medskip\vskip-\parskip
  \firstdescribe@false\else\par\fi
  \noindent@@\hangindent2pc\hangafter\@ne
  {\bf##1}\hskip.5em}}

\Invalid@\pullin
\Invalid@\pullinmore
\newif\iffirstpull@
\def\margins{\par\begingroup\firstpull@true
 \def\pullin##1##2{\par
  \iffirstpull@\firstpull@false\else\endgroup\fi
  \begingroup\DN@{##1}%
  \ifx\next@\empty\leftskip\z@\else\ifx\next@\space\leftskip\z@
  \else\leftskip##1\fi\fi
  \DN@{##2}\ifx\next@\empty\rightskip\z@\else\ifx\next@\space
  \rightskip\z@\else\rightskip##2\fi\fi\ignorespaces}%
 \def\pullinmore##1##2{\par
  \xdef\Next@{\leftskip\the\leftskip\relax\rightskip\the\rightskip\relax}%
  \iffirstpull@\firstpull@false\else\endgroup\fi
  \begingroup\Next@
  \DN@{##1}%
  \ifx\next@\empty\else\ifx\next@\space\else\advance\leftskip##1\fi\fi
  \DN@{##2}\ifx\next@\empty\else\ifx\next@\space\else
  \advance\rightskip##2\fi\fi\ignorespaces}}

\newif\ifnopunct@
\newif\ifnospace@
\newif\ifoverlong@
\let\nofrillslist@\empty
\let\overlonglist@\empty
\def\nopunct{\nopunct@true\FN@\nopunct@}
\def\nospace{\nospace@true\FN@\nospace@}
\def\overlong{\overlong@true\FN@\overlong@}
\def\nopunct@{\ifx\next\nospace
 \DN@\nospace{\nospace@true\FN@\nopnos@}\else\ifx\next\overlong
 \DN@\overlong{\overlong@true\FN@\nopol@}\else
 \let\next@\nopunct@@\fi\fi\next@}
\def\nopunct@@#1{\ismember@\nofrillslist@#1%
 \iftest@\let\next@#1\else
 \DN@{\nopunct@false\Err@{\noexpand\nopunct can't be used with
 \string#1}#1}\fi\next@}
\def\nospace@{\ifx\next\nopunct
 \DN@\nopunct{\nopunct@true\FN@\nopnos@}\else\ifx\next\overlong
 \DN@\overlong{\overlong@true\FN@\nosol@}\else
 \let\next@\nospace@@\fi\fi\next@}
\def\nospace@@#1{\ismember@\nofrillslist@#1%
 \iftest@\let\next@#1\else
 \DN@{\nospace@false\Err@{\noexpand\nospace can't be used with
 \string#1}#1}\fi\next@}
\def\overlong@{\ifx\next\nopunct
 \DN@\nopunct{\nopunct@true\FN@\nopol@}\else\ifx\next\nospace
 \DN@\nospace{\nospace@true\FN@\nosol@}\else
 \let\next@\overlong@@\fi\fi\next@}
\def\overlong@@#1{\ismember@\overlonglist@#1%
 \iftest@\let\next@#1\else
 \DN@{\overlong@false\Err@{\noexpand\overlong can't be used with
 \string#1}#1}\fi\next@}
\def\nopnos@{\ifx\next\overlong
 \DN@\overlong{\overlong@true\nopnosol@}\else
 \let\next@\nopnos@@\fi\next@}
\def\nopol@{\ifx\next\nospace
 \DN@\nospace{\nospace@true\nopnosol@}\else
 \let\next@\nopol@@\fi\next@}
\def\nosol@{\ifx\next\nopunct
 \DN@\nopunct{\nopunct@true\nopnosol@}\else
 \let\next@\nosol@@\fi\next@}
\def\nopnos@@#1{\ismember@\nofrillslist@#1%
 \iftest@\let\next@#1\else
 \DN@{\nopunct@false\nospace@false
  \Err@{\noexpand\nopunct\noexpand\nospace
   can't be used with \string#1}#1}\fi\next@}
\def\testii@#1{\ismember@\nofrillslist@#1%
 \iftest@\let\nextiii@ T\else\let\nextiii@ F\fi
 \ismember@\overlonglist@#1%
 \iftest@\let\nextiv@ T\else\let\nextiv@ F\fi
 \test@false\if\nextiii@ T\if\nextiv@ T\test@true\fi\fi}
\def\nopol@@#1{\testii@{#1}%
 \iftest@\let\next@#1%
 \else\DN@{\if\nextiii@ T\else\nopunct@false\fi
  \if\nextiv@ T\else\overlong@false\fi
  \Err@{\if\nextiii@ T\else\noexpand\nopunct\fi
  \if\nextiv@ T\else\noexpand\overlong\fi can't be used
  with \string#1}#1}\fi\next@}
\def\nosol@@#1{\testii@{#1}%
 \iftest@\let\next@#1%
 \else\DN@{\if\nextiii@ T\else\nospace@false\fi
  \if\nextiv@ T\else\overlong@false\fi
  \Err@{\if\nextiii@ T\else\noexpand\nospace\fi
  \if\nextiv@ T\else\noexpand\overlong\fi can't be used
  with \string#1}#1}\fi\next@}
\def\nopnosol@#1{\testii@{#1}%
 \iftest@\let\next@#1%
 \else\DN@{\if\nextiii@ T\else\nopunct@false\nospace@false\fi
  \if\nextiv@ T\else\overlong@false\fi
  \Err@{\if\nextiii@ T\else\noexpand\nopunct\noexpand\nospace\fi
  \if\nextiv@ T\else\noexpand\overlong\fi can't be used
  with \string#1}#1}\fi\next@}
\def\punct@#1{\ifnopunct@\else#1\fi}
\def\addspace@#1{\ifnospace@\else#1\fi}
\def\hss@{\ifoverlong@\z@ plus\@m\p@ minus\@m\p@
 \else \z@ plus\@m\p@\fi}
\rightadd@\demo\to\nofrillslist@
\newif\ifclaim@
\def\exxx@{\expandafter\expandafter\expandafter\eat@\expandafter\string}
\let\colon@:
\def\demo#1{\ifclaim@
 \Err@{Previous \expandafter\noexpand\claimtype@ has
  no matching \string\end\exxx@\claimtype@}%
 \let\next@\relax
 \else
  \par
  \ifdim\lastskip<\smallskipamount\removelastskip\smallskip\fi
  \begingroup
  \noindent@@{\smc\ignorespaces#1\unskip
   \punct@{\null\colon@}\addspace@\enspace}%
  \nopunct@false\nospace@false
  \rm
  \DN@{\FNSSP@}%
 \fi
 \next@}
\def\enddemo{\par\endgroup\nopunct@false\nospace@false\smallskip}
\rightadd@\claim\to\nofrillslist@
\def\claim@F{\smc}
\def\claim@@@F{\csname\exxx@\claimtype@ @F\endcsname}
\def\claimformat@#1#2#3{%
 \medbreak\noindent@@{\smc#1 {\claim@@@F#2} #3%
 \punct@{\null.}\addspace@\enspace}\sl}
\def\claimformat@@#1#2{\claimformat@{\ignorespaces#1\unskip}%
 {\ifx\thelabel@@\empty\unskip\else\thelabel@@\fi}%
 {\ignorespaces#2\unskip}%
 \let\Claimformat@@\claimformat@@\FNSSP@}
\let\Claimformat@@\claimformat@@
\def\claim@@@P{\csname\exxx@\claimtype@ @P\endcsname}
\def\claim@@@Q{\csname\exxx@\claimtype@ @Q\endcsname}
\def\claim@@@S{\csname\exxx@\claimtype@ @S\endcsname}
\def\claim@@@N{\csname\exxx@\claimtype@ @N\endcsname}
\def\claim@@@C{\csname claim@C\claimclass@\endcsname}
\newcount\claim@C
\claim@C\z@
\let\claim@P\empty
\let\claim@Q\empty
\def\claim@S#1{#1\/}
\let\claim@N\arabic
\def\claim{\claim@true\let\claimclass@\empty
 \def\claimtype@{\claim}\FN@\claim@}
\def\claim@{%
 \ifx\next\c
  \let\next@\claim@c
 \else
  \ifx\next"%
   \let\next@\claim@q
  \else
   \begingroup\global\advance\claim@C\@ne
   {\noexpands@
    \xdef\Thelabel@@@{\number\claim@C}%
    \xdefThelabel@\claim@N
    \xdef\Thelabel@@@@{\claim@P\Thelabel@\claim@Q}%
    \xdefThelabel@@\claim@S
   }%
   \locallabel@
   \let\next@\Claimformat@@
  \fi
 \fi
 \next@}
\def\claim@c\c#1{\claim@true\begingroup
 \expandafter
 \ifx\csname claim@C#1\endcsname\relax
  \expandafter\newcount@\csname claim@C#1\endcsname
  \global\csname claim@C#1\endcsname\@ne
 \else
  \global\advance\csname claim@C#1\endcsname\@ne
 \fi
 \def\claimclass@{#1}%
 {\noexpands@
  \xdef\Thelabel@@@{\number\claim@@@C}%
  \xdefThelabel@\claim@@@N
  \xdef\Thelabel@@@@{\claim@@@P\Thelabel@\claim@@@Q}%
  \xdefThelabel@@\claim@@@S
 }%
 \locallabel@
 \FNSS@\claim@c@}
\def\claim@q"#1"{\begingroup
 {\let\pre\claim@@@P\let\post\claim@@@Q
  \let\style\claim@@@S\let\numstyle\claim@@@N
  \noexpands@
  \Qlabel@{#1}}%
 \locallabel@
 \FNSS@\claim@q@}
\def\claim@c@{\ifx\next"%
 \global\advance\claim@@@C\m@ne\let\next@\claim@cq
 \else\let\next@\Claimformat@@\fi\next@}
\def\claim@cq"#1"{{\let\pre\claim@@@P\let\post\claim@@@Q
 \let\style\claim@@@S\let\numstyle\claim@@@N
 \noexpands@
 \Qlabel@{#1}}%
 \locallabel@
 \FNSS@\Claimformat@@}
\def\claim@q@{\ifx\next\c\expandafter\claim@qc
 \else\expandafter\Claimformat@@\fi}
\def\claim@qc\c#1{\expandafter\ifx\csname claim@C#1\endcsname\relax
 \expandafter\newcount@\csname claim@C#1\endcsname
 \global\csname claim@C#1\endcsname\z@\fi
 \FNSS@\Claimformat@@}
\def\endclaim{\endgroup\claim@false\nopunct@false\nospace@false
 \let\Claimformat@@\claimformat@@\medbreak}
\Invalid@\claimclause
\def\newclaim{\FN@\newclaim@}
\def\newclaim@{\ifx\next\claimclause
 \DN@\claimclause##1{\newclaim@@{##1}}\else
 \DN@{\newclaim@@\relax}\fi\next@}
\def\claimlist@{\\\claim}
\newtoks\claim@i
\newtoks\claim@v
\let\noclaimclause@=F
\def\newclaim@@#1#2#3\c#4#5{\define#2{}%
 \rightadd@#2\to\claimlist@\rightadd@#2\to\nofrillslist@%
 \expandafter\def\csname\exstring@#2@P\endcsname{\claim@P}%
 \expandafter\def\csname\exstring@#2@Q\endcsname{\claim@Q}%
 \expandafter\def\csname\exstring@#2@S\endcsname{\claim@S}%
 \expandafter\def\csname\exstring@#2@N\endcsname{\claim@N}%
 \expandafter\def\csname\exstring@#2@F\endcsname{\claim@F}%
 \expandafter\def\csname end\exstring@#2\endcsname{\endclaim}%
 \expandafter\ifx\csname claim@C#4\endcsname\relax
  \expandafter\newcount@\csname claim@C#4\endcsname
  \global\csname claim@C#4\endcsname\z@\fi
 \edef\next@{\let\csname\exstring@#2@C\endcsname
   \csname claim@C#4\endcsname}\next@
 \def#2{\ifx\noclaimclause@ T\else#1\fi
  \global\claim@i{#1}\gdef\claim@iv{#4}\global\claim@v{#5}%
  \def\claimtype@{#2}\def\Claimformat@@{\claimformat@@{#5}}\claim@c\c{#4}}}
\def\shortenclaim#1#2{\define#2{}%
 \ismember@\claimlist@#1%
 \iftest@
  \rightadd@#2\to\nofrillslist@%
  \expandafter\def\csname\exstring@#2@P\endcsname
   {\csname\exstring@#1@P\endcsname}%
  \expandafter\def\csname\exstring@#2@Q\endcsname
   {\csname\exstring@#1@Q\endcsname}%
  \expandafter\def\csname\exstring@#2@S\endcsname
   {\csname\exstring@#1@S\endcsname}%
  \expandafter\def\csname\exstring@#2@N\endcsname
   {\csname\exstring@#1@N\endcsname}%
  \expandafter\def\csname\exstring@#2@F\endcsname
   {\csname\exstring@#1@F\endcsname}%
  \expandafter\def\csname end\exstring@#2\endcsname{\endclaim}%
  \edef\next@{\let\csname\exstring@#2@C\endcsname
    \csname claim\exstring@#1C\endcsname}\next@
  \setbox\z@\vbox{\let\noclaimclause@ T#1""\relax\endgroup}%
  \edef#2{\the\claim@i
   \def\noexpand\claimtype@{\noexpand#2}%
   \def\noexpand\Claimformat@@{\noexpand\claimformat@@{\the\claim@v}\relax}%
   \noexpand\claim@c\noexpand\c{\claim@iv}}%
 \else
  \Err@{\noexpand#1not yet created by \string\newclaim}%
 \fi}
\def\classtest@#1{\DN@{#1}\ifx\next@\claimclass@
 \test@true\else\test@false\fi}
\def\typetest@#1{\DN@{#1}\ifx\next@\claimtype@\test@true\else
  \test@false\fi}
\newif\iftoc@
\def\tocfile{\iftoc@\else\alloc@@7\write\chardef\sixt@@n\toc@
 \immediate\openout\toc@=\jobname.toc
 \alloc@@7\write\chardef\sixt@@n\tic@
 \immediate\openout\tic@=\jobname.tic
 \global\toc@true\fi}
\rightadd@\hl\to\nofrillslist@
\rightadd@\HL\to\overlonglist@
\def\HL@@C{\csname HL@C\HLlevel@\endcsname}
\def\HL@@P{\csname HL@P\HLlevel@\endcsname}
\def\HL@@Q{\csname HL@Q\HLlevel@\endcsname}
\def\HL@@S{\csname HL@S\HLlevel@\endcsname}
\def\HL@@N{\csname HL@N\HLlevel@\endcsname}
\def\HL@@F{\csname HL@F\HLlevel@\endcsname}
\def\HL@@@C{\csname\exxx@\HLtype@ @C\endcsname}
\def\HL@@@P{\csname\exxx@\HLtype@ @P\endcsname}
\def\HL@@@Q{\csname\exxx@\HLtype@ @Q\endcsname}
\def\HL@@@S{\csname\exxx@\HLtype@ @S\endcsname}
\def\HL@@@N{\csname\exxx@\HLtype@ @N\endcsname}
\def\HL#1{\expandafter
 \ifx\csname HL@C#1\endcsname\relax
  \DN@{\Err@{\string\HL#1 not defined in this style}}%
 \else
  \DN@{\gdef\HLlevel@{#1}\def\HLname@{\HL{#1}}\let\HLtype@\relax\FNSS@\HL@}%
 \fi
 \next@}%
\newif\ifquoted@
\let\aftertoc@\relax
\def\HL@{%
 \DN@"##1"##2\endHL{\def\entry@{##2}\quoted@true
  {\noexpands@
  \ifx\HLtype@\relax
   \let\pre\HL@@P\let\post\HL@@Q\let\style\HL@@S\let\numstyle\HL@@N
  \else
   \let\pre\HL@@@P\let\post\HL@@@Q\let\style\HL@@@S\let\numstyle\HL@@@N
  \fi
  \Qlabel@{##1}\let\style\relax\xdef\Qlabel@@@@{##1}%
  \xdef\Thepref@{\Thelabel@@@@}}%
  \csname HL@\HLlevel@\endcsname##2\endHL
  \let\pref\Thepref@
  \csname HL@I\HLlevel@\endcsname
  \csname HL@J\HLlevel@\endcsname
  \let\pref\pref@
  \HLtoc@
  \aftertoc@
  \let\aftertoc@\relax\overlong@false}%
 \DNii@##1\endHL{\def\entry@{##1}\quoted@false
  {\noexpands@
  \ifx\HLtype@\relax
   \global\advance\HL@@C\@ne
   \xdef\Thelabel@@@{\number\HL@@C}%
   \xdefThelabel@{\HL@@N}%
   \xdef\Thelabel@@@@{\HL@@P\Thelabel@\HL@@Q}%
   \xdefThelabel@@{\HL@@S}%
  \else
   \global\advance\HL@@@C\@ne
   \xdef\Thelabel@@@{\number\HL@@@C}%
   \xdefThelabel@{\HL@@@N}%
   \xdef\Thelabel@@@@{\HL@@@P\Thelabel@\HL@@@Q}%
   \xdefThelabel@@{\HL@@@S}%
  \fi
  \xdef\Thepref@{\Thelabel@@@@}}%
  \csname HL@\HLlevel@\endcsname##1\endHL
  \let\pref\Thepref@
  \csname HL@I\HLlevel@\endcsname
  \csname HL@J\HLlevel@\endcsname
  \let\pref\pref@
  \HLtoc@
  \aftertoc@
  \let\aftertoc@\relax\overlong@false}%
 \ifx\next"\expandafter\next@\else\expandafter\nextii@\fi}%
\Invalid@\endHL
\def\hl@@C{\csname hl@C\hllevel@\endcsname}
\def\hl@@P{\csname hl@P\hllevel@\endcsname}
\def\hl@@Q{\csname hl@Q\hllevel@\endcsname}
\def\hl@@S{\csname hl@S\hllevel@\endcsname}
\def\hl@@N{\csname hl@N\hllevel@\endcsname}
\def\hl@@F{\csname hl@F\hllevel@\endcsname}
\def\hl@@@C{\csname\exxx@\hltype@ @C\endcsname}
\def\hl@@@P{\csname\exxx@\hltype@ @P\endcsname}
\def\hl@@@Q{\csname\exxx@\hltype@ @Q\endcsname}
\def\hl@@@S{\csname\exxx@\hltype@ @S\endcsname}
\def\hl@@@N{\csname\exxx@\hltype@ @N\endcsname}
\def\hl#1{\expandafter
 \ifx\csname hl@C#1\endcsname\relax
  \DN@{\Err@{\string\hl#1 not defined in this style}}%
 \else
  \DN@{\gdef\hllevel@{#1}\def\hlname@{\hl{#1}}\let\hltype@\relax\FNSS@\hl@}%
 \fi
 \next@}
\def\hl@{%
 \DN@"##1"##2{\def\entry@{##2}\quoted@true
  {\noexpands@
  \ifx\hltype@\relax
   \let\pre\hl@@P\let\post\hl@@Q\let\style\hl@@S\let\numstyle\hl@@N
  \else
   \let\pre\hl@@@P\let\post\hl@@@Q\let\style\hl@@@S\let\numstyle\hl@@@N
  \fi
  \Qlabel@{##1}\let\style\relax\xdef\Qlabel@@@@{##1}%
  \xdef\Thepref@{\Thelabel@@@@}}%
  \csname hl@\hllevel@\endcsname{##2}%
  \let\pref\Thepref@
  \csname hl@I\hllevel@\endcsname
  \csname hl@J\hllevel@\endcsname
  \let\pref\pref@
  \hltoc@
  \aftertoc@
  \let\aftertoc@\relax\nopunct@false\nospace@false\FNSSP@}%
 \DNii@##1{\def\entry@{##1}\quoted@false
  {\noexpands@
  \ifx\hltype@\relax
   \global\advance\hl@@C\@ne
   \xdef\Thelabel@@@{\number\hl@@C}%
   \xdefThelabel@{\hl@@N}%
   \xdef\Thelabel@@@@{\hl@@P\Thelabel@\hl@@Q}%
   \xdefThelabel@@{\hl@@S}%
  \else
   \global\advance\hl@@@C\@ne
   \xdef\Thelabel@@@{\number\hl@@@C}%
   \xdefThelabel@{\hl@@@N}%
   \xdef\Thelabel@@@@{\hl@@@P\Thelabel@\hl@@@Q}%
   \xdefThelabel@@{\hl@@@S}%
  \fi
  \xdef\Thepref@{\Thelabel@@@@}}%
  \csname hl@\hllevel@\endcsname{##1}%
  \let\pref\Thepref@
  \csname hl@I\hllevel@\endcsname
  \csname hl@J\hllevel@\endcsname
  \let\pref\pref@
  \hltoc@
  \aftertoc@
  \let\aftertoc@\relax\nopunct@false\nospace@false\FNSSP@}%
 \ifx\next"\expandafter\next@\else\expandafter\nextii@\fi}%
\def\six@#1#2 #3 #4 #5 #6 #7 {\DN@{#2}\ifx\next@\empty
 \DN@##1\six@{}\else
 \write#1{ #2 #3 #4 #5 #6 #7}\DN@{\six@#1}\fi
 \next@}
\def\Sixtoc@{\ifx\macdef@\empty\else
 \DN@##1##2\next@{\def\macdef@{##1##2}}%
 \expandafter\next@\macdef@\next@
 \edef\next@
  {\noexpand\six@\toc@\macdef@
  \space\space\space\space\space\space\space\space\space\space\space\space
  \noexpand\six@}%
 \next@\let\macdef@\relax\fi}
\def\QorThelabel@@@@{\ifquoted@
 \noexpand\noexpand\noexpand"\Qlabel@@@@\noexpand\noexpand\noexpand"\else
 \Thelabel@@@@\fi}
\def\HLtoc@{%
 \iftoc@
 \expandafter\expandafter\expandafter\unmacro@
  \expandafter\meaning\csname HL@W\HLlevel@\endcsname\unmacro@
  {\noexpands@\let\style\relax
   \edef\next@{\write\toc@{\noexpand\noexpand\expandafter\noexpand\HLname@
   {\macdef@}{\QorThelabel@@@@}}}%
  \next@}%
  \expandafter\unmacro@\meaning\entry@\unmacro@
  \Sixtoc@
  \write\toc@{\noexpand\Page{\number\pageno}{\page@N}%
   {\page@P}{\page@Q}^^J}%
 \fi}
\def\hltoc@{%
 \iftoc@
 \expandafter\expandafter\expandafter\unmacro@
  \expandafter\meaning\csname hl@W\hllevel@\endcsname\unmacro@
  {\noexpands@\let\style\relax
  \edef\next@{\write\toc@{%
   \ifnopunct@\noexpand\noexpand\noexpand\nopunct\fi
   \ifnospace@\noexpand\noexpand\noexpand\nospace\fi
   \noexpand\noexpand\expandafter\noexpand\hlname@
   {\macdef@}{\QorThelabel@@@@}}}%
  \next@}%
  \expandafter\unmacro@\meaning\entry@\unmacro@
  \Sixtoc@
  \write\toc@{\noexpand\Page{\number\pageno}{\page@N}%
   {\page@P}{\page@Q}^^J}%
 \fi}
\def\mainfile#1{\def\mainfile@{#1}}
\def\checkmainfile@{\ifx\mainfile@\undefined
 \Err@{No \noexpand\mainfile specified}\fi}
\expandafter\newcount@\csname HL@C1\endcsname
\csname HL@C1\endcsname\z@
\expandafter\def\csname HL@S1\endcsname#1{#1\null.}
\expandafter\let\csname HL@N1\endcsname\arabic
\expandafter\let\csname HL@P1\endcsname\empty
\expandafter\let\csname HL@Q1\endcsname\empty
\expandafter\def\csname HL@F1\endcsname{\bf}
\expandafter\let\csname HL@W1\endcsname\empty
\expandafter\newcount@\csname hl@C1\endcsname
\csname hl@C1\endcsname\z@
\expandafter\def\csname hl@S1\endcsname#1{#1\/}
\expandafter\let\csname hl@N1\endcsname\arabic
\expandafter\let\csname hl@P1\endcsname\empty
\expandafter\let\csname hl@Q1\endcsname\empty
\expandafter\def\csname hl@F1\endcsname{\bf}
\expandafter\let\csname hl@W1\endcsname\empty
\expandafter\def\csname HL@1\endcsname#1\endHL{\bigbreak
 {\locallabel@
  \global\setbox\@ne\vbox{\Let@\tabskip\hss@
  \halign to\hsize{\bf\hfil\ignorespaces##\unskip\hfil\cr
  \expandafter\ifx\csname HL@W1\endcsname\empty\else
   \csname HL@W1\endcsname\space\fi
  {\HL@@F\ifx\thelabel@@\empty\else\thelabel@@\space\fi}%
  \ignorespaces#1\crcr}}%
  }%
 \unvbox\@ne\nobreak\medskip}
\expandafter\def\csname hl@1\endcsname#1{\medbreak\noindent@@
 {\locallabel@
 \bf{\hl@@F\ifx\thelabel@@\empty\else\thelabel@@\space\fi}%
 \ignorespaces#1\unskip\punct@{\null.}\addspace@\enspace}}
\expandafter\def\csname HL@I1\endcsname{\Reset\hl1{1}%
 \ifx\pref\empty\newpre\hl1{}\else\newpre\hl1{\pref.}\fi}
\def\NameHL#1#2{\define#2{}%
 \expandafter\ifx\csname HL@R#1\endcsname\relax
 \else
  \def\nextiv@{\let\nextiii@}%
  \expandafter\nextiv@\csname HL@R#1\endcsname
  \expandafter\let\nextiii@\undefined
  \expandafter\let\csname\exxx@\nextiii@ @C\endcsname\relax
  \expandafter\let\csname\exxx@\nextiii@ @P\endcsname\relax
  \expandafter\let\csname\exxx@\nextiii@ @Q\endcsname\relax
  \expandafter\let\csname\exxx@\nextiii@ @S\endcsname\relax
  \expandafter\let\csname\exxx@\nextiii@ @N\endcsname\relax
  \expandafter\let\csname\exxx@\nextiii@ @F\endcsname\relax
  \expandafter\let\csname\exxx@\nextiii@ @W\endcsname\relax
  \expandafter\let\csname end\exxx@\nextiii@\endcsname\undefined
 \fi
 \expandafter\gdef\csname HL@R#1\endcsname{#2}%
 \expandafter\gdef\csname\exstring@#2@R\endcsname{{HL}{#1}}%
 \iftoc@\write\toc@{\noexpand\NameHL#1\noexpand#2^^J}\fi
 \rightadd@#2\to\overlonglist@
 \edef\next@{\let\csname\exstring@#2@C\endcsname\expandafter\noexpand
  \csname HL@C#1\endcsname}\next@
 \edef\next@{\let\csname\exstring@#2@P\endcsname\expandafter\noexpand
  \csname HL@P#1\endcsname}\next@
 \edef\next@{\let\csname\exstring@#2@Q\endcsname\expandafter\noexpand
  \csname HL@Q#1\endcsname}\next@
 \edef\next@{\let\csname\exstring@#2@S\endcsname\expandafter\noexpand
  \csname HL@S#1\endcsname}\next@
 \edef\next@{\let\csname\exstring@#2@N\endcsname\expandafter\noexpand
  \csname HL@N#1\endcsname}\next@
 \edef\next@{\let\csname\exstring@#2@F\endcsname\expandafter\noexpand
  \csname HL@F#1\endcsname}\next@
 \edef\next@{\let\csname\exstring@#2@W\endcsname\expandafter\noexpand
  \csname HL@W#1\endcsname}\next@
 \edef\next@{\def\noexpand#2####1\expandafter\noexpand
  \csname end\exstring@#2\endcsname
  {\def\noexpand\HLtype@{\noexpand#2}%
   \def\noexpand\HLname@{\noexpand#2}%
   \gdef\noexpand\HLlevel@{#1}%
   \noexpand\FNSS@\noexpand\HL@####1\noexpand\endHL}}%
  \next@
 \edef\next@{\noexpand\Invalid@\expandafter\noexpand
  \csname end\exstring@#2\endcsname}%
 \next@}
\def\Namehl#1#2{\define#2{}%
 \expandafter\ifx\csname hl@R#1\endcsname\relax
 \else
  \def\nextiv@{\let\nextiii@}%
  \expandafter\nextiv@\csname hl@R#1\endcsname
  \expandafter\let\nextiii@\undefined
  \expandafter\let\csname\exxx@\nextiii@ @C\endcsname\relax
  \expandafter\let\csname\exxx@\nextiii@ @P\endcsname\relax
  \expandafter\let\csname\exxx@\nextiii@ @Q\endcsname\relax
  \expandafter\let\csname\exxx@\nextiii@ @S\endcsname\relax
  \expandafter\let\csname\exxx@\nextiii@ @N\endcsname\relax
  \expandafter\let\csname\exxx@\nextiii@ @F\endcsname\relax
  \expandafter\let\csname\exxx@\nextiii@ @W\endcsname\relax
 \fi
 \expandafter\gdef\csname hl@R#1\endcsname{#2}%
 \expandafter\gdef\csname\exstring@#2@R\endcsname{{hl}{#1}}%
 \iftoc@\write\toc@{\noexpand\Namehl#1\noexpand#2^^J}\fi
 \rightadd@#2\to\nofrillslist@%
 \edef\next@{\let\csname\exstring@#2@C\endcsname\expandafter\noexpand
  \csname hl@C#1\endcsname}\next@
 \edef\next@{\let\csname\exstring@#2@P\endcsname\expandafter\noexpand
  \csname hl@P#1\endcsname}\next@
 \edef\next@{\let\csname\exstring@#2@Q\endcsname\expandafter\noexpand
  \csname hl@Q#1\endcsname}\next@
 \edef\next@{\let\csname\exstring@#2@S\endcsname\expandafter\noexpand
  \csname hl@S#1\endcsname}\next@
 \edef\next@{\let\csname\exstring@#2@N\endcsname\expandafter\noexpand
  \csname hl@N#1\endcsname}\next@
 \edef\next@{\let\csname\exstring@#2@F\endcsname\expandafter\noexpand
  \csname hl@F#1\endcsname}\next@
 \edef\next@{\let\csname\exstring@#2@W\endcsname\expandafter\noexpand
  \csname hl@W#1\endcsname}\next@
 \edef\next@{\def\noexpand#2{%
  \def\noexpand\hltype@{\noexpand#2}%
  \def\noexpand\hlname@{\noexpand#2}%
  \gdef\noexpand\hllevel@{#1}%
  \noexpand\FNSS@\noexpand\hl@}}%
 \next@}%
\def\Initialize{\FN@\Init@}
\def\Init@{\ifx\next\HL\let\next@\InitH@\else\ifx\next\hl\let\next@\InitH@
  \else\let\next@\InitS@\fi\fi\next@}
\def\InitH@#1#2{\expandafter\ifx\csname\exstring@#1@C#2\endcsname\relax
 \DN@{\Err@{\noexpand#1level #2 not defined in this style}}\else
 \DN@{\expandafter\gdef\csname\exstring@#1@J#2\endcsname}\fi\next@}
\def\InitC@#1#2{\edef\nextii@{\expandafter\noexpand\csname#1\endcsname{#2}}}
\def\InitS@#1{\expandafter\ifx\csname\exstring@#1@R\endcsname\relax
 \Err@{\noexpand#1not defined in this style}\let\next@\relax\else
 \DN@{\let\next@}\expandafter\next@\csname\exstring@#1@R\endcsname
 \expandafter\InitC@\next@
 \DN@{\expandafter\InitH@\nextii@}\fi\next@}
\def\value#1{\expandafter
 \ifx\csname\exstring@#1@C\endcsname\relax
  \expandafter\ifx\csname\exstring@#1@C1\endcsname\relax
   \DN@{\Err@{\noexpand\value can't be used with \string#1}}%
  \else
   \DN@{\value@#1}%
  \fi
 \else
  \DN@{\number\csname\exstring@#1@C\endcsname\relax}%
 \fi
 \next@}
\def\value@#1#2{\expandafter
 \ifx\csname\exstring@#1@C#2\endcsname\relax
  \DN@{\Err@{\string\value\string#1 can't be followed by \string#2}}%
 \else
  \DN@{\number\csname\exstring@#1@C#2\endcsname\relax}%
 \fi
 \next@}
\newcount\Value
\def\Evaluate#1{\expandafter
 \ifx\csname\exstring@#1@C\endcsname\relax
  \expandafter\ifx\csname\exstring@#1@C1\endcsname\relax
   \DN@{\Err@{\noexpand\Evaluate can't be used with \string#1}}%
  \else
   \DN@{\Evaluate@#1}%
  \fi
 \else
  \DN@{\global\Value\csname\exstring@#1@C\endcsname}%
 \fi
 \next@}
\def\Evaluate@#1#2{\expandafter
 \ifx\csname\exstring@#1@C#2\endcsname\relax
  \DN@{\Err@{\string\Evaluate\string#1 can't be followed by \string#2}}%
 \else
  \DN@{\global\Value\csname\exstring@#1@C#2\endcsname}%
 \fi\next@}
\def\pre#1{\expandafter
 \ifx\csname\exstring@#1@P\endcsname\relax
  \expandafter\ifx\csname\exstring@#1@P1\endcsname\relax
   \DN@{\Err@{\noexpand\pre can't be used with \string#1}}%
  \else
   \DN@{\pre@#1}%
  \fi
 \else
  \DN@{{\csname\exstring@#1@P\endcsname}}%
 \fi
 \next@}
\def\pre@#1#2{\expandafter
 \ifx\csname\exstring@#1@P#2\endcsname\relax
  \DN@{\Err@{\string\pre\string#1 can't be followed by \string#2}}%
 \else
  \DN@{{\csname\exstring@#1@P#2\endcsname}}%
 \fi
 \next@}
\def\post#1{\expandafter
 \ifx\csname\exstring@#1@Q\endcsname\relax
  \expandafter\ifx\csname\exstring@#1@Q1\endcsname\relax
   \DN@{\Err@{\noexpand\post can't be used with \string#1}}%
  \else
   \DN@{\post@#1}%
  \fi
 \else
  \DN@{{\csname\exstring@#1@Q\endcsname}}%
 \fi
 \next@}
\def\post@#1#2{\expandafter
 \ifx\csname\exstring@#1@Q#2\endcsname\relax
  \DN@{\Err@{\string\post\string#1 can't be followed by \string#2}}%
 \else
  \DN@{{\csname\exstring@#1@Q#2\endcsname}}%
 \fi
 \next@}
\def\style#1{\expandafter
 \ifx\csname\exstring@#1@S\endcsname\relax
  \expandafter\ifx\csname\exstring@#1@S1\endcsname\relax
   \DN@{\Err@{\noexpand\style can't be used with \string#1}}%
  \else
   \DN@{\style@#1}%
  \fi
 \else
  \DN@{\csname\exstring@#1@S\endcsname}%
 \fi
 \next@}
\def\style@#1#2{\expandafter
 \ifx\csname\exstring@#1@S#2\endcsname\relax
  \DN@{\Err@{\string\style\string#1 can't be followed by \string#2}}%
 \else
  \DN@{\csname\exstring@#1@S#2\endcsname}%
 \fi
 \next@}
\def\fontstyle#1{\expandafter
 \ifx\csname\exstring@#1@F\endcsname\relax
  \expandafter\ifx\csname\exstring@#1@F1\endcsname\relax
   \DN@{\Err@{\noexpand\fontstyle can't be used with \string#1}}%
  \else
   \DN@{\fontstyle@#1}%
  \fi
 \else
  \DN@##1{{\csname\exstring@#1@F\endcsname##1}}%
 \fi
 \next@}
\def\fontstyle@#1#2{\expandafter
 \ifx\csname\exstring@#1@F#2\endcsname\relax
  \DN@{\Err@{\string\fontstyle\string#1 can't be followed by \string#2}}%
 \else
  \DN@##1{{\csname\exstring@#1@F#2\endcsname##1}}%
 \fi
 \next@}
\def\Reset#1{\expandafter
 \ifx\csname\exstring@#1@C\endcsname\relax
  \expandafter\ifx\csname\exstring@#1@C1\endcsname\relax
   \DN@{\Err@{\noexpand\Reset can't be used with \string#1}}%
  \else
   \DN@{\Reset@#1}%
  \fi
 \else
  \DN@##1{\count@##1\relax\ifx#1\page\else\advance\count@\m@ne\fi
   \global\csname\exstring@#1@C\endcsname\count@}%
 \fi
 \next@}
\def\Reset@#1#2{\expandafter
 \ifx\csname\exstring@#1@C#2\endcsname\relax
  \DN@{\Err@{\string\Reset\string#1 can't be followed by \string#2}}%
 \else
  \DN@##1{\count@##1\relax\advance\count@\m@ne
   \global\csname\exstring@#1@C#2\endcsname\count@}%
 \fi
 \next@}
\def\Offset#1{\expandafter
 \ifx\csname\exstring@#1@C\endcsname\relax
  \expandafter\ifx\csname\exstring@#1@C1\endcsname\relax
   \DN@{\Err@{\noexpand\Offset can't be used with \string#1}}%
  \else
   \DN@{\Offset@#1}%
  \fi
 \else
  \DN@##1{\count@##1\relax\advance\count@\m@ne\global\advance
   \csname\exstring@#1@C\endcsname\count@}%
 \fi
 \next@}
\def\Offset@#1#2{\expandafter
 \ifx\csname\exstring@#1@C#2\endcsname\relax
  \DN@{\Err@{\string\Offset\string#1 can't be followed by \string#2}}%
 \else
  \DN@##1{\count@##1\relax\advance\count@\m@ne
   \global\advance\csname\exstring@#1@C#2\endcsname\count@}%
 \fi
 \next@}
\def\getR@#1#2{\def\nextiv@{\let\nextiii@}\expandafter\nextiv@
 \csname\exstring@#1@R#2\endcsname}
\def\letR@#1#2#3{\expandafter\let\csname#1@#3#2\endcsname\Next@}
\def\letR@@#1#2{\expandafter\let\csname\exstring@#1@#2\endcsname\Next@}
\def\newpre#1{\expandafter
 \ifx\csname\exstring@#1@P\endcsname\relax
  \expandafter\ifx\csname\exstring@#1@P1\endcsname\relax
   \DN@{\Err@{\noexpand\newpre can't be used with \string#1}}%
  \else
   \DN@{\newpre@#1}%
  \fi
 \else
  \DN@{%
   \DNii@{%
    \endgroup
    \expandafter\let\csname\exstring@#1@P\endcsname\Next@
    \expandafter\ifx\csname\exstring@#1@R\endcsname\relax\else
    \getR@#1{}\expandafter\letR@\nextiii@ P\fi
    }%
   \begingroup\noexpands@\afterassignment\nextii@\xdef\Next@}%
 \fi
 \next@}
\def\newpre@#1#2{\expandafter
 \ifx\csname\exstring@#1@P#2\endcsname\relax
  \DN@{\Err@{\string\newpre\string#1 can't be followed by \string#2}}%
 \else
  \DN@{%
   \DNii@{%
    \endgroup
    \expandafter\let\csname\exstring@#1@P#2\endcsname\Next@
    \expandafter\ifx\csname\exstring@#1@R#2\endcsname\relax\else
    \getR@#1{#2}\expandafter\letR@@\nextiii@ P\fi
    }%
   \begingroup\noexpands@\afterassignment\nextii@\xdef\Next@}%
 \fi
 \next@}
\def\newpost#1{\expandafter
 \ifx\csname\exstring@#1@Q\endcsname\relax
  \expandafter\ifx\csname\exstring@#1@Q1\endcsname\relax
   \DN@{\Err@{\noexpand\newpost can't be used with \string#1}}%
  \else
   \DN@{\newpost@#1}%
  \fi
 \else
  \DN@{%
   \DNii@{%
    \endgroup
    \expandafter\let\csname\exstring@#1@Q\endcsname\Next@
    \expandafter\ifx\csname\exstring@#1@R\endcsname\relax\else
    \getR@#1{}\expandafter\letR@\nextiii@ Q\fi
    }%
   \begingroup\noexpands@\afterassignment\nextii@\xdef\Next@}%
 \fi
 \next@}
\def\newpost@#1#2{\expandafter
 \ifx\csname\exstring@#1@Q#2\endcsname\relax
  \DN@{\Err@{\string\newpost\string#1 can't be followed by \string#2}}%
 \else
  \DN@{%
   \DNii@{%
    \endgroup
    \expandafter\let\csname\exstring@#1@Q#2\endcsname\Next@
    \expandafter\ifx\csname\exstring@#1@R#2\endcsname\relax\else
    \getR@#1{#2}\expandafter\letR@@\nextiii@ Q\fi
    }%
   \begingroup\noexpands@\afterassignment\nextii@\xdef\Next@}%
 \fi
 \next@}
\def\newstyle#1{\expandafter
 \ifx\csname\exstring@#1@S\endcsname\relax
  \expandafter\ifx\csname\exstring@#1@S1\endcsname\relax
   \DN@{\Err@{\noexpand\newstyle can't be used
    with \string#1}}%
  \else
   \DN@{\newstyle@#1}%
  \fi
 \else
  \DN@{%
   \DNii@{%
    \expandafter\let\csname\exstring@#1@S\endcsname\Next@
    \expandafter\ifx\csname\exstring@#1@R\endcsname\relax\else
    \getR@#1{}\expandafter\letR@\nextiii@ S\fi
    }%
   \afterassignment\nextii@\gdef\Next@}%
 \fi
 \next@}
\def\newstyle@#1#2{\expandafter
 \ifx\csname\exstring@#1@S#2\endcsname\relax
  \DN@{\Err@{\string\newstyle\string#1 can't be followed by
   \string#2}}%
 \else
  \DN@{%
   \DNii@{%
    \expandafter\let\csname\exstring@#1@S#2\endcsname\Next@
    \expandafter\ifx\csname\exstring@#1@R#2\endcsname\relax\else
    \getR@#1{#2}\expandafter\letR@@\nextiii@ S\fi
    }%
   \afterassignment\nextii@\gdef\Next@}%
 \fi
 \next@}
\def\newnumstyle#1{\expandafter
 \ifx\csname\exstring@#1@N\endcsname\relax
  \expandafter\ifx\csname\exstring@#1@N1\endcsname\relax
   \DN@{\Err@{\noexpand\newnumstyle can't be used with
    \string#1}}%
  \else
   \DN@{\newnumstyle@#1}%
  \fi
 \else
  \DN@##1{%
   \gdef\Next@{##1}%
    \expandafter\let\csname\exstring@#1@N\endcsname\Next@
    \expandafter\ifx\csname\exstring@#1@R\endcsname\relax\else
    \getR@#1{}\expandafter\letR@\nextiii@ N\fi
    }%
 \fi
 \next@}
\def\newnumstyle@#1#2{\expandafter
 \ifx\csname\exstring@#1@N#2\endcsname\relax
  \DN@{\Err@{\string\newnumstyle\string#1 can't be followed by
   \string#2}}%
 \else
  \DN@##1{%
   \gdef\Next@{##1}%
    \expandafter\let\csname\exstring@#1@N#2\endcsname\Next@
    \expandafter\ifx\csname\exstring@#1@R#2\endcsname\relax\else
    \getR@#1{#2}\expandafter\letR@@\nextiii@ N\fi
    }%
  \fi
 \next@}
\def\newfontstyle#1{\expandafter
 \ifx\csname\exstring@#1@F\endcsname\relax
  \expandafter\ifx\csname\exstring@#1@F1\endcsname\relax
   \DN@{\Err@{\noexpand\newfontstyle can't be used with
    \string#1}}%
  \else
   \DN@{\newfontstyle@#1}%
  \fi
 \else
  \DN@##1{%
   \gdef\Next@{##1}%
    \expandafter\let\csname\exstring@#1@F\endcsname\Next@
    \expandafter\ifx\csname\exstring@#1@R\endcsname\relax\else
    \getR@#1{}\expandafter\letR@\nextiii@ F\fi
    }%
 \fi
 \next@}
\def\newfontstyle@#1#2{\expandafter
 \ifx\csname\exstring@#1@F#2\endcsname\relax
  \DN@{\Err@{\string\newfontstyle\string#1 can't be followed by
   \string#2}}%
 \else
  \DN@##1{%
   \gdef\Next@{##1}%
    \expandafter\let\csname\exstring@#1@F#2\endcsname\Next@
    \expandafter\ifx\csname\exstring@#1@R#2\endcsname\relax\else
    \getR@#1{#2}\expandafter\letR@@\nextiii@ F\fi
    }%
 \fi
 \next@}
\def\word#1{\expandafter
 \ifx\csname\exstring@#1@W\endcsname\relax
  \expandafter\ifx\csname\exstring@#1@W1\endcsname\relax
   \DN@{\Err@{\noexpand\word can't be used with \string#1}}%
  \else
   \DN@{\word@#1}%
  \fi
 \else
  \DN@{{\csname\exstring@#1@W\endcsname}}%
 \fi
 \next@}
\def\word@#1#2{\expandafter
 \ifx\csname\exstring@#1@W#2\endcsname\relax
  \DN@{\Err@{\string\word\noexpand#1can't be followed by \string#2}}%
 \else
  \DN@{{\csname\exstring@#1@W#2\endcsname}}%
 \fi
 \next@}
\def\newword#1{\expandafter
 \ifx\csname\exstring@#1@W\endcsname\relax
  \expandafter\ifx\csname\exstring@#1@W1\endcsname\relax
   \DN@{\Err@{\noexpand\newword can't be used  with \string#1}}%
  \else
   \DN@{\newword@#1}%
  \fi
 \else
  \DN@{%
   \DNii@{%
    \expandafter\let\csname\exstring@#1@W\endcsname\Next@
    \expandafter\ifx\csname\exstring@#1@R\endcsname\relax\else
     \getR@#1{}\expandafter\letR@\nextiii@ W\fi
    }%
   \afterassignment\nextii@\gdef\Next@}%
 \fi
 \next@}
\def\newword@#1#2{\expandafter
 \ifx\csname\exstring@#1@W#2\endcsname\relax
  \DN@{\Err@{\string\newword\noexpand#1can't be followed by \string#2}}%
 \else
  \DN@{%
   \DNii@{%
    \expandafter\let\csname\exstring@#1@W#2\endcsname\Next@
    \expandafter\ifx\csname\exstring@#1@R#2\endcsname\relax\else
     \getR@#1{#2}\expandafter\letR@@\nextiii@ W\fi
    }%
   \afterassignment\nextii@\gdef\Next@}%
 \fi
 \next@}
\newif\iffn@
\newcount\footmark@C
\footmark@C\z@
\def\footmark@S#1{$^{#1}$}
\let\footmark@N\arabic
\def\footmark@F{\rm}
\def\foottext@S#1{$^{#1}$}
\def\foottext@F{\rm}
\let\modifyfootnote@\relax
\def\modifyfootnote#1{\def\modifyfootnote@{#1}}
\def\vfootnote@#1{\insert\footins
 \bgroup
 \floatingpenalty\@MM\interlinepenalty\interfootnotelinepenalty
 \leftskip\z@\rightskip\z@\spaceskip\z@\xspaceskip\z@
 \rm\splittopskip\ht\strutbox\splitmaxdepth\dp\strutbox
 \locallabel@\noindent@@{\foottext@F#1}\modifyfootnote@
 \footstrut\FN@\fo@t}
\def\fo@t{\ifcat\bgroup\noexpand\next\expandafter\f@@t\else
 \expandafter\f@t\fi}
\def\f@t#1{#1\@foot}
\def\f@@t{\bgroup\aftergroup\@foot\afterassignment\FNSSP@\let\next@}
\def\@foot{\unskip\lower\dp\strutbox\vbox to\dp\strutbox{}\egroup
 \iffn@\expandafter\fn@false\else
 \expandafter\postvanish@\fi}
\newif\ifplainfn@
\plainfn@true
\def\fancyfootnotes{\plainfn@false}
\newcount\fancyfootmarkcount@
\fancyfootmarkcount@\z@
\newcount\lastfnpage@
\lastfnpage@-\@M
\let\justfootmarklist@\empty
\def\footmark{\let\@sf\empty
 \ifhmode\edef\@sf{\spacefactor\the\spacefactor}\/\fi
 \DN@{\ifx"\next\expandafter\nextii@\else\expandafter\footmark@\fi}%
 \DNii@"##1"{%
  \iffirstchoice@
   {\let\style\footmark@S\let\numstyle\footmark@N
   \footmark@F##1%
   \noexpands@
   \let\style\foottext@S
   \Qlabel@{##1}%
   }%
   \iffn@\else
    {\noexpands@
    \xdef\Next@{{\Thelabel@}{\Thelabel@@}{\Thelabel@@@}{\Thelabel@@@@}}%
    }%
    \expandafter\rightappend@\Next@\to\justfootmarklist@
   \fi
  \fi
  \@sf\relax}%
 \FN@\next@}
\def\footmark@{%
 \iffirstchoice@
  \global\advance\footmark@C\@ne
  \ifplainfn@
   \xdef\adjustedfootmark@{\number\footmark@C}%
  \else
   {\let\\\or\xdef\Next@{\ifcase\number\footmark@C\fnpages@\else
     -\@M\fi}}%
   \ifnum\Next@=-\@M
    \xdef\adjustedfootmark@{\number\footmark@C}%
   \else
    \ifnum\Next@=\lastfnpage@
     \global\advance\fancyfootmarkcount@\@ne
    \else
     \global\fancyfootmarkcount@\@ne
     \global\lastfnpage@\Next@
    \fi
    \xdef\adjustedfootmark@{\number\fancyfootmarkcount@}%
   \fi
  \fi
  {\noexpands@
  \xdef\Thelabel@@@{\adjustedfootmark@}%
  \xdefThelabel@\footmark@N
  \xdef\Thelabel@@@@{\Thelabel@}%
  \xdefThelabel@@\foottext@S
  }%
  \iffn@\else
   {\noexpands@
   \xdef\Next@{{\Thelabel@}{\Thelabel@@}{\Thelabel@@@}{\Thelabel@@@@}}%
   }%
   \expandafter\rightappend@\Next@\to\justfootmarklist@
  \fi
  \ifplainfn@
  \else
   \edef\next@{\write\laxwrite@{F\noexpand\the\pageno}}\next@
  \fi
 \fi
 \footmark@S{\footmark@N{\adjustedfootmark@}}%
 \@sf\relax}
\def\foottext{\prevanish@
 \ifx\justfootmarklist@\empty
  \Err@{There is no \noexpand\footmark for this \string\foottext}\fi
 \DN@\\##1##2\next@{\DN@{##1}\gdef\justfootmarklist@{##2}}%
 \expandafter\next@\justfootmarklist@\next@
 \expandafter\foottext@\next@}
\def\foottext@#1#2#3#4{{\noexpands@
  \xdef\Thelabel@{#1}\xdef\Thelabel@@{#2}%
  \xdef\Thelabel@@@{#3}\xdef\Thelabel@@@@{#4}}%
  \vfootnote@{\thelabel@@}}
\rightadd@\foottext\to\vanishlist@
\def\footnote{\fn@true
 \let\@sf\empty
 \ifhmode\edef\@sf{\spacefactor\the\spacefactor}\/\fi
 \DN@{\ifx"\next\expandafter\nextii@\else\expandafter\nextiii@\fi}%
 \DNii@"##1"{\footmark"##1"\vfootnote@{\let\style\foottext@S
  \let\numstyle\footmark@N##1}}%
 \def\nextiii@{\footmark\vfootnote@{\foottext@S{\footmark@N
  {\adjustedfootmark@}}}}%
 \FN@\next@}
\newdimen\litindent
\litindent20\p@
\newbox\litbox@
\newbox\Litbox@
\newcount\interlitpenalty@
\interlitpenalty@\@M
\newcount\litlines@
{\obeyspaces\gdef\defspace@{\def {\allowbreak\hskip.5emminus.15em}}}
{\obeylines\gdef\letM@{\let^^M\CtrlM@}}
\def\CtrlM@{\egroup
 \ifcase\litlines@\advance\litlines@\@ne\or
 \box\litbox@\advance\litlines@\@ne\else
 \penalty\interlitpenalty@\box\litbox@\fi
 \Lit@}
\def\Lit@{\setbox\litbox@\hbox\bgroup\litdefs@\hskip\litindent}
\newcount\littab@
\littab@8
\def\littab#1{\littab@#1\relax}
{\catcode`\^^I=\active\gdef\letTAB@{\let^^I\TAB@}}
\def\TAB@{\egroup
 \dimen@\wd\litbox@
 \advance\dimen@-\litindent
 \setboxz@h{\tt0}%
 \dimen@ii\littab@\wdz@
 \divide\dimen@\dimen@ii
 \multiply\dimen@\dimen@ii
 \advance\dimen@\littab@\wdz@
 \advance\dimen@\litindent
 \setbox\litbox@\hbox\bgroup\litdefs@\hbox to\dimen@{\unhbox\litbox@\hfil}}
{\catcode`\`=\active\gdef`{\relax\lq}}
\let\litbs@\relax
\let\litbs@@\relax
\def\litbackslash#1{%
 \edef\litbs@{\catcode`\string#1=\z@
 \def\noexpand\litbs@@{\def\expandafter\noexpand\csname\string#1\endcsname
  {\char`\string#1}}}}
\def\litcodes@{\catcode`\\=12
 \catcode`\{=12 \catcode`\}=12
 \catcode`\$=12 \catcode`\&=12
 \catcode`\#=12
 \catcode`\^=12 \catcode`\_=12
 \catcode`\@=12 \catcode`\~=12 \catcode`\"=12
 \catcode`\;=12 \catcode`\:=12 \catcode`\!=12 \catcode`\?=12
 \catcode`\%=12 \litbs@\catcode`\`=\active\obeyspaces\defspace@}
\def\activate@#1#2{{\lccode`\~=`#2%
 \lowercase{%
  \if0#1%
  \gdef\Next@{\def~{\egroup\endgroup\bigskip\vskip-\parskip
   \def\next@{\noindent@@\FN@\pretendspace@}\FNSS@\next@}}\else
  \gdef\Next@{\def~{\egroup\egroup\endgroup}}\fi
  }%
 }}
\def\litdefs@{\let\0\empty\let\1\litdelim@\def\ {\char32 }\litbs@@}%
\def\litdelimiter#1{%
 \edef\litdelim@{\char`#1}%
 \def\lit#1{\leavevmode\begingroup\litcodes@\litdefs@
  \tt\hyphenchar\tentt\m@ne\lit@}%
 \def\lit@##1#1{##1\endgroup\null}%
 \def\Lit#1{\ifhmode$$\abovedisplayskip\bigskipamount
  \abovedisplayshortskip\bigskipamount
  \belowdisplayskip\z@\belowdisplayshortskip\z@
  \postdisplaypenalty\@M
  $$\vskip-\baselineskip\else\bigskip\fi
  \begingroup\litlines@\z@
  \catcode`#1=\active\activate@0#1\Next@
  \def\displaybreak{\egroup\break\litlines@\z@\Lit@}%
  \def\allowdisplaybreak{\egroup\allowbreak\litlines@\z@\Lit@}%
  \def\allowdisplaybreaks{\egroup\allowbreak\interlitpenalty@\z@
   \litlines@\z@\Lit@}%
  \litcodes@\tt\catcode`\^^I=\active\letTAB@
  \obeylines\letM@\Lit@}%
 \def\Litbox##1=#1{\begingroup\ifodd##1\relax\aftergroup\global\fi
  \aftergroup\setbox\aftergroup##1\aftergroup\box\aftergroup\Litbox@
  \def\allowdisplaybreak{\egroup\allowbreak\litlines@\z@\Lit@}%
  \def\allowdisplaybreaks{\egroup\allowbreak\interlitpenalty@\z@
   \litlines@\z@\Lit@}%
  \catcode`#1=\active\activate@1#1\Next@
  \litcodes@\tt\catcode`\^^I=\active\letTAB@
  \obeylines\letM@\global\setbox\Litbox@\vbox\bgroup\litindent\z@%
  \litlines@\z@\Lit@}%
}
\newbox\titlebox@
\setbox\titlebox@\vbox{}
\rightadd@\title\to\overlonglist@
\def\title{\begingroup\Let@
 \global\setbox\titlebox@\vbox\bgroup\tabskip\hss@
 \halign to\hsize\bgroup\bf\hfil\ignorespaces##\unskip\hfil\cr}
\def\endtitle{\crcr\egroup\egroup\endgroup\overlong@false}
\newbox\authorbox@
\rightadd@\author\to\overlonglist@
\def\author{\begingroup\Let@
 \global\setbox\authorbox@\vbox\bgroup\tabskip\hss@
 \halign to\hsize\bgroup\rm\hfil\ignorespaces##\unskip\hfil\cr}
\def\endauthor{\crcr\egroup\egroup\endgroup\overlong@false}
\newbox\affilbox@
\def\affil{\begingroup\Let@
 \global\setbox\affilbox@\vbox\bgroup\tabskip\hss@
 \halign to\hsize\bgroup\rm\hfil\ignorespaces##\unskip\hfil\cr}%
\def\endaffil{\crcr\egroup\egroup\endgroup\overlong@false}
\let\date@\relax
\def\date#1{\gdef\date@{\ignorespaces#1\unskip}}
\def\today{\ifcase\month\or January\or February\or March\or April\or May\or
 June\or July\or August\or September\or October\or November\or December\fi
 \space\number\day, \number\year}
\def\maketitle{\hrule\height\z@\vskip-\topskip
 \vskip24\p@ plus12\p@ minus12\p@
 \unvbox\titlebox@
 \ifvoid\authorbox@\else\vskip12\p@ plus6\p@ minus3\p@\unvbox\authorbox@\fi
 \ifvoid\affilbox@\else\vskip10\p@ plus5\p@ minus2\p@\unvbox\affilbox@\fi
 \ifx\date@\relax\else\vskip6\p@ plus2\p@ minus\p@\centerline{\rm\date@}\fi
 \vskip18\p@ plus12\p@ minus6\p@}
\def\cite{%
 \DNii@(##1)##2{{\rm[}{##2}, {##1\/}{\rm]}}%
 \def\nextiii@##1{{\rm[}{##1\/}{\rm]}}%
 \DN@{\ifx\next(\expandafter\nextii@\else\expandafter\nextiii@\fi}%
 \FN@\next@}
\def\makebib@W{Bibliography}
\def\makebibp@W{References}
\def\makebib{\begingroup\rm\bigbreak\centerline{\smc\makebib@W}%
 \nobreak\medskip
 \sfcode`\.=\@m\everypar{}\parindent\z@
 \def\nopunct{\nopunct@true}\def\nospace{\nospace@true}%
 \nopunct@false\nospace@false
 \def\lkerns@{\null\kern\m@ne sp\kern\@ne sp}%
 \def\nkerns@{\null\kern-\tw@ sp\kern\tw@ sp}%
}

\newif\ifnoprepunct@
\newif\ifnoprespace@
\newif\ifnoquotes@
\def\noprepunct{\noprepunct@true}
\def\noprespace{\noprespace@true}
\def\noquotes{\noquotes@true}
\newbox\nobox@
\newbox\keybox@
\newbox\bybox@
\newbox\paperbox@
\newbox\paperinfobox@
\newbox\jourbox@
\newbox\volbox@
\newbox\issuebox@
\newbox\yrbox@
\newbox\pgbox@
\newbox\ppbox@
\newbox\bookbox@
\newbox\inbookbox@
\newbox\bookinfobox@
\newbox\publbox@
\newbox\publaddrbox@
\newbox\edbox@
\newbox\edsbox@
\newbox\langbox@
\newbox\translbox@
\newbox\finalinfobox@
\def\setbibinfo@#1{\edef\next@{\ifnopunct@1\else0\fi
 \ifnospace@1\else0\fi\ifnoprepunct@1\else0\fi\ifnoprespace@1\else0\fi
 \ifnoquotes@1\else0\fi}%
 \DNii@{00000}%
 \ifx\next@\nextii@\else\xdef\bibinfo@{\bibinfo@\the#1,\next@}%
 \fi}
\def\getbibinfo@#1{\ifx\bibinfo@\empty
 \let\next@0\let\nextii@0\let\nextiii@0\let\nextiv@0\let\nextv@0\else
 \edef\next@{\def
  \noexpand\next@####1\the#1,####2####3####4####5####6####7\noexpand\next@
  {\let\noexpand\next@####2\let\noexpand\nextii@####3%
  \let\noexpand\nextiii@####4\let\noexpand\nextiv@####5%
  \let\noexpand\nextv@####6}%
  \noexpand\next@\bibinfo@\the#1,00000\noexpand\next@}\next@
 \fi}
\newif\ifbookinquotes@
\def\bookinquotes{\bookinquotes@true}
\newif\ifpaperinquotes@
\def\paperinquotes{\paperinquotes@true}
\newif\ifininbook@
\def\ininbook{\ininbook@true}
\newif\ifopenquotes@
\def\closequotes@{\ifopenquotes@''\openquotes@false\fi}
\newif\ifbeginbib@
\newif\ifendbib@
\newif\ifprevjour@
\newif\ifprevbook@
\newdimen\bibindent@
\bibindent@20\p@
\def\bib{\global\let\bibinfo@\empty\global\let\translinfo@\relax\beginbib@true
 \begingroup\noindent@
 \hangindent\bibindent@\hangafter\@ne\bib@}
\def\v@id#1{\setbox#1\box\voidb@x}
\def\bib@{\v@id\nobox@\v@id\keybox@\v@id\bybox@\v@id\paperbox@
 \v@id\paperinfobox@\v@id\jourbox@\v@id\volbox@\v@id\issuebox@
 \v@id\yrbox@\v@id\pgbox@\v@id\ppbox@\v@id\bookbox@\v@id\inbookbox@
 \v@id\bookinfobox@\v@id\publbox@\v@id\publaddrbox@\v@id\edbox@
 \v@id\edsbox@\v@id\langbox@\v@id\translbox@\v@id\finalinfobox@
 \bgroup}
\def\Setnonemptybox@#1#2{\unskip\setbibinfo@#1\egroup#2%
 \def\aftergroup@{\ifdim\wd#1=\z@\setbox#1\box\voidb@x\fi}%
 \setbox#1\vbox\bgroup\aftergroup\aftergroup@\hsize\maxdimen\leftskip\z@
 \rightskip\z@\hbadness\@M\hfuzz\maxdimen\noindent}
\def\setnonemptybox@#1{\Setnonemptybox@#1\relax}
\def\no{\setnonemptybox@\nobox@}
\def\key{\setnonemptybox@\keybox@\bf}
\def\by{\setnonemptybox@\bybox@}
\def\bysame{\setnonemptybox@\bybox@\leaders\hrule\hskip3em\null}
\def\paper{\setnonemptybox@\paperbox@
 \ifpaperinquotes@\getbibinfo@\paperbox@
 \if\nextv@1\else``\fi\else\it\fi}
\def\paperinfo{\setnonemptybox@\paperinfobox@}
\def\jour{\Setnonemptybox@\jourbox@\prevjour@true}
\def\vol{\setnonemptybox@\volbox@\bf}
\def\issue{\setnonemptybox@\issuebox@}
\def\yr{\setnonemptybox@\yrbox@}

\def\pg{\setnonemptybox@\pgbox@}
\def\pp{\setnonemptybox@\ppbox@}
\def\book{\Setnonemptybox@\bookbox@\prevbook@true
 \ifbookinquotes@\getbibinfo@\bookbox@
 \if\nextv@1\else``\fi\else\it\fi}
\def\inbook{\Setnonemptybox@\inbookbox@\prevbook@true
 \ifininbook@ in \fi\ifbookinquotes@\getbibinfo@\inbookbox@
 \if\nextv@1\else``\fi\fi}
\def\bookinfo{\setnonemptybox@\bookinfobox@}
\def\publ{\setnonemptybox@\publbox@}
\def\publaddr{\setnonemptybox@\publaddrbox@}
\def\ed{\setnonemptybox@\edbox@}
\def\eds{\setnonemptybox@\edsbox@}
\def\lang{\setnonemptybox@\langbox@}
\def\finalinfo{\setnonemptybox@\finalinfobox@}
\def\setboxzl@{\setbox\z@\lastbox}
\def\getbox@#1{\setbox\z@\vbox{\vskip-\@M\p@
 \unvbox#1%
 \setboxzl@
 \global\setbox\@ne\hbox{\unhbox\z@\unskip\unskip\unpenalty}%
 \ifdim\lastskip=-\@M\p@\else
 \loop\ifdim\lastskip=-\@M\p@
 \else\unskip\unpenalty\setboxzl@
 \global\setbox\@ne\hbox{\unhbox\z@\unhbox\@ne}%
 \repeat\fi}%
 \unhbox\@ne}
\def\adjustpunct@#1{\count@\lastkern
 \ifnum\count@=\z@#1\closequotes@\else
 \ifnum\count@>\tw@#1\closequotes@\else
 \ifnum\count@<-\tw@#1\closequotes@\else
  \unkern\unkern\setboxzl@
  \skip@\lastskip\unskip
  \count@@\lastpenalty\unpenalty
  \ifnum\count@=\tw@\unskip\setboxzl@\fi
  \ifdim\skip@=\z@\else\hskip\skip@\fi
  #1\closequotes@
  \ifnum\count@=\tw@\null\hfill\fi
  \penalty\count@@
 \fi\fi\fi}
\def\prepunct@#1#2{\getbibinfo@#2%
 \ifnopunct@
 \else
  \if\nextiii@0\adjustpunct@#1\fi
 \fi
 \closequotes@
 \ifnospace@
 \else
  \if\nextiv@0\space\else\fi
 \fi
 \nopunct@false\nospace@false
 \if\next@1\nopunct@true\fi
 \if\nextii@1\nospace@true\fi}
\def\ppunbox@#1#2{\prepunct@{#1}#2%
 \getbox@#2}
\let\semicolon@;
\def\endbib@{%
 \ifbeginbib@
  \ifvoid\nobox@
   \ifvoid\keybox@\else\hbox to\bibindent@{[\getbox@\keybox@]\hss}\fi
  \else\hbox to\bibindent@{\hss\getbox@\nobox@. }\fi
  \ifvoid\bybox@\else\getbox@\bybox@\fi
 \else
  \nopunct@true
  \ifvoid\bybox@\else\ppunbox@\relax\bybox@\fi
 \fi
 \ifvoid\translbox@\else\ppunbox@,\translbox@\fi
 \ifvoid\paperbox@\else\ppunbox@,\paperbox@\ifpaperinquotes@
  \if\nextv@1\else\openquotes@true\fi\fi
 \fi
 \ifvoid\paperinfobox@\else\ppunbox@,\paperinfobox@\fi
 \test@false
 \ifvoid\jourbox@\else\test@true\ppunbox@,\jourbox@\fi
 \ifprevjour@\test@true\fi
 \iftest@
  \ifvoid\volbox@\else\ppunbox@\relax\volbox@\fi
  \ifvoid\issuebox@
   \else\prepunct@\relax\issuebox@ no.~\getbox@\issuebox@\fi
  \ifvoid\yrbox@\else\prepunct@\relax\yrbox@(\getbox@\yrbox@)\fi
  \ifvoid\ppbox@\else\ppunbox@,\ppbox@\fi
  \ifvoid\pgbox@\else\prepunct@,\pgbox@ p.~\getbox@\pgbox@\fi
 \fi
 \test@false
 \ifvoid\bookbox@\else\test@true\ppunbox@,\bookbox@\ifbookinquotes@
  \if\nextv@1\else\openquotes@true\fi\fi\fi
 \ifvoid\inbookbox@\else\test@true\ppunbox@,\inbookbox@\ifbookinquotes@
  \if\nextv@1\else\openquotes@true\fi\fi\fi
 \ifprevbook@\test@true\fi
 \iftest@
  \ifvoid\edbox@\else\prepunct@\relax\edbox@(\getbox@\edbox@, ed.)\fi
  \ifvoid\edsbox@\else\prepunct@\relax\edsbox@(\getbox@\edsbox@, eds.)\fi
  \ifvoid\bookinfobox@\else\ppunbox@,\bookinfobox@\fi
  \ifvoid\publbox@\else\ppunbox@,\publbox@\fi
  \ifvoid\publaddrbox@\else\ppunbox@,\publaddrbox@\fi
  \ifvoid\yrbox@\else\ppunbox@,\yrbox@\fi
  \ifvoid\ppbox@\else\prepunct@,\ppbox@ pp.~\getbox@\ppbox@\fi
  \ifvoid\pgbox@\else\prepunct@,\pgbox@ p.~\getbox@\pgbox@\fi
 \fi
 \ifvoid\finalinfobox@
  \ifendbib@
   \ifnopunct@\else.\closequotes@\fi
  \else
  \ifvoid\langbox@\else\space(\getbox@\langbox@)\fi
   \/\semicolon@\closequotes@
  \fi
 \else
  \ifendbib@
   \ppunbox@{.\spacefactor3000\relax}\finalinfobox@
    \ifnopunct@\else.\fi
  \else
   \ppunbox@,\finalinfobox@\/\semicolon@\fi
 \fi
 \ifvoid\langbox@\else\space(\getbox@\langbox@)\fi
}
\def\endbib{\unskip\egroup\endbib@true\endbib@\par\endgroup}
\def\morebib{\unskip\egroup
 \endbib@false\endbib@
 \global\let\bibinfo@\empty\beginbib@false
 \bib@}
\def\anotherbib{\unskip\egroup
 \endbib@false\endbib@
 \global\let\bibinfo@\empty\beginbib@false
 \prevjour@false\prevbook@false\bib@}
\def\transl{\unskip
 \xdef\translinfo@{\the\translbox@,\ifnopunct@1\else0\fi
 \ifnospace@1\else0\fi\ifnoprepunct@1\else0\fi\ifnoprespace@1\else0\fi0}%
 \egroup\endbib@false\endbib@
 \global\let\bibinfo@\translinfo@\beginbib@false
 \bib@
 \egroup
 \def\aftergroup@{\ifdim\wd\translbox@=\z@\setbox\translbox@\box\voidb@x\fi}%
 \setbox\translbox@\vbox\bgroup\aftergroup\aftergroup@
 \hsize\maxdimen\leftskip\z@\rightskip\z@\hbadness\@M\hfuzz\maxdimen
 \noindent}
\newwrite\auxwrite@
\newread\bbl@
\def\UseBibTeX{\immediate\openout\auxwrite@=\jobname.aux
 \let\cite\BTcite@
 \def\nocite##1{\immediate\write\auxwrite@{\string\citation{##1}}}%
 \def\bibliographystyle##1{\immediate\write\auxwrite@{\string
  \bibstyle{##1}}}%
 \def\bibliography@W{Bibliography}%
 \def\bibliography##1{\immediate\write\auxwrite@{\string\bibdata{##1}}%
  \immediate\openin\bbl@=\jobname.bbl
  \ifeof\bbl@
   \W@{No .bbl file}%
  \else
   \immediate\closein\bbl@
   \begingroup\input bibtex \input\jobname.bbl \endgroup
  \fi}%
 }
\def\BTcite@{%
 \DNii@(##1)##2{{\rm[}\BTcite@@##2,\BTcite@@{\rm, }{##1\/}{\rm]}%
  \immediate\write\auxwrite@{\string\citation{##2}}}%
 \def\nextiii@##1{{\rm[}\BTcite@@##1,\BTcite@@\/{\rm]}%
  \immediate\write\auxwrite@{\string\citation{##1}}}%
 \DN@{\ifx\next(\expandafter\nextii@\else\expandafter\nextiii@\fi}%
 \FN@\next@}%
\def\BTcite@@#1,{\BTcite@@@{#1}\FN@\BTcite@@@@}
\def\BTcite@@@@{\ifx\next\BTcite@@
 \expandafter\eat@\else{\rm, }\expandafter\BTcite@@\fi}
\catcode`\~=11
\def\BTcite@@@#1{\nolabel@\cite{#1}\relax
 \DNii@##1~##2\nextii@{##1}%
 \csL@{#1}\expandafter\nextii@\Next@\nextii@\fi}
\catcode`\~=\active

\def\beginthebibliography@#1{\rm\setboxz@h{#1\ }\bibindent@\wdz@
 \bigbreak\centerline{\smc\bibliography@W}\nobreak\medskip
 \sfcode`\.=\@m\everypar{}\parindent\z@}

\newif\iffigproofing@
\def\Figureproofing{\figproofing@true}
\def\noFigureproofing{\figproofing@false}
\newif\ifHby@
\def\Hbyw#1{\global\Hby@true\hbyw\vsize{#1}}
\def\hbyw#1#2{%
 \hbox{%
  \ifHby@
  \else
   \iffigproofing@
    \setbox\z@\vbox{\hrule\width5\p@}\ht\z@\z@
    \vbox to#1{\hrule\height5\p@\width.4\p@\vfil\hrule\height5\p@\width.4\p@}%
    \kern-.4\p@\rlap{\copy\z@}\raise#1\hbox{\rlap{\copy\z@}}%
   \fi
  \fi
  \vbox to#1{\hbox to#2{}\vfil}%
  \ifHby@
  \else
   \iffigproofing@
    \vbox to#1{\hrule\height5\p@\width.4\p@\vfil\hrule\height5\p@\width.4\p@}%
    \kern-.4\p@\llap{\copy\z@}\raise#1\hbox{\llap{\boxz@}}%
   \fi
  \fi}}
\newcount\island@C
\let\island@P\empty
\let\island@Q\empty
\def\island@S#1{#1\null.}
\let\island@N\arabic
\def\island@F{\rm}
\def\island@@@P{\csname\exxx@\islandtype@ @P\endcsname}
\def\island@@@Q{\csname\exxx@\islandtype@ @Q\endcsname}
\def\island@@@S{\csname\exxx@\islandtype@ @S\endcsname}
\def\island@@@N{\csname\exxx@\islandtype@ @N\endcsname}
\def\island@@@F{\csname\exxx@\islandtype@ @F\endcsname}
\def\island@@@C{\csname island@C\islandclass@\endcsname}
\newif\ifplace@
\newif\ifisland@
\def\island{%
 \ifplace@
  \DN@{\let\islandclass@\empty\def\islandtype@{\island}\FN@\island@}%
 \else
  \long\DN@##1\endisland{\Err@{\noexpand\island must be used after some
   type of \string\...place}}%
 \fi
 \next@}
\def\island@{\ifx\next\c\let\next@\island@c\else
 \DN@{\FN@\island@@}\fi\next@}
\def\island@@{\ifcat\bgroup\noexpand\next\let\next@\island@@@\else
 \DN@{\Err@{\noexpand\island must be followed by a {prefix} for
 \string\caption's}}\fi\next@}
\newbox\islandbox@
\newcount\captioncount@
\def\island@@@#1{\def\captionprefix@{#1}\captioncount@\z@
 \global\setbox\islandbox@\vbox\bgroup}
\def\island@c\c#1{%
 \ifplace@
 \DN@{\def\islandclass@{#1}%
  \expandafter\ifx\csname island@C#1\endcsname\relax
  \expandafter\newcount@\csname island@C#1\endcsname
   \global\csname island@C#1\endcsname\z@\fi
  \FNSS@\island@c@}%
 \else
 \DN@{\edef\next@{\long\def\noexpand\next@########1\expandafter\noexpand
  \csname end\exxx@\islandtype@\endcsname{\noexpand\Err@{\noexpand\noexpand
  \expandafter\noexpand
  \islandtype@ must be used after some type of \noexpand\string
   \noexpand\...place}}}\next@\next@}%
 \fi
 \next@}
\def\island@c@{%
 \ifcat\bgroup\noexpand\next
  \let\next@\island@c@@
 \else
  \DN@{\Err@{\noexpand\island\string\c{\expandafter\string\islandclass@} must
   be followed by a {prefix} for \string\caption's}}%
 \fi\next@}
\def\island@c@@#1{\def\captionprefix@{#1}%
 \captioncount@\z@\global\setbox\islandbox@\vbox\bgroup}
\rightadd@\caption\to\nofrillslist@
\newbox\captionbox@
\newbox\Captionbox@
\def\caption{%
 \ifnum\captioncount@=\z@
  \ifnopunct@
   \DN@{\egroup\nopunct@true}%
  \else
   \let\next@\egroup
  \fi
 \else
  \let\next@\relax
 \fi
 \next@
 \advance\captioncount@\@ne
 \FN@\caption@}
\def\caption@{\ifx\next"\expandafter\caption@q\else\expandafter\caption@@\fi}
\def\caption@q"#1"{\quoted@true
 {\noexpands@
 \let\pre\island@@@P\let\post\island@@@Q
 \let\style\island@@@S\let\numstyle\island@@@N
 \Qlabel@{#1}\let\style\relax\xdef\Qlabel@@@@{#1}}%
 \finishcaption@}
\def\caption@@{\quoted@false
 \global\advance\island@@@C\@ne
 {\noexpands@
 \xdef\Thelabel@@@{\number\island@@@C}%
 \xdefThelabel@\island@@@N
 \xdef\Thelabel@@@@{\island@@@P\Thelabel@\island@@@Q}%
 \xdefThelabel@@\island@@@S
 \xdef\Thepref@{\Thelabel@@@@}}%
 \finishcaption@}
\long\def\captionformat@#1#2#3{\rm\strut#1 {\island@@@F#2} #3%
 \punct@.\strut}
\long\def\widerthanisland@#1#2#3{\test@true\setbox\z@\vbox{\hsize\maxdimen
 \noindent@@\captionformat@{#1}{#2}{#3}\par\setboxzl@}%
 \ifdim\wdz@=\z@
  \global\setbox\captionbox@\hbox{\noset@\unlabel@
   \captionformat@{#1}{#2}{#3}}%
  \ifdim\wd\captionbox@>\wd\islandbox@\else\test@false\fi
 \fi}
\long\def\captionformat@@#1#2#3{\widerthanisland@{#1}{#2}{#3}%
 \iftest@
  \global\setbox\captionbox@\vbox{\hsize\wd\islandbox@
   \vskip-\parskip\noindent@@\noset@\unlabel@
   \captionformat@{#1}{#2}{#3}\par}%
 \else
  \global\setbox\captionbox@
   \hbox to\wd\islandbox@{\hfil\box\captionbox@\hfil}%
 \fi}
\long\def\finishcaption@#1{\def\entry@{#1}%
 {\locallabel@
 \captionformat@@
  {\expandafter\ignorespaces\captionprefix@\unskip}%
  {\ifx\thelabel@@\empty\unskip\else\thelabel@@\fi}%
  {\ignorespaces#1\unskip}%
 \ifnum\captioncount@=\@ne
  \global\setbox\islandbox@\vbox{\ticwrite@\vbox{\box\islandbox@}}%
  \global\setbox\Captionbox@\vbox{\box\captionbox@}%
 \else
  \global\setbox\islandbox@\vbox{\unvbox\islandbox@\setboxzl@
   \ticwrite@\boxz@}%
  \global\setbox\Captionbox@\vbox{\unvbox\Captionbox@
   \smallskip\box\captionbox@}%
 \fi}%
 \nopunct@false\nospace@false\ignorespaces}
\def\Sixtic@{\ifx\macdef@\empty\else
 \DN@##1##2\next@{\def\macdef@{##1##2}}%
 \expandafter\next@\macdef@\next@
 \edef\next@
  {\noexpand\six@\tic@\macdef@
  \space\space\space\space\space\space\space\space\space\space\space\space
  \noexpand\six@}%
 \next@\let\macdef@\relax\fi}
\def\ticwrite@{%
 \iftoc@
  {\noexpands@\let\style\relax
  \DN@{\island}%
  \edef\next@{\write\tic@{%
   \ifnopunct@\noexpand\noexpand\noexpand\nopunct\fi
   \ifx\islandtype@\next@\noexpand\noexpand\noexpand\island
    \noexpand\string\noexpand\c{\islandclass@}{\captionprefix@}%
     {\QorThelabel@@@@}\else\noexpand\noexpand\expandafter\noexpand
     \islandtype@{\QorThelabel@@@@}}\fi}%
  \next@}%
  \expandafter\unmacro@\meaning\entry@\unmacro@
  \Sixtic@
  \write\tic@{\noexpand\Page{\number\pageno}{\page@N}{\page@P}{\page@Q}^^J}%
 \fi}
\def\Htrim@#1{%
 \ifHby@
  \dimen@\vsize
  \ifnum\captioncount@=\z@
  \else
   \advance\dimen@-\ht\Captionbox@
   \advance\dimen@-#1%
  \fi
  \global\Hby@false
  \dimen@ii\wd\islandbox@
  \global\setbox\islandbox@\vbox
   {\unvbox\islandbox@\setboxzl@
   \vbox to\z@{\vss\boxz@}\nointerlineskip\hbyw\dimen@\dimen@ii}%
  \global\Hby@true
 \fi}
\newif\ifdata@
\def\iclasstest@#1{\DN@{#1}\ifx\next@\islandclass@
 \test@true\else\test@false\fi}
\skipdef\skipi@=1
\def\endisland{\ifnum\captioncount@=\z@\expandafter\egroup\fi
 \ifdata@
 \else
  \iclasstest@{T}%
  \iftest@
   {\rm\global\skipi@-\dp\strutbox}\global\advance\skipi@\bigskipamount
   \Htrim@\skipi@
   \global\setbox\islandbox@\vbox
    {\ifnum\captioncount@=\z@\else
     \box\Captionbox@
     \nointerlineskip
     \vskip\skipi@\fi
     \box\islandbox@}%
  \else
   {\rm\global\skipi@\dp\strutbox}\global\advance\skipi@\medskipamount
   \Htrim@\skipi@
   \global\setbox\islandbox@\vbox
    {\box\islandbox@
     \ifnum\captioncount@=\z@\else
     \nointerlineskip
     \vskip\skipi@
     \box\Captionbox@
     \fi}%
  \fi
  \ifHby@
  \else
   \dimen@\ht\islandbox@\advance\dimen@\dp\islandbox@
   \ifdim\dimen@>\vsize
    \DN@{\island}%
    \Err@{%
     \ifx\islandtype@\next@\noexpand\island\else
      \expandafter\noexpand\islandtype@\fi
     \ifnum\captioncount@=\z@\else
       with \noexpand\caption\fi
      is larger than page}%
     \ht\islandbox@=\vsize
   \fi
  \fi
 \fi
 \global\Hby@false\island@true}
\def\newisland#1\c#2#3{\define#1{}%
 \iftoc@\immediate\write\tic@{\noexpand\newisland\noexpand#1%
  \string\c{#2}{#3}^^J}\fi
 \expandafter\def\csname\exstring@#1@S\endcsname{\island@S}%
 \expandafter\def\csname\exstring@#1@N\endcsname{\island@N}%
 \expandafter\def\csname\exstring@#1@P\endcsname{\island@P}%
 \expandafter\def\csname\exstring@#1@Q\endcsname{\island@Q}%
 \expandafter\def\csname\exstring@#1@F\endcsname{\island@F}%
 \expandafter\def\csname end\exstring@#1\endcsname{\endisland}%
 \expandafter
 \ifx\csname island@C#2\endcsname\relax
  \expandafter\newcount@\csname island@C#2\endcsname
  \global\csname island@C#2\endcsname\z@
 \fi
 \edef\next@{\noexpand\expandafter\noexpand\let\noexpand
  \csname\exstring@#1@C\noexpand\endcsname
  \csname island@C#2\endcsname}%
 \next@
 \def#1{\def\islandtype@{#1}\island@c\c{#2}{#3}}}
\newisland\Figure\c{F}{Figure}
\newisland\Table\c{T}{Table}
\newbox\islandboxi
\newbox\islandboxii
\newbox\islandboxiii
\newbox\captionboxi
\newbox\captionboxii
\newbox\captionboxiii
\long\def\islandpairdata#1#2{{\data@true
 \place@true
 #1%
 \global\setbox\islandboxi\box\islandbox@
 \global\setbox\captionboxi\box\Captionbox@
 #2%
 \global\setbox\islandboxii\box\islandbox@
 \global\setbox\captionboxii\box\Captionbox@
 }}
\long\def\islandpairbox#1#2{\islandpairdata{#1}{#2}%
 \dimen@\ht\captionboxi
 \ifdim\ht\captionboxii>\dimen@\dimen@\ht\captionboxii\fi
 \ifdim\dimen@>\z@
  \ifdim\ht\captionboxi<\dimen@
   \global\setbox\captionboxi\vbox to\dimen@{\unvbox\captionboxi\vfil}\fi
  \ifdim\ht\captionboxii<\dimen@
   \global\setbox\captionboxii\vbox to\dimen@{\unvbox\captionboxii\vfil}\fi
 \fi
 \global\setbox\islandbox@\vbox
 {\hbox to\hsize{\hfil\box\islandboxi\hfil\box\islandboxii\hfil}%
 \ifdim\dimen@>\z@\nointerlineskip
 {\rm\global\skipi@\dp\strutbox}\global\advance\skipi@\medskipamount
  \vskip\skipi@
  \hbox to\hsize{\hfil\box\captionboxi\hfil\box\captionboxii\hfil}\fi}}
\long\def\islandpairboxa#1#2{\islandpairdata{#1}{#2}%
 \dimen@\ht\captionboxi
 \ifdim\ht\captionboxii>\dimen@\dimen@\ht\captionboxii\fi
 \ifdim\dimen@>\z@
  \ifdim\ht\captionboxi<\dimen@
   \global\setbox\captionboxi\vbox to\dimen@{\vfil\unvbox\captionboxi}\fi
  \ifdim\ht\captionboxii<\dimen@
   \global\setbox\captionboxii\vbox to\dimen@{\vfil\unvbox\captionboxii}\fi
 \fi
 \dimen@ii\ht\islandboxi
 \ifdim\ht\islandboxii>\dimen@ii \dimen@ii\ht\islandboxii\fi
 \ifdim\dimen@ii>\z@
  \ifdim\ht\islandboxi<\dimen@ii
   \global\setbox\islandboxi\vbox to\dimen@ii{\box\islandboxi\vfil}\fi
  \ifdim\ht\islandboxii<\dimen@ii
   \global\setbox\islandboxii\vbox to\dimen@ii{\box\islandboxii\vfil}\fi
 \fi
 \global\setbox\islandbox@\vbox{\ifdim\dimen@>\z@
  \hbox to\hsize{\hfil\box\captionboxi\hfil\box\captionboxii\hfil}%
  \nointerlineskip{\rm\global\skipi@-\dp\strutbox}%
  \global\advance\skipi@\bigskipamount\vskip\skipi@\fi
  \hbox to\hsize{\hfil\box\islandboxi\hfil\box\islandboxii\hfil}}}
\long\def\islandtripledata#1#2#3{{\data@true\place@true
 #1%
 \global\setbox\islandboxi\box\islandbox@
 \global\setbox\captionboxi\box\Captionbox@
 #2%
 \global\setbox\islandboxii\box\islandbox@
 \global\setbox\captionboxii\box\Captionbox@
 #3%
 \global\setbox\islandboxiii\box\islandbox@
 \global\setbox\captionboxiii\box\Captionbox@
 }}
\long\def\islandtriplebox#1#2#3{\islandtripledata{#1}{#2}{#3}%
 \dimen@\ht\captionboxi
 \ifdim\ht\captionboxii>\dimen@ \dimen@\ht\captionboxii\fi
 \ifdim\ht\captionboxiii>\dimen@ \dimen@\ht\captionboxiii\fi
 \ifdim\dimen@>\z@
  \ifdim\ht\captionboxi<\dimen@
   \global\setbox\captionboxi\vbox to\dimen@{\unvbox\captionboxi\vfil}\fi
  \ifdim\ht\captionboxii<\dimen@
   \global\setbox\captionboxii\vbox to\dimen@{\unvbox\captionboxii\vfil}\fi
  \ifdim\ht\captionboxiii<\dimen@
   \global\setbox\captionboxiii\vbox to\dimen@{\unvbox\captionboxiii\vfil}\fi
 \fi
 \global\setbox\islandbox@\vbox
  {\hbox to\hsize{\hfil\box\islandboxi\hfil\box\islandboxii\hfil
   \box\islandboxiii\hfil}%
 \ifdim\dimen@>\z@\nointerlineskip
  {\rm\global\skipi@\dp\strutbox}\global\advance\skipi@\medskipamount
  \vskip\skipi@
  \hbox to\hsize{\hfil\box\captionboxi\hfil\box\captionboxii\hfil
   \box\captionboxiii\hfil}\fi}}
\def\islandtripleboxa#1#2#3{\islandtripledata{#1}{#2}{#3}%
 \dimen@\ht\captionboxi
 \ifdim\ht\captionboxii>\dimen@ \dimen@\ht\captionboxii\fi
 \ifdim\ht\captionboxiii>\dimen@ \dimen@\ht\captionboxiii\fi
 \ifdim\dimen@>\z@
  \ifdim\ht\captionboxi<\dimen@
   \global\setbox\captionboxi\vbox to\dimen@{\vfil\unvbox\captionboxi}\fi
  \ifdim\ht\captionboxii<\dimen@
   \global\setbox\captionboxii\vbox to\dimen@{\vfil\unvbox\captionboxii}\fi
  \ifdim\ht\captionboxiii<\dimen@
   \global\setbox\captionboxiii\vbox to\dimen@{\vfil\unvbox\captionboxiii}\fi
 \fi
 \dimen@ii\ht\islandboxi
 \ifdim\ht\islandboxii>\dimen@ii \dimen@ii\ht\islandboxii\fi
 \ifdim\ht\islandboxiii>\dimen@ii \dimen@ii\ht\islandboxiii\fi
 \ifdim\dimen@ii>\z@
  \ifdim\ht\islandboxi<\dimen@ii
   \global\setbox\islandboxi\vbox to\dimen@ii{\box\islandboxi\vfil}\fi
  \ifdim\ht\islandboxii<\dimen@ii
   \global\setbox\islandboxii\vbox to\dimen@ii{\box\islandboxii\vfil}\fi
  \ifdim\ht\islandboxiii<\dimen@ii
   \global\setbox\islandboxiii\vbox to\dimen@ii{\box\islandboxiii\vfil}\fi
 \fi
 \global\setbox\islandbox@\vbox
  {\ifdim\dimen@>\z@
  \hbox to\hsize{\hfil\box\captionboxi\hfil\box\captionboxii\hfil
   \box\captionboxiii\hfil}%
  \nointerlineskip{\rm\global\skipi@-\dp\strutbox}%
  \global\advance\skipi@\bigskipamount\vskip\skipi@\fi
  \hbox to\hsize{\hfil\box\islandboxi\hfil\box\islandboxii\hfil
   \box\islandboxiii\hfil}}}
\def\Figurepair#1\and#2\endFigurepair{\island@true
 \islandpairbox{\Figure#1\endFigure}{\Figure#2\endFigure}}
\def\Figuretriple#1\and#2\and#3\endFiguretriple{\island@true
 \islandtriplebox{\Figure#1\endFigure}{\Figure#2\endFigure}%
  {\Figure#3\endFigure}}
\def\Tablepair#1\and#2\endTablepair{\island@true
 \islandpairboxa{\Table#1\endTable}{\Table#2\endTable}}
\def\Tabletriple#1\and#2\and#3\endTabletriple{\island@true
 \islandtripleboxa{\Table#1\endTable}{\Table#2\endTable}%
 {\Table#3\endTable}}
\def\place#1{\place@true\island@false
 #1%
 \ifisland@
  \box\islandbox@
 \else
  \Err@{Whoa ... there's no \string\Figure, \string\Table,
   etc., here}%
 \fi
 \place@false}
\newskip\belowtopfigskip
\belowtopfigskip 15\p@ plus 5\p@ minus5\p@
\newskip\abovebotfigskip
\abovebotfigskip 18\p@ plus 6\p@ minus6\p@
\newdimen\minpagesize
\minpagesize 5pc
\dimen@\belowtopfigskip
\advance\dimen@-\abovebotfigskip
\skip\topins\dimen@
\dimen\topins\z@
\newcount\topinscount@
\newbox\topinsdims@
\def\storedim@{\global\setbox\topinsdims@
 \vbox{\hbox to\dimen@{}\unvbox\topinsdims@}}
\def\advancedimtopins@{%
 \ifnum\pageno=\@ne
 \else
   \advance\dimen@\dimen\topins
   \global\dimen\topins\dimen@
 \fi}
\newcount\flipcount@
\def\fliptopins@{%
 \global\flipcount@\z@
 \ifvoid\topins\else
 \setbox\z@\vbox
  {\vskip\p@
   \unvbox\topins
   \global\setbox\topins\vbox{}%
   \loop
    \test@false
    \ifdim\lastskip=\z@\unskip
     \ifdim\lastskip=\z@
      \test@true\fi\fi
    \iftest@
    \global\advance\flipcount@\@ne
    \setboxzl@
    \global\setbox\topins\vbox{\unvbox\topins\boxz@}%
    \unpenalty
   \repeat}\fi}
\newif\ifPar@
\newcount\Parcount@
\newbox\Parbox@
\expandafter\newbox\csname Parfigbox1\endcsname
\expandafter\newbox\csname Parfigbox2\endcsname
\expandafter\newbox\csname Parfigbox3\endcsname
\expandafter\newbox\csname Parfigbox4\endcsname
\expandafter\newbox\csname Parfigbox5\endcsname
\expandafter\newdimen\csname Parprev1\endcsname
\expandafter\newdimen\csname Parprev2\endcsname
\expandafter\newdimen\csname Parprev3\endcsname
\expandafter\newdimen\csname Parprev4\endcsname
\expandafter\newdimen\csname Parprev5\endcsname
\expandafter\newdimen\csname Parprev6\endcsname
\def\Par{\par\global\csname Parprev1\endcsname\prevdepth
 \global\Parcount@\@ne
 \global\Par@true\global\let\Parlist@\empty
 \global\setbox\Parbox@\vbox\bgroup\break}
\def\place@#1#2{%
 \ifisland@
  \ifhmode
   \ifPar@
    \ifnum\Parcount@>5
     \Err@{Only 5 \string\place's allowed per
      \string\Par...\noexpand\endPar paragraph}%
    \else
     \expandafter\expandafter\expandafter
      \global\expandafter\setbox
       \csname Parfigbox\number\Parcount@\endcsname\box\islandbox@
     \global\advance\Parcount@\@ne
     \xdef\Parlist@{\Parlist@#1}%
    \fi
   \else
    \vadjust{#2}%
   \fi
  \else
   #2%
  \fi
 \else
  \Err@{Whoa ... there's no \string\Figure,
   \string\Table, etc., here}%
 \fi
 \place@false}
\long\def\Aplace#1{\prevanish@
 \place@true\island@false
 #1%
 \place@ a\Aplace@
 \postvanish@}
\long\def\AAplace#1{\prevanish@\place@true\island@false
 #1%
 \place@ A\AAplace@
 \postvanish@}
\newif\ifAA@
\def\AAplace@{\AA@true\Aplace@\AA@false}
\let\AAlist@\empty
\def\Aplace@{\allowbreak
 \dimen@=\ht\islandbox@
 \advance\dimen@\abovebotfigskip
 \ht\islandbox@\dimen@
 \advance\dimen@\dp\islandbox@
 \storedim@
 \ifAA@
  \xdef\AAlist@{\AAlist@1}%
  \advancedimtopins@
 \else
  \xdef\AAlist@{\AAlist@0}%
  \ifnum\topinscount@>\@ne\else\advancedimtopins@\fi
 \fi
 \insert\topins{\penalty\z@\splittopskip\z@\floatingpenalty\z@
  \box\islandbox@}%
 \global\advance\topinscount@\@ne}
\long\def\Bplace#1{\prevanish@\place@true\island@false
 #1%
 \place@ b\Bplace@
 \postvanish@}
\def\Bplace@{\allowbreak
 \ifnum\topinscount@=\z@
  \setbox\z@\vbox{\vbox to-\belowtopfigskip{}}%
  \dimen@-\skip\topins
  \ht\z@\dimen@
  \storedim@
  \advancedimtopins@
  \insert\topins{\boxz@}%
  \global\advance\topinscount@\@ne
  \xdef\AAlist@{\AAlist@0}%
 \fi
 \dimen@\ht\islandbox@
 \advance\dimen@\abovebotfigskip
 \ht\islandbox@\dimen@
 \advance\dimen@\dp\islandbox@
 \storedim@
 \xdef\AAlist@{\AAlist@0}%
 \ifnum\topinscount@>\@ne\else\advancedimtopins@\fi
 \insert\topins{\penalty\z@\splittopskip\z@
  \floatingpenalty\z@
  \box\islandbox@}%
 \global\advance\topinscount@\@ne}
\def\breakisland@{\global\setbox\@ne\lastbox\global\skipi@\lastskip\unskip
 \global\setbox\thr@@\lastbox}%
\def\printisland@{\centerline{\box\thr@@}\nobreak\nointerlineskip
 \vskip\skipi@
 \ifdim\ht\@ne<\z@\box\@ne\else\centerline{\box\@ne}\fi}
\def\bottomfigs@{%
 \count@\@ne
 \loop
  \ifnum\count@<\flipcount@
  \nointerlineskip
  \vskip\abovebotfigskip
  \global\setbox\topins\vbox{\unvbox\topins\setboxzl@
   \unvbox\z@
   \breakisland@}%
  \printisland@
  \advance\count@\@ne
  \repeat}
\def\resetdimtopins@{%
 \global\advance\topinscount@-\flipcount@
 \global\setbox\topinsdims@\vbox
  {\unvbox\topinsdims@
   \count@\z@
   \DN@##1##2\next@{\gdef\AAlist@{##2}}%
   \loop
    \ifnum\count@<\flipcount@\setboxzl@
    \expandafter\next@\AAlist@\next@
    \advance\count@\@ne
    \repeat
   \dimen@\z@
   \count@\z@
   \setbox\tw@\vbox{}%
   \edef\nextiii@{\AAlist@}%
   \DN@##1##2\next@{\DNii@{##1}\def\nextiii@{##2}}%
   \loop
    \test@false
    \ifnum\count@<\topinscount@
    \expandafter\next@\nextiii@\next@
     \ifnum\count@<\tw@
      \test@true
     \else
      \if\nextii@ 1\test@true\fi
     \fi
    \fi
    \iftest@
     \setboxzl@
     \advance\dimen@\wdz@
     \setbox\tw@\vbox{\boxz@\unvbox\tw@}%
     \advance\count@\@ne
    \repeat
    \unvbox\tw@
    \global\dimen\topins\dimen@}}
\def\Place@#1#2{%
 \ifisland@
  \ifhmode
   \ifPar@
    \ifnum\Parcount@>5
     \Err@{Only 5 \string\place's allowed per
       \string\Par...\noexpand\endPar paragraph}%
    \else
     \expandafter\expandafter\expandafter\global\expandafter\setbox
      \csname Parfigbox\number\Parcount@\endcsname\box\islandbox@
     \global\advance\Parcount@\@ne
     \xdef\Parlist@{\Parlist@#1}%
     \vadjust{\break}%
    \fi
   \else
    \Err@{\noexpand#2allowed only in a \string\Par...\noexpand\endPar
     paragraph}%
   \fi
  \else
   #2%
  \fi
 \else
  \Err@{Who ... there's no \string\Figure, \string\Table,
   etc., here}%
 \fi
 \place@false}
\newif\ifC@
\newdimen\Cdim@
\long\def\Cplace#1{\prevanish@\place@true\island@false
 #1%
 \Place@ c\Cplace@
 \postvanish@}
\def\Cplace@{\allowbreak
 \ifnum\topinscount@>\z@\else
  \global\C@true\global\Cdim@\pagetotal\fi
 \Aplace@}
\long\def\Mplace#1{\prevanish@\place@true\island@false
 #1%
 \Place@ m\Mplace@
 \postvanish@}
\long\def\MXplace#1{\prevanish@\place@true\island@false
 #1%
 \Place@ M\MXplace@
 \postvanish@}
\newif\ifMX@
\def\MXplace@{\MX@true\Mplace@\MX@false}
\def\Mplace@{\allowbreak
 \dimen@\ht\islandbox@\advance\dimen@\dp\islandbox@
 \ifdim\pagetotal=\z@\else
  \ifdim\lastskip<\abovebotfigskip\advance\dimen@\abovebotfigskip
  \advance\dimen@-\lastskip\fi
 \fi
 \advance\dimen@\pagetotal
 \ifdim\dimen@>\pagegoal
  \Aplace@
 \else
  \nointerlineskip
  \ifdim\lastskip<\abovebotfigskip\removelastskip\vskip\abovebotfigskip\fi
  \setbox\z@\vbox{\unvbox\islandbox@
   \breakisland@}%
  \printisland@
  \ifnum\topinscount@=\z@
   \setbox\z@\vbox{\vbox to-\belowtopfigskip{}}%
   \dimen@-\skip\topins
   \ht\z@\dimen@
   \storedim@
   \advancedimtopins@
   \insert\topins{\boxz@}%
   \global\advance\topinscount@\@ne
   \xdef\AAlist@{\AAlist@0}%
  \fi
  \ifMX@
   \ifnum\topinscount@=\@ne
    \setbox\z@\vbox{\vbox to-\abovebotfigskip{}}%
    \ht\z@\z@
    \dimen@\z@
    \storedim@
    \advancedimtopins@
    \insert\topins{\boxz@}%
    \global\advance\topinscount@\@ne
    \xdef\AAlist@{\AAlist@0}%
   \fi
  \fi
  \nointerlineskip
  \vskip\belowtopfigskip
 \fi}
\expandafter\newbox\csname Parbox1\endcsname
\expandafter\newbox\csname Parbox2\endcsname
\expandafter\newbox\csname Parbox3\endcsname
\expandafter\newbox\csname Parbox4\endcsname
\expandafter\newbox\csname Parbox5\endcsname
\def\endPar{\egroup
 \count@\@ne
 {\vbadness\@M\vfuzz\maxdimen\splitmaxdepth\maxdimen\splittopskip\ht\strutbox
 \setbox\z@\vsplit\Parbox@ to\ht\Parbox@
 \loop
  \ifnum\count@<\Parcount@
  \expandafter\expandafter\expandafter\global\expandafter\setbox
   \csname Parbox\number\count@\endcsname\vsplit\Parbox@ to\ht\Parbox@
  \count@@\count@\advance\count@@\@ne
  \global\csname Parprev\number\count@@\endcsname
   \dp\csname Parbox\number\count@\endcsname
  \advance\count@\@ne
  \repeat}%
 \vskip\parskip
 \count@\@ne
 \def\nextv@##1##2\nextv@{\DN@{##1}\gdef\Parlist@{##2}}%
 \loop
  \ifnum\count@<\Parcount@
   \dimen@\csname Parprev\number\count@\endcsname
   \advance\dimen@\ht\strutbox
   \ifdim\dimen@<\baselineskip
    \advance\dimen@-\baselineskip\vskip-\dimen@
   \else
    \vskip\lineskip
   \fi
   \unvbox\csname Parbox\number\count@\endcsname
   \global\setbox\islandbox@\box\csname Parfigbox\number\count@\endcsname
   \expandafter\nextv@\Parlist@\nextv@
   \if a\next@\Aplace@\else
   \if A\next@\AAplace@\else
   \if b\next@\Bplace@\else
   \if c\next@\Cplace@\else
   \if m\next@\Mplace@\else
   \if M\next@\MXplace@\fi\fi\fi\fi\fi\fi
  \advance\count@\@ne
  \repeat
 \global\Par@false
 \ifvoid\Parbox@
  \prevdepth\csname Parprev\number\count@\endcsname
 \else
  \dimen@\csname Parprev\number\count@\endcsname\advance\dimen@\ht\strutbox
  \ifdim\dimen@<\baselineskip
    \advance\dimen@-\baselineskip\vskip-\dimen@
  \else
    \vskip\lineskip
  \fi
  \dimen@\dp\Parbox@
  \unvbox\Parbox@
  \prevdepth\dimen@
 \fi}
\def\folio{{\page@F\page@S{\page@P\page@N{\number\page@C}\page@Q}}}
\def\advancepageno{\global\advance\pageno\@ne}
\newif\ifspecialsplit@
\newbox\outbox@
\let\shipout@\shipout
\def\plainoutput{\specialsplit@false\ifvoid\topins\else\ifdim\ht\topins=\z@
 \specialsplit@true\advance\minpagesize-\skip\topins\fi\fi
 \fliptopins@
 \setbox\outbox@\vbox{\makeheadline\pagebody\makefootline}%
 {\noexpands@\let\style\relax
 \shipout@\box\outbox@}%
 \advancepageno
 \resetdimtopins@
 \ifvoid\@cclv\else\unvbox\@cclv\penalty\outputpenalty\fi
 \ifnum\outputpenalty>-\@MM\else\dosupereject\fi}
\def\pagebody{\vbox to\vsize{\boxmaxdepth\maxdepth
 \ifvoid\margin@\else
 \rlap{\kern\hsize\vbox to\z@{\kern4\p@\box\margin@\vss}}\fi
 \pagecontents}}
\newif\ifonlytop@
\def\pagecontents{%
 \onlytop@false
 \ifdim\ht\@cclv<\minpagesize\ifnum\flipcount@<\tw@\ifvoid\footins
  \onlytop@true\fi\fi\fi
 \test@false
 \ifC@
  \ifnum\flipcount@=\@ne
   \global\multiply\Cdim@\tw@
   \ifdim\Cdim@>\ht\@cclv
    \test@true
   \fi
  \fi
 \fi
 \global\C@false
 \iftest@
  \dimen@\ht\@cclv
  \advance\dimen@\skip\topins
  {\vfuzz\maxdimen\vbadness\@M
  \splitmaxdepth\maxdepth\splittopskip\topskip
  \setbox\z@\vsplit\@cclv to\dimen@
  \unvbox\z@}%
  \global\setbox\topins\vbox{\unvbox\topins
   \global\setbox\@ne\lastbox}%
  \setbox\z@\vbox{\unvbox\@ne
   \breakisland@}%
  \nointerlineskip
  \vskip\abovebotfigskip
  \printisland@
 \else
  \ifnum\flipcount@>\z@
   \global\setbox\topins\vbox{\unvbox\topins\global\setbox\@ne\lastbox}%
   \setbox\z@\vbox{\unvbox\@ne
    \breakisland@}%
   \printisland@
   \ifonlytop@\kern-\prevdepth\vfill\else\vskip\belowtopfigskip\fi
  \fi
 \fi
 \ifdim\ht\@cclv<\minpagesize
  \ifonlytop@\else\vfill\fi
 \else
  \ifspecialsplit@
   {\vfuzz\maxdimen\vbadness\@M
   \splitmaxdepth\maxdepth\splittopskip\topskip
   \dimen@ii\ht\@cclv \advance\dimen@ii\skip\topins
   \setbox\z@\vsplit\@cclv to\dimen@ii
   \unvbox\z@}%
  \else
   \unvbox\@cclv
  \fi
 \fi
 \bottomfigs@
 \ifvoid\footins\else\vskip\skip\footins\footnoterule\unvbox\footins\fi}
\newread\readdata@
\def\readthedata@#1{\expandafter
 \ifx\csname#1@D\endcsname\relax
  \immediate\openin\readdata@=#1.dat
  \ifeof\readdata@
   \Err@{No file #1.dat}%
  \else
   {\endlinechar\m@ne\gdef\Next@{}%
   \DNii@##1 ##2 ##3pt{\global\data@ht##1\global\data@dp##2%
    \global\data@wd##3pt}%
   \loop
    \ifeof\readdata@
    \else
    \read\readdata@ to\next@
    \ifx\next@\empty\else
     \edef\next@{\expandafter\nextii@\next@}%
     \expandafter\rightadd@\next@\to\Next@
    \fi
    \repeat}%
   \immediate\closein\readdata@
   \expandafter\expandafter\expandafter\global\expandafter
    \let\csname#1@D\endcsname\Next@\global\let\Next@\relax
  \fi
 \fi}
\newdimen\data@ht
\newdimen\data@dp
\newdimen\data@wd
\newif\ifgetdata@
\def\getdata@#1#2{\global\getdata@true\count@#2\relax
 {\let\\\or\xdef\Next@{\ifcase\number\count@#1\else
 \global\noexpand\getdata@false\fi}}\Next@}
\def\paste#1#2{\readthedata@{#1}%
 \getdata@{\csname#1@D\endcsname}{#2}%
 \ifgetdata@
 \dimen@\data@ht \advance\dimen@\data@dp
  \hbox{\special{dvipaste: #1 #2}%
   \lower\data@dp\vbox to\dimen@{\hbox to\data@wd{}\vfil}}%
 \else
  {\lccode`\Z=`\#\lccode`\N=`\N\lccode`\F=`\F%
   \lowercase{\Err@{No data for File [#1], Z#2}}}%
 \fi}
\newdimen\httable
\newdimen\dptable
\newdimen\wdtable
\def\measuretable#1#2{\readthedata@{#1}%
 \getdata@{\csname#1@D\endcsname}{#2}%
 \ifgetdata@
  \httable\data@ht \dptable\data@dp \wdtable\data@wd
 \else
  {\lccode`\Z=`\#\lccode`\N=`\N\lccode`\F=`\F%
  \lowercase{\Err@{No data for File [#1], Z#2}}}%
 \fi}
\def\East#1#2{\setboxz@h{$\m@th\ssize\;{#1}\;\;$}%
 \setbox\tw@\hbox{$\m@th\ssize\;{#2}\;\;$}\setbox4=\hbox{$\m@th#2$}%
 \dimen@\minaw@
 \ifdim\wdz@>\dimen@\dimen@\wdz@\fi\ifdim\wd\tw@>\dimen@\dimen@\wd\tw@\fi
 \ifdim\wd4 >\z@
  \mathrel{\mathop{\hbox to\dimen@{\rightarrowfill}}\limits^{#1}_{#2}}%
 \else
  \mathrel{\mathop{\hbox to\dimen@{\rightarrowfill}}\limits^{#1}}%
 \fi}
\def\West#1#2{\setboxz@h{$\m@th\ssize\;\;{#1}\;$}%
 \setbox\tw@\hbox{$\m@th\ssize\;\;{#2}\;$}\setbox4=\hbox{$\m@th#2$}%
 \dimen@\minaw@
 \ifdim\wdz@>\dimen@\dimen@\wdz@\fi\ifdim\wd\tw@>\dimen@\dimen@\wd\tw@\fi
 \ifdim\wd4 >\z@
  \mathrel{\mathop{\hbox to\dimen@{\leftarrowfill}}\limits^{#1}_{#2}}%
 \else
  \mathrel{\mathop{\hbox to\dimen@{\leftarrowfill}}\limits^{#1}}%
 \fi}
\font\arrow@i=lams1
\font\arrow@ii=lams2
\font\arrow@iii=lams3
\font\arrow@iv=lams4
\font\arrow@v=lams5
\newdimen\standardcgap
\standardcgap40\p@
\newdimen\hunit
\hunit\tw@\p@
\newdimen\standardrgap
\standardrgap32\p@
\newdimen\vunit
\vunit1.6\p@
\def\Cgaps#1{\RIfM@
 \standardcgap#1\standardcgap\relax\hunit#1\hunit\relax
 \else\nonmatherr@\Cgaps\fi}
\def\Rgaps#1{\RIfM@
 \standardrgap#1\standardrgap\relax\vunit#1\vunit\relax
 \else\nonmatherr@\Rgaps\fi}
\newdimen\getdim@
\def\getcgap@#1{\ifcase#1\or\getdim@\z@\else\getdim@\standardcgap\fi}
\def\getrgap@#1{\ifcase#1\getdim@\z@\else\getdim@\standardrgap\fi}
\def\cgaps{\RIfM@\expandafter\cgaps@\else\expandafter\nonmatherr@
 \expandafter\cgaps\fi}
\def\cgaps@{\ifnum\catcode`\;=\active\expandafter\cgapsA@\else
 \expandafter\cgapsO@\fi}
\def\cgapsO@#1{\toks@{\ifcase\i@\or\getdim@=\z@}%
 \gaps@@\standardcgap#1;\gaps@@\gaps@@
 \edef\next@{\the\toks@\noexpand\else\noexpand\getdim@\noexpand\standardcgap
  \noexpand\fi}%
 \toks@=\expandafter{\next@}%
 \edef\getcgap@##1{\i@##1\relax\the\toks@}\toks@{}}
{\catcode`\;=\active
 \gdef\cgapsA@#1{\toks@{\ifcase\i@\or\getdim@=\z@}%
 \gaps@@\standardcgap#1;\gaps@@\gaps@@
 \edef\next@{\the\toks@\noexpand\else\noexpand\getdim@\noexpand\standardcgap
  \noexpand\fi}%
 \toks@=\expandafter{\next@}%
 \edef\getcgap@##1{\i@##1\relax\the\toks@}\toks@{}}
}
\def\Gaps@@{\gaps@@}
\def\gaps@@#1#2;#3{\mgaps@#1#2\mgaps@
 \edef\next@{\the\toks@\noexpand\or\noexpand\getdim@
  \noexpand#1\the\mgapstoks@@}%
 \toks@\expandafter{\next@}%
 \DN@{#3}%
 \ifx\next@\Gaps@@\def\next@##1\gaps@@{}\else
  \def\next@{\gaps@@#1#3}\fi\next@}
{\catcode`\;=\active
 \gdef\rgaps#1{\RIfM@{\ifnum\catcode`\;=\active\def;{\string;}\fi
   \xdef\Next@{\noexpand\rgaps@{#1}}}%
  \Next@\edef\getrgap@##1{\i@##1\relax\the\toks@}\toks@{}\else
  \nonmatherr@\rgaps\fi}
}
\def\rgaps@#1{\toks@{\ifcase\i@\getdim@=\z@}%
 \gaps@@\standardrgap#1;\gaps@@\gaps@@
 \edef\next@{\the\toks@\noexpand\else\noexpand\getdim@\noexpand\standardrgap
  \noexpand\fi}%
 \toks@=\expandafter{\next@}}
\newbox\ZER@
\def\mgaps@#1{\let\mgapsnext@#1\FNSS@\mgaps@@}
\def\mgaps@@{\ifx\next\w\expandafter\mgaps@@@\else
 \expandafter\mgaps@@@@\fi}
\newtoks\mgapstoks@@
\def\mgaps@@@@#1\mgaps@{\getdim@\mgapsnext@\getdim@#1\getdim@
 \edef\next@{\noexpand\getdim@\the\getdim@}%
 \mgapstoks@@\expandafter{\next@}}
\def\mgaps@@@\w#1#2\mgaps@{\mgaps@@@@#2\mgaps@
 \setbox\ZER@\hbox{$\m@th\hskip15\p@\tsize@#1$}%
 \dimen@\wd\ZER@
 \ifdim\dimen@>\getdim@\getdim@\dimen@\fi
 \edef\next@{\noexpand\getdim@\the\getdim@}%
 \mgapstoks@@\expandafter{\next@}}
\def\changewidth#1#2{\setbox\ZER@{$\m@th#2}%
 \hbox to\wd\ZER@{\hss$\m@th#1$\hss}}
\atdef@({\FN@\ARROW@}
\def\ARROW@{\ifx\next)\let\next@\OPTIONS@\else
 \DN@{\csname\string @(\endcsname}\fi\next@}
\newif\ifoptions@
\def\OPTIONS@){\ifoptions@\let\next@\relax\else
 \DN@{\global\options@true\begingroup\optioncodes@}\fi\next@}
\newif\ifN@
\newif\ifE@
\newif\ifNESW@
\newif\ifH@
\newif\ifV@
\newif\ifHshort@
\expandafter\def\csname\string @(\endcsname #1,#2){%
 \ifoptions@\expandafter\endgroup\fi
 \N@false\E@false\H@false\V@false\Hshort@false
 \ifnum#1>\z@\E@true\fi
 \ifnum#1=\z@\V@true\global\tX@false\global\tY@false\global\a@false\fi
 \ifnum#2>\z@\N@true\fi
 \ifnum#2=\z@\H@true\global\tX@false\global\tY@false\global\a@false
  \ifshort@\Hshort@true\fi\fi
 \NESW@false
 \ifN@\ifE@\NESW@true\fi\else\ifE@\else\NESW@true\fi\fi
 \arrow@{#1}{#2}%
 \global\options@false
 \global\scount@\z@\global\tcount@\z@\global\arrcount@\z@
 \global\s@false\global\sxdimen@\z@\global\sydimen@\z@
 \global\tX@false\global\tXdimen@i\z@\global\tXdimen@ii\z@
 \global\tY@false\global\tYdimen@i\z@\global\tYdimen@ii\z@
 \global\a@false\global\exacount@\z@
 \global\x@false\global\xdimen@\z@
 \global\X@false\global\Xdimen@\z@
 \global\y@false\global\ydimen@\z@
 \global\Y@false\global\Ydimen@\z@
 \global\p@false\global\pdimen@\z@
 \global\label@ifalse\global\label@iifalse
 \global\dl@ifalse\global\ldimen@i\z@
 \global\dl@iifalse\global\ldimen@ii\z@
 \global\short@false\global\unshort@false}
\newif\iflabel@i
\newif\iflabel@ii
\newcount\scount@
\newcount\tcount@
\newcount\arrcount@
\newif\ifs@
\newdimen\sxdimen@
\newdimen\sydimen@
\newif\iftX@
\newdimen\tXdimen@i
\newdimen\tXdimen@ii
\newif\iftY@
\newdimen\tYdimen@i
\newdimen\tYdimen@ii
\newif\ifa@
\newcount\exacount@
\newif\ifx@
\newdimen\xdimen@
\newif\ifX@
\newdimen\Xdimen@
\newif\ify@
\newdimen\ydimen@
\newif\ifY@
\newdimen\Ydimen@
\newif\ifp@
\newdimen\pdimen@
\newif\ifdl@i
\newif\ifdl@ii
\newdimen\ldimen@i
\newdimen\ldimen@ii
\newif\ifshort@
\newif\ifunshort@
\def\zero@#1{\ifnum\scount@=\z@
 \if#1e\global\scount@\m@ne\else
 \if#1t\global\scount@\tw@\else
 \if#1h\global\scount@\thr@@\else
 \if#1'\global\scount@6 \else
 \if#1`\global\scount@7 \else
 \if#1(\global\scount@8 \else
 \if#1)\global\scount@9 \else
 \if#1s\global\scount@12 \else
 \if#1H\global\scount@13 \else
 \Err@{\Invalid@@ option \string\0}\fi\fi\fi\fi\fi\fi\fi\fi\fi
 \fi}
\def\one@#1{\ifnum\tcount@=\z@
 \if#1e\global\tcount@\m@ne\else
 \if#1h\global\tcount@\tw@\else
 \if#1t\global\tcount@\thr@@\else
 \if#1'\global\tcount@4 \else
 \if#1`\global\tcount@5 \else
 \if#1(\global\tcount@\ten@ \else
 \if#1)\global\tcount@11 \else
 \if#1s\global\tcount@12 \else
 \if#1H\global\tcount@13 \else
 \Err@{\Invalid@@ option \string\1}\fi\fi\fi\fi\fi\fi\fi\fi\fi
 \fi}
\def\a@#1{\ifnum\arrcount@=\z@
 \if#10\global\arrcount@\m@ne\else
 \if#1+\global\arrcount@\@ne\else
 \if#1-\global\arrcount@\tw@\else
 \if#1=\global\arrcount@\thr@@\else
 \Err@{\Invalid@@ option \string\a}\fi\fi\fi\fi
 \fi}
\def\ds@{\ifnum\catcode`\;=\active\expandafter\dsA@\else
 \expandafter\dsO@\fi}
\def\dsO@(#1;#2){\ds@@{#1}{#2}}
\def\ds@@#1#2{\ifs@\else
 \global\s@true
 \global\sxdimen@\hunit\global\sxdimen@#1\sxdimen@\relax
 \global\sydimen@\vunit\global\sydimen@#2\sydimen@\relax
 \fi}
\def\dtX@{\ifnum\catcode`\;=\active\expandafter\dtXA@\else
 \expandafter\dtXO@\fi}
\def\dtXO@(#1;#2){\dtX@@{#1}{#2}}
\def\dtX@@#1#2{\iftX@\else
 \global\tX@true
 \global\tXdimen@i\hunit\global\tXdimen@i#1\tXdimen@i\relax
 \global\tXdimen@ii\vunit\global\tXdimen@ii#2\tXdimen@ii\relax
 \fi}
\def\dtY@{\ifnum\catcode`\;=\active\expandafter\dtYA@\else
 \expandafter\dtYO@\fi}
\def\dtYO@(#1;#2){\dtY@@{#1}{#2}}
\def\dtY@@#1#2{\iftY@\else
 \global\tY@true
 \global\tYdimen@i\hunit\global\tYdimen@i#1\tYdimen@i\relax
 \global\tYdimen@ii\vunit\global\tYdimen@ii#2\tYdimen@ii\relax
 \fi}
{\catcode`\;=\active
 \gdef\dsA@(#1;#2){\ds@@{#1}{#2}}
 \gdef\dtXA@(#1;#2){\dtX@@{#1}{#2}}
 \gdef\dtYA@(#1;#2){\dtY@@{#1}{#2}}
}
\def\da@#1{\ifa@\else\global\a@true\global\exacount@#1\relax\fi}
\def\dx@#1{\ifx@\else
 \global\x@true
 \global\xdimen@\hunit\global\xdimen@#1\xdimen@\relax
 \fi}
\def\dX@#1{\ifX@\else
 \global\X@true
 \global\Xdimen@\hunit\global\Xdimen@#1\Xdimen@\relax
 \fi}
\def\dy@#1{\ify@\else
 \global\y@true
 \global\ydimen@\vunit\global\ydimen@#1\ydimen@\relax
 \fi}
\def\dY@#1{\ifY@\else
 \global\Y@true
 \global\Ydimen@\vunit\global\Ydimen@#1\Ydimen@\relax
 \fi}
\def\p@@#1{\ifp@\else
 \global\p@true
 \global\pdimen@\hunit\global\divide\pdimen@\tw@
 \global\pdimen@#1\pdimen@\relax
 \fi}
\def\L@#1{\iflabel@i\else
 \global\label@itrue\gdef\label@i{#1}%
 \fi}
\def\l@#1{\iflabel@ii\else
 \global\label@iitrue\gdef\label@ii{#1}%
 \fi}
\def\dL@#1{\ifdl@i\else
 \global\dl@itrue\global\ldimen@i\hunit\global\ldimen@i#1\ldimen@i\relax
 \fi}
\def\dl@#1{\ifdl@ii\else
 \global\dl@iitrue\global\ldimen@ii\hunit\global\ldimen@ii#1\ldimen@ii\relax
 \fi}
\def\s@{\ifunshort@\else\global\short@true\fi}
\def\uns@{\ifshort@\else\global\unshort@true\global\short@false\fi}
\def\optioncodes@{\let\0\zero@\let\1\one@\let\a\a@\let\ds\ds@\let\dtX\dtX@
 \let\dtY\dtY@\let\da\da@\let\dx\dx@\let\dX\dX@\let\dY\dY@\let\dy\dy@
 \let\p\p@@\let\L\L@\let\l\l@\let\dL\dL@\let\dl\dl@\let\s\s@\let\uns\uns@}
\def\slopes@{\\161\\152\\143\\134\\255\\126\\357\\238\\349\\45{10}\\56{11}%
 \\11{12}\\65{13}\\54{14}\\43{15}\\32{16}\\53{17}\\21{18}\\52{19}\\31{20}%
 \\41{21}\\51{22}\\61{23}}
\newcount\tan@i
\newcount\tan@ip
\newcount\tan@ii
\newcount\tan@iip
\newdimen\slope@i
\newdimen\slope@ip
\newdimen\slope@ii
\newdimen\slope@iip
\newcount\angcount@
\newcount\extracount@
\def\slope@{{\slope@i\secondy@\advance\slope@i-\firsty@
 \ifN@\else\multiply\slope@i\m@ne\fi
 \slope@ii\secondx@\advance\slope@ii-\firstx@
 \ifE@\else\multiply\slope@ii\m@ne\fi
 \ifdim\slope@ii<\z@
  \global\tan@i6 \global\tan@ii\@ne\global\angcount@23
 \else
  \dimen@\slope@i\multiply\dimen@6
  \ifdim\dimen@<\slope@ii
   \global\tan@i\@ne\global\tan@ii6 \global\angcount@\@ne
  \else
   \dimen@\slope@ii\multiply\dimen@6
   \ifdim\dimen@<\slope@i
    \global\tan@i6 \global\tan@ii\@ne\global\angcount@23
   \else
    \global\tan@ip\z@\global\tan@iip\@ne
    \def\\##1##2##3{\global\angcount@##3\relax
     \slope@ip\slope@i\slope@iip\slope@ii
     \multiply\slope@iip##1\relax\multiply\slope@ip##2\relax
     \ifdim\slope@iip<\slope@ip
      \global\tan@ip##1\relax\global\tan@iip##2\relax
     \else
      \global\tan@i##1\relax\global\tan@ii##2\relax
      \def\\####1####2####3{}%
     \fi}%
    \slopes@
    \slope@i\secondy@\advance\slope@i-\firsty@
    \ifN@\else\multiply\slope@i\m@ne\fi
    \multiply\slope@i\tan@ii\multiply\slope@i\tan@iip\multiply\slope@i\tw@
    \count@\tan@i\multiply\count@\tan@iip
    \extracount@\tan@ip\multiply\extracount@\tan@ii
    \advance\count@\extracount@
    \slope@ii\secondx@\advance\slope@ii-\firstx@
    \ifE@\else\multiply\slope@ii\m@ne\fi
    \multiply\slope@ii\count@
    \ifdim\slope@i<\slope@ii
     \global\tan@i\tan@ip\global\tan@ii\tan@iip
     \global\advance\angcount@\m@ne
    \fi
   \fi
  \fi
 \fi}%
}
\def\slope@a#1{{\def\\##1##2##3{\ifnum##3=#1\global\tan@i##1\relax
 \global\tan@ii##2\relax\fi}\slopes@}}
\newcount\i@
\newcount\j@
\newcount\colcount@
\newcount\Colcount@
\newcount\tcolcount@
\newdimen\rowht@
\newdimen\rowdp@
\newcount\rowcount@
\newcount\Rowcount@
\newcount\maxcolrow@
\newtoks\colwidthtoks@
\newtoks\Rowheighttoks@
\newtoks\Rowdepthtoks@
\newtoks\widthtoks@
\newtoks\Widthtoks@
\newtoks\heighttoks@
\newtoks\Heighttoks@
\newtoks\depthtoks@
\newtoks\Depthtoks@
\newif\iffirstCDcr@
\def\dotoks@i{%
 \global\widthtoks@\expandafter{\the\widthtoks@\else\getdim@\z@\fi}%
 \global\heighttoks@\expandafter{\the\heighttoks@\else\getdim@\z@\fi}%
 \global\depthtoks@\expandafter{\the\depthtoks@\else\getdim@\z@\fi}}
\def\dotoks@ii{%
 \global\widthtoks@{\ifcase\j@}%
 \global\heighttoks@{\ifcase\j@}%
 \global\depthtoks@{\ifcase\j@}}
\def\preCD@#1\endCD{\setbox\ZER@
 \vbox{%
  \def\arrow@##1##2{{}}%
  \global\rowcount@\m@ne\global\colcount@\z@\global\Colcount@\z@
  \global\firstCDcr@true\toks@{}%
  \global\widthtoks@{\ifcase\j@}%
  \global\Widthtoks@{\ifcase\i@}%
  \global\heighttoks@{\ifcase\j@}%
  \global\Heighttoks@{\ifcase\i@}%
  \global\depthtoks@{\ifcase\j@}%
  \global\Depthtoks@{\ifcase\i@}%
  \global\Rowheighttoks@{\ifcase\i@}%
  \global\Rowdepthtoks@{\ifcase\i@}%
  \Let@
  \everycr{%
   \noalign{%
    \global\advance\rowcount@\@ne
    \ifnum\colcount@<\Colcount@
    \else
     \global\Colcount@\colcount@\global\maxcolrow@\rowcount@
    \fi
    \global\colcount@\z@
    \iffirstCDcr@
     \global\firstCDcr@false
    \else
     \edef\next@{\the\Rowheighttoks@\noexpand\or\noexpand\getdim@\the\rowht@}%
      \global\Rowheighttoks@\expandafter{\next@}%
     \edef\next@{\the\Rowdepthtoks@\noexpand\or\noexpand\getdim@\the\rowdp@}%
      \global\Rowdepthtoks@\expandafter{\next@}%
     \global\rowht@\z@\global\rowdp@\z@
     \dotoks@i
     \edef\next@{\the\Widthtoks@\noexpand\or\the\widthtoks@}%
      \global\Widthtoks@\expandafter{\next@}%
     \edef\next@{\the\Heighttoks@\noexpand\or\the\heighttoks@}%
      \global\Heighttoks@\expandafter{\next@}%
     \edef\next@{\the\Depthtoks@\noexpand\or\the\depthtoks@}%
      \global\Depthtoks@\expandafter{\next@}%
     \dotoks@ii
    \fi}}%
  \tabskip\z@
  \halign{&\setbox\ZER@\hbox{\vrule\height\ten@\p@\width\z@\depth\z@     %1
   $\m@th\displaystyle{##}$}\copy\ZER@
   \ifdim\ht\ZER@>\rowht@\global\rowht@\ht\ZER@\fi
   \ifdim\dp\ZER@>\rowdp@\global\rowdp@\dp\ZER@\fi
   \global\advance\colcount@\@ne
   \edef\next@{\the\widthtoks@\noexpand\or\noexpand\getdim@\the\wd\ZER@}%
    \global\widthtoks@\expandafter{\next@}%
   \edef\next@{\the\heighttoks@\noexpand\or\noexpand\getdim@\the\ht\ZER@}%
    \global\heighttoks@\expandafter{\next@}%
   \edef\next@{\the\depthtoks@\noexpand\or\noexpand\getdim@\the\dp\ZER@}%
    \global\depthtoks@\expandafter{\next@}%
   \cr#1\crcr}}%
 \Rowcount@\rowcount@
 \global\Widthtoks@\expandafter{\the\Widthtoks@\fi\relax}%
 \edef\Width@##1##2{\i@##1\relax\j@##2\relax\the\Widthtoks@}%
 \global\Heighttoks@\expandafter{\the\Heighttoks@\fi\relax}%
 \edef\Height@##1##2{\i@##1\relax\j@##2\relax\the\Heighttoks@}%
 \global\Depthtoks@\expandafter{\the\Depthtoks@\fi\relax}%
 \edef\Depth@##1##2{\i@##1\relax\j@##2\relax\the\Depthtoks@}%
 \edef\next@{\the\Rowheighttoks@\noexpand\fi\relax}%
 \global\Rowheighttoks@\expandafter{\next@}%
 \edef\Rowheight@##1{\i@##1\relax\the\Rowheighttoks@}%
 \edef\next@{\the\Rowdepthtoks@\noexpand\fi\relax}%
 \global\Rowdepthtoks@\expandafter{\next@}%
 \edef\Rowdepth@##1{\i@##1\relax\the\Rowdepthtoks@}%
 \global\colwidthtoks@{\fi}%
 \setbox\ZER@\vbox{%
  \unvbox\ZER@
  \count@\rowcount@
  \loop
   \unskip\unpenalty
   \setbox\ZER@\lastbox
   \ifnum\count@>\maxcolrow@\advance\count@\m@ne
   \repeat
  \hbox{%
   \unhbox\ZER@
   \count@\z@
   \loop
    \unskip
    \setbox\ZER@\lastbox
    \edef\next@{\noexpand\or\noexpand\getdim@\the\wd\ZER@\the\colwidthtoks@}%
     \global\colwidthtoks@\expandafter{\next@}%
    \advance\count@\@ne
    \ifnum\count@<\Colcount@
    \repeat}}%
 \edef\next@{\noexpand\ifcase\noexpand\i@\the\colwidthtoks@}%
  \global\colwidthtoks@\expandafter{\next@}%
 \edef\Colwidth@##1{\i@##1\relax\the\colwidthtoks@}%
 \global\colwidthtoks@{}\global\Rowheighttoks@{}\global\Rowdepthtoks@{}%
 \global\widthtoks@{}\global\Widthtoks@{}\global\heighttoks@{}%
 \global\Heighttoks@{}\global\depthtoks@{}\global\Depthtoks@{}%
}
\newcount\xoff@
\newcount\yoff@
\newcount\endcount@
\newcount\rcount@
\newdimen\firstx@
\newdimen\firsty@
\newdimen\secondx@
\newdimen\secondy@
\newdimen\tocenter@
\newdimen\charht@
\newdimen\charwd@
\def\outside@{\Err@{This arrow points outside the \string\CD}}
\newif\ifsvertex@
\newif\iftvertex@
\def\arrow@#1#2{\global\xoff@#1\relax\global\yoff@#2\relax
 \count@\rowcount@\advance\count@-\yoff@
 \ifnum\count@<\@ne\outside@\else\ifnum\count@>\Rowcount@\outside@\fi\fi
 \count@\colcount@\advance\count@\xoff@
 \ifnum\count@<\@ne\outside@\else\ifnum\count@>\Colcount@\outside@\fi\fi
 \tcolcount@\colcount@\advance\tcolcount@\xoff@
 \Width@\rowcount@\colcount@\divide\getdim@\tw@\tocenter@-\getdim@
 \ifdim\getdim@=\z@
  \firstx@\z@\firsty@\mathaxis@\svertex@true
 \else
  \svertex@false
  \ifHshort@
   \Colwidth@\colcount@\divide\getdim@\tw@
   \ifE@ \firstx@\getdim@ \else \firstx@-\getdim@ \fi
  \else
   \ifE@ \firstx@\getdim@ \else \firstx@-\getdim@ \fi
  \fi
  \ifE@
   \ifH@ \advance\firstx@\thr@@\p@ \else \advance\firstx@-\thr@@\p@ \fi  %2
  \else
   \ifH@ \advance\firstx@-\thr@@\p@ \else \advance\firstx@\thr@@\p@ \fi  %3
  \fi
  \ifN@
   \Height@\rowcount@\colcount@ \firsty@\getdim@                         %4
   \ifV@ \advance\firsty@\thr@@\p@ \fi                                   %5
  \else
   \ifV@
    \Depth@\rowcount@\colcount@ \firsty@-\getdim@                        %6
    \advance\firsty@-\thr@@\p@                                           %7
   \else
    \firsty@\z@                                                          %8
   \fi
  \fi
 \fi
 \ifV@
 \else
  \Colwidth@\colcount@\divide\getdim@\tw@
  \ifE@\secondx@\getdim@\else\secondx@-\getdim@\fi
  \ifE@\else\getcgap@\colcount@\advance\secondx@-\getdim@\fi
  \endcount@\colcount@\advance\endcount@\xoff@
  \count@\colcount@
  \ifE@
   \advance\count@\@ne
   \loop
    \ifnum\count@<\endcount@
    \Colwidth@\count@\advance\secondx@\getdim@
    \getcgap@\count@\advance\secondx@\getdim@
    \advance\count@\@ne
    \repeat
  \else
   \advance\count@\m@ne
   \loop
    \ifnum\count@>\endcount@
    \Colwidth@\count@\advance\secondx@-\getdim@
    \getcgap@\count@\advance\secondx@-\getdim@
    \advance\count@\m@ne
    \repeat
  \fi
  \Colwidth@\count@\divide\getdim@\tw@
  \ifHshort@
  \else
   \ifE@\advance\secondx@\getdim@\else\advance\secondx@-\getdim@\fi
  \fi
  \ifE@\getcgap@\count@\advance\secondx@\getdim@\fi
  \rcount@\rowcount@\advance\rcount@-\yoff@
  \Width@\rcount@\count@\divide\getdim@\tw@
  \tvertex@false
  \ifH@\ifdim\getdim@=\z@\tvertex@true\Hshort@false\fi\fi
  \ifHshort@
  \else
   \ifE@\advance\secondx@-\getdim@\else\advance\secondx@\getdim@\fi
  \fi
  \iftvertex@
   \advance\secondx@.4\p@
  \else
   \ifE@\advance\secondx@-\thr@@\p@\else\advance\secondx@\thr@@\p@\fi    %9
  \fi
 \fi
 \ifH@
 \else
  \ifN@
   \Rowheight@\rowcount@\secondy@\getdim@
  \else
   \Rowdepth@\rowcount@\secondy@-\getdim@
   \getrgap@\rowcount@\advance\secondy@-\getdim@
  \fi
  \endcount@\rowcount@\advance\endcount@-\yoff@
  \count@\rowcount@
  \ifN@
   \advance\count@\m@ne
   \loop
    \ifnum\count@>\endcount@
    \Rowheight@\count@\advance\secondy@\getdim@
    \Rowdepth@\count@\advance\secondy@\getdim@
    \getrgap@\count@\advance\secondy@\getdim@
    \advance\count@\m@ne
    \repeat
  \else
   \advance\count@\@ne
   \loop
    \ifnum\count@<\endcount@
    \Rowheight@\count@\advance\secondy@-\getdim@
    \Rowdepth@\count@\advance\secondy@-\getdim@
    \getrgap@\count@\advance\secondy@-\getdim@
    \advance\count@\@ne
    \repeat
  \fi
  \tvertex@false
  \ifV@\Width@\count@\colcount@\ifdim\getdim@=\z@\tvertex@true\fi\fi
  \ifN@
   \getrgap@\count@\advance\secondy@\getdim@
   \Rowdepth@\count@\advance\secondy@\getdim@
   \iftvertex@
    \advance\secondy@\mathaxis@
   \else
    \Depth@\count@\tcolcount@\advance\secondy@-\getdim@
    \advance\secondy@-\thr@@\p@                                          %10
   \fi
  \else
   \Rowheight@\count@\advance\secondy@-\getdim@
   \iftvertex@
    \advance\secondy@\mathaxis@
   \else
    \Height@\count@\tcolcount@\advance\secondy@\getdim@
    \advance\secondy@\thr@@\p@                                           %11
   \fi
  \fi
 \fi
 \ifV@\else\advance\firstx@\sxdimen@\fi
 \ifH@\else\advance\firsty@\sydimen@\fi
 \iftX@
  \advance\secondy@\tXdimen@ii
  \advance\secondx@\tXdimen@i
  \slope@
 \else
  \iftY@
   \advance\secondy@\tYdimen@ii
   \advance\secondx@\tYdimen@i
   \slope@
   \secondy@\secondx@\advance\secondy@-\firstx@
   \ifNESW@\else\multiply\secondy@\m@ne\fi
   \multiply\secondy@\tan@i\divide\secondy@\tan@ii\advance\secondy@\firsty@
  \else
   \ifa@
    \slope@
    \ifNESW@\global\advance\angcount@\exacount@\else
     \global\advance\angcount@-\exacount@\fi
    \ifnum\angcount@>23 \global\angcount@23 \fi
    \ifnum\angcount@<\@ne\global\angcount@\@ne\fi
    \slope@a\angcount@
    \ifY@
     \advance\secondy@\Ydimen@
    \else
     \ifX@
      \advance\secondx@\Xdimen@
      \dimen@\secondx@\advance\dimen@-\firstx@
      \ifNESW@\else\multiply\dimen@\m@ne\fi
      \multiply\dimen@\tan@i\divide\dimen@\tan@ii
      \advance\dimen@\firsty@\secondy@\dimen@
     \fi
    \fi
   \else
    \ifH@\else\ifV@\else\slope@\fi\fi
   \fi
  \fi
 \fi
 \ifH@\else\ifV@\else\ifsvertex@\else
  \dimen@6\p@\multiply\dimen@\tan@ii
  \count@\tan@i\advance\count@\tan@ii\divide\dimen@\count@
  \ifE@\advance\firstx@\dimen@\else\advance\firstx@-\dimen@\fi
  \multiply\dimen@\tan@i\divide\dimen@\tan@ii
  \ifN@\advance\firsty@\dimen@\else\advance\firsty@-\dimen@\fi
 \fi\fi\fi
 \ifp@
  \ifH@\else\ifV@\else
   \getcos@\pdimen@\advance\firsty@\dimen@\advance\secondy@\dimen@
   \ifNESW@\advance\firstx@-\dimen@ii\else\advance\firstx@\dimen@ii\fi
  \fi\fi
 \fi
 \ifH@\else\ifV@\else
  \ifnum\tan@i>\tan@ii
   \charht@\ten@\p@\charwd@\ten@\p@
   \multiply\charwd@\tan@ii\divide\charwd@\tan@i
  \else
   \charwd@\ten@\p@\charht@\ten@\p@
   \divide\charht@\tan@ii\multiply\charht@\tan@i
  \fi
  \ifnum\tcount@=\thr@@
   \ifN@\advance\secondy@-.3\charht@\else\advance\secondy@.3\charht@\fi
  \fi
  \ifnum\scount@=\tw@
   \ifE@\advance\firstx@.3\charht@\else\advance\firstx@-.3\charht@\fi
  \fi
  \ifnum\tcount@=12
   \ifN@\advance\secondy@-\charht@\else\advance\secondy@\charht@\fi
  \fi
  \iftY@
  \else
   \ifa@
    \ifX@
    \else
     \secondx@\secondy@\advance\secondx@-\firsty@
     \ifNESW@\else\multiply\secondx@\m@ne\fi
     \multiply\secondx@\tan@ii\divide\secondx@\tan@i
     \advance\secondx@\firstx@
    \fi
   \fi
  \fi
 \fi\fi
 \ifH@\harrow@\else\ifV@\varrow@\else\arrow@@\fi\fi}
\newdimen\mathaxis@
\mathaxis@90\p@\divide\mathaxis@36
\def\harrow@b{\ifE@\hskip\tocenter@\hskip\firstx@\fi}
\def\harrow@bb{\ifE@\hskip\xdimen@\else\hskip\Xdimen@\fi}
\def\harrow@e{\ifE@\else\hskip-\firstx@\hskip-\tocenter@\fi}
\def\harrow@ee{\ifE@\hskip-\Xdimen@\else\hskip-\xdimen@\fi}
\def\harrow@{\dimen@\secondx@\advance\dimen@-\firstx@
 \ifE@\let\next@\rlap\else\multiply\dimen@\m@ne\let\next@\llap\fi
 \next@{%
  \harrow@b
  \smash{\raise\pdimen@\hbox to\dimen@
   {\harrow@bb\arrow@ii
    \ifnum\arrcount@=\m@ne\else\ifnum\arrcount@=\thr@@\else
     \ifE@
      \ifnum\scount@=\m@ne
      \else
       \ifcase\scount@\or\or\char118 \or\char117 \or\or\or\char119 \or
       \char120 \or\char121 \or\char122 \or\or\or\arrow@i\char125 \or
       \char117 \hskip\thr@@\p@\char117 \hskip-\thr@@\p@\fi
      \fi
     \else
      \ifnum\tcount@=\m@ne
      \else
       \ifcase\tcount@\char117 \or\or\char117 \or\char118 \or\char119 \or
       \char120 \or\or\or\or\or\char121 \or\char122 \or\arrow@i\char125
       \or\char117 \hskip\thr@@\p@\char117 \hskip-\thr@@\p@\fi
      \fi
     \fi
    \fi\fi
    \dimen@\mathaxis@\advance\dimen@.2\p@
    \dimen@ii\mathaxis@\advance\dimen@ii-.2\p@
    \ifnum\arrcount@=\m@ne
     \let\leads@\null
    \else
     \ifcase\arrcount@
      \def\leads@{\hrule\height\dimen@\depth-\dimen@ii}\or
      \def\leads@{\hrule\height\dimen@\depth-\dimen@ii}\or
      \def\leads@{\hbox to\ten@\p@{%
       \leaders\hrule\height\dimen@\depth-\dimen@ii\hfil
       \hfil
      \leaders\hrule\height\dimen@\depth-\dimen@ii\hskip\z@ plus2fil\relax
       \hfil
       \leaders\hrule\height\dimen@\depth-\dimen@ii\hfil}}\or
     \def\leads@{\hbox{\hbox to\ten@\p@{\dimen@\mathaxis@\advance\dimen@1.2\p@
       \dimen@ii\dimen@\advance\dimen@ii-.4\p@
       \leaders\hrule\height\dimen@\depth-\dimen@ii\hfil}%
       \kern-\ten@\p@
       \hbox to\ten@\p@{\dimen@\mathaxis@\advance\dimen@-1.2\p@
       \dimen@ii\dimen@\advance\dimen@ii-.4\p@
       \leaders\hrule\height\dimen@\depth-\dimen@ii\hfil}}}\fi
    \fi
    \cleaders\leads@\hfil
    \ifnum\arrcount@=\m@ne\else\ifnum\arrcount@=\thr@@\else
     \arrow@i
     \ifE@
      \ifnum\tcount@=\m@ne
      \else
       \ifcase\tcount@\char119 \or\or\char119 \or\char120 \or\char121 \or
       \char122 \or \or\or\or\or\char123 \or\char124 \or
       \char125 \or\char119 \hskip-\thr@@\p@\char119 \hskip\thr@@\p@\fi
      \fi
     \else
      \ifcase\scount@\or\or\char120 \or\char119 \or\or\or\char121 \or\char122
      \or\char123 \or\char124 \or\or\or\char125 \or
      \char119 \hskip-\thr@@\p@\char119 \hskip\thr@@\p@\fi
     \fi
    \fi\fi
    \harrow@ee}}%
  \harrow@e}%
 \iflabel@i
  \dimen@ii\z@\setbox\ZER@\hbox{$\m@th\tsize@@\label@i$}%
  \ifnum\arrcount@=\m@ne
  \else
   \advance\dimen@ii\mathaxis@
   \advance\dimen@ii\dp\ZER@\advance\dimen@ii\tw@\p@
   \ifnum\arrcount@=\thr@@\advance\dimen@ii\tw@\p@\fi
  \fi
  \advance\dimen@ii\pdimen@
  \next@{\harrow@b\smash{\raise\dimen@ii\hbox to\dimen@
   {\harrow@bb\hskip\tw@\ldimen@i\hfil\box\ZER@\hfil\harrow@ee}}\harrow@e}%
 \fi
 \iflabel@ii
  \ifnum\arrcount@=\m@ne
  \else
   \setbox\ZER@\hbox{$\m@th\tsize@\label@ii$}%
   \dimen@ii-\ht\ZER@\advance\dimen@ii-\tw@\p@
   \ifnum\arrcount@=\thr@@\advance\dimen@ii-\tw@\p@\fi
   \advance\dimen@ii\mathaxis@\advance\dimen@ii\pdimen@
   \next@{\harrow@b\smash{\raise\dimen@ii\hbox to\dimen@
    {\harrow@bb\hskip\tw@\ldimen@ii\hfil\box\ZER@\hfil\harrow@ee}}\harrow@e}%
  \fi
 \fi}
\let\tsize@\tsize
\def\tsizeCDlabels{\let\tsize@\tsize}
\def\ssizeCDlabels{\let\tsize@\ssize}
\def\tsize@@{\ifnum\arrcount@=\m@ne\else\tsize@\fi}
\def\varrow@{\dimen@\secondy@\advance\dimen@-\firsty@
 \ifN@\else\multiply\dimen@\m@ne\fi
 \setbox\ZER@\vbox to\dimen@
  {\ifN@\vskip-\Ydimen@\else\vskip\ydimen@\fi
   \ifnum\arrcount@=\m@ne\else\ifnum\arrcount@=\thr@@\else
    \hbox{\arrow@iii
     \ifN@
      \ifnum\tcount@=\m@ne
      \else
       \ifcase\tcount@\char117 \or\or\char117 \or\char118 \or\char119 \or
       \char120 \or\or\or\or\or\char121 \or\char122 \or\char123 \or
       \vbox{\hbox{\char117}\nointerlineskip\vskip\thr@@\p@
       \hbox{\char117}\vskip-\thr@@\p@}\fi
      \fi
     \else
      \ifcase\scount@\or\or\char118 \or\char117 \or\or\or\char119 \or
      \char120 \or\char121 \or\char122 \or\or\or\char123 \or
      \vbox{\hbox{\char117}\nointerlineskip\vskip\thr@@\p@
      \hbox{\char117}\vskip-\thr@@\p@}\fi
     \fi}%
    \nointerlineskip
   \fi\fi
   \ifnum\arrcount@=\m@ne
    \let\leads@\null
   \else
    \ifcase\arrcount@\let\leads@\vrule\or\let\leads@\vrule\or
    \def\leads@{\vbox to\ten@\p@{%
     \hrule\height1.67\p@\depth\z@\width.4\p@
     \vfil
     \hrule\height3.33\p@\depth\z@\width.4\p@
     \vfil
     \hrule\height1.67\p@\depth\z@\width.4\p@}}\or
    \def\leads@{\hbox{\vrule\height\p@\hskip\tw@\p@\vrule}}\fi
   \fi
  \cleaders\leads@\vfill\nointerlineskip
   \ifnum\arrcount@=\m@ne\else\ifnum\arrcount@=\thr@@\else
    \hbox{\arrow@iv
     \ifN@
      \ifcase\scount@\or\or\char118 \or\char117 \or\or\or\char119 \or
      \char120 \or\char121 \or\char122 \or\or\or\arrow@iii\char123 \or
      \vbox{\hbox{\char117}\nointerlineskip\vskip-\thr@@\p@
      \hbox{\char117}\vskip\thr@@\p@}\fi
     \else
      \ifnum\tcount@=\m@ne
      \else
       \ifcase\tcount@\char117 \or\or\char117 \or\char118 \or\char119 \or
       \char120 \or\or\or\or\or\char121 \or\char122 \or\arrow@iii\char123 \or
       \vbox{\hbox{\char117}\nointerlineskip\vskip-\thr@@\p@
       \hbox{\char117}\vskip\thr@@\p@}\fi
      \fi
     \fi}%
   \fi\fi
   \ifN@\vskip\ydimen@\else\vskip-\Ydimen@\fi}%
 \ifN@
  \dimen@ii\firsty@
 \else
  \dimen@ii-\firsty@\advance\dimen@ii\ht\ZER@\multiply\dimen@ii\m@ne
 \fi
 \rlap{\smash{\hskip\tocenter@\hskip\pdimen@\raise\dimen@ii\box\ZER@}}%
 \iflabel@i
  \setbox\ZER@\vbox to\dimen@{\vfil
   \hbox{$\m@th\tsize@@\label@i$}\vskip\tw@\ldimen@i\vfil}%
  \rlap{\smash{\hskip\tocenter@\hskip\pdimen@
  \ifnum\arrcount@=\m@ne\let\next@\relax\else\let\next@\llap\fi
  \next@{\raise\dimen@ii\hbox{\ifnum\arrcount@=\m@ne\hskip-.5\wd\ZER@\fi
   \box\ZER@\ifnum\arrcount@=\m@ne\else\hskip\tw@\p@\fi}}}}%
 \fi
 \iflabel@ii
  \ifnum\arrcount@=\m@ne
  \else
   \setbox\ZER@\vbox to\dimen@{\vfil
    \hbox{$\m@th\tsize@\label@ii$}\vskip\tw@\ldimen@ii\vfil}%
   \rlap{\smash{\hskip\tocenter@\hskip\pdimen@
   \rlap{\raise\dimen@ii\hbox{\ifnum\arrcount@=\thr@@\hskip4.5\p@\else
    \hskip2.5\p@\fi\box\ZER@}}}}%
  \fi
 \fi
}
\newdimen\goal@
\newdimen\shifted@
\newcount\Tcount@
\newcount\Scount@
\newbox\shaft@
\newcount\slcount@
\def\getcos@#1{%
 \ifnum\tan@i<\tan@ii
  \dimen@#1%
  \ifnum\slcount@<8 \count@9 \else \ifnum\slcount@<12 \count@8 \else
   \count@7 \fi\fi
  \multiply\dimen@\count@\divide\dimen@\ten@
  \dimen@ii\dimen@\multiply\dimen@ii\tan@i\divide\dimen@ii\tan@ii
 \else
  \dimen@ii#1%
  \count@-\slcount@\advance\count@24
  \ifnum\count@<8 \count@9 \else \ifnum\count@<12 \count@8
   \else\count@7 \fi\fi
  \multiply\dimen@ii\count@\divide\dimen@ii\ten@
  \dimen@\dimen@ii\multiply\dimen@\tan@ii\divide\dimen@\tan@i
 \fi}
\newdimen\adjust@
\def\Nnext@{\ifN@\let\next@\raise\else\let\next@\lower\fi}
\def\arrow@@{\slcount@\angcount@
 \ifNESW@
  \ifnum\angcount@<\ten@
   \let\arrowfont@\arrow@i\global\advance\angcount@\m@ne
   \global\multiply\angcount@13
  \else
   \ifnum\angcount@<19
    \let\arrowfont@\arrow@ii\global\advance\angcount@-\ten@
    \global\multiply\angcount@13
   \else
    \let\arrowfont@\arrow@iii\global\advance\angcount@-19
    \global\multiply\angcount@13
  \fi\fi
  \Tcount@\angcount@
 \else
  \ifnum\angcount@<5
   \let\arrowfont@\arrow@iii\global\advance\angcount@\m@ne
   \global\multiply\angcount@13 \global\advance\angcount@65
  \else
   \ifnum\angcount@<14
    \let\arrowfont@\arrow@iv\global\advance\angcount@-5
    \global\multiply\angcount@13
   \else
    \ifnum\angcount@<23
     \let\arrowfont@\arrow@v\global\advance\angcount@-14
     \global\multiply\angcount@13
    \else
     \let\arrowfont@\arrow@i\global\angcount@117
  \fi\fi\fi
  \ifnum\angcount@=117 \Tcount@115 \else\Tcount@\angcount@\fi
 \fi
 \Scount@\Tcount@
 \ifE@
  \ifnum\tcount@=\z@\advance\Tcount@\tw@\else\ifnum\tcount@=13
   \advance\Tcount@\tw@\else\advance\Tcount@\tcount@\fi\fi
  \ifnum\scount@=\z@\else\ifnum\scount@=13 \advance\Scount@\thr@@\else
   \advance\Scount@\scount@\fi\fi
 \else
  \ifcase\tcount@\advance\Tcount@\thr@@\or\or\advance\Tcount@\thr@@\or
  \advance\Tcount@\tw@\or\advance\Tcount@6 \or\advance\Tcount@7
  \or\or\or\or\or\advance\Tcount@8 \or\advance\Tcount@9 \or
  \advance\Tcount@12 \or\advance\Tcount@\thr@@\fi
  \ifcase\scount@\or\or\advance\Scount@\thr@@\or\advance\Scount@\tw@\or
  \or\or\advance\Scount@4 \or\advance\Scount@5 \or\advance\Scount@\ten@
  \or\advance\Scount@11 \or\or\or\advance\Scount@12 \or\advance
  \Scount@\tw@\fi
 \fi
 \ifcase\arrcount@\or\or\global\advance\angcount@\@ne\else\fi
 \ifN@\shifted@\firsty@\else\shifted@-\firsty@\fi
 \ifE@\else\advance\shifted@\charht@\fi
 \goal@\secondy@\advance\goal@-\firsty@
 \ifN@\else\multiply\goal@\m@ne\fi
 \setbox\shaft@\hbox{\arrowfont@\char\angcount@}%
 \ifnum\arrcount@=\thr@@
  \getcos@{1.5\p@}%
  \setbox\shaft@\hbox to\wd\shaft@{\arrowfont@
   \rlap{\hskip\dimen@ii
    \smash{\ifNESW@\let\next@\lower\else\let\next@\raise\fi
     \next@\dimen@\hbox{\arrowfont@\char\angcount@}}}%
   \rlap{\hskip-\dimen@ii
    \smash{\ifNESW@\let\next@\raise\else\let\next@\lower\fi
      \next@\dimen@\hbox{\arrowfont@\char\angcount@}}}\hfil}%
 \fi
 \rlap{\smash{\hskip\tocenter@\hskip\firstx@
  \ifnum\arrcount@=\m@ne
  \else
   \ifnum\arrcount@=\thr@@
   \else
    \ifnum\scount@=\m@ne
    \else
     \ifnum\scount@=\z@
     \else
      \setbox\ZER@\hbox{\ifnum\angcount@=117 \arrow@v\else\arrowfont@\fi
       \char\Scount@}%
      \ifNESW@
       \ifnum\scount@=\tw@
        \dimen@\shifted@\advance\dimen@-\charht@
        \ifN@\hskip-\wd\ZER@\fi
        \Nnext@
        \next@\dimen@\copy\ZER@
        \ifN@\else\hskip-\wd\ZER@\fi
       \else
        \Nnext@
        \ifN@\else\hskip-\wd\ZER@\fi
        \next@\shifted@\copy\ZER@
        \ifN@\hskip-\wd\ZER@\fi
       \fi
       \ifnum\scount@=12
        \advance\shifted@\charht@\advance\goal@-\charht@
        \ifN@\hskip\wd\ZER@\else\hskip-\wd\ZER@\fi
       \fi
       \ifnum\scount@=13
        \getcos@{\thr@@\p@}%
        \ifN@\hskip\dimen@\else\hskip-\wd\ZER@\hskip-\dimen@\fi
        \adjust@\shifted@\advance\adjust@\dimen@ii
        \Nnext@
        \next@\adjust@\copy\ZER@
        \ifN@\hskip-\dimen@\hskip-\wd\ZER@\else\hskip\dimen@\fi
       \fi
      \else
       \ifN@\hskip-\wd\ZER@\fi
       \ifnum\scount@=\tw@
        \ifN@\hskip\wd\ZER@\else\hskip-\wd\ZER@\fi
        \dimen@\shifted@\advance\dimen@-\charht@
        \Nnext@
        \next@\dimen@\copy\ZER@
        \ifN@\hskip-\wd\ZER@\fi
       \else
        \Nnext@
        \next@\shifted@\copy\ZER@
        \ifN@\else\hskip-\wd\ZER@\fi
       \fi
       \ifnum\scount@=12
        \advance\shifted@\charht@\advance\goal@-\charht@
        \ifN@\hskip-\wd\ZER@\else\hskip\wd\ZER@\fi
       \fi
       \ifnum\scount@=13
        \getcos@{\thr@@\p@}%
        \ifN@\hskip-\wd\ZER@\hskip-\dimen@\else\hskip\dimen@\fi
        \adjust@\shifted@\advance\adjust@\dimen@ii
        \Nnext@
        \next@\adjust@\copy\ZER@
        \ifN@\hskip\dimen@\else\hskip-\dimen@\hskip-\wd\ZER@\fi
       \fi
      \fi
  \fi\fi\fi\fi
  \ifnum\arrcount@=\m@ne
  \else
   \loop
    \ifdim\goal@>\charht@
    \ifE@\else\hskip-\charwd@\fi
    \Nnext@
    \next@\shifted@\copy\shaft@
    \ifE@\else\hskip-\charwd@\fi
    \advance\shifted@\charht@\advance\goal@-\charht@
    \repeat
   \ifdim\goal@>\z@
    \dimen@\charht@\advance\dimen@-\goal@
    \divide\dimen@\tan@i\multiply\dimen@\tan@ii
    \ifE@\hskip-\dimen@\else\hskip-\charwd@\hskip\dimen@\fi
    \adjust@\shifted@\advance\adjust@-\charht@\advance\adjust@\goal@
    \Nnext@
    \next@\adjust@\copy\shaft@
    \ifE@\else\hskip-\charwd@\fi
   \else
    \adjust@\shifted@\advance\adjust@-\charht@
   \fi
  \fi
  \ifnum\arrcount@=\m@ne
  \else
   \ifnum\arrcount@=\thr@@
   \else
    \ifnum\tcount@=\m@ne
    \else
     \setbox\ZER@
      \hbox{\ifnum\angcount@=117 \arrow@v\else\arrowfont@\fi\char\Tcount@}%
     \ifnum\tcount@=\thr@@
      \advance\adjust@\charht@
      \ifE@\else\ifN@\hskip-\charwd@\else\hskip-\wd\ZER@\fi\fi
     \else
      \ifnum\tcount@=12
       \advance\adjust@\charht@
       \ifE@\else\ifN@\hskip-\charwd@\else\hskip-\wd\ZER@\fi\fi
      \else
       \ifE@\hskip-\wd\ZER@\fi
     \fi\fi
     \Nnext@
     \next@\adjust@\copy\ZER@
     \ifnum\tcount@=13
      \hskip-\wd\ZER@
      \getcos@{\thr@@\p@}%
      \ifE@\hskip-\dimen@\else\hskip\dimen@\fi
      \advance\adjust@-\dimen@ii
      \Nnext@
      \next@\adjust@\box\ZER@
     \fi
  \fi\fi\fi}}%
 \iflabel@i
  \rlap{\hskip\tocenter@
  \dimen@\firstx@\advance\dimen@\secondx@\divide\dimen@\tw@
  \advance\dimen@\ldimen@i
  \dimen@ii\firsty@\advance\dimen@ii\secondy@\divide\dimen@ii\tw@
  \global\multiply\ldimen@i\tan@i\global\divide\ldimen@i\tan@ii
  \ifNESW@\advance\dimen@ii\ldimen@i\else\advance\dimen@ii-\ldimen@i\fi
  \setbox\ZER@\hbox{\ifNESW@\else\ifnum\arrcount@=\thr@@\hskip4\p@\else
   \hskip\tw@\p@\fi\fi
   $\m@th\tsize@@\label@i$\ifNESW@\ifnum\arrcount@=\thr@@\hskip4\p@\else
   \hskip\tw@\p@\fi\fi}%
  \ifnum\arrcount@=\m@ne
   \ifNESW@\advance\dimen@.5\wd\ZER@\advance\dimen@\p@\else
    \advance\dimen@-.5\wd\ZER@\advance\dimen@-\p@\fi
   \advance\dimen@ii-.5\ht\ZER@
  \else
   \advance\dimen@ii\dp\ZER@
   \ifnum\slcount@<6 \advance\dimen@ii\tw@\p@\fi
  \fi
  \hskip\dimen@
  \ifNESW@\let\next@\llap\else\let\next@\rlap\fi
  \next@{\smash{\raise\dimen@ii\box\ZER@}}}%
 \fi
 \iflabel@ii
  \ifnum\arrcount@=\m@ne
  \else
   \rlap{\hskip\tocenter@
   \dimen@\firstx@\advance\dimen@\secondx@\divide\dimen@\tw@
   \ifNESW@\advance\dimen@\ldimen@ii\else\advance\dimen@-\ldimen@ii\fi
   \dimen@ii\firsty@\advance\dimen@ii\secondy@\divide\dimen@ii\tw@
   \global\multiply\ldimen@ii\tan@i\global\divide\ldimen@ii\tan@ii
   \advance\dimen@ii\ldimen@ii
   \setbox\ZER@\hbox{\ifNESW@\ifnum\arrcount@=\thr@@\hskip4\p@\else
    \hskip\tw@\p@\fi\fi
    $\m@th\tsize@\label@ii$\ifNESW@\else\ifnum\arrcount@=\thr@@\hskip4\p@
    \else\hskip\tw@\p@\fi\fi}%
   \advance\dimen@ii-\ht\ZER@
   \ifnum\slcount@<9 \advance\dimen@ii-\thr@@\p@\fi
   \ifNESW@\let\next@\rlap\else\let\next@\llap\fi
   \hskip\dimen@\next@{\smash{\raise\dimen@ii\box\ZER@}}}%
  \fi
 \fi
}
\def\outCD@#1{\def#1{\Err@{\noexpand#1must not be used within \string\CD}}}
\newskip\preCDskip@
\newskip\postCDskip@
\preCDskip@\z@
\postCDskip@\z@
\def\preCDspace#1{\RIfMIfI@
 \onlydmatherr@\preCDspace\else\advance\preCDskip@#1\relax\fi\else
 \onlydmatherr@\preCDspace\fi}
\def\postCDspace#1{\RIfMIfI@
 \onlydmatherr@\postCDspace\else\advance\postCDskip@#1\relax\fi\else
 \onlydmatherr@\postCDspace\fi}
\def\predisplayspace#1{\RIfMIfI@
 \onlydmatherr@\predisplayspace\else
 \advance\abovedisplayskip#1\relax
 \advance\abovedisplayshortskip#1\relax\fi
 \else\onlydmatherr@\preCDspace\fi}
\def\postdisplayspace#1{\RIfMIfI@
 \onlydmatherr@\postdisplayspace\else
 \advance\belowdisplayskip#1\relax
 \advance\belowdisplayshortskip#1\relax\fi
 \else\onlydmatherr@\postdisplayspace\fi}
\def\PreCDSpace#1{\global\preCDskip@#1\relax}
\def\PostCDSpace#1{\global\postCDskip@#1\relax}
\def\CD#1\endCD{%
 \outCD@\cgaps\outCD@\rgaps\outCD@\Cgaps\outCD@\Rgaps
 \preCD@#1\endCD
 \advance\abovedisplayskip\preCDskip@
 \advance\abovedisplayshortskip\preCDskip@
 \advance\belowdisplayskip\postCDskip@
 \advance\belowdisplayshortskip\postCDskip@
 \vcenter{\offinterlineskip
  \vskip\preCDskip@\Let@\global\colcount@\@ne\global\rowcount@\z@
  \everycr{%
   \noalign{%
    \ifnum\rowcount@=\Rowcount@
    \else
     \getrgap@\rowcount@\vskip\getdim@
     \global\advance\rowcount@\@ne\global\colcount@\@ne
    \fi}}%
  \tabskip\z@
  \halign{&\global\xoff@\z@\global\yoff@\z@
   \getcgap@\colcount@\hskip\getdim@
   \hfil\vrule\height\ten@\p@\width\z@\depth\z@
   $\m@th\displaystyle{##}$\hfil
   \global\advance\colcount@\@ne\cr
   #1\crcr}\vskip\postCDskip@}%
 \preCDskip@\z@\postCDskip@\z@
 \def\getcgap@##1{\ifcase##1\or\getdim@\z@\else\getdim@\standardcgap\fi}%
 \def\getrgap@##1{\ifcase##1\getdim@\z@\else\getdim@\standardrgap\fi}%
 \let\Width@\relax\let\Height@\relax\let\Depth@\relax\let\Rowheight@\relax
 \let\Rowdepth@\relax\let\Colwidth@\relax
}
\let\enddocument\bye
\def\alloc@#1#2#3#4#5{\global\advance\count1#1by\@ne
  \ch@ck#1#4#2%
  \allocationnumber=\count1#1%
  \global#3#5=\allocationnumber
  \wlog{\string#5=\string#2\the\allocationnumber}}
\catcode`\@=\active

\catcode`\"=12
\font\black=cmbx10
\font\sblack=cmbx7
\font\ssblack=cmbx5
\font\sanss=cmss10
\font\ssanss=cmss8 scaled 900
\font\sssanss=cmss8 scaled 600
\font\blackboard=msbm10
\font\sblackboard=msbm7
\font\ssblackboard=msbm5

\def\tx#1{{\fam0\relax#1}}

\newfam\blfam
\textfont\blfam=\black
\scriptfont\blfam=\sblack
\scriptscriptfont\blfam=\ssblack

\newfam\bbfam
\textfont\bbfam=\blackboard
\scriptfont\bbfam=\sblackboard
\scriptscriptfont\bbfam=\ssblackboard
\def\bb#1{{\fam\bbfam\relax#1}}

\newfam\ssfam
\textfont\ssfam=\sanss
\scriptfont\ssfam=\ssanss
\scriptscriptfont\ssfam=\sssanss
\def\ss#1{{\fam\ssfam\relax#1}}

\font\bsymb=cmsy10 scaled\magstep2
\def\all#1{\setbox0=\hbox{\lower1.5pt\hbox{\bsymb \char"38}}\setbox1=\hbox{$_{#1}$} \box0\lower2pt\box1}
\def\exi#1{\setbox0=\hbox{\lower1.5pt\hbox{\bsymb \char"39}}\setbox1=\hbox{$_{#1}$} \box0\lower2pt\box1}

\mathchardef\za="710B  %\alpha
\mathchardef\zb="710C  %\beta
\mathchardef\zg="710D  %\gamma
\mathchardef\zd="710E  %\delta
\mathchardef\zve="710F %\epsilon
\mathchardef\zz="7110  %\zeta
\mathchardef\zh="7111  %\eta
\mathchardef\zvy="7112 %\theta
\mathchardef\zi="7113  %\iota
\mathchardef\zk="7114  %\kappa
\mathchardef\zl="7115  %\lambda
\mathchardef\zm="7116  %\mu
\mathchardef\zn="7117  %\nu
\mathchardef\zx="7118  %\xi
\mathchardef\zp="7119  %\pi
\mathchardef\zr="711A  %\rho
\mathchardef\zs="711B  %\sigma
\mathchardef\zt="711C  %\tau
\mathchardef\zu="711D  %\upsilon
\mathchardef\zvf="711E %\phi
\mathchardef\zq="711F  %\chi
\mathchardef\zc="7120  %\psi
\mathchardef\zw="7121  %\omega
\mathchardef\ze="7122  %\varepsilon
\mathchardef\zy="7123  %\vartheta
\mathchardef\zvp="7124  %\varpi
\mathchardef\zvr="7125 %\varrho
\mathchardef\zvs="7126 %\varsigma
\mathchardef\zf="7127  %\varphi
\mathchardef\zG="7000  %\Gamma
\mathchardef\zD="7001  %\Delta
\mathchardef\zY="7002  %\Theta
\mathchardef\zL="7003  %\Lambda
\mathchardef\zX="7004  %\Xi
\mathchardef\zP="7005  %\Pi
\mathchardef\zS="7006  %\Sigma
\mathchardef\zU="7007  %\Upsilon
\mathchardef\zF="7008  %\Phi
\mathchardef\zC="7009  %\Psi
\mathchardef\zW="700A  %\Omega

\catcode`\"=\active

\loadmsam
\newsymbol\leqslant 1336
\newsymbol\geqslant 133E

\def\*{{\textstyle *}}

\def\R{{\bb R}}

\def\sA{{\ss A}}
\def\sP{{\ss P}}
\def\sT{{\ss T}}
\def\st{{\ss t}}
\def\xd{\tx{d}}
\def\xi{\tx{i}}
\def\xD{\tx{D}}

\def\Cr{\operatorname{Cr}}
\def\im{\operatorname{im}}

\input paper.st\relax
\hsize=37pc
\hoffset=-10pt
\vsize=53pc
\voffset=6pt
\TagsOnRight
\document
\input xy
\xyoption{all}
\def\oT{\overset\circ\to\sT}
\newsymbol\blacktriangle 104E

    \title
        A slow and careful Legendre transformation for singular Lagrangians
    \endtitle

    \author
        W\l odzimierz M. Tulczyjew \\
        Istituto Nazionale di Fisica Nucleare,
        Sezione di Napoli, Italy \\
        {\tt tulczy\@camserv.unicam.it} \\
                \\
        Pawe\l\ Urba\'nski \\
        Division of Mathematical Methods in Physics \\
        University of Warsaw \\
        Ho\D{z}a 74, 00-682 Warszawa \\
        {\tt urbanski\@fuw.edu.pl}

    \endauthor

        \thanks{Supported by KBN, grant No 2 PO3A 074 10}

    \maketitle

    \vskip5mm
    \centerline{Dedicated to the memory of Leopold Infeld, our teacher.}
    \vskip5mm

    \vskip3mm
        \leftline{\black 0. Introduction.}
    \vskip1mm

    In the present note we address the issue of singular Lagrangians in analytical
mechanics.  Deriving Hamiltonian formulations of physical systems with singular
Lagrangians was attempted by Dirac and Bergmann [1].  The aim was to obtain
Hamiltonian formulations of relativistic field theories although Dirac formulated
his theory in terms of finite dimensional geometry.  Applying Dirac procedures to
relativistic mechanical systems we find that in most cases the resulting Hamiltonian
description contains less information than was available in the Lagrangian
formulation.  We propose a version of the Legendre transformation without this
defect.

    In a recent paper Cendra, Holm, Hoyle, and Marsden [2] express the opinion that
Lagrangian systems and Hamiltonian systems offer different representations of the
same object.  The Legendre transformation is the passage from one of these
representations to the other.  We agree with these concepts.  We also agree with the
statement that ``one should {\it do the Legendre transformation slowly and
carefully} when there are degeneracies''.  We think that our Legendre transformation
is slow and careful enough to provide the correct Hamiltonian representation of
relativistic mechanical systems.

    We provide an almost complete although somewhat superficial review of the
geometric background for analytical mechanics.  Complete coordinate
characterizations of all structures are provided.  Intrinsic constructions of most
of the object are given.  A more rigorous version of this material is in preparation.
Related material can be found in [10] [12] [13].

    Much of the material was developed in collaboration with G. Marmo at Istituto
Nazionale di Fisica Nucleare, Sezione di Napoli.

    \vskip3mm
        \leftline{\black 1. Geometry of tangent and cotangent bundles.}
    \vskip1mm

    Let $Q$ be a differential manifold of dimension $m$.  We use a coordinate system
or a chart
        $$\align
    (q^\zk) &\colon Q \rightarrow \R^m \\
    &\colon x \mapsto (q^\zk)(x) = (q^1(x),\ldots ,q^m(x)).
                                                                                \tag \label{Fxx1}\endalign$$
    Each individual coordinate is a function
        $$q^\zk \colon Q \rightarrow \R.
                                                                                \tag \label{Fxx2}$$
    We ignore the fact that the domain of a chart could be an open submanifold of $Q$
and not all of $Q$.

    Let $F$ be a differentiable function on $Q$.  The function
        $$F \circ (q^\zk)^{-1} \colon \R^m \rightarrow \R
                                                                                \tag \label{Fxx3}$$
    is the coordinate expression of the function $F$.  It is a function of the
coordinates $(q^\zk(q)) \in \R^m$ of a point $q \in Q$.  We define partial
derivatives
        $$\partial_\zk F = \frac{\partial(F \circ (q^\zm)^{-1})}{\partial
q^\zk(v)} \circ (q^\zm)
                                                                                \tag \label{Fxx4}$$
    These partial derivatives are functions on $Q$.

    The {\it tangent bundle} of a manifold $Q$ is a manifold $\sT Q$.  There is a
mapping
        $$\zt_Q \colon \sT Q \rightarrow Q
                                                                                \tag \label{Fxx5}$$
     called the {\it tangent fibration}.  Tangent vectors (elements of $\sT Q$) are
equivalence classes of curves in $Q$.  Two curves $\zg \colon \R \rightarrow Q$ and
$\zg' \colon \R \rightarrow Q$ are equivalent if $\zg'(0) = \zg(0)$ and $\xD(f \circ
\zg')(0) = \xD(f \circ \zg)(0)$ for each function $f \colon Q \rightarrow \R$.  The
equivalence class of a curve $\zg \colon \R \rightarrow Q$ will be denoted by
$\st\zg(0)$.  Coordinates
        $$\align
    (q^\zk,\zd q^\zl) &\colon \sT Q \rightarrow \R^{2m} \\
    &\colon v \mapsto (q^1(v),\ldots ,q^m(v),\zd q^1(v),\ldots ,\zd q^m(v))
                                                                                \tag \label{Fxx6}\endalign$$
    are induced by coordinates $(q^\zk)$ in $Q$.  If $\zg$ is a representative of a
vector $v$, then $q^\zk(v) = q^\zk(\zg(0))$ and $\zd q^\zl(v) = \xD(q^\zl \circ
\zg)(0)$.  The tangent fibration is defined by
        $$\zt_Q(\st\zg(0)) = \zg(0).
                                                                                \tag \label{Fxx7}$$
    Fibres of this fibration are vector spaces.  We have operations
        $$+ \colon \sT Q \underset{(\zt_Q,\zt_Q)}\to\times \sT Q \rightarrow \sT Q
                                                                                \tag \label{Fxx8}$$
    and
        $$\cdot\, \colon \R \times \sT Q \rightarrow \sT Q
                                                                                \tag \label{Fxx9}$$
    with coordinate representations
        $$(q^\zk,\zd q^\zl)(v_1 + v_2) = (q^\zk(v_1),\zd q^\zl(v_1) + \zd q^\zl(v_2))
                                                                                \tag \label{Fxx10}$$
    and
        $$(q^\zk,\zd q^\zl)(k\cdot v) = (q^\zk(v),k\zd q^\zl(v)).
                                                                                \tag \label{Fxx11}$$
    We denote by $\sT Q \underset{(\zt_Q,\zt_Q)}\to\times \sT Q$ the set
        $$\{(v_1,v_2) \in \sT Q \times \sT Q ;\; \zt_Q(v_1) = \zt_Q(v_2)\}
                                                                                \tag \label{Fxx12}$$
    Since representatives of vectors (curves in $Q$) can not be added the
construction of linear operations in fibres of $\zt_Q$ is somewhat indirect.  Let $v
= \st\zg(0)$, $v_1 = \st\zg_1(0)$, and $v_2 = \st\zg_2(0)$ be elements of the same
fibre $\sT_q Q = \zt_Q^{-1}(q)$.  We write
        $$v = v_1 + v_2
                                                                                \tag \label{Fxx13}$$
    if
        $$\xD(f \circ \zg)(0) = \xD(f \circ \zg_1)(0) + \xD(f \circ \zg_2)(0)
                                                                                \tag \label{Fxx14}$$
    for each function $f$ on $Q$.  We have defined a relation between three elements
of a fibre $\sT_q Q$.  This relation will turn into a binary operation if we show
that for each pair $(v_1,v_2) \in \sT_q Q \times \sT_q Q$ there is an unique vector
$v \in \sT_q Q$ such that $v = v_1 + v_2$.  The coordinate construction
        $$(q^\zk \circ \zg)(s) = (q^\zk(v_1) + (\zd q^\zk(v_1) + \zd q^\zk(v_2))s)
                                                                                \tag \label{Fxx15}$$
    of a representative $\zg$ of $v$ proves existence.  Let $v = \st\zg(0)$ and $v'
= \st\zg'(0)$ be in relations $v = v_1 + v_2$ and $v' = v_1 + v_2$ with $v_1 =
\st\zg_1(0)$ and $v_2 = \st\zg_2(0)$.  Then
        $$\xD(f \circ \zg')(0) = \xD(f \circ \zg)(0) = \xD(f \circ \zg_1)(0) + \xD(f
\circ \zg_2)(0)
                                                                                \tag \label{Fxx16}$$
    for each function $f$ on $Q$.  It follows that $\zg'$ and $\zg$ represent the
same vector $v' = v$.  This proves uniqueness.  Let $v = \st\zg(0)$ and $u =
\st\zl(0)$ be elements of $\sT_q Q$ and let $k$ be a number.  We write
        $$v = ku
                                                                                \tag \label{Fxx17}$$
    if
        $$\xD(f \circ \zg)(0) = k\xD(f \circ \zl)(0)
                                                                                \tag \label{Fxx18}$$
    for each function $f$ on $Q$.  The coordinate construction
        $$(q^\zk \circ \zg)(s) = (q^\zk(u) + k\zd q^\zk(u)s)
                                                                                \tag \label{Fxx19}$$
    shows that for each $k \in \R$ and $u \in \sT_q Q$ there is a vector $v \in
\sT_q Q$ such that $v = ku$.  If $v = \st\zg(0)$ and $v' = \st\zg'(0)$ are two such
vectors, then
        $$\xD(f \circ \zg')(0) = \xD(f \circ \zg)(0) = k\xD(f \circ \zl)(0).
                                                                                \tag \label{Fxx20}$$
    It follows that the vector $v$ is unique.

    Each curve $\zg \colon \R \rightarrow Q$ has a {\it tangent prolongation}
        $$\align
    \st\zg&\colon \R \rightarrow \sT Q \\
    &\colon s \mapsto \st\zg(\cdot + s)(0).
                                                                                \tag \label{Fxx21}\endalign$$
    The curve $\zg(\cdot + s)$ is the mapping
        $$\align
    \zg(\cdot + s) &\colon \R \rightarrow Q \\
    &\colon s' \mapsto \zg(s' + s)
                                                                                \tag \label{Fxx22}\endalign$$
    The vector $\st\zg(s)$ is the vector tangent to $\zg$ at $\zg(s)$.  The
coordinate description of the prolongation is given by
        $$(q^\zk,\zd q^\zl) \circ \st\zg = (q^\zk \circ \zg,\xD(q^\zl \circ \zg)).
                                                                                \tag \label{Fxx23}$$

    A mapping $X \colon Q \rightarrow  \sT Q$ such that $\zt_Q \circ X \colon Q
\rightarrow Q$ is the identity mapping is called a {\it section} of the fibration
$\zt_Q$.  A section of the tangent fibration is called a {\it vector field}.

    Let $P$ be a differential manifold with coordinates
        $$(p^i) \colon P \rightarrow \R^n
                                                                                \tag \label{Fxx24}$$
    For each differentiable mapping
        $$\za \colon Q \rightarrow P
                                                                                \tag \label{Fxx25}$$
    we have the {\it tangent mapping}
        $$\sT\za \colon \sT Q \rightarrow \sT P.
                                                                                \tag \label{Fxx26}$$
    If $\zg \colon \R \rightarrow Q$ is a representative of a vector $v \in \sT Q$,
then $\za \circ \zg \colon \R \rightarrow P$ is a representative of the vector
$\sT\za(v) \in \sT P$:
        $$\sT\za(\st\zg(0)) = \st(\za \circ \zg)(0).
                                                                                \tag \label{Fxx27}$$
    The coordinate definition of the tangent mapping is given by
        $$(p^i,\zd p^j) \circ \sT\za = (\za^i \circ \zt_Q,(\partial_\zk\za^j \circ
\zt_Q)\zd q^\zk)
                                                                                \tag \label{Fxx28}$$
    with $\za^i = p^i \circ \za$ or by a simplified formula
        $$(p^i,\zd p^j) \circ \sT\za = (\za^i,\partial_\zk\za^j \zd q^\zk).
                                                                                \tag \label{Fxx29}$$
    Einstein's summation convention is used.  The commutative diagram
    \vskip1mm
        $$\xymatrix@C+4mm{{\sT Q} \ar[d]_*{\zt_Q} \ar[r]^*{\sT\za} &
            \sT P \ar[d]_*{\zt_P} \\
            Q \ar[r]^*{\za} & P}
                                                                                \tag \label{Fxx30}$$
    \vskip2mm
    \noindent is a vector fibration morphism.

    A differentiable mapping $\zs \colon T \rightarrow Q$ is called an immersion if
at each point $t \in T$ the linear mapping $\sT_t\zs \colon \sT_t T \rightarrow
\sT_{\zs(t)} Q$ obtained by restricting the mapping $\sT\zs$ to the fibre $\sT_t T =
\zt_T^{-1}(t)$ is injective.  If
        $$(t^i) \colon T \rightarrow \R^k
                                                                                \tag \label{Fxx31}$$
    are coordinates in $T$ and $\zs^\zk = q^\zk \circ \zs$, then $\zs$ is an
immersion if the matrix $(\partial_i\zs^\zk)$ is of maximal rank $k$.  The image $S
= \im(\zs) \subset Q$ is called an (immersed) {\it submanifold} of $Q$ of dimension
$k$.  A submanifold $S \subset Q$ is frequently given as a set
        $$S = \left\{q \in Q ;\; \all{A}F_A(q) = 0 \right\},
                                                                                \tag \label{Fxx32}$$
    where $F_A$ are $m-k$ functions on $Q$ such that the matrix $(\partial_\zk F_A)$
is of maximal rank $m-k$ at points of $S$.  A set $S$ specified in this way is
called an {\it embedded submanifold}.  Submanifolds are usually assumed to be
embedded.  We will adopt the standard practice of not distinguishing elements of
geometric spaces from their coordinates.  Functions defined on these geometric spaces
will be considered functions of coordinates.  Instead of writing a formula \Ref{Fxx32}
we will say that $S$ satisfies equations $F_A(q^\zk) = 0$.  The {\it tangent set} of
a subset $S \subset Q$ (not necessarily a submanifold) is a subset of $\sT Q$.  A
vector $v$ is in $\sT S$ if there is a curve $\zg \colon \R \rightarrow Q$ such that
$v = \st\zg(0)$ and $\zg(s) \in S$ for each $s$ in a neighbourhood of $0 \in \R$.
We have $\zt_Q(\sT S) = S$.  If $S$ is the image of an immersion $\zs \colon T
\rightarrow Q$, then $\sT S$ is the image of $\sT\zs \colon \sT T \rightarrow \sT
Q$.  The coordinates $(q^\zk,\zd q^\zl)$ of elements of $\sT S$ are related to
coordinates $(t^i,\zd t^j)$ by

        $$q^\zk = \zs^\zk(t^i),\; \zd q^\zl = \partial_j \zs^\zl(t^i) \zd t^j.
                                                                                \tag \label{Fxx33}$$
    If $S$ satisfies equations $F_A(q^\zk) = 0$, then $\sT S$ satisfies equations
$\partial_\zk F_A \zd q^\zk = 0$ in addition to $F_A(q^\zk) = 0$.

    A 0-{\it form} on $Q$ is a function on $Q$.  A 1-{\it form} on $Q$ is a mapping
        $$\align
    A&\colon \sT Q \rightarrow \R \\
    &\colon v \mapsto \langle A, v \rangle
                                                                                \tag \label{Fxx34}\endalign$$
    linear on fibres of $\zt_Q$.  The product of a 0-form $F$ with a 1-form $A$ is a
1-form $FA$ defined by
        $$\langle FA, v \rangle = F(\zt_Q(v))\langle A, v \rangle.
                                                                                \tag \label{Fxx35}$$
    The {\it differential} $\xd F$ of a function $F$ on $Q$ is 1-form defined by
        $$\langle \xd F,\st\zg(0) \rangle = \xD(F \circ \zg)(0).
                                                                                \tag \label{Fxx36}$$
    The differential of the product $FG$ of two functions is the 1-form $F\xd G +
G\xd F$.  Coordinates $(\zd q^\zk)$ in $\sT Q$ are 1-forms.  They are the
differentials $(\xd q^\zk)$ of coordinates $(q^\zk)$ in $Q$.  Each 1-form $A$ can be
expressed as a combination
        $$A = A_\zk \xd q^\zk
                                                                                \tag \label{Fxx37}$$
    of these differentials.  The coefficients $A_\zk$ are 0-forms  obtained from
        $$\langle A, v \rangle = A_\zk(v) \zd q^\zk(v)
                                                                                \tag \label{Fxx38}$$
    for each $v \in \sT Q$.  The differential of a function $F(q^\zk)$ is the 1-form
        $$\xd F = \partial_\zl F(q^\zk)\xd q^\zl.
                                                                                \tag \label{Fxx39}$$

    A 2-{\it form} on $Q$ is a function
        $$\align
    B&\colon \sT Q \underset{(\zt_Q,\zt_Q)}\to\times \sT Q \rightarrow \R \\
    &\colon (v_1,v_2) \mapsto \langle B,v_1 \wedge v_2 \rangle,
                                                                                \tag \label{Fxx40}\endalign$$
    which is antisymmetric:
        $$\langle B,v_1 \wedge v_2 \rangle + \langle B,v_2 \wedge v_1 \rangle = 0
                                                                                \tag \label{Fxx41}$$
    and linear in its first argument:
        $$\langle B,(k v_1 + k' v'_1) \wedge v_2 \rangle = k\langle B,v_1 \wedge
v_2\rangle  + k'\langle B,v'_1 \wedge v_2 \rangle .
                                                                                \tag \label{Fxx42}$$
    Linearity in the first argument and antisymmetry imply linearity in the second
argument.  The product of 0-form with a 2-form is a 2-form.  The {\it exterior
product} of 1-forms $A^1$ and $A^2$ is a 2-form $A^1 \wedge A^2$ defined by
        $$\langle A^1 \wedge A^2,v_1 \wedge v_2 \rangle = \langle A^1,v_1\rangle
\langle A^2, v_2\rangle - \langle A^1,v_2\rangle \langle A^2, v_1\rangle.
                                                                                \tag \label{Fxx43}$$
    Each 2-form $B$ is a combination
        $$B = \frac{1}{2} B_{\zk\zl} \xd q^\zk \wedge \xd q^\zl.
                                                                                \tag \label{Fxx44}$$
    The coefficients $B_{\zk\zl}$ are 0-forms characterized by
        $$\langle B, v_1 \wedge v_2 \rangle = \frac{1}{2} B_{\zk\zl}(\zd
q^\zk(v_1)\zd q^\zl(v_2) - \zd q^\zk(v_2)\zd q^\zl(v_1))
                                                                                \tag \label{Fxx45}$$
    and
        $$B_{\zk\zl} + B_{\zl\zk} = 0.
                                                                                \tag \label{Fxx46}$$
    The {\it exterior differential} of a 1-form $A$ is a 2-form $\xd A$.  In order
to construct the exterior differential we associate with each pair $(v_1,v_2) \in
\sT Q \underset{(\zt_Q,\zt_Q)}\to\times \sT Q$ a mapping $\zq \colon \R^2
\rightarrow Q$ such that $v_1 = \st\zq(\cdot,0)$ and $v_2 = \st\zq(0,\cdot)$.  The
coordinate construction
        $$\zq^\zk(s_1,s_2) = q^\zk(\zq(s_1,s_2)) = q^\zk(v_1) + \zd q^\zk(v_1)s_1 +
\zd q^\zk(v_2)s_2
                                                                                \tag \label{Fxx47}$$
    proves the existence of such mappings.  We define curves
        $$\align
    \zx_1&\colon \R \rightarrow \sT Q \\
    &\colon s \mapsto \st\zq(\cdot,s)(0)
                                                                                \tag \label{Fxx48}\endalign$$
    and
        $$\align
    \zx_2&\colon \R \rightarrow \sT Q \\
    &\colon s \mapsto \st\zq(s,\cdot)(0)
                                                                                \tag \label{Fxx49}\endalign$$
    with coordinate representations
        $$(\zx^\zk_1(s_2),\zd\zx^\zl_1(s_2)) = (q^\zk(\zx_1(s_2)),\zd
q^\zl(\zx_1(s_2))) = (\zq^\zk(0,s_2),\partial_{s_1}\zq^\zl(0,s_2))
                                                                                \tag \label{Fxx50}$$
    and
        $$(\zx^\zk_2(s_1),\zd\zx^\zl_2(s_1)) = (q^\zk(\zx_2(s_1)),\zd
q^\zl(\zx_2(s_1))) = (\zq^\zk(s_1,0),\partial_{s_2}\zq^\zl(s_1,0)).
                                                                                \tag \label{Fxx51}$$
    For the mapping defined in \Ref{Fxx47} we have
        $$(\zx^\zk_1(s),\zd\zx^\zl_1(s)) = (q^\zk(v_1) + \zd q^\zk(v_2)s,\zd
q^\zl(v_1))
                                                                                \tag \label{Fxx52}$$
    and
        $$(\zx^\zk_2(s),\zd\zx^\zl_2(s)) = (q^\zk(v_1) + \zd q^\zk(v_1)s,\zd
q^\zl(v_2))
                                                                                \tag \label{Fxx53}$$
    The exterior differential is defined by
        $$\langle \xd A, v_1 \wedge v_2 \rangle = \xD\langle A, \zx_2 \rangle(0) -
\xD\langle A, \zx_1 \rangle(0)
                                                                                \tag \label{Fxx54}$$
    Relations
        $$A^1 \wedge A^2 + A^2 \wedge A^1 = 0,
                                                                                \tag \label{Fxx55}$$
        $$\xd(FA) = \xd F \wedge A + F\xd A,
                                                                                \tag \label{Fxx56}$$
    and
        $$\xd\xd F = 0
                                                                                \tag \label{Fxx57}$$
    are easily established for an arbitrary 0-form $F$ and arbitrary 1-forms $A$,
$A^1$, and $A^2$.  The exterior differential of a 1-form $A = A_\zl \xd q^\zl$ is the
2-form
        $$\xd A = \xd A_\zl \wedge \xd q^\zl = \partial_\zk A_\zl \xd q^\zk \wedge
\xd q^\zl = \frac{1}{2} (\partial_\zk A_\zl - \partial_\zl A_\zk)\xd q^\zk \wedge
\xd q^\zl.
                                                                                \tag \label{Fxx58}$$
    A 2-form which is the differential of a 1-form is said to be {\it exact}.

    A 1-form $A$ is said to be {\it closed} if $\xd A = 0$.  If $A$ is closed, then
there is a neighbourhood $V$ of each point $q_0$ and a 0-form $F$ on $V$ such that
$A|V = \xd F$.  This is as a consequence of the Poincar\'e lemma.

    Let $P$ be a differential manifold with coordinates $(p^i)$ and let $\za \colon
Q \rightarrow P$ be a differentiable mapping.  Let $\za^i = p^i \circ \za$.  The
{\it pull back} of a 0-form $F$ on $P$ is the 0-form $F \circ \za$ on $Q$.  The {\it
pull back} of a 1-form $A$ on $P$ is the 1-form $\za^\* A$ on $Q$ defined by
        $$\langle \za^\* A, v \rangle = \langle A, \sT\za(v) \rangle.
                                                                                \tag \label{Fxx59}$$
    If
        $$A = A_i \xd p^i,
                                                                                \tag \label{Fxx60}$$
    then
        $$\za^\* A = A_i \partial_\zk \za^i \xd q^\zk
                                                                                \tag \label{Fxx61}$$
    The {\it pull back} of a 2-form $B$ on $P$ is the 2-form $\za^\* B$ on $Q$
defined by
        $$\langle \za^\* B, v_1 \wedge v_2 \rangle = \langle B, \sT\za(v_1) \wedge
\sT\za(v_1) \rangle.
                                                                                \tag \label{Fxx62}$$
    If
        $$B = \frac{1}{2} B_{ij} \xd p^i \wedge \xd p^j,
                                                                                \tag \label{Fxx63}$$
    then
        $$\za^\* B = \frac{1}{2} B_{ij} \partial_\zk \za^i \partial_\zl \za^j \xd
q^\zk \wedge \xd q^\zl.
                                                                                \tag \label{Fxx64}$$
    The relations
        $$\xd(\za^\*F) = \za^\*\xd F
                                                                                \tag \label{Fxx65}$$
    and
        $$\xd(\za^\*A) = \za^\*\xd A
                                                                                \tag \label{Fxx66}$$
    hold for a 0-form $F$ and a 1-form $A$.  Let $C \subset Q$ be a submanifold.
The mapping
        $$\align
    \zi_C &\colon C \rightarrow Q \\
    &\colon q \mapsto q
                                                                                \tag \label{Fxx67}\endalign$$
    is the {\it canonical injection}.  The pull backs $\zi_C^\*F$, $\zi_C^\*A$, and
$\zi_C^\*B$ are denoted by $F|C$, $A|C$, and $B|C$ respectively.

    The {\it cotangent bundle} of a manifold $Q$ is a manifold $\sT^\* Q$.  The {\it
cotangent fibration}
        $$\zp_Q \colon \sT^\*Q \rightarrow Q
                                                                                \tag \label{Fxx68}$$
     is the vector fibration dual to the tangent fibration $\zt_Q$.  The {\it
canonical pairing} is a bilinear mapping
        $$\align
    \langle\,,\,\rangle&\colon  \sT^\*Q \underset{(\zp_Q,\zt_Q)}\to\times \sT Q \rightarrow \R \\
    &\colon (f,v) \mapsto \langle f, v \rangle
                                                                                \tag \label{Fxx69}\endalign$$
    defined on the set
        $$\sT^\*Q \underset{(\zp_Q,\zt_Q)}\to\times \sT Q = \left\{(f,v) \in \sT^\*Q
\times \sT Q ;\; \zp_Q(f) = \zt_Q(v)\right\}
                                                                                \tag \label{Fxx70}$$
    Each {\it covector} $f \in \sT^\*_q Q = \zp_Q^{-1}(q)$ is the differential $\xd
F(q)$ of a function $F \colon Q \rightarrow \R$.  Differentials $(\xd q^\zk(q))$
form a basis of the vector space $\sT^\*_q Q$.  Let $(e_\zk(q))$ be the basis of the
vector space $\sT_q Q$ dual to the base $(\xd q^\zk(q))$ in the sense that
        $$\langle \xd q^\zk(q), e_\zl(q) \rangle = \zd^\zk{}_\zl.
                                                                                \tag \label{Fxx71}$$
    Coordinates
        $$(q^\zk,f_\zl) \colon \sT^\*Q \rightarrow \R^{2m}
                                                                                \tag \label{Fxx72}$$
    are defined by
        $$(q^\zk,f_\zl)(f) = (q^\zk(\zp_Q(f)),\langle f, e_\zl(\zp_Q(f))\rangle)
                                                                                \tag \label{Fxx73}$$
    The canonical pairing has the coordinate expression
        $$\langle f, v \rangle = f_\zl(f) \zd q^\zl(v).
                                                                                \tag \label{Fxx74}$$

    For the tangent bundle $\sT\sT^\*Q$ of the cotangent bundle $\sT^\*Q$ we have
the tangent fibration
        $$\zt_{\sT^\*Q} \colon \sT\sT^\*Q \rightarrow \sT^\*Q
                                                                                \tag \label{Fxx75}$$
    and the tangent mapping
        $$\sT\zp_Q \colon \sT\sT^\*Q \rightarrow \sT Q
                                                                                \tag \label{Fxx76}$$
     of the cotangent fibration $\zp_Q \colon \sT^\*Q \rightarrow Q$.  The diagram
    \vskip1mm
        $$\xymatrix@C+6mm{{\sT\sT^\*Q} \ar[d]_*{\zt_{\sT^\*Q}} \ar[r]^*{\sT\zp_Q} &
            \sT Q \ar[d]_*{\zt_Q} \\
            {\sT^\*Q} \ar[r]^*{\zp_Q} & Q}
                                                                                \tag \label{Fxx77}$$
    \vskip2mm
    \noindent
    is commutative.  Hence, $(\zt_{\sT^\*Q}(w),\sT\zp_Q(w)) \in \sT^\*Q
\underset{(\zp_Q,\zt_Q)}\to\times \sT Q$ for each $w \in \sT\sT^\*Q$.  A canonical
1-form $\zy_Q$ on $\sT^\*Q$, called the {\it Liouville form}, is defined by
        $$\langle \zy_Q, w \rangle = \langle \zt_{\sT^\*Q}(w),\sT\zp_Q(w) \rangle.
                                                                                \tag \label{Fxx78}$$
    In the manifold $\sT\sT^\*Q$ we have coordinates
        $$(q^\zk,p_\zl,\zd q^\zm,\zd p_\zn) \colon \sT\sT^\*Q \rightarrow \R^{4m}
                                                                                \tag \label{Fxx79}$$
    related to the coordinates $(q^\zk,f_\zl)$ as the coordinates $(q^\zk,\zd
q^\zl)$ in $\sT Q$ are related to the coordinates $(q^\zk)$ in $Q$.  In terms o
these coordinates, coordinates $(q^\zk,f_\zl)$ in $\sT^\*Q$, and coordinates
$(q^\zk,\zd q^\zl)$ in $\sT Q$ we have the coordinate definitions of the fibrations
$\zt_{\sT^\*Q}$ and $\sT\zp_Q$:
        $$(q^\zk,f_\zl) \circ \zt_{\sT^\*Q} = (q^\zk,f_\zl)
                                                                                \tag \label{Fxx80}$$
    and
        $$(q^\zk,\zd q^\zl) \circ \sT\zp_Q = (q^\zk,\zd q^\zl).
                                                                                \tag \label{Fxx81}$$
    It follows that
        $$\langle \zy_Q, w \rangle = f_\zk(w) \zd q^\zk(w).
                                                                                \tag \label{Fxx82}$$
    Hence,
        $$\zy_Q = f_\zk \xd q^\zk.
                                                                                \tag \label{Fxx83}$$
    A 1-form $A$ on $Q$ is a function on $\sT Q$ but it can be interpreted as a
section $A \colon Q \rightarrow \sT^\*Q$ of the cotangent fibration.  In terms of
this dual interpretation we state the following fundamental property of the
Liouville form:
        $$A^\* \zy_Q = A.
                                                                                \tag \label{Fxx84}$$

    A manifold $P$ and an exact, non degenerate 2-form $\zw$ form an (exact) {\it
symplectic manifold} $(P,\zw)$.  The 2-form $\zw$ defines a mapping $\zb_{(P,\zw)}
\colon \sT P \rightarrow \sT^\*P$ characterized by the equality
        $$\langle \zb_{(P,\zw)}(u), v \rangle = \langle \zw, u \wedge v \rangle
                                                                                \tag \label{Fxx85}$$
    for vectors $u \in \sT P$ and $v \in \sT P$ such that $\zt_P(v) = \zt_P(u)$. The
2-form $\zw$ is said to be {\it non degenerate} if the mapping $\zb_{(P,\zw)}$ is
invertible.  The cotangent bundle $\sT^\*Q$ together with the 2-form
        $$\zw_Q = \xd\zy_Q = \xd f_\zk \wedge \xd q^\zk
                                                                                \tag \label{Fxx86}$$
    form a symplectic manifold $(\sT^\*Q,\zw_Q)$.  In the cotangent bundle
$\sT^\*\sT^\*Q$ we use coordinates
        $$(q^\zk,f_\zl,a_\zm,b^\zn) \colon \sT^\*\sT^\*Q \rightarrow \R^{4m}
                                                                                \tag \label{Fxx87}$$
    induced by coordinates $(q^\zk,f_\zl)$ in $\sT^\*Q$.  The coordinate definition
of the mapping
        $$\zb_{(\sT^\*Q,\zw_Q)} \colon \sT\sT^\*Q \rightarrow \sT^\*\sT^\*Q
                                                                                \tag \label{Fxx88}$$
    is given by
        $$(q^\zk,f_\zl,a_\zm,b^\zn) \circ \zb_{(\sT^\*Q,\zw_Q)} = (q^\zk,f_\zl,\zd
f_\zm,- \zd q^\zn).
                                                                                \tag \label{Fxx89}$$
    This mapping is invertible.  Its inverse
        $$\zb^{-1}_{(\sT^\*Q,\zw_Q)} \colon \sT^\*\sT^\*Q \rightarrow \sT\sT^\*Q
                                                                                \tag \label{Fxx90}$$
    is defined by
        $$(q^\zk,f_\zl,\zd q^\zm,\zd f_\zn) \circ \zb^{-1}_{(\sT^\*Q,\zw_Q)} =
(q^\zk,f_\zl,-b^\zm,a_\zn).
                                                                                \tag \label{Fxx91}$$

    The {\it Poisson bracket}
        $$\{F,G\} \colon \sT^\*Q \rightarrow \R
                                                                                \tag \label{Fxx92}$$
    of two functions $F$ and $G$ on $\sT^\*Q$ is defined by
        $$\{F,G\}(f) = \langle \xd G(f), \zb^{-1}_{(\sT^\*Q,\zw_Q)}(\xd F(f)) \rangle.
                                                                                \tag \label{Fxx93}$$
    It follows from the coordinate relation \Ref{Fxx91} that the Poisson bracket
$\{F,G\}$ of two functions $F(q^\zk,f_\zl)$ and $G(q^\zk,f_\zl)$ is the function
        $$\frac{\partial F}{\partial q^\zk}\frac{\partial G}{\partial f_\zk} -
\frac{\partial G}{\partial q^\zk}\frac{\partial F}{\partial f_\zk}
                                                                                \tag \label{Fxx94}$$
    or
        $$\partial_\zk F \partial^\zk G - \partial_\zk G \partial^\zk F
                                                                                \tag \label{Fxx95}$$
    with the symbol $\partial^\zk$ used to denote the partial derivative with
respect to $f_\zk$.

    \vskip3mm
        \leftline{\black 2. Lagrangian submanifolds.}
    \vskip1mm

    A {\it Lagrangian submanifold} of a general symplectic manifold $(P,\zw)$ is a
submanifold $S \subset P$ of dimension $\dim(S) = \frac{1}{2}\dim(P)$ such that
$\zw|S = 0$.  This last condition means that the symplectic form $\zw$ evaluated on
two vectors tangent to $S$ vanishes.  If $S$ is the image of an immersion $\zs
\colon T \rightarrow P$, then $\zw|S = 0$ is equivalent to $\zs^\*\zw = 0$.

    A Lagrangian submanifold of $(\sT^\*Q,\zw_Q)$ is a submanifold $S \subset
\sT^\*Q$ of dimension $m$ such that $\zw_Q|S = 0$.  If $S$ is the image of an
immersion $\zs \colon T \rightarrow \sT^\*Q$ from a manifold $T$ with coordinates
$(t^\za)$ and
        $$(q^\zk,f_\zl) \circ \zs = (\zs^\zk,\zs_\zl),
                                                                                \tag \label{Fxx96}$$
    then
        $$\zs^\* \zw_Q = \partial_\za\zs_\zk \partial_\zb\zs^\zk \xd t^\za \wedge
\xd t^\zb = \frac{1}{2} \left(\partial_\za\zs_\zk \partial_\zb\zs^\zk -
\partial_\zb\zs_\zk \partial_\za\zs^\zk \right) \xd t^\za \wedge \xd t^\zb.
                                                                                \tag \label{Fxx97}$$
    If $S$ is a Lagrangian submanifold, then the {\it Lagrange brackets}
        $$\partial_\za\zs_\zk \partial_\zb\zs^\zk - \partial_\zb\zs_\zk
\partial_\za\zs^\zk
                                                                                \tag \label{Fxx98}$$
    vanish.  Let $f \in S$ and let $\sT_f S \subset \sT_f\sT^\*Q$ denote the space of
vectors tangent to $S$ at $f$.  Let
        $$\sT^\circ_f S = \left\{a \in \sT^\*_f\sT^\*Q ;\; \all{w \in \sT_f S}
\langle a, w \rangle = 0 \right\}
                                                                                \tag \label{Fxx99}$$
    be the {\it polar} of $\sT_f S$.  If $u \in \sT_f S$, then
        $$\langle \zb_{(\sT^\*Q,\zw_Q)}(u), w \rangle = \langle \zw_Q, u \wedge w
\rangle = 0
                                                                                \tag \label{Fxx100}$$
    for each $w \in \sT_f S$.  Hence, $\zb_{(\sT^\*Q,\zw_Q)}(\sT_f S) \subset
\sT^\circ_f S$.  Since $\dim(\zb_{(\sT^\*Q,\zw_Q)}(\sT_f S)) = \dim(\sT_f S) = m$ and
$\dim(\sT^\circ_f S) = \dim(\sT^\*_f\sT^\*Q) - \dim(\sT_f S) = m$, the spaces
$\sT^\circ_f S$ and $\zb_{(\sT^\*Q,\zw_Q)}(\sT_f S)$ are equal.  If $F$ and $G$ are
functions on $\sT^\*Q$ constant on $S$, then $\xd F(f)$ and $\xd G(f)$ are in
$\sT^\circ_f S$ for each $f \in S$.  It follows that
        $$\{F,G\}|S = 0.
                                                                                \tag \label{Fxx101}$$
    If $S$ is specified by equations $F_A = 0$, where $F_A$ are $m$ independent
functions on $\sT^\*Q$, then
        $$\{F_A,F_B\}|S = 0.
                                                                                \tag \label{Fxx102}$$

    There are three categories of Lagrangian submanifolds of cotangent bundles
generated by increasingly complex objects.

\vskip3mm
\noindent I. {\black Lagrangian submanifolds generated by functions.}
\vskip1mm

    Let $U$ be a function on $Q$.  The image $S$ of the differential $\xd U \colon Q
\rightarrow \sT^\*Q$ is a Lagrangian submanifold of $(\sT^\*Q,\zw_Q)$ since $\dim(S)
= m$ and
        $$(\xd U)^\* \zw_Q = (\xd U)^\* \xd\zy_Q = \xd(\xd U)^\* \zy_Q = \xd\xd U = 0.
                                                                                \tag \label{Fxx103}$$
    The submanifold $S$ is said to be {\it generated} by $U$.  In terms of
coordinates $(q^\zk,f_\zl)$ the set $S$ is described by equations
        $$f_\zl = \partial_\zl U(q^\zk),
                                                                                \tag \label{Fxx104}$$
    equivalent to the simple version of the {\it principle of virtual work}
        $$f_\zl\zd q^\zl = \zd U(q^\zk) = \partial_\zl U(q^\zk)\zd q^\zl,
                                                                                \tag \label{Fxx105}$$
    where the {\it virtual displacements} $\zd q^\zl$ are coordinates of a vector $v
\in \sT Q$.

    Let $S = \im(\zs) \subset \sT^\*Q$ be the image of a 1-form interpreted as a
section $\zs \colon Q \rightarrow \sT^\*Q$ of the cotangent fibration.  From
        $$(\zs)^\* \zw_Q = (\zs)^\* \xd\zy_Q = \xd(\zs)^\* \zy_Q = \xd \zs
                                                                                \tag \label{Fxx106}$$
    it follows that if $S$ is a Lagrangian submanifold, then for each element $f_0
\in S$ there is a neighbourhood $W \subset \sT^\*Q$ of $f_0$ and a function
$U$ on $Q$ such that $S \cap W = \im(\xd U) \cap W$.  This is a version of the
Poincar\'e lemma.

\vskip3mm
\noindent II. {\black Lagrangian submanifolds generated by constrained functions.}
\vskip1mm

    Let $C \subset Q$ be a submanifold of dimension $k$ and let $U \colon C
\rightarrow \R$ be a differentiable function.  The set
        $$S = \left\{f \in \sT^\*Q ;\; q = \zp_Q(f) \in C, \all{v \in \sT_q C
\subset \sT_q Q} \langle f, v\rangle = \langle \xd U, v \rangle \right\}
                                                                                \tag \label{Fxx107}$$
    is an affine subbundle of the cotangent bundle $\sT^\*Q$ restricted to $C$.  At
each point $q \in C$ the fibre $S_q = S \cap \sT_q^\*Q$ is an affine subspace of
$\sT_q^\*Q$ modeled on the vector subspace $\sT^\circ_q C \subset \sT_q^\*Q$ of
dimension $m-k$.  It follows that $S$ is a submanifold of $\sT^\*Q$ of dimension
$m$. We choose a function $\overline U \colon Q \rightarrow \R$ such that $\overline
U|C = U$ and define functions $\widetilde{\overline U} = \overline U \circ \zp_Q$
on $\sT^\*Q$ and $\widetilde{U} = \widetilde{\overline U}|S$ on $S$.  The function
$\widetilde{U}$ does not depend on the choice of the function $\overline U$, it can
be defined directly by $\widetilde U(f) = U(\zp_Q(f))$ for each $f \in S$.  If $w \in
\sT S$, then $\sT\zp_Q(w) \in \sT C$ since $\zp_Q(S) = C$.  From
        $$\align
    \langle \zy_Q, w \rangle &= \langle \zt_{\sT^\*Q}(w),\sT\zp_Q(w) \rangle \\
            &= \langle \xd U,\sT\zp_Q(w) \rangle \\
            &= \langle \xd \overline U,\sT\zp_Q(w) \rangle \\
            &= \langle \xd \widetilde{\overline U}, w \rangle \\
            &= \langle \xd \widetilde{U}, w \rangle
                                                                                \tag \label{Fxx108}\endalign$$
    it follows that
        $$\zy_Q|S = \xd \widetilde{U}
                                                                                \tag \label{Fxx109}$$
    and
        $$\zw_Q|S = \xd\zy_Q|S = \xd(\zy_Q|S) = \xd\xd\widetilde{U} = 0.
                                                                                \tag \label{Fxx110}$$
    Hence, $S$ is a Lagrangian submanifold of $(\sT^\*Q,\zw_Q)$.

    Given a function $\overline U(q^\zk)$ and $m-k$ independent functions
$F_A(q^\zk)$ such that the set $C$ is described by the equations $F_A(q^\zk) = 0$ we
write the principle of virtual work
        $$\aligned
            F_A(q^\zk) &= 0 \\
            f_\zl\zd q^\zl &= \partial_\zl \overline U(q^\zk)\zd q^\zl \\
            \partial_\zl F_A(q^\zk)\zd q^\zl &= 0
        \endaligned
                                                                                \tag \label{Fxx111}$$
    for the set $S$.  Coordinates $(q^\zk,f_\zl)$ of elements of $S$ satisfy the
variational principle with arbitrary virtual displacements $\zd q^\zl$ satisfying the
last equality.  This last equality indicate that the virtual displacements are
coordinates of vectors tangent to $C$.  Using Lagrange multipliers $\zl^A$ we write the
equations for $S$ in the form
        $$\aligned
            F_A(q^\zk) &= 0 \\
            f_\zl &= \partial_\zl \overline U(q^\zk) + \partial_\zl F_A(q^\zk)\zl^A.
        \endaligned
                                                                                \tag \label{Fxx112}$$

    Let $(t^i)$ be the coordinates in $C$ and let $q^\zk = \zs^\zk(t^i)$ be the
coordinate expression of the canonical injection of $C$ in $Q$.  If $U(t^i)$ is the
internal energy, then $S$ is represented by
        $$\aligned
            q^\zk &= \zs^\zk(t^i) \\
            f_\zl \partial_j\zs^\zl(t^i) &= \partial_j U(t^i).
        \endaligned
                                                                                \tag \label{Fxx113}$$

    Let $C \subset Q$ be a submanifold and let $S$ be an affine subbundle of the
cotangent bundle $\sT^\*Q$ restricted to $C$ modeled on the vector subbundle
$\sT^\circ C$ of $\sT^\*Q$ restricted to $C$.  If $S$ is a Lagrangian submanifold of
$(\sT^\*Q,\zw_Q)$, then $\zy_Q|S$ is closed.  Let $f_0$ be an element of $S$ and let
$W \subset \sT^\*Q$ be a neighbourhood of $f_0$ and $\widetilde U$ a function on $S
\cap W$ such that $\zy_Q|S \cap W = \xd \widetilde{U}$.  We choose the neighbourhood
$W$ to have a connected intersection $W_q = S_q \cap W$ with the fibre $S_q = S \cap
\sT_q^\*Q$ for each $q$ in $V = \zp_Q(S \cap W)$.  The restriction of $\zy_Q$ to the
fibre $\sT_q^\*Q$ is the zero form since $\langle \zy_Q, w \rangle = 0$ if
$\sT\zp_Q(w) = 0$.  Consequently,
        $$\xd \widetilde{U}|W_q = \zy_Q|W_q = 0
                                                                                \tag \label{Fxx114}$$
    and the function $\widetilde{U}$ is constant on the connected set $W_q$.  This
permits the introduction of a function $U$ on $C$ such that $\widetilde
U(f) = U(\zp_Q(f))$ for each $f \in S \cap W$.  The set
        $$\left\{f \in \sT^\*Q ;\; q = \zp_Q(f) \in C, \all{v \in \sT_q C
\subset \sT_q Q} \langle f, v\rangle = \langle \xd U, v \rangle \right\}
                                                                                \tag \label{Fxx115}$$
    intersected with $W$ is the intersection of $S$ with $W$.  We have obtained an extension
of the Poincar\'e lemma to constrained Lagrangian submanifolds.

\vskip3mm
\noindent III. {\black Lagrangian submanifolds generated by Morse families.}
\vskip1mm

    Let $\zh \colon Y \rightarrow Q$ be a differential fibration with coordinates
$(q^\zk,y^A)$ adapted in the sense that
        $$(q^\zk) \circ \zh = (q^\zk),
                                                                                \tag \label{Fxx116}$$
    where the coordinates $(q^\zk)$ on the right hand side are coordinates in $Y$.
Let $U \colon Y \rightarrow \R$ be a function interpreted as a family of functions
defined on fibres of the fibration $\zh$.  The family is called a {\it Morse family}
if the $k \times (m + k)$ matrix
        $$\pmatrix
    \dfrac{\partial^2 U}{\partial y^A \partial y^B} & \dfrac{\partial^2 U}{\partial
y^A \partial q^\zk}
            \endpmatrix
                                                                                \tag \label{Fxx117}$$
    is of maximal rank.  A Morse family generates a set
        $$S = \left\{f \in \sT^\*Q ;\; \exi{y \in Y_{\zp_q(f)}} \all{z \in \sT_y Y}
\langle f, \sT\zh(z) \rangle = \langle \xd U, z \rangle \right\}.
                                                                                \tag \label{Fxx118}$$
    The {\it critical set}
        $$\Cr(U,\zh) = \left\{y \in Y;\; \all{w \in \sT_y Y} \sT\zh(w) = 0
\Rightarrow \langle \xd U, w \rangle = 0 \right\}
                                                                                \tag \label{Fxx119}$$
    of the Morse family is a submanifold of $Y$ of dimension $m$.  A mapping
        $$\zk \colon \Cr(U,\zh) \rightarrow \sT^\*Q
                                                                                \tag \label{Fxx120}$$
    such that $\zp_Q(\zk(y)) = \zh(y)$ is defined by
        $$\langle \zk(y), v \rangle = \langle \xd U, w \rangle,
                                                                                \tag \label{Fxx121}$$
    where $v$ is any vector in $\sT_{\zh(y)}$ and $w \in \sT_y Y$ such that
$\sT\zh(w) = v$.  This mapping is an immersion and $S = \im(\zk)$.  Let $y \in
\Cr(U,\zh)$ and $w \in \sT_y \Cr(U,\zh)$.  From
        $$\align
    \langle \zk^\*\zy_Q, w \rangle &= \langle \zy_Q, \sT\zk(w) \rangle \\
    &= \langle \zt_{\sT^\*Q}(\sT\zk(w)), \sT\zp_Q(\sT\zk(w)) \rangle \\
    &= \langle \zk(y), \sT\zh(w) \rangle \\
    &= \langle \xd U, w \rangle
                                                                                \tag \label{Fxx122}\endalign$$
    it follows that
        $$\zk^\*\zy_Q = \xd U|\Cr(U,\zh).
                                                                                \tag \label{Fxx123}$$
    The set $S$ is an immersed Lagrangian submanifold of $(\sT^\*Q,\zw_Q)$ since
        $$\zk^\*\zw_Q = 0
                                                                                \tag \label{Fxx124}$$
    and $\dim(S) = \dim(\Cr(U,\zh)) = m$.

    It follows from a theorem of H\"ormander [4][7] that for each element $f_0$ of a
Lagrangian submanifold of $(\sT^\*Q,\zw_Q)$ there is a neighbourhood $W \subset
\sT^\*Q$ and a Morse family $U \colon Y \rightarrow \R$ of functions on fibres of a
fibration $\zh \colon Y \rightarrow Q$ such that $S$ and the Lagrangian submanifold
generated by $U$ coincide in $W$.  This is an extension of the Poincar\'e lemma.

    The coordinates $(q^\zk,f_\zl)$ of elements of $S$ satisfy equations
        $$\align
    f_\zl &= \partial_\zl U(q^\zk,y^A) \\
    0 &= \partial_B U(q^\zk,y^A)
                                                                                \tag \label{Fxx125}\endalign$$
    derived from the variational principle of virtual work
        $$f_\zl\zd q^\zl = \zd U(q^\zk,y^A) = \partial_\zl U(q^\zk,y^A)\zd q^\zl +
\partial_B U(q^\zk,y^A)\zd y^A
                                                                                \tag \label{Fxx126}$$
    with some values of the variables $(y^A)$ and all variations $(\zd q^\zl,\zd
y^B)$.  The symbol $\partial_A$ stands for the partial derivative with respect to
$y^A$.  Equations \Ref{Fxx125} imply the equalities
        $$\align
    \xd f_\zl &= \partial_\zm\partial_\zl U(q^\zk,y^A)\xd q^\zm + \partial_B\partial_\zl U(q^\zk,y^A)\xd y^B \\
    0 &= \partial_\zl\partial_B U(q^\zk,y^A).
                                                                                \tag \label{Fxx127}\endalign$$
    Consequently,
        $$\zw_Q|S = \xd f_\zl \wedge \xd q^\zl|S = \partial_\zm\partial_\zl
U(q^\zk,y^A)\xd q^\zm \wedge \xd q^\zl = 0.
                                                                                \tag \label{Fxx128}$$
    It follows from the maximality of the rank of the matrix \Ref{Fxx117} that
$\dim(S) = m$.

    Note that the affine subbundle \Ref{Fxx107} is generated by the Morse family
        $$U(q^\zk,y^A) = \overline U(q^\zk) + F_A(q^\zk)y^A.
                                                                                \tag \label{Fxx129}$$
    The rank of the matrix
        $$\pmatrix
    \dfrac{\partial^2 U}{\partial y^A \partial y^B} & \dfrac{\partial^2 U}{\partial
y^A \partial q^\zk}
            \endpmatrix
            = \pmatrix
    0 & \dfrac{\partial F_A}{\partial q^\zk}
            \endpmatrix
                                                                                \tag \label{Fxx130}$$
    is maximal due to independence of the functions $F_A(q^\zk)$.  The function
\Ref{Fxx129} depends linearly on the unrestricted variables $(y^A)$.  This is the
characteristic feature of a Morse family equivalent to a constrained generating
function.  There is little difference between the variables $(y^A)$ and the Lagrange
multipliers $(\zl^A)$.

    A Morse family generating a Lagrangian submanifold is not unique.  It is
frequently possible to reduce the dimension of the fibration $\zh$.  Reductions are
based on the following observation.  Let $S$ be a Lagrangian submanifold of
$(\sT^\*Q,\zw_Q)$ generated by a Morse family $U \colon Y \rightarrow \R$ of
functions defined on fibres of a fibration $\zh \colon Y \rightarrow Q$.  If the
critical set $\Cr(U,\zh)$ is the image of a section $\zs \colon Q \rightarrow Y$ of
$\zh$, then $S$ is generated by the function $U \circ \zs \colon Q \rightarrow \R$.
If $f \in S$ and $q = \zp_Q(f)$, then $f = \zk(\zs(q))$ and
        $$\langle f, v \rangle = \langle \zk(\zs(q)), v \rangle = \langle \xd U,
\sT\zs(v) \rangle = \langle \xd(U \circ \zs), v \rangle
                                                                                \tag \label{Fxx131}$$
    for each $v \in \sT_q Q$.  Hence, $f = \xd(U \circ \zs)(q)$.  This shows that $S
\subset \im(U \circ \zs)$.  If $f = \xd(U \circ \zs)(q)$, then
        $$\langle f, v \rangle = \langle \xd(U \circ \zs), v \rangle = \langle \xd U,
\sT\zs(v) \rangle = \langle \zk(\zs(q)), v \rangle
                                                                                \tag \label{Fxx132}$$
    for each $v \in \sT_q Q$.  Hence, $f = \zk(\zs(q))$.  It follows that $\im(U
\circ \zs) \subset S$.  It may happen that the fibration $\zh$ is the composition
$\zh'' \circ \zh'$ of fibrations $\zh' \colon Y \rightarrow Y'$ and $\zh'' \colon Y'
\rightarrow Q$ and that the critical set $\Cr(U,\zh')$ is the image of a section
$\zs \colon Y' \rightarrow Y$ of $\zh'$.  In this case the Lagrangian submanifold
$S$ is generated by the Morse family $U \circ \zs \colon Y' \rightarrow \R$ of
functions on fibres of $\zh''$.

    \vskip3mm
        \leftline{\black 3. Statics of mechanical systems.}
    \vskip1mm

    Let $Q$ be the {\it configuration space} of a static mechanical system. Elements
of the cotangent bundle $\sT^\*Q$ are the {\it generalized forces} applied to the
system.  The {\it constitutive set} of a static system is subset $S$ (usually a
submanifold) of the cotangent bundle.  An element $f \in S$ is the generalized force
which when applied by an external controlling device will maintain the system in
equilibrium at the configuration $q = \zp_Q(f)$.  The constitutive set provides a
complete characterization of the response of the static system to external control
represented by generalized forces applied to it.  The knowledge of equilibrium
configurations of an isolated system does not characterize the system completely.
Two systems may have the same equilibrium configurations and yet respond differently
to external control.

    The system is said to be {\it reciprocal} if $\zw_Q|S = 0$.  Let $w_1$ and $w_2$
be vectors tangent to $S$ such that $\zt_{\sT^\*Q}(w_2) = \zt_{\sT^\*Q}(w_1)$.  Let
$\zd_1 q^\zk = \zd q^\zk(w_1)$, $\zd_1 f_\zk = \zd f_\zk(w_1)$, $\zd_2 q^\zk = \zd
q^\zk(w_2)$, and $\zd_2 f_\zk = \zd f_\zk(w_2)$.  The equality
        $$\zd_1 f_\zk \zd_2 q^\zk = \zd_2 f_\zk \zd_1 q^\zk
                                                                                \tag \label{Fxx133}$$
    derived from $\langle \zw_Q, w_1 \wedge w_2 \rangle = 0$ is a reciprocity
relation.  The system is said to be {\it potential} if $S$ is a Lagrangian
submanifold generated globally by a generating function, a constrained function or a
Morse family.  The generating function is interpreted as the {\it internal energy}
of the system.  A potential system is reciprocal.

    In the following three examples the configuration space is an affine Euclidean
plane with Cartesian coordinates $(x,y)$.  Coordinates $(x,y,f,g)$ are used in $\sT^\*Q$.

        \claim \c{e}{Example}{}{\rm                                                     \label{Cxx1}
    The function
        $$U(x,y) = \frac{k}{2} (x^2 + y^2)
                                                                                \tag \label{Fxx134}$$
    is the internal energy of an elastically suspended material point.  The
constitutive set $S$ is the Lagrangian submanifold generated by $U$.  It is
described by equations
        $$f = kx,\;\; g = ky
                                                                                \tag \label{Fxx135}$$
    derived from the principle of virtual work
        $$f\zd x + g\zd y = \partial_x U(x,y) \zd x + \partial_y U(x,y) \zd y.
                                                                                \tag \label{Fxx136}$$
        }{\hfill    $\blacktriangle$}\endclaim

        \claim \c{e}{Example}{}{\rm                                                     \label{Cxx2}
    Let $C \subset Q$ be the circle
        $$x^2 + y^2 = a^2.
                                                                                \tag \label{Fxx137}$$
    Let
        $$\overline U(x,y) = ky
                                                                                \tag \label{Fxx138}$$
    represent the internal energy of a material point constrained to the circle
        $$x^2 + y^2 = a^2.
                                                                                \tag \label{Fxx139}$$
    From the variational principle
        $$\aligned
            x^2 + y^2 &= 0 \\
            f\zd x + g\zd y &= k\zd y \\
            x\zd x + y\zd y &= 0
        \endaligned
                                                                                \tag \label{Fxx140}$$
    we derive equations
        $$\aligned
            x^2 + y^2 &= 0 \\
            f &= \zl x \\
            g &= k + \zl y
        \endaligned
                                                                                \tag \label{Fxx141}$$
    for the constitutive set $S$ with a Lagrange multiplier $\zl$.  With the
parametric representation
        $$\aligned
            x &= a\cos\zy \\
            y &= a\sin\zy
        \endaligned
                                                                                \tag \label{Fxx142}$$
    we obtain the expression $U(\zy) = ka\sin\zy$ for the internal energy and the
variational principle
        $$\aligned
            x &= a\cos\zy \\
            y &= a\sin\zy \\
            - fa\sin\zy\zd\zy + ga\cos\zy\zd\zy &= ka\cos\zy\zd\zy
        \endaligned
                                                                                \tag \label{Fxx143}$$
    equivalent to \Ref{Fxx140}.  The constitutive set is generated by the Morse family
        $$U(x,y,\zl) = ky + \frac{\zl}{2} (x^2 + y^2 - a^2).
                                                                                \tag \label{Fxx144}$$
        }{\hfill    $\blacktriangle$}\endclaim

        \claim \c{e}{Example}{}{\rm                                                     \label{Cxx3}
    The function
        $$U(x,y,\zy) = \frac{k}{2} ((x - a\cos \zy)^2 + (y - a\sin \zy)^2)
                                                                                \tag \label{Fxx145}$$
    is the internal energy of a material point tied elastically to a point left to
move freely on the circle
        $$x = a\cos \zy,\;\; y = a\sin \zy.
                                                                                \tag \label{Fxx146}$$
    The function $U$ is a Morse family of functions of the variable $\zy$ since the
rank of the $1 \times 3$ matrix
        $$\pmatrix
    \dfrac{\partial^2 U}{\partial\zy \partial\zy} & \dfrac{\partial^2 U}{\partial\zy \partial
x} & \dfrac{\partial^2 U}{\partial\zy \partial y}
            \endpmatrix = \pmatrix ka(x\cos\zy + y\sin\zy),& ka\sin\zy,& -ka\cos\zy
        \endpmatrix
                                                                                \tag \label{Fxx147}$$
    is 1.  From the principle of virtual work
        $$f\zd x + g\zd y = \zd U(x,y,\zy) = k(x - a\cos\zy)\zd x + k(y -
a)\sin\zy)\zd y + ka(x\sin\zy - y\cos\zy)\zd\zy
                                                                                \tag \label{Fxx148}$$
    we obtain equations
        $$\aligned
            f &= k(x - a\cos\zy) \\
            g &= k(y - a\sin\zy) \\
            0 &= ka(x\sin\zy - y\cos\zy)
        \endaligned
                                                                                \tag \label{Fxx149}$$
    for the constitutive set $S$.  Equations
        $$\aligned
            x &= \zr\cos\zy \\
            y &= \zr\sin\zy \\
            f &= k(\zr - a)\cos\zy \\
            g &= k(\zr - a)\sin\zy
        \endaligned
                                                                                \tag \label{Fxx150}$$
    represent a mapping $\zs$ from $\R^2$ to $\sT^\*Q$.  The set $S$ is the image of this
mapping.  The matrix
        $$\pmatrix
    \dfrac{\partial x}{\partial\zr} & \dfrac{\partial y}{\partial\zr} &
\dfrac{\partial f}{\partial\zr} & \dfrac{\partial g}{\partial\zr} \\
    \dfrac{\partial x}{\partial\zy} & \dfrac{\partial y}{\partial\zy} &
\dfrac{\partial f}{\partial\zy} & \dfrac{\partial g}{\partial\zy}
            \endpmatrix = \pmatrix \cos\zy & \sin\zy & k\cos\zy & k\sin\zy \\
    - \zr\sin\zy & \zr\cos\zy & - k(\zr - a)\sin\zy & k(\zr - a)\cos\zy
        \endpmatrix
                                                                                \tag \label{Fxx151}$$
    is of rank 2.  This indicates that $S$ is an immersed submanifold.  With the
exclusion of points corresponding to $\zr = 0$ the set $S$ is the union of images of
two sections of $\zp_Q$ corresponding to the two different signs in the formulae
        $$\aligned
            f &= \frac{kx}{\sqrt{x^2 + y^2}}\left(\sqrt{x^2 + y^2} \pm a \right) \\
            g &= \frac{ky}{\sqrt{x^2 + y^2}}\left(\sqrt{x^2 + y^2} \pm a \right).
        \endaligned
                                                                                \tag \label{Fxx152}$$
    With the exclusion of points corresponding to $x^2 + y^2 \geq a^2$, $S$ is  the set  of
points satisfying equations
        $$\aligned
            F^0_x(x,y,f,g)  &= x - \frac{f}{k\sqrt{f^2 + g^2}}\left(\sqrt{f^2 + g^2}
- ka\right) = 0 \\
            F^0_y(x,y,f,g)  &= y - \frac{g}{k\sqrt{f^2 + g^2}}\left(\sqrt{f^2 + g^2}
- ka\right) = 0.
        \endaligned
                                                                                \tag \label{Fxx153}$$
    The functions $F^0_x$ and $F^0_y$ are obviously independent.  It follows that $S$
is an embedded submanifold.  The rank of the Jacobian
        $$\pmatrix
    \dfrac{\partial x}{\partial\zr} & \dfrac{\partial y}{\partial\zr} \\
    \dfrac{\partial x}{\partial\zr} & \dfrac{\partial y}{\partial\zr} \\
            \endpmatrix = \pmatrix \cos\zy & \sin\zy \\
    - \zr\sin\zy & \zr\cos\zy
        \endpmatrix
                                                                                \tag \label{Fxx154}$$
    of the mapping $\zp_Q \circ \zs$ represented by
        $$\aligned
            x &= \zr\cos\zy \\
            y &= \zr\sin\zy
        \endaligned
                                                                                \tag \label{Fxx155}$$
    changes from 2 to 1 at $\zr = 0$.  This indicates the presence of a {\it
Lagrangian singularity} above the point with coordinates $(x,y) = (0,0)$.
        }{\hfill    $\blacktriangle$}\endclaim

    \vskip3mm
        \leftline{\black 4. Differential equations.}
    \vskip1mm

    The tangent fibration $\zt_q \colon \sT Q \rightarrow Q$ was introduced in
Section 1.  Elements of the tangent bundle $\sT Q$ were interpreted as virtual
displacements.  We return to the topic of tangent vectors this time interpreted as
velocities.  Coordinates $(q^\zk,\zd q^\zl) \colon \sT Q \rightarrow \R^{2m}$
introduced in Section 1 will be now denoted by $(q^\zk,\dot q^\zl)$.

    A {\it first order differential equation} in $Q$ is a submanifold $D \subset \sT
Q$.  A curve $\zg \colon I \rightarrow Q$ defined on an open interval $I \subset \R$
is said to be a {\it solution} of $D$ if for each $t \in I$ the vector $\st\zg(t)$
tangent to $\zg$ at $\zg(t)$ is an element of $D$.  A differential equation $D$ is
said to be {\it integrable} if for each $v \in D$ there is a solution $\zg \colon I
\rightarrow Q$ of $D$ such that $v = \st\zg(t_0)$ for some $t_0 \in I$.

    Not all differential equations are integrable.  Let $D \subset \sT Q$ be a
differential equation and let $C$ be the set $\zt_Q(D)$. If $v \in D$ and $D$ is
integrable, then there is a solution $\zg \colon I \rightarrow Q$ of $D$ such that
$v = \st\zg(t_0)$ for some $t_0 \in I$.  Since $\st\zg(t) \in D$ for each $t \in I$,
it follows that $\zg(t) \in C$ for each $t \in I$.  Consequently $\st\zg(t) \in \sT
C$ for each $t \in I$ and $v = \st\zg(t_0) \in \sT C$.  We have shown that the
condition $D \subset \sT C$ is necessary for integrability of the equation $D$.  This
condition is sufficient for a class of differential equations described below.

    The image $D = \im(X)$ of a vector field $X \colon Q \rightarrow \sT Q$ is an
integrable differential equation.  Let $C \subset Q$ be a submanifold and let $D$ be
the union
        $$\bigcup_{\za \in \sA}\{\im(X_\za|C)\}
                                                                                \tag \label{Fxx156}$$
    of a family of vector fields
        $$X_\za \colon Q \rightarrow \sT Q
                                                                                \tag \label{Fxx157}$$
    restricted to $C$.  If $D \subset \sT C$, then each field $X_\za$ induces a
vector field
        $$\align
    \overline{X}_\za&\colon C \rightarrow \sT C \\
    &\colon q \mapsto X_\za(q)
                                                                                \tag \label{Fxx158}\endalign$$
    since $\im(X_\za|C) \subset D \subset \sT C$.  The differential equation
$\im(\overline{X}_\za)$ is integrable for each $\za \in \sA$ and
        $$D = \bigcup_{\za \in \sA}\{\im(\overline{X}_\za)\}.
                                                                                \tag \label{Fxx159}$$
    Hence, $D$ is integrable.

    If the necessary condition $D \subset \sT\zt_Q(D)$ is not satisfied, then the
reduced equation $D \cap \sT\zt_Q(D)$ is closer to being integrable although the
condition $D \cap \sT\zt_Q(D) \subset \sT\zt_Q(D \cap \sT\zt_Q(D))$ is not
necessarily satisfied.  This observation suggests the following algorithm for
extracting the integrable part of a differential equation.  We consider the sequence
of sets
        $$\overline{C}^0 = \zt_Q(D), \overline{C}^1 = \zt_Q(D \cap \sT
\overline{C}^0), \dots, \overline{C}^k = \zt_Q(D \cap \sT \overline{C}^{k-1}), \ldots
                                                                                \tag \label{Fxx160}$$
    and the sequence of differential equations
        $$\overline D^0 = D, \overline D^1 = D \cap \sT \overline{C}^0, \dots,
\overline D^k = D \cap \sT \overline{C}^{k-1}, \ldots
                                                                                \tag \label{Fxx161}$$
    It may happen that after a finite number of steps the sets in the sequence
\Ref{Fxx160} are all equal to a set $\overline C$.  This set satisfies the equality
        $$\overline C = \zt_Q(D \cap \sT\overline C).
                                                                                 \tag \label{Fxx162}$$
    If the differential equation $\overline D = D \cap \sT\overline C$ is
integrable, then it is the integrable part of $D$.

    In Section 11 we give an example of a version of the above algorithm applied to
a Hamiltonian system.

    Other algorithms for extracting the integrable part of a differential equations
have been designed.  They require the use of higher order tangent vectors.

    The {\it second tangent bundle} of a manifold $Q$ is the set $\sT^2 Q$ of
equivalence classes of curves in $Q$.  Two curves $\zg \colon \R \rightarrow Q$ and
$\zg' \colon \R \rightarrow Q$ are equivalent if $\zg'(0) = \zg(0)$, $\xD(f \circ
\zg')(0) = \xD(f \circ \zg)(0)$, and $\xD^2(f \circ \zg')(0) = \xD^2(f \circ \zg)(0)$
for each function $f \colon Q \rightarrow \R$.  We use coordinates
        $$(q^\zk,\dot q^\zl,\ddot q^\zm) \colon \sT^2 Q \rightarrow \R^{3m}
                                                                                \tag \label{Fxx163}$$
    in $\sT^2 Q$.  If $\zg$ is a representative of a second tangent vector $a \in
\sT^2 Q$, then $q^\zk(a) = q^\zk(\zg(0))$, $\dot q^\zl(a) = \xD(q^\zl \circ
\zg)(0)$, and $\ddot q^\zm(a) = \xD^2(q^\zm \circ \zg)(0)$.  The equivalence class
of a curve $\zg \colon \R \rightarrow Q$ will be denoted by $\st^2\zg(0)$.  Each
curve $\zg \colon \R \rightarrow Q$ has a {\it second tangent prolongation}
        $$\align
    \st^2\zg&\colon \R \rightarrow \sT^2 Q \\
    &\colon t \mapsto \st^2\zg(\cdot + t)(0)
                                                                                \tag \label{Fxx164}\endalign$$
    The coordinate description of the prolongation is given by
        $$(q^\zk,\dot q^\zl,\ddot q^\zm) \circ \st^2\zg = (q^\zk \circ \zg,\xD(q^\zl
\circ \zg),\xD^2(q^\zm \circ \zg)).
                                                                                \tag \label{Fxx165}$$
    The {\it second tangent fibration} is the mapping
        $$\align
    \zt_{2\,Q}&\colon \sT^2 Q \rightarrow Q \\
    &\colon \st\zg(0) \mapsto \zg(0).
                                                                                \tag \label{Fxx166}\endalign$$
    There is also the fibration
        $$\align
    \zt^1{}_{2\,Q}&\colon \sT^2 Q \rightarrow \sT Q \\
    &\colon \st^2\zg(0) \mapsto \st\zg(0).
                                                                                \tag \label{Fxx167}\endalign$$

    A {\it second order differential equation} in $Q$ is a submanifold $E \subset
\sT^2 Q$.  A {\it solution} is a curve $\zg \colon I \rightarrow Q$ such that
$\st^2\zg(t) \in E$ for each $t$ in the open interval $I \subset \R$.  The concept of
integrability is easily extended to second order equations.  The image $\im(X)$ of a
section
        $$X \colon \sT Q \rightarrow \sT^2 Q
                                                                                \tag \label{Fxx168}$$
    of the fibration $\zt^1{}_{2\,Q}$ is an integrable differential equation.

    Elements of the iterated tangent bundle $\sT\sT Q$ are equivalence classes of
curves in $\sT Q$.  Coordinates
        $$(q^\zk,\dot q^\zl,q'{}^\zm,\dot q'{}^\zn) \colon \sT Q \rightarrow \R^{4m}
                                                                                \tag \label{Fxx169}$$
    will be used.  These coordinates are related to coordinates $(q^\zk,\dot q^\zl)$
as coordinates $(q^\zk,\dot q^\zl)$ are related to coordinates $(q^\zk)$.  We have
fibrations
        $$\zt_{\sT Q} \colon \sT\sT Q \rightarrow \sT Q
                                                                                \tag \label{Fxx170}$$
    and
        $$\sT\zt_Q \colon \sT\sT Q \rightarrow \sT Q
                                                                                \tag \label{Fxx171}$$
    with coordinate representations
        $$(q^\zk,\dot q^\zl) \circ \zt_{\sT Q} = (q^\zk,\dot q^\zl)
                                                                                \tag \label{Fxx172}$$
    and
        $$(q^\zk,\dot q^\zl) \circ \sT\zt_Q = (q^\zk,q'{}^\zl).
                                                                                \tag \label{Fxx173}$$

    There is an useful immersion $\zl_Q$ of $\sT^2 Q$ in $\sT\sT Q$.  This immersion
associates with a second tangent vector $a = \st^2\zg(0)$ the vector $w = \st\st\zg(0)$
tangent to the prolongation $\st\zg$ of the curve $\zg$ at $\st\zg(0)$.  The formal
definition is expressed in
        $$\align
    \zl_Q &\colon \sT^2 Q \rightarrow \sT\sT Q \\
    &\colon \st^2\zg(0) \mapsto \st\st\zg(0).
                                                                                \tag \label{Fxx174}\endalign$$
    From $\zt_{\sT Q}(\st\st\zg(0)) = \st\zg(0)$, $\sT\zt_Q(\st\st\zg(0)) =
\st\zg(0)$, and $\zt^1{}_{2\,Q}(\st^2\zg(0)) = \st\zg(0)$ it follows that
        $$\zt_{\sT Q} \circ \zl_Q = \sT\zt_Q \circ \zl_Q = \zt^1{}_{2\,Q}.
                                                                                \tag \label{Fxx175}$$

    Let $D \subset \sT Q$ be a differential equation.  The set
        $$\sP D = \zl_Q^{-1}(\sT D) \subset \sT^2 Q
                                                                                \tag \label{Fxx176}$$
    is a second order differential equation called the {\it prolongation} of $D$.  If
the differential equation is given in the form
        $$D = \left\{v \in \sT Q ;\; \all{i}f_i(v) = 0 \right\},
                                                                                \tag \label{Fxx177}$$
    where $f_i$ are functions on $\sT Q$, then $\sT D$ is the set
        $$\left\{w \in \sT\sT Q ;\; \all{i}f_i(\zt_{\sT Q}(w)) = 0, \partial_\zm
f_i(\zt_{\sT Q}(w))q'{}^\zm(a) + \partial_{\dot\zm} f_i(\zt_{\sT Q}(w))\dot
q'{}^\zm(a) = 0 \right\}
                                                                                \tag \label{Fxx178}$$
    and $\sP D$ is the set
        $$\left\{a \in \sT^2 Q ;\; \all{i}f_i(\zt^1{}_{2\,Q}(a)) = 0, \partial_\zm
f_i(\zt^1{}_{2\,Q}(a))\dot q^\zm(a) + \partial_{\dot\zm} f_i(\zt^1{}_{2\,Q}(a))\ddot
q^\zm(a) = 0 \right\}.
                                                                                \tag \label{Fxx179}$$
    The symbol $\partial_{\dot\zm}$ stands for the partial derivative with respect
to $\dot q^\zm$.

    The inclusion
        $$\zt^1{}_{2\,Q}(\sP D) \subset D
                                                                                \tag \label{Fxx180}$$
    follows from
        $$\zt^1{}_{2\,Q}(\sP D) = \zt_{\sT Q}(\zl_Q(\sP D)) = \zt_{\sT
Q}(\zl_Q(\zl_Q^{-1}(\sT D))) \subset \zt_{\sT Q}(\sT D) = D.
                                                                                \tag \label{Fxx181}$$
    Let $D \subset \sT Q$ be an integrable equation.  If $v \in D$, then there is a
solution $\zg \colon I \rightarrow Q$ of $D$ such that $\st\zg(0) = v$.  We have
$\zl_Q(\st^2\zg(0)) = \st\st\zg(0) \in \sT D$ since $\st\zg(t) \in D$ for each $t
\in I$.  It follows that $v = \st\zg(0) = \zt^1{}_{2\,Q}(\st^2\zg(0)) \in
\zt^1{}_{2\,Q}(\sP D)$.  Hence,
        $$D \subset \zt^1{}_{2\,Q}(\sP D)
                                                                                \tag \label{Fxx182}$$
     if $D$ is integrable.  We have established a necessary condition
        $$\zt^1{}_{2\,Q}(\sP D) = D
                                                                                \tag \label{Fxx183}$$
    for integrability of a differential equation $D \subset \sT Q$.  If this
condition is not satisfied, then the integrable part of $D$ is a subset of the set
$\zt^1{}_{2\,Q}(\sP D)$.  The set $\zt^1{}_{2\,Q}(\sP D)$ is a subset of $D$ closer
to the integrable part without being necessarily integrable.  These observations
suggest a new algorithm for extracting the integrable part of a differential
equation.  We introduce the sequence of differential equations
        $$\widetilde D^0 = D, \widetilde D^1 = \zt^1{}_{2\,Q}(\sP \widetilde D^0),
\dots, \widetilde D^k = \zt^1{}_{2\,Q}(\sP \widetilde{D}^{k-1}), \ldots
                                                                                \tag \label{Fxx184}$$
    It may happen that after a finite number of steps the sets in the sequence
\Ref{Fxx184} are all equal to a set $\widetilde D$.  It may happen that $\widetilde D$
is the integrable part of $D$.

    \vskip3mm
        \leftline{\black 5. The iterated tangent bundle.}
    \vskip1mm

    We have already introduced the iterated tangent bundle $\sT\sT Q$ and the
coordinates
        $$(q^\zk,\dot q^\zl,q'{}^\zm,\dot q'{}^\zn) \colon \sT Q \rightarrow \R^{4m}.
                                                                                \tag \label{Fxx185}$$
    The fibration
        $$\zt_{\sT Q} \colon \sT\sT Q \rightarrow \sT Q
                                                                                \tag \label{Fxx186}$$
    is a vector fibration.  We have the operations
        $$+ \colon \sT\sT Q \underset{(\zt_{\sT Q},\zt_{\sT Q})}\to\times  \sT\sT Q
\rightarrow \sT\sT Q
                                                                                \tag \label{Fxx187}$$
    and
        $$\cdot\, \colon \R \times \sT\sT Q \rightarrow \sT\sT Q
                                                                                \tag \label{Fxx188}$$
    with coordinate representations
        $$(q^\zk,\dot q^\zl,q'{}^\zm,\dot q'{}^\zn)(w_1 + w_2) = (q^\zk(w_1),\dot
q^\zl(w_1),q'{}^\zm(w_1) + q'{}^\zm(w_2),\dot q'{}^\zn(w_1) + \dot q'{}^\zn(w_2))
                                                                                \tag \label{Fxx189}$$
    and
        $$(q^\zk,\dot q^\zl,q'{}^\zm,\dot q'{}^\zn)(k\cdot w) = (q^\zk(w),\dot
q^\zl(w),kq'{}^\zm(w),k\dot q'{}^\zn(w)).
                                                                                \tag \label{Fxx190}$$
    The diagram
    \vskip1mm
        $$\xymatrix@R+3mm @C+6mm{{\sT\sT Q} \ar[d]_*{\zt_{\sT Q}} \ar[r]^*{\sT\zt_Q} &
            \sT Q \ar[d]_*{\zt_Q} \\
            {\sT Q} \ar[r]^*{\zt_Q} & Q}
                                                                                \tag \label{Fxx191}$$
    \vskip2mm
    \noindent is a vector fibration morphism.

    We show that the mapping
        $$\sT\zt_Q \colon \sT\sT Q \rightarrow \sT Q
                                                                                \tag \label{Fxx192}$$
    is a vector fibration by constructing operations
        $$\tilde+ \colon \sT\sT Q \underset{(\sT\zt_Q,\sT\zt_Q)}\to\times  \sT\sT Q
\rightarrow \sT\sT Q
                                                                                \tag \label{Fxx193}$$
    and
        $$\tilde\cdot\, \colon \R \times \sT\sT Q \rightarrow \sT\sT Q.
                                                                                \tag \label{Fxx194}$$
    Let $w_1$ and $w_2$ be elements of $\sT\sT Q$ such that $\sT\zt_Q(w_2) =
\sT\zt_Q(w_1)$.  It is possible to choose curves $\zx_1 \colon \R \rightarrow \sT Q$
and $\zx_2 \colon \R \rightarrow \sT Q$ such that $w_1 = \st\zx_1(0)$, $w_2 =
\st\zx_2(0)$ and $\zt_Q \circ \zx_2 = \zt_Q \circ \zx_1$.  The coordinate
constructions
        $$(q^\zk,\dot q^\zl) \circ \zx_1 = (q^\zk(w_1) + q'{}^\zk(w_1)s,\dot
q^\zl(w_1) + \dot q'{}^\zl(w_1)s)
                                                                                \tag \label{Fxx195}$$
    and
        $$(q^\zk,\dot q^\zl) \circ \zx_2 = (q^\zk(w_2) + q'{}^\zk(w_2)s,\dot
q^\zl(w_2) + \dot q'{}^\zl(w_2)s)
                                                                                \tag \label{Fxx196}$$
    provide an example.  The operation $\,\tilde+\,$ is defined by
        $$w_1 \,\tilde+\, w_2 = \st(\zx_1 + \zx_2)(0).
                                                                                \tag \label{Fxx197}$$
    The operation $\,\tilde\cdot\,$ is defined by
        $$k \,\tilde\cdot\, \st\zx(0) = \st(k\zx)(0).
                                                                                \tag \label{Fxx198}$$
    Coordinate representations of these operations are given by
        $$(q^\zk,\dot q^\zl,q'{}^\zm,\dot q'{}^\zn)(w_1 \,\tilde+\, w_2) =
(q^\zk(w_1),\dot q^\zl(w_1) + \dot q^\zl(w_2),q'{}^\zm(w_1),\dot q'{}^\zn(w_1) +
\dot q'{}^\zn(w_2))
                                                                                \tag \label{Fxx199}$$
    and
        $$(q^\zk,\dot q^\zl,q'{}^\zm,\dot q'{}^\zn)(k\,\tilde\cdot\, w) =
(q^\zk(w),k\dot q^\zl(w),q'{}^\zm(w),k\dot q'{}^\zn(w)).
                                                                                \tag \label{Fxx200}$$
    The diagram
    \vskip1mm
        $$\xymatrix@R+3mm @C+6mm{{\sT\sT Q} \ar[r]^*{\zt_{\sT Q}} \ar[d]_*{\sT\zt_Q} &
            \sT Q \ar[d]_*{\zt_Q} \\
            {\sT Q} \ar[r]^*{\zt_Q} & Q}
                                                                                \tag \label{Fxx201}$$
    \vskip2mm
    \noindent is a vector fibration morphism.

    Elements of the iterated bundle $\sT\sT Q$ are equivalence classes of curves in
a set of equivalence classes of curves in $Q$.  A simpler representation of these
elements is needed.  Let $\zq \colon \R^2 \rightarrow Q$ be a differentiable
mapping.  For each $s \in \R$ we denote by $\st^{(0,1)}\zq(s,0)$ the vector
$\st\zq(s,\cdot)(0) \in \sT Q$. For each $t \in \R$ we denote by
$\st^{(1,0)}\zq(0,t)$ the vector $\st\zq(\cdot,t)(0) \in \sT Q$.  We have curves
        $$\st^{(0,1)}\zq(\cdot,0) \colon \R \rightarrow \sT Q
                                                                                \tag \label{Fxx202}$$
    and
        $$\st^{(1,0)}\zq(0,\cdot) \colon \R \rightarrow \sT Q.
                                                                                \tag \label{Fxx203}$$
    Vectors $\st\st^{(0,1)}\zq(\cdot,0)(0) \in \sT\sT Q$ and
$\st\st^{(1,0)}\zq(0,\cdot)(0) \in \sT\sT Q$ will be denoted by
$\st\st^{(0,1)}\zq(0,0)$ and $\st\st^{(1,0)}\zq(0,0)$ respectively.  For each $w \in
\sT\sT Q$ there is a mapping $\zq \colon \R^2 \rightarrow Q$ such that $w =
\st\st^{(0,1)}\zq(0,0)$.  The mapping specified by coordinate relations
        $$(q^\zk \circ \zq)(s,t) = (q^\zk(w) + \dot q^\zk(w)t + q'{}^\zk(w)s + \dot
q'{}^\zk(w)st)
                                                                                \tag \label{Fxx204}$$
    has the required property.  We consider mappings $\zq \colon \R^2 \rightarrow Q$
and $\zq' \colon \R^2 \rightarrow Q$ equivalent if
        $$\st\st^{(0,1)}\zq'(0,0) =
\st\st^{(0,1)}\zq(0,0).
                                                                                \tag \label{Fxx205}$$
    These mappings are equivalent if
        $$\zq'(0,0) = \zq(0,0),
                                                                                \tag \label{Fxx206}$$
        $$\xD^{(1,0)}\zq'(0,0) = \xD^{(1,0)}\zq(0,0),
                                                                                \tag \label{Fxx207}$$
        $$\xD^{(0,1)}\zq'(0,0) = \xD^{(0,1)}\zq(0,0),
                                                                                \tag \label{Fxx208}$$
    and
        $$\xD^{(1,1)}\zq'(0,0) = \xD^{(1,1)}\zq(0,0).
                                                                                \tag \label{Fxx209}$$
    We have obtained an efficient representation of elements of $\sT\sT Q$.  In
terms of this representation we define the {\it canonical involution}
        $$\align
    \zk_Q &\colon \sT\sT Q \rightarrow \sT\sT Q \\
    &\colon \st\st^{(0,1)}\zq(0,0) \mapsto \st\st^{(1,0)}\zq(0,0) =
\st\st^{(0,1)}\widetilde\zq(0,0),
                                                                                \tag \label{Fxx210}\endalign$$
    with
        $$\align
    \widetilde\zq &\colon \R^2 \rightarrow Q \\
    &\colon (s,t) \mapsto \zq(t,s).
                                                                                \tag \label{Fxx211}\endalign$$
    The coordinate expression of this involution is given by
        $$(q^\zk,\dot q^\zl,q'{}^\zm,\dot q'{}^\zn) \circ \zk_Q =
(q^\zk,q'{}^\zl,\dot q^\zm,\dot q'{}^\zn).
                                                                                \tag \label{Fxx212}$$
    The commutative diagram
    \vskip1mm
        $$\xymatrix@R+3mm @C+6mm{{\sT\sT Q} \ar[d]_*{\sT\zt_Q} \ar[r]^*{\zk_Q} &
            \sT\sT Q \ar[d]_*{\zt_{\sT Q}} \\
            {\sT Q} \ar@{=}[r] & \sT Q}
                                                                                \tag \label{Fxx213}$$
    \vskip2mm
    \noindent is a vector fibration isomorphism.  The diagram
    \vskip1mm
        $$\xymatrix@R+3mm @C+6mm{{\sT\sT Q} \ar[d]_*{\zt_{\sT Q}} \ar[r]^*{\zk_Q} &
            \sT\sT Q \ar[d]_*{\sT\zt_Q} \\
            {\sT Q} \ar@{=}[r] & \sT Q}
                                                                                \tag \label{Fxx214}$$
    \vskip2mm
    \noindent is the inverse isomorphism.  For a differentiable mapping $\za \colon
Q \rightarrow P$ we have
        $$\sT\sT\za(\st\st^{(0,1)}\zq(0,0)) = \st\st^{(0,1)}(\za \circ
\zq)(0,0)
                                                                                \tag \label{Fxx215}$$
    and
        $$\zk_P \circ \sT\sT\za = \sT\sT\za \circ \zk_Q.
                                                                                \tag \label{Fxx216}$$

    Let $A$ be a 1-form on $Q$.  A 0-form $\xi_T A$ on $\sT Q$ is defined as the
function
        $$\xi_T A(v) = \langle A, v \rangle.
                                                                                \tag \label{Fxx217}$$
    Let $B$ be a 2-form on $Q$.  If $w \in \sT\sT Q$, then $(\zt_{\sT
Q}(w),\sT\zt_Q(w)) \in \sT Q \underset{(\zt_Q,\zt_Q)}\to\times \sT Q$ since $\zt_Q
\circ \sT\zt_Q = \zt_Q \circ \zt_{\sT Q}$.  A 1-form $\xi_T B$ on $\sT Q$ is defined
by
        $$\langle \xi_T B, w \rangle = \langle B, \zt_{\sT Q}(w) \wedge \sT\zt_Q(w)
\rangle.
                                                                                \tag \label{Fxx218}$$
    Let $F$, $A$, and $B = \xd A$ be a 0-form, a 1-form, and an exact 2-form on $Q$
respectively.  We define a 0-form $\xd_T F$, a 1-form $\xd_T A$, and a 2-form $\xd_T
B$ on $\sT Q$ by
        $$\xd_T F = \xi_T\xd F,
                                                                                \tag \label{Fxx219}$$
        $$\xd_T A = \xi_T\xd A + \xd\xi_T A,
                                                                                \tag \label{Fxx220}$$
    and
        $$\xd_T B = \xd\xi_T B = \xd\xi_T\xd A = \xd\xd_T A.
                                                                                \tag \label{Fxx221}$$
    The coordinate expression of the function $\xd_T F$ is
        $$\xd_T F(q^\zk,\dot q^\zl) = \partial_\zl F(q^\zk)\dot q^\zl.
                                                                                \tag \label{Fxx222}$$
    If
        $$A = A_\zk(q^\zm,p_\zn)\xd q^\zk
                                                                                \tag \label{Fxx223}$$
    and
        $$B = \frac{1}{2} B_{\zk\zl}(q^\zm,p_\zn)\xd q^\zk \wedge \xd q^\zl,
                                                                                \tag \label{Fxx224}$$
    then
        $$\xi_T A = A_\zk(q^\zm,p_\zn)\dot q^\zk,
                                                                                \tag \label{Fxx225}$$
        $$\xd_T A = \partial_\zl A_\zk \dot q^\zl \xd q^\zk,
                                                                                \tag \label{Fxx226}$$
        $$\xi_T B = B_{\zk\zl}(q^\zm,p_\zn)\dot q^\zk \xd q^\zl,
                                                                                \tag \label{Fxx227}$$
    and
        $$\xd\xi_T\xd A = \frac{1}{2} \partial_\zm(\partial\zk A_\zl - \partial\zl
A_\zk)\dot q^\zm \xd q^\zk \wedge \xd q^\zl + (\partial\zk A_\zl - \partial\zl
A_\zk)\xd\dot q^\zk \wedge \xd q^\zl.
                                                                                \tag \label{Fxx228}$$

    Each 1-form on $Q$ can be expressed as a sum of products $F\xd G$ and from
        $$\langle \xd_T (F\xd G), \st\st^{(0,1)}\zq(0,0) \rangle =
\frac{\partial}{\partial t} \left(F(\zq(0,t))\frac{\partial}{\partial
s}G(\zq(s,t))\right)_{|s=0,t=0}
                                                                                \tag \label{Fxx229}$$
    and
        $$\frac{\xd}{\xd t} \langle F\xd G, \st\zq(\cdot,t)(0) \rangle_{|t=0} =
\frac{\partial}{\partial t} \left(F(\zq(0,t))\frac{\partial}{\partial
s}G(\zq(s,t))\right)_{|s=0,t=0}
                                                                                \tag \label{Fxx230}$$
    it follows that
        $$\langle \xd_T (F\xd G), \st\st^{(0,1)}\zq(0,0) \rangle = \frac{\xd}{\xd t}
\langle F\xd G, \st\zq(\cdot,t)(0) \rangle_{|t=0}.
                                                                                \tag \label{Fxx231}$$
    Hence,
        $$\langle \xd_T A, \st\st^{(0,1)}\zq(0,0) \rangle = \frac{\xd}{\xd t}
\langle A, \st\zq(\cdot,t)(0) \rangle_{|t=0}
                                                                                \tag \label{Fxx232}$$
    for each 1-form $A$.

    \vskip3mm
        \leftline{\black 6. A geometric framework for analytical mechanics.}
    \vskip1mm
    Let $Q$ be a manifold of dimension $m$.  We have already described the geometry
of the tangent bundle $\sT Q$, the cotangent bundle $\sT^\*Q$ and the tangent bundle
$\sT\sT^\*Q$ of the cotangent bundle $\sT^\*Q$.  The present section is devoted to
the study of the canonical symplectic structure of the bundle $\sT\sT^\*Q$.  We will
use coordinates
        $$(q^\zk,\dot q^\zl) \colon \sT Q \rightarrow \R^{2m},
                                                                                \tag \label{Fxx233}$$
        $$(q^\zk,p_\zl) \colon \sT^\* Q \rightarrow \R^{2m},
                                                                                \tag \label{Fxx234}$$
    and
        $$(q^\zk,p_\zl,\dot q^\zm,\dot p_\zn) \colon \sT\sT^\*Q \rightarrow \R^{4m}
                                                                                \tag \label{Fxx235}$$
    in the manifolds $\sT Q$, $\sT^\*Q$ and $\sT\sT^\*Q$.  The mappings
$\zt_{\sT^\*Q}$ and $\sT\zp_Q$ have now the coordinate expressions
        $$(q^\zk,p_\zl) \circ \zt_{\sT^\*Q} = (q^\zk,p_\zl)
                                                                                \tag \label{Fxx236}$$
    and
        $$(q^\zk,\dot q^\zl) \circ \sT\zp_Q = (q^\zk,\dot q^\zl).
                                                                                \tag \label{Fxx237}$$

    We introduce the exact 2-form
        $$\xd_T\zw_Q = \xd\xi_T\zw_Q.
                                                                                \tag \label{Fxx238}$$
    It will be shown that this 2-form is non degenerate.  The manifold $\sT\sT^\*Q$
with the form $\xd_T\zw_Q$ form a symplectic manifold $(\sT\sT^\*Q,\xd_T\zw_Q)$.
We believe that the symplectic form $\xd_T\zw_Q$ is the only natural symplectic form
in $\sT\sT^\*Q$.  The discovery of a second symplectic structure in $\sT\sT^\*Q$ was
announced in a recent Springer-Verlag publication [9].  We have not been able to
identify the second symplectic structure.  We strongly suspect that this announcement
is false.  The formula
        $$\xd\zY_\zi = \xd\dot q^\zk \wedge \xd q^\zk + \xd\dot p_\zk \wedge \xd
p_\zk
                                                                                \tag \label{Fxx239}$$
    for the Marsden-Ratiu symplectic form does not seem to have an intrinsic meaning
since elementary rules of tensor calculus have been violated.  We have the
coordinate expressions
        $$\xd_T\zy_Q = \dot p_\zk\xd q^\zk + p_\zk\xd\dot q^\zk
                                                                                \tag \label{Fxx240}$$
    and
        $$\xd_T\zw_Q = \xd\dot p_\zk \wedge \xd q^\zk + \xd p_\zk \wedge \xd\dot q^\zk.
                                                                                \tag \label{Fxx241}$$

    The fibration $\zt_{\sT^\*Q} \colon \sT\sT^\*Q \rightarrow \sT^\*Q$ is a vector
fibration.  We will construct a vector fibration structure for the fibration
$\sT\zp_Q \colon \sT\sT^\*Q \rightarrow \sT Q$.  For two vectors $z_1 \in \sT\sT^\*Q$
and $z_2 \in \sT\sT^\*Q$ such that $\sT\zp_Q(z_2) = \sT\zp_Q(z_1)$ it is possible to
choose representatives $\zz_1 \colon \R \rightarrow \sT^\*Q$ and $\zz_2 \colon \R
\rightarrow \sT^\*Q$ such that $z_1 = \st\zz_1(0)$, $z_2 = \st\zz_2(0)$ and $\zp_Q
\circ \zz_2 = \zp_Q \circ \zz_1$.  An example is provided by the coordinate
constructions
        $$(q^\zk,p_\zl) \circ \zz_1 = (q^\zk(z_1) + \dot q^\zk(z_1)s,p_\zl(z_1) +
\dot p_\zl(z_1)s)
                                                                                \tag \label{Fxx242}$$
    and
        $$(q^\zk,p_\zl) \circ \zz_2 = (q^\zk(z_2) + \dot q^\zk(z_2)s,p_\zl(z_2) +
\dot p_\zl(z_2)s).
                                                                                \tag \label{Fxx243}$$
    An operation
        $$\tilde+ \colon \sT\sT^\*Q \underset{(\sT\zp_Q,\sT\zp_Q)}\to\times
\sT\sT^\*Q \rightarrow \sT\sT^\*Q
                                                                                \tag \label{Fxx244}$$
    is defined by
        $$z_1 \,\tilde+\, z_2 = \st(\zz_1 + \zz_2)(0).
                                                                                \tag \label{Fxx245}$$
    An operation
        $$\tilde\cdot\, \colon \R \times \sT\sT Q \rightarrow \sT\sT Q.
                                                                                \tag \label{Fxx246}$$
    is defined by
        $$k \,\tilde\cdot\, \st\zz(0) = \st(k\zz)(0).
                                                                                \tag \label{Fxx247}$$
    Coordinate representations of these operations are given by
        $$(q^\zk,p_\zl,\dot q^\zm,\dot p_\zn)(z_1 \,\tilde+\, z_2) =
(q^\zk(z_1),p_\zl(z_1) + p_\zl(z_2),\dot q^\zm(z_1),\dot p_\zn(z_1) +
\dot p_\zn(z_2))
                                                                                \tag \label{Fxx248}$$
    and
        $$(q^\zk,p_\zl,\dot q^\zm,\dot p_\zn)(k\,\tilde\cdot\, z) =
(q^\zk(z),kp_\zl(z),\dot q^\zm(z),k\dot p_\zn(z)).
                                                                                \tag \label{Fxx249}$$
    The diagram
    \vskip1mm
        $$\xymatrix@R+3mm @C+6mm{{\sT\sT^\*Q} \ar[r]^*{\zt_{\sT^\*Q}} \ar[d]_*{\sT\zp_Q} &
            \sT^\*Q \ar[d]_*{\zp_Q} \\
            {\sT Q} \ar[r]^*{\zt_Q} & Q}
                                                                                \tag \label{Fxx250}$$
    \vskip2mm
    \noindent is a vector fibration morphism.  The vector fibration $\sT\zp_Q \colon
\sT\sT^\*Q \rightarrow \sT Q$ is dual to the vector fibration $\sT\zt_Q \colon
\sT\sT Q \rightarrow \sT Q$.  The pairing
        $$\langle \,,\,\rangle^\sim \colon \sT\sT^\*Q
\underset(\sT\zp_Q,\sT\zt_Q)\to\times \sT\sT Q \rightarrow \R
                                                                                \tag \label{Fxx251}$$
    is defined by
        $$\langle z, w \rangle^\sim = \frac{\xd}{\xd s} \langle \zz(s),
\zx(s)\rangle_{|s=0},
                                                                                \tag \label{Fxx252}$$
    where $\zz \colon \R \rightarrow \sT^\*Q$ and $\zx \colon \R \rightarrow \sT Q$
are curves such that $z = \st\zz(0)$, $w = \st\zx(0)$ and $\zp_Q \circ \zz = \zt_Q
\circ \zx$.  Such curves are provided by the coordinate constructions
        $$(q^\zk,p_\zl) \circ \zz = (q^\zk(z) + \dot q^\zk(z)s,p_\zl(z) +
\dot p_\zl(z)s)
                                                                                \tag \label{Fxx253}$$
    and
        $$(q^\zk,\dot q^\zl) \circ \zx = (q^\zk(w) + q'{}^\zk(w)s,\dot q^\zl(w) +
\dot q'{}^\zl(w)s).
                                                                                \tag \label{Fxx254}$$
    The coordinate expression of the pairing is
        $$\langle z, w \rangle^\sim = p_\zk(z) \dot q'{}^\zk(w) + \dot p_\zk(z) \dot
q^\zk(w).
                                                                                \tag \label{Fxx255}$$

    A mapping
        $$\zc_Q \colon \sT\sT^\*Q \underset{(\zp_Q \circ
\zt_{\sT^\*Q},\zp_Q)}\to\times \sT^\*Q \rightarrow \sT\sT^\*Q
                                                                                \tag \label{Fxx256}$$
    is defined by
        $$\zc_Q(w,f) = w - \st\zh(0)
                                                                                \tag \label{Fxx257}$$
    with $\zh \colon \R \rightarrow \sT^\*Q$ defined by $\zh(s) = \zt_{\sT^\*Q}(w) +
sf$.  The coordinate expression of the mapping in terms of coordinates
$(q^\zk,p_\zl,\dot q^\zm,\dot p_\zn)$ in $\sT\sT^\*Q$ and coordinates $(q^\zk,f_\zl)$
in $\sT^\*Q$ is given by
        $$(q^\zk,p_\zl,\dot q^\zm,\dot p_\zn) \circ \zc_Q = (q^\zk,p_\zl,\dot
q^\zm,\dot p_\zn - f_\zn).
                                                                                \tag \label{Fxx258}$$

    In the cotangent bundle $\sT^\*\sT Q$ we use coordinates
        $$(q^\zk,\dot q^\zl,a_\zm,b_\zn) \colon \sT^\*\sT Q \rightarrow \R^{4m}
                                                                                \tag \label{Fxx259}$$
    induced by coordinates $(q^\zk,\dot q^\zl)$ in $\sT Q$.  The Liouville form is
the 1-form
        $$\zy_{\sT Q} = a_\zk\xd q^\zk + b_\zk\xd\dot q^\zk.
                                                                                \tag \label{Fxx260}$$
    The 2-form
        $$\zw_{\sT Q} = \xd a_\zk \wedge \xd q^\zk + \xd b_\zk \wedge \xd\dot q^\zk
                                                                                \tag \label{Fxx261}$$
    is the symplectic form on $\sT^\*\sT Q$.  A vector fibration isomorphism
    \vskip1mm
        $$\xymatrix@+5mm{{\sT\sT^\*Q} \ar[d]_*{\sT\zp_Q} \ar[r]^*{\za_Q} &
            \sT^\*\sT Q \ar[d]_*{\zp_{\sT Q}} \\
            {\sT Q} \ar@{=}[r] & {\sT Q}}
                                                                                \tag \label{Fxx262}$$
    \vskip2mm
    \noindent is defined as dual to the vector fibration isomorphism
    \vskip1mm
        $$\xymatrix@R+3mm @C+6mm{{\sT\sT Q} \ar[d]_*{\sT\zt_Q} &
            \sT\sT Q \ar[l]_*{\zk_Q} \ar[d]_*{\zt_{\sT Q}} \\
            {\sT Q} \ar@{=}[r] & \sT Q}
                                                                                \tag \label{Fxx263}$$
    \vskip2mm
    \noindent in the sense that
        $$\langle \za_Q(z), w \rangle = \langle z, \zk_Q(w) \rangle^\sim
                                                                                \tag \label{Fxx264}$$
    for $z \in \sT\sT^\*Q$ and $w \in \sT\sT Q$ such that $\sT\zp_Q(z) = \zt_{\sT
Q}(w)$.  We have the coordinate characterization
        $$(q^\zk,\dot q^\zl,a_\zm,b_\zn) \circ \za_Q = (q^\zk,\dot q^\zl,\dot
p_\zm,p_\zn)
                                                                                \tag \label{Fxx265}$$
    of the mapping $\za_Q$.

    For a vector $z = \st\st^{(0,1)}\zq(0,0) \in \sT\sT\sT^\*Q$ represented by a
mapping $\zq \colon \R^2 \rightarrow \sT^\*Q$ we have
        $$\align
    \langle \za_Q^\*\zy_{\sT Q}, z \rangle &= \langle \zy_{\sT Q}, \sT\za_Q(z) \rangle \\
            &= \langle \zt_{\sT^\*\sT Q}(\sT\za_Q(z)), \sT\zp_{\sT Q}(\sT\za_Q(z))
\rangle \\
            &= \langle \za_Q(\zt_{\sT\sT^\*Q}(z)), \sT\sT\zp_Q(z) \rangle \\
            &= \langle \zt_{\sT\sT^\*Q}(z), \zk_Q(\sT\sT\zp_Q(z)) \rangle^\sim \\
            &= \langle \zt_{\sT\sT^\*Q}(\st\st^{(0,1)}\zq(0,0)),
\zk_Q(\sT\sT\zp_Q(\st\st^{(0,1)}\zq(0,0))) \rangle^\sim \\
            &= \langle \st^{(0,1)}\zq(0,0), \zk_Q(\st\st^{(0,1)}(\zp_Q \circ
\zq)(0,0)) \rangle^\sim \\
            &= \langle \st\zq(0,\cdot)(0), \st\st^{(1,0)}(\zp_Q \circ \zq)(0,0)
\rangle^\sim \\
            &= \frac{\xd}{\xd t} \langle \zq(0,t), \st^{(1,0)}(\zp_Q \circ \zq)(0,t)
\rangle_{|t=0} \\
            &= \frac{\xd}{\xd t} \langle \zq(0,t), \st(\zp_Q \circ \zq)(\cdot,t)(0)
\rangle_{|t=0} \\
            &= \frac{\xd}{\xd t} \langle \zt_{\sT^\*Q}(\st\zq(\cdot,t)(0)),
\sT\zp_Q(\st\zq(\cdot,t)(0)) \rangle_{|t=0} \\
            &= \frac{\xd}{\xd t} \langle \zy_Q, \st\zq(\cdot,t)(0) \rangle_{|t=0} \\
            &= \frac{\xd}{\xd t} \langle \zy_Q, \st^{(1,0)}\zq(0,t) \rangle_{|t=0} \\
            &= \langle \xd_T\zy_Q, \st\st^{(0,1)}\zq(0,0) \rangle \\
            &= \langle \xd_T\zy_Q, z \rangle.
                                                                                \tag \label{Fxx266}\endalign$$
    We have used the formula \Ref{Fxx232} and relations
        $$\zt_{\sT^\*\sT Q} \circ \sT\za_Q = \za_Q \circ \zt_{\sT\sT^\*Q}
                                                                                \tag \label{Fxx267}$$
    and
        $$\sT\zp_{\sT Q} \circ \sT\za_Q = \sT\sT\zp_Q
                                                                                \tag \label{Fxx268}$$
    derived from
        $$\zp_{\sT Q} \circ \za_Q = \sT\zp_Q.
                                                                                \tag \label{Fxx269}$$
    We have shown that
        $$\za_Q^\*\zy_{\sT Q} = \xd_T\zy_Q.
                                                                                \tag \label{Fxx270}$$
    It follows that the 2-form $\xd_T\zw_Q$ is non degenerate and that the mapping
$\za_Q \colon \sT\sT^\*Q \rightarrow \sT^\*\sT Q$ is a symplectomorphism from
$(\sT\sT^\*Q,\xd_T\zw_Q)$ to $(\sT^\*\sT Q,\zw_{\sT Q})$ since
        $$\xd_T\zw_Q = \xd_T\xd\zy_Q = \xd\xd_T\zy_Q = \xd\za_Q^\*\zy_{\sT Q} =
\za_Q^\*\xd\zy_{\sT Q} = \za_Q^\*\zw_{\sT Q}.
                                                                                \tag \label{Fxx271}$$
    These results are confirmed by the coordinate calculations
        $$\za_Q^\*\zy_{\sT Q} = \dot p_\zk \xd q^\zk + p_\zk \xd\dot q^\zk = \xd_T\zy_Q
                                                                                \tag \label{Fxx272}$$
    and
        $$\za_Q^\*\zw_{\sT Q} = \xd\dot p_\zk \wedge \xd q^\zk + \xd p_\zk \wedge
\xd\dot q^\zk = \xd_T\zw_Q.
                                                                                \tag \label{Fxx273}$$

    In the cotangent bundle $\sT^\*\sT^\*Q$ we use coordinates
        $$(q^\zk,p_\zl,u_\zm,v^\zn) \colon \sT^\*\sT^\*Q \rightarrow \R^{4m}
                                                                                \tag \label{Fxx274}$$
    induced by coordinates $(q^\zk,p_\zl)$ in $\sT^\*Q$.  We have the Liouville form
        $$\zy_{\sT^\*Q} = u_\zk\xd q^\zk + v^\zk\xd p_\zk.
                                                                                \tag \label{Fxx275}$$
    and the symplectic 2-form
        $$\zw_{\sT^\*Q} = \xd u_\zk \wedge \xd q^\zk + \xd v^\zk \wedge \xd p_\zk.
                                                                                \tag \label{Fxx276}$$
    on $\sT^\*\sT^\*Q$.  We have already introduced the mapping
        $$\zb_{(\sT^\*Q,\zw_Q)} \colon \sT\sT^\*Q \rightarrow \sT^\*\sT^\*Q
                                                                                \tag \label{Fxx277}$$
    characterized by the equality
        $$\langle \zb_{(\sT^\*Q,\zw_Q)}(u), v \rangle = \langle \zw_Q, u \wedge v
\rangle
                                                                                \tag \label{Fxx278}$$
    for vectors $u \in \sT\sT^\*Q$ and $v \in \sT\sT^\*Q$ such that
$\zt_{\sT^\*Q}(v) = \zt_{\sT^\*Q}(u)$.  The diagram
        $$\xymatrix@R+5mm @C+12mm{{\sT\sT^\*Q} \ar[d]_*{\zt_{\sT^\*Q}}
\ar[r]^*{\zb_{(\sT^\*Q,\zw_Q)}} &
            \sT^\*\sT^\*Q \ar[d]_*{\zp_{\sT^\*Q}} \\
            {\sT^\*Q} \ar@{=}[r] & {\sT^\*Q}}
                                                                                \tag \label{Fxx279}$$
    \vskip2mm
    \noindent is a vector fibration isomorphism.  For each $z \in \sT\sT^\*\sT^\*Q$
we have
        $$\align
    \langle \zb_{(\sT^\*Q,\zw_Q)}^\*\zy_{\sT^\*Q}, z \rangle &= \langle
\zy_{\sT^\*Q}, \sT\zb_{(\sT^\*Q,\zw_Q)}(z) \rangle \\
            &= \langle \zt_{\sT^\*\sT^\*Q}(\sT\zb_{(\sT^\*Q,\zw_Q)}(z)), \sT\zp_{\sT^\*Q}(\sT\zb_{(\sT^\*Q,\zw_Q)}(z)) \rangle \\
            &= \langle \zb_{(\sT^\*Q,\zw_Q)}(\zt_{\sT\sT^\*Q}(z)), \sT\zt_{\sT^\*Q}(z) \rangle \\
            &= \langle \zw_Q, \zt_{\sT\sT^\*Q}(z) \wedge \sT\zt_{\sT^\*Q}(z) \rangle \\
            &= \langle \xi_T\zw_Q, z \rangle.
                                                                                \tag \label{Fxx280}\endalign$$
    The formula \Ref{Fxx218} and relations
        $$\zt_{\sT^\*\sT^\*Q} \circ \sT\zb_{(\sT^\*Q,\zw_Q)} = \zb_{(\sT^\*Q,\zw_Q)}
\circ \zt_{\sT\sT^\*Q}
                                                                                \tag \label{Fxx281}$$
    and
        $$\sT\zp_{\sT^\*Q} \circ \sT\zb_{(\sT^\*Q,\zw_Q)} = \sT\zt_{\sT^\*Q}
                                                                                \tag \label{Fxx282}$$
    derived from
        $$\zp_{\sT^\*Q} \circ \zb_{(\sT^\*Q,\zw_Q)} = \zt_{\sT^\*Q}
                                                                                \tag \label{Fxx283}$$
    were used.  We have shown that
        $$\zb_{(\sT^\*Q,\zw_Q)}^\*\zy_{\sT^\*Q} = \xi_T\zw_Q.
                                                                                \tag \label{Fxx284}$$
    It follows that the mapping $\zb_{(\sT^\*Q,\zw_Q)} \colon \sT\sT^\*Q \rightarrow
\sT^\*\sT^\*Q$ is a symplectomorphism from $(\sT\sT^\*Q,\xd_T\zw_Q)$ to
$(\sT^\*\sT^\*Q,\zw_{\sT^\*Q})$ since
        $$\xd_T\zw_Q = \xd\xi_T\zw_Q = \xd\zb_{(\sT^\*Q,\zw_Q)}^\*\zy_{\sT^\*Q} =
\zb_{(\sT^\*Q,\zw_Q)}^\*\xd\zy_{\sT^\*Q} = \zb_{(\sT^\*Q,\zw_Q)}^\*\zw_{\sT^\*Q}.
                                                                                \tag \label{Fxx285}$$
    Coordinate calculations
        $$\zb_{(\sT^\*Q,\zw_Q)}^\* \zy_{\sT^\*Q} = \dot p_\zk \xd q^\zk -
\dot q^\zk \xd p_\zk = \xi_T\zw_Q
                                                                                \tag \label{Fxx286}$$
and
        $$\zb_{(\sT^\*Q,\zw_Q)}^\* \zw_{\sT^\*Q} = \xd\dot p_\zk \wedge \xd q^\zk +
\xd p_\zk \wedge \xd\dot q^\zk = \xd_T\zw_Q.
                                                                                \tag \label{Fxx287}$$
    confirm these results.

    \vskip3mm
        \leftline{\black 7. Dynamics of mechanical systems.}
    \vskip1mm

    Let $Q$ be the {\it configuration manifold} of a mechanical system.  The
cotangent bundle $\sT^\*Q$ is the {\it phase space} of the system.  Elements of the
phase space are {\it momenta}.  The commutative diagram
    \vskip2mm
        $$\xymatrix{{\sT^\*\sT^\*Q} \ar[rd]_*{\zp_{\sT^\*Q}}
            && {\sT\sT^\*Q} \ar[rr]^*{\za_Q} \ar[ll]_*{\zb_{(\sT^\*Q,\zw_Q)}}
            \ar[ld]^*{\zt_{\sT^\*Q}} \ar[rd]_*{\sT\zp_Q}
            && {\sT^\*\sT Q} \ar[ld]^*{\zp_{\sT Q}}\\
        & {\sT^\*Q} \ar[rd]_*{\zp_Q} && {\sT Q} \ar[ld]^*{\zt_Q} & \\
        && Q && }
                                                                                \tag \label{Fxx288}$$
    \vskip2mm
    \noindent contains the geometric structures used to formulate the dynamics of
the system.  The dynamics is a differential equation $D \subset \sT\sT^\*Q$.  A
solution $\zp \colon I \rightarrow \sT^\*Q$ of this equation is a phase space
trajectory of the system.  External forces have to be included in a complete
description of dynamics.  The dynamics of the system with external forces is the
differential equation
        $$D_f = \zc_Q^{-1}(D) \subset \sT\sT^\*Q \underset{(\zp_Q \circ
\zt_{\sT^\*Q},\zp_Q)}\to\times \sT^\*Q.
                                                                                \tag \label{Fxx289}$$
    A solution is a curve $(\zp,\zf) \colon I \rightarrow \sT^\*Q
\underset{(\zp_Q,\zp_Q)}\to\times \sT^\*Q$.  The values of this curve represent the
momenta and external forces.  The differential equation $D_f$ is of first order for
the momentum component $\zp$ and of zero order for the force component $\zf$.  This
treatment of external forces is suitable for non relativistic systems.  Relativistic
systems described by homogeneous Lagrangians require a modification of the concept
of force.  We will deal with dynamics without external forces.

    Trajectories of the system in the configuration manifold $Q$ are solutions of the
second order Euler-Lagrange equation
        $$E = \sT^2\zp_Q(\sP D).
                                                                                \tag \label{Fxx290}$$

    We have recognized the presence of a canonical symplectic structure in
$\sT\sT^\*Q$ with the symplectic form $\xd_T\zw_Q$.  In most cases of interest in
relativistic physics the dynamics is a Lagrangian submanifold of
$(\sT\sT^\*Q,\xd_T\zw_Q)$.  Morphisms $\za_Q$ and $\zb_{(\sT^\*Q,\zw_Q)}$ are
canonical symplectomorphisms from $(\sT\sT^\*Q,\xd_T\zw_Q)$ to $(\sT^\*\sT
Q,\zw_{\sT Q})$ and to $(\sT^\*\sT^\*Q,\zw_{\sT^\*Q})$.  These symplectomorphisms
with cotangent bundles create the possibility of generating the dynamics from
(generalized) Lagrangians associated with $\sT Q$ or (generalized) Hamiltonians
associated with $\sT^\*Q$.

    We will present a number of examples of mechanical systems in Lagrangian and
Hamiltonian formulations.  We will perform the Legendre transformations and test the
integrability criteria for these systems.

    \vskip3mm
        \leftline{\black 8. Lagrangian systems.}
    \vskip1mm
    Let
        $$L \colon \sT Q \rightarrow \R
                                                                                \tag \label{Fxx291}$$
    be the Lagrangian of a mechanical system with configuration space $Q$.  The
Lagrangian may be defined on all of $\sT Q$ or on an open subset.  The image $N =
\im(\xd L)$ of the mapping
        $$\xd L \colon \sT Q \rightarrow \sT^\*\sT Q
                                                                                \tag \label{Fxx292}$$
    is a Lagrangian submanifold of $(\sT^\*\sT Q,\zw_{\sT Q})$ and the set $D =
\za_Q^{-1}(N) \subset \sT\sT^\*Q$ is a Lagrangian submanifold of
$(\sT\sT^\*Q,\xd_T\zw_Q)$.  Coordinates $(q^\zk,\dot q^\zl,a_\zm,b_\zn)$ of $N$
satisfy equations
        $$\aligned
            a_\zm&= \partial_\zm L(q^\zk,\dot q^\zl) \\
            b_\zm&= \partial_{\dot\zm} L(q^\zk,\dot q^\zl)
        \endaligned
                                                                                \tag \label{Fxx293}$$
    and coordinates $(q^\zk,p_\zl,\dot q^\zm,\dot p_\zn)$ of elements of $D$ satisfy
equations
        $$\aligned
            \dot p_\zm&= \partial_\zm L(q^\zk,\dot q^\zl) \\
            p_\zm&= \partial_{\dot\zm} L(q^\zk,\dot q^\zl)
        \endaligned
                                                                                \tag \label{Fxx294}$$
    derived from the variational principle
        $$\dot p_\zk\zd q^\zk + p_\zk\zd\dot q^\zk = \zd L(q^\zk,\dot q^\zl) =
\partial_\zm L(q^\zk,\dot q^\zl)\zd q^\zm + \partial_{\dot\zm} L(q^\zk,\dot
q^\zl)\zd\dot q^\zm.
                                                                                \tag \label{Fxx295}$$
    Substituting the equalities
        $$\xd\dot p_\zm = \partial_\zn\partial_\zm L \xd q^\zn +
\partial_{\dot\zn}\partial_\zm L \xd\dot q^\zn
                                                                                \tag \label{Fxx296}$$
    and
        $$\xd p_\zm = \partial_\zn\partial_{\dot\zm} L \xd q^\zn +
\partial_{\dot\zn}\partial_{\dot\zm} L \xd\dot q^\zn
                                                                                \tag \label{Fxx297}$$
    in
        $$\xd_T\zw_Q = \xd\dot p_\zm \wedge \xd q^\zm + \xd p_\zm \wedge \xd\dot q^\zm
                                                                                \tag \label{Fxx298}$$
    we obtain the equality
        $$\xd_T\zw_Q|D = \partial_\zn\partial_\zm L \xd q^\zn \wedge \xd q^\zm +
\partial_{\dot\zn}\partial_\zm L \xd\dot q^\zn \wedge \xd q^\zm +
        \partial_\zn\partial_{\dot\zm} L \xd q^\zn \wedge \xd\dot q^\zm +
\partial_{\dot\zn}\partial_{\dot\zm} L \xd\dot q^\zn \wedge \xd\dot q^\zm = 0.
                                                                                \tag \label{Fxx299}$$
    This equality together with $\dim(D) = 2m = 1/2 \dim(\sT\sT^\*Q)$ confirms that
$D$ is a Lagrangian submanifold of $(\sT\sT^\*Q,\xd_T\zw_Q)$.  The set $N$ is a
Lagrangian submanifold since it is the image of the differential of a function and
$D$ is a Lagrangian submanifold since it is obtained from $N$ by applying the
symplectomorphism $\za_Q$.  We have confirmed this fact by direct calculation.

    The set $D \subset \sT\sT^\*Q$ is a differential equation.  A solution is a curve
$\zg \colon \R \rightarrow \sT^\*Q$ such that vectors tangent to $\zg$ are in $D$.
Equations
        $$\aligned
            \dot\zg_\zm&= \partial_\zm L(\zg^\zk,\dot\zg^\zl) \\
            \zg_\zn&= \partial_{\dot\zn} L(\zg^\zk,\dot\zg^\zl)
        \endaligned
                                                                                \tag \label{Fxx300}$$
    are differential equations for the coordinate expression
        $$(\zg^\zk,\zg_\zl) = (q^\zk,p_\zl) \circ \zg
                                                                                \tag \label{Fxx301}$$
    of a curve $\zg$ derived from equations \Ref{Fxx294}.  Dots indicate derivatives.
The differential equation $D$ represents dynamics in the sense that solution curves
are phase space trajectories of the mechanical system.  We say that $D$ is a {\it
Lagrangian system} since it was obtained from a Lagrangian function \Ref{Fxx291}.
Equations \Ref{Fxx294} and equations \Ref{Fxx300} will be called the {\it Lagrange
equations}.

    The second order equations
        $$\aligned
            \dot p_\zm&= \partial_\zm L(q^\zk,\dot q^\zl) \\
            p_\zm&= \partial_{\dot\zm} L(q^\zk,\dot q^\zl) \\
            \ddot p_\zm&= \partial_\zn\partial_\zm L(q^\zk,\dot q^\zl)\dot q^\zn +
\partial_{\dot\zn}\partial_\zm L(q^\zk,\dot q^\zl)\ddot q^\zn \\
            \dot p_\zm&= \partial_\zn\partial_{\dot\zm} L(q^\zk,\dot q^\zl)\dot
q^\zn + \partial_{\dot\zn}\partial_{\dot\zm} L(q^\zk,\dot q^\zl)\ddot q^\zn
        \endaligned
                                                                                \tag \label{Fxx302}$$
    represent the prolongation $\sP D$ of the Lagrange equations.  The equations
        $$\partial_\zn\partial_{\dot\zm} L(q^\zk,\dot q^\zl)\dot q^\zn +
\partial_{\dot\zn}\partial_{\dot\zm} L(q^\zk,\dot q^\zl)\ddot q^\zn - \partial_\zm
L(q^\zk,\dot q^\zl) = 0
                                                                                \tag \label{Fxx303}$$
    are the Euler-Lagrange equation $\sT^2\zp_Q(\sP D)$ in coordinate form.

        \claim \c{e}{Example}{}{\rm                                                     \label{Cxx4}
    Let $Q$ be a manifold of dimension 3 with coordinates $(q^i) = (q^1,q^2,q^3)$.
There is a Riemannian metric tensor $g_{ij}$, a function $\zf$, and a 1-form $A =
A_i \xd q^i$ on $Q$.  The function
        $$L(q^i,\dot q^j) = \frac{m}{2} g_{ij}\dot q^i\dot q^j - e\zf + eA_i \dot q^i
                                                                                \tag \label{Fxx304}$$
    is the Lagrangian of a particle of mass $m$ and charge $e$ in an electric field
        $$E = E_i \xd q^i = - \xd\zf = - \partial_i\zf\xd q^i
                                                                                \tag \label{Fxx305}$$
     and a magnetic field (induction)
        $$B = \frac{1}{2} B_{ij}\xd q^i \wedge \xd q^j = \xd A = \xd A_j \wedge \xd
q^j = \partial_i A_j \xd q^i \wedge \xd q^j = \frac{1}{2} (\partial_i A_j -
\partial_j A_i) \xd q^i \wedge \xd q^j.
                                                                                \tag \label{Fxx306}$$
    The dynamics $D$ of the particle is described by equations
        $$\aligned
            \dot p_i&= \frac{m}{2} \partial_i g_{jk}\dot q^j \dot q^k - e
\partial_i\zf + e \partial_i A_j \dot q^j \\
            p_i&= mg_{ij}\dot q^j + eA_i
        \endaligned
                                                                                \tag \label{Fxx307}$$
    Gauge independent covariant second order Euler-Lagrange differential equations
        $$mg_{ij}(\ddot q^j + \zG_{kl}^{\,j}\dot q^k \dot q^l) = eE_i + B_{ij}\dot q^j
                                                                                \tag \label{Fxx308}$$
    are easily derived.  The symbol $\zG_{kl}^{\,j}$ is the Christoffel symbol
        $$\zG_{kl}^{\,j} = \frac{1}{2} g^{ji}\left(\partial_k g_{li} + \partial_l
g_{ki} - \partial_i g_{kl} \right).
                                                                                \tag \label{Fxx309}$$
    Solution curves of \Ref{Fxx308} are motions in the configuration space $Q$.
These equations provide a partial description of dynamics.  The complete description
of dynamics is obtained by complementing these equations with the gauge dependent
velocity-momentum relation
        $$p_i= mg_{ij}\dot q^j + eA_i.
                                                                                \tag \label{Fxx310}$$
    A gauge transformation
        $$A_i \mapsto A_i + \partial_i\zc
                                                                                \tag \label{Fxx311}$$
    will not modify equations \Ref{Fxx308} but will change the velocity-momentum relation.
    }{\hfill    $\blacktriangle$}\endclaim

        \claim \c{e}{Example}{}{\rm                                                     \label{Cxx5}
    A gauge independent formulation of dynamics of a charged particle is obtained by
extending the configuration space $Q$ to a manifold $\overline Q$ of four dimensions
with coordinates $(q,q^i)$.  A {\it gauge transformation} is a coordinate
transformation $(q,q^i) \mapsto (q + \zc(q^k),q^i)$.  The dynamics
is derived from the gauge independent Lagrangian
        $$L(q,q^i,\dot q,\dot q^j) = \frac{m}{2} g_{ij}\dot q^i\dot q^j - e\zf +
eA_i \dot q^i + e\dot q.
                                                                                \tag \label{Fxx312}$$
    The equations
        $$\aligned
            \dot p&= 0 \\
            \dot p_i&= \frac{m}{2} \partial_i g_{jk}\dot
q^j \dot q^k - e \partial_i\zf + e \partial_i A_j \dot q^j \\
                p&= e \\
            p_i&= mg_{ij}\dot q^j + eA_i
        \endaligned
                                                                                \tag \label{Fxx313}$$
    provide a description of dynamics in terms of coordinates $(q,q^i,p,p_j,\dot
q,\dot q^k,\dot p,\dot p_l)$ in $\sT\sT^\*\overline Q$.  These equations are gauge
independent and can be given an explicitly covariant and gauge independent form
        $$\aligned
            \dot p&= 0 \\
            mg_{ij}(\ddot q^j + \zG_{kl}^{\,j}\dot q^k \dot q^l) &= eE_i +
B_{ij}\dot q^j \\
                p&= e \\
            p_i - eA_i&= mg_{ij}\dot q^j.
        \endaligned
                                                                                \tag \label{Fxx314}$$
    The gauge invariant quantity $(p_i - eA_i)$ is the momentum of the particle.
        }{\hfill    $\blacktriangle$}\endclaim

        \claim \c{e}{Example}{}{\rm                                                     \label{Cxx6}
    Let $Q$ be the space-time of general relativity with coordinates $(q^\zk) =
(q^0,q^1,q^2,q^3)$.  The gravitational field is represented by a Minkowski metric
$g_{\zk\zl}$ and the electromagnetic field is a 2-form
        $$F = \frac{1}{2} F_{\zk\zl}\xd q^\zk \wedge \xd q^\zl = - \xd A = - \xd
A_\zl \wedge \xd q^\zl = -\partial_\zk A_\zl \xd q^\zk \wedge \xd q^\zl = \frac{1}{2}
(\partial_\zl A_\zk - \partial_\zk A_\zl) \xd q^\zk \wedge \xd q^\zl
                                                                                \tag \label{Fxx315}$$
    derived from a potential
        $$A = A_\zk \xd q^\zk.
                                                                                \tag \label{Fxx316}$$
    The dynamics of a relativistic particle of mass $m$ and charge $e$ is derived
from the Lagrangian
        $$L(q^\zk,\dot q^\zl) = m \sqrt{g_{\zk\zl}\dot q^\zk
\dot q^\zl} + eA_\zk\dot q^\zk
                                                                                \tag \label{Fxx317}$$
    defined for time-like vectors -- vectors satisfying $g_{\zk\zl}\dot q^\zk \dot
q^\zl > 0$.  Dynamics $D \subset \sT\sT^\*Q$ is described by the Lagrange equations
        $$\aligned
            \dot p_\zk &= m\partial_\zk
g_{\zl\zm}\frac{\dot q^\zl\dot q^\zm}{2\|\dot q\|} + e\partial_\zk
A_\zl\dot q^\zl = m g_{\zl\zn}\zG^{\,\zn}_{\zk\zm}
\frac{\dot q^\zl\dot q^\zm}{\|\dot q\|} + e\partial_\zk
A_\zl\dot q^\zl \\
            p_\zk &= mg_{\zk\zl}\frac{\dot q^\zl}{\|\dot q\|} + e A_\zk
        \endaligned
                                                                                \tag \label{Fxx318}$$
    with $\|\dot q\| = \sqrt{g_{\zk\zl}\dot q^\zk \dot q^\zl}$.  Note that these
equations are reparametrization invariant: if $\zg \colon \R \rightarrow \sT^\*Q$ is
a solution and $\zs \colon \R \rightarrow \R$ is a diffeomorphism with positive
derivative, then $\zg \circ \zs$ is a solution.  One can say that solutions are
one-dimensional oriented but not parametrized submanifolds of $\sT^\*Q$.

    The Euler-Lagrange equations
        $$\frac{m}{\|\dot q\|} g_{\zk\zl}\left(\ddot q^\zl +
\zG_{\zm\zn}^{\,\zl}\dot q^\zm \dot q^\zn\right) + eF_{\zk\zl}\dot q^\zl =
\frac{m}{\|\dot q\|^3} g_{\zk\zl}\dot q^\zl g_{\zm\zn}\left(\ddot q^\zm +
\zG_{\zr\zs}^{\,\zl}\dot q^\zr \dot q^\zs\right)\dot q^\zn
                                                                                \tag \label{Fxx319}$$
    are reparametrization invariant and gauge invariant.  If proper time is chosen
as the parameter, then $\|\dot q\| = 1$ and the world line of the particle in
space-time is a solution of simplified gauge invariant second order differential
equations
        $$m(\ddot q^\zk + \zG_{\zl\zm}^{\,\zk}\dot q^\zl \dot q^\zm) = - eg^{\zk\zl}
F_{\zl\zm}\dot q^\zm.
                                                                                \tag \label{Fxx320}$$
    The complete dynamics is gauge dependent.
        }{\hfill    $\blacktriangle$}\endclaim

        \claim \c{e}{Example}{}{\rm                                                     \label{Cxx7}
    Let $Q$ be the space-time of general relativity with coordinates $(q^\zk)$, a
Minkowski metric $g_{\zk\zl}$ and an electromagnetic potential $A = A_\zk \xd
q^\zk$.  Let $\overline Q$ be a manifold of dimension 5 with coordinates
$(q,q^\zk)$.  A {\it gauge transformation} is a coordinate transformation $(q,q^\zk)
\mapsto (q + \zc(q^\zm),q^\zk)$.  We use coordinates $(q,q^\zk,p,p_\zl)$ in
$\sT^\*\overline Q$.  Two gauge invariant quantities are derived from the 5-{\it
momentum} $(p,p_\zl)$.  These are the {\it charge} $p$ and the {\it energy momentum}
$(p_\zl - pA_\zl)$.  There are two equivalent interpretations of the manifold
$\overline Q$ [6].  This manifold is interpreted as a pseudoriemannian manifold
(Kaluza [5]) with a metric tensor
        $$\pmatrix
            1  & A_\zl \\
            A_\zk & g_{\zk\zl} + A_\zk A_\zl
        \endpmatrix
                                                                                \tag \label{Fxx321}$$
    or as the total space of a principal fibration (Utiyama [14])
        $$\zq \colon \overline Q \rightarrow Q
                                                                                \tag \label{Fxx322}$$
    characterized by
        $$(q^\zk) \circ \zq = (q^\zk)
                                                                                \tag \label{Fxx323}$$
    The electromagnetic potential is used to introduce the connection form
        $$\za = \xd q + A_\zk\xd q^\zk
                                                                                \tag \label{Fxx324}$$
    in the principal bundle $\overline Q$.  The curvature form
        $$\zb = \xd\za = -\frac{1}{2} F_{\zk\zl}\xd q^\zk \xd q^\zl = \xd A_\zl
\wedge \xd q^\zl = \partial_\zk A_\zl \xd q^\zk \wedge \xd q^\zl = -\frac{1}{2}
(\partial_\zl A_\zk - \partial_\zk A_\zl) \xd q^\zk \wedge \xd q^\zl
                                                                                \tag \label{Fxx325}$$
    represents the electromagnetic field.

    The Lagrangian of a particle with mass $m$ and charge $e$ is the gauge invariant
function
        $$L(q,q^\zk,\dot q,\dot q^\zl) = m \sqrt{g_{\zk\zl}\dot q^\zk
\dot q^\zl} + eA_\zk\dot q^\zk + e\dot q.
                                                                                \tag \label{Fxx326}$$
    Coordinates $(q,q^\zk,p,p_\zl,\dot q,\dot q^\zm,\dot p,\dot p_\zn)$ are used in
$\sT\sT^\*\overline Q$.  The Lagrange equations
        $$\aligned
            \dot p &= 0 \\
            \dot p_\zk &= m\partial_\zk g_{\zl\zm}\frac{\dot q^\zl\dot
q^\zm}{2\|\dot q\|} + e\partial_\zk A_\zl\dot q^\zl = m
g_{\zl\zn}\zG^{\,\zn}_{\zk\zm} \frac{\dot q^\zl\dot q^\zm}{\|\dot q\|} +
e\partial_\zk A_\zl\dot q^\zl \\
            p &= e \\
            p_\zk &= mg_{\zk\zl}\frac{\dot q^\zl}{\|\dot q\|} + e A_\zk
        \endaligned
                                                                                \tag \label{Fxx327}$$
    are equivalent to the explicitly covariant and gauge independent second order
equations
        $$\aligned
            \dot p &= 0 \\
            m(\ddot q^\zk + \zG_{\zl\zm}^{\,\zk}\dot q^\zl \dot q^\zm) &=
eg^{\zk\zl}F_{\zl\zm}\dot q^\zm \\
            p &= e \\
            p_\zk - e A_\zk &= mg_{\zk\zl}\dot q^\zl.
        \endaligned
                                                                                \tag \label{Fxx328}$$
    These equations are obtained by adopting the simplifying condition $\|\dot q\| =
1$.  Trajectories in $\overline Q$ satisfy the second order equations
        $$m(\ddot q^\zk + \zG_{\zl\zm}^{\,\zk}\dot q^\zl \dot q^\zm) =
eg^{\zk\zl}F_{\zl\zm}\dot q^\zm
                                                                                \tag \label{Fxx329}$$
    with no conditions on $q$.  Compatibility with field equations requires that
trajectories in $\overline Q$ be two dimensional.  The dynamics of a charge particle
has to be suitably modified for correct description of interaction with the
electromagnetic field.
            }{\hfill    $\blacktriangle$}\endclaim

    Not all mechanical systems are Lagrangian systems derived from a Lagrangian
defined on the tangent bundle $\sT Q$.  The dynamics could be generated by a Morse
family of functions defined on fibres of a fibration
        $$\zh \colon Y \rightarrow \sT Q.
                                                                                \tag \label{Fxx330}$$
    We have coordinates $(q^\zk,\dot q^\zl)$ in $\sT
Q$.  In the space $Y$ we use adapted coordinates
        $$(q^\zk,\dot q^\zl,y^A) \colon Y \rightarrow \R^{2m+k}
                                                                                \tag \label{Fxx331}$$
    such that
        $$(q^\zk,\dot q^\zl) \circ \zh = (q^\zk,\dot q^\zl).
                                                                                \tag \label{Fxx332}$$
    Let $L \colon Y \rightarrow \R$ be a Morse family of functions defined on fibres
of $\zh$.  The $k \times (2m + k)$ matrix
        $$\pmatrix
    \dfrac{\partial^2 L}{\partial y^A \partial y^B} & \dfrac{\partial^2 L}{\partial
y^A \partial q^\zk} & \dfrac{\partial^2 L}{\partial y^A \partial\dot q^\zl}
            \endpmatrix
                                                                                \tag \label{Fxx333}$$
    is of maximal rank.  The Lagrangian submanifold $N \subset \sT^\*\sT Q$
generated by the family is the set of elements of $\sT^\*\sT Q$ with coordinates
$(q^\zk,\dot q^\zl,a_\zm,b_\zn)$ satisfying equations
        $$\aligned
            a_\zm&= \partial_\zm L(q^\zk,\dot q^\zl,y^A) \\
            b_\zn&= \partial_{\dot\zn} L(q^\zk,\dot q^\zl,y^A) \\
                0&= \partial_A L(q^\zk,\dot q^\zl,y^A),
        \endaligned
                                                                                \tag \label{Fxx334}$$
    for some values of the variables $y^A$.  The set $D = \za_Q^{-1}(N) \subset
\sT\sT^\*Q$ is a Lagrangian submanifold of $(\sT\sT^\*Q,\xd_T\zw_Q)$.  It is the set
of elements of $\sT\sT^\*Q$ with coordinates $(q^\zk,p_\zl,\dot q^\zm,\dot p_\zn)$
satisfying equations
        $$\aligned
            \dot p_\zm&= \partial_\zm L(q^\zk,\dot q^\zl,y^A) \\
            p_\zn&= \partial_{\dot\zn} L(q^\zk,\dot q^\zl,y^A) \\
                0&= \partial_A L(q^\zk,\dot q^\zl,y^A)
        \endaligned
                                                                                \tag \label{Fxx335}$$
    for some values of the variables $y^A$.  These equations can be derived from a
variational principle
        $$\dot p_\zk\zd q^\zk + p_\zk\zd\dot q^\zk = \zd L(q^\zk,\dot q^\zl,y^A) =
\partial_\zm L(q^\zk,\dot q^\zl,y^A)\zd q^\zm + \partial_{\dot\zm} L(q^\zk,\dot
q^\zl,y^A)\zd\dot q^\zm + \partial_B L(q^\zk,\dot q^\zl,y^A)\zd y^B.
                                                                                \tag \label{Fxx336}$$
    Equations \Ref{Fxx335} present the set $D$ parametrized by variables $(q^\zk,\dot
q^\zl,y^A)$ subject to $\partial_A L(q^\zk,\dot q^\zl,y^A) = 0$.  From
        $$\xd\dot p_\zm = \partial_\zn\partial_\zm L \xd q^\zn +
\partial_{\dot\zn}\partial_\zm L \xd\dot q^\zn + \partial_A\partial_\zm L \xd y^A,
                                                                                \tag \label{Fxx337}$$
        $$\xd p_\zm = \partial_\zn\partial_{\dot\zm} L \xd q^\zn +
\partial_{\dot\zn}\partial_{\dot\zm} L \xd\dot q^\zn + \partial_A\partial_{\dot\zm}
L \xd y^A,
                                                                                \tag \label{Fxx338}$$
        $$\partial_A\partial_\zm L = 0,
                                                                                \tag \label{Fxx339}$$
        $$\partial_A\partial_{\dot\zm} L = 0
                                                                                \tag \label{Fxx340}$$
    we obtain the equality
        $$\xd_T\zw_Q|D = 0.
                                                                                \tag \label{Fxx341}$$
    Equations $\partial_A L(q^\zk,\dot q^\zl,y^A) = 0$ leave only $2m$ out of the
$2m + k$ variables $(q^\zk,\dot q^\zl,y^A)$ independent.  This is a consequence of
maximality of the rank of the matrix \Ref{Fxx333}.  It follows that $\dim(D) = 2m$.
We have thus confirmed that $D$ is a Lagrangian submanifold of
$(\sT\sT^\*Q,\xd_T\zw_Q)$.  The set $D$ is a differential equation and may represent
the dynamics of a mechanical system.

        \claim \c{e}{Example}{}{\rm                                                     \label{Cxx8}
    Let $Q$ be the space-time of general relativity with coordinates $(q^\zk) =
(q^0,q^1,q^2,q^3)$ and a Minkowski metric $g_{\zk\zl}$.  Let
        $$L(q^\zk,\dot q^\zl,y) = \frac{1}{2y} g_{\zk\zl}\dot q^\zk \dot q^\zl
                                                                                \tag \label{Fxx342}$$
    be a function on $\oT Q \times \R_+$, where $\oT Q$ is the tangent bundle with
the zero vectors removed.  This function is a Morse family of functions of the
variable $y$ with the coordinates $(q^\zk,\dot q^\zl)$ treated as parameters.  It
represents the Lagrangian of a particle of mass zero.  The dynamics of the particle
is governed by the equations
        $$\aligned
            \dot p_\zk&= \frac{1}{2y}\partial_k g_{\zl\zm}\dot q^\zl \dot q^\zm \\
            p_\zk&= \frac{1}{y}g_{\zk\zl}\dot q^\zl \\
                0&= g_{\zk\zl}\dot q^\zk \dot q^\zl
        \endaligned
                                                                                \tag \label{Fxx343}$$
    satisfied for some value of the variable $y$.  The variable $y$ can be eliminated
from the equation for $\dot p_\zk$.  It follows from the resulting equation
        $$\dot p_\zk - \zG^{\,\zl}_{\zk\zm}p_\zl\dot q^\zm = 0
                                                                                \tag \label{Fxx344}$$
    that the covector $p_\zk$ is covariant constant along the world line.  If an
affine parameter is chosen then $y$ is constant and the dynamics satisfies equations
        $$\aligned
            \ddot q^\zl + \zG^{\,\zl}_{\zk\zm}\dot q^\zk\dot q^\zm &= 0 \\
                g_{\zk\zl}q^\zk q^\zl&= 0 \\
            p_\zk&= \frac{1}{y}g_{\zk\zl}q^\zl
        \endaligned
                                                                                \tag \label{Fxx345}$$
    for some constant $y > 0$.
        }{\hfill    $\blacktriangle$}\endclaim

    \vskip3mm
        \leftline{\black 9. Hamiltonian systems.}
    \vskip1mm
    Let
        $$H \colon \sT^\*Q \rightarrow \R
                                                                                \tag \label{Fxx346}$$
    be the Hamiltonian of a mechanical system with configuration space $Q$.  The
mapping
        $$- \xd H \colon \sT^\*Q \rightarrow \sT^\*\sT^\*Q
                                                                                \tag \label{Fxx347}$$
    is a section of the fibration $\zp_{\sT^\*Q}$.  Consequently
        $$X = \zb_{(\sT^\*Q,\zw_Q)}^{-1} \circ (- \xd H) \colon \sT^\*Q \rightarrow
\sT\sT^\*Q
                                                                                \tag \label{Fxx348}$$
    is a vector field.  The image $M = \im(- \xd H)$ is a Lagrangian submanifold of
$(\sT^\*\sT^\*Q,\zw_{\sT^\*Q})$ and $D = \im(X)$ is a Lagrangian submanifold of
$(\sT\sT^\*Q,\xd_T\zw_Q)$.  The equations describing $D$ are the {\it Hamilton
equations}
        $$\aligned
            \dot q^\zk&= \partial^\zk H  \\
            \dot p_\zk&= -\partial_\zk H
        \endaligned
                                                                                \tag \label{Fxx349}$$
    derived from the variational principle
        $$\dot p_\zm\zd q^\zm - \dot q^\zm\zd p_\zm = \zd H(q^\zk,p_\zl) = -
\partial_\zm H(q^\zk,p_\zl) \zd q^\zm - \partial^\zm H(q^\zk,p_\zl) \zd p_\zk.
                                                                                \tag \label{Fxx350}$$
    The symbol $\partial^\zk$ denotes the partial derivative
$\frac{\partial}{\partial p_\zk}$.  The dimension of $D$ is $2m$ and by using
equalities
        $$\xd\dot q^\zm = \partial_\zn\partial^\zm H \xd q^\zn +
\partial^\zn\partial^\zm H \xd p_\zn
                                                                                \tag \label{Fxx351}$$
    and
        $$\xd\dot p_\zm = - \partial_\zn\partial_\zm H \xd q^\zn -
\partial^\zn\partial_\zm H \xd p_\zn
                                                                                \tag \label{Fxx352}$$
    in
        $$\xd_T\zw_Q = \xd\dot p_\zm \wedge \xd q^\zm + \xd p_\zm \wedge \xd\dot q^\zm
                                                                                \tag \label{Fxx353}$$
    we obtain the equality
        $$\xd_T\zw_Q|D =  - \partial_\zn\partial_\zm H \xd q^\zn \wedge \xd q^\zm -
\partial^\zn\partial_\zm H \xd p_\zn \wedge \xd q^\zm + \partial_\zn\partial^\zm H
\xd p_\zm \wedge \xd q^\zn + \partial^\zn\partial_\zm H \xd p_\zm \wedge \xd p_\zn =
0.
                                                                                \tag \label{Fxx354}$$
    It follows that $D$ is a Lagrangian submanifold of $(\sT\sT^\*Q,\xd_T\zw_Q)$.
The set $D$ is a differential equation and may represent the dynamics of a
mechanical system.

        \claim \c{e}{Example}{}{\rm                                                     \label{Cxx9}
    Equations \Ref{Fxx307} of Example \Ref{Cxx4} can be rewritten in the form
        $$\aligned
            \dot q^i&= \frac{1}{m} g^{ij}(p_j - eA_j) \\
            \dot p_i&= \frac{1}{2m} \partial_i g_{jk}g^{jl}g^{km}(p_l - eA_l)(p_m -
eA_m) - e \partial_i\zf + \frac{e}{m}  \partial_i A_j g^{jl}(p_l - eA_l).
        \endaligned
                                                                                \tag \label{Fxx355}$$
    These equations describe a Hamiltonian vector field.  They are the Hamilton
equations for the Hamiltonian
        $$H(q^i,p_j) = \frac{1}{2m} g^{ij}(p_i - eA_i)(p_j - eA_j) + e\zf.
                                                                                \tag \label{Fxx356}$$
        }{\hfill    $\blacktriangle$}\endclaim

    Gauge independent dynamics of charged particles and the dynamics of relativistic
particles are not images of Hamiltonian vector fields.  Dirac [1] introduced {\it
generalized Hamiltonian systems} in order to be able to deal with similar cases.  In
the original construction of Dirac a generalized Hamiltonian system is a family of
Hamiltonian vector fields on the phase space $\sT^\*Q$ restricted to a {\it
constraint set} $C \subset \sT^\*Q$.  We have translated this construction in an
equivalent construction of a differential equation $D \subset \sT\sT^\*Q$.

    Let $C \subset \sT^\*Q$ be a submanifold and let
        $$H \colon C \rightarrow \R
                                                                                \tag \label{Fxx357}$$
    be a differentiable function.  The set
        $$M = \left\{b \in \sT^\*\sT^\*Q ;\; a = \zp_{\sT^\*Q}(b) \in C, \all{u \in
\sT_a C \subset \sT_a\sT^\*Q} \langle b,u \rangle = -\langle \xd H,u \rangle \right\}
                                                                                \tag \label{Fxx358}$$
    is a Lagrangian submanifold of $(\sT^\*\sT^\*Q,\zw_{\sT^\*Q})$ and
        $$D = \zb^{-1}_{(\sT^\*Q,\zw_Q)}(M) = \left\{w \in \sT\sT^\*Q ;\; a =
\zt_{\sT^\*Q}(w) \in C, \all{u \in \sT_a C \subset \sT_a\sT^\*Q} \langle \zw_Q,u
\wedge w \rangle = \langle \xd H,u \rangle \right\}
                                                                                \tag \label{Fxx359}$$
    is a Lagrangian submanifold of $(\sT\sT^\*Q,\xd_T\zw_Q)$.  If $(\zF_A)$ is a set
of $k$ independent functions on $\sT^\*Q$ such that
        $$C = \left\{a \in \sT^\*Q ;\; \zF_A(a) = 0 \text{ for } A = 1,\ldots,k
\right\}
                                                                                \tag \label{Fxx360}$$
    and $\overline H$ is a function on $\sT^\*Q$ such that $\overline H|C = H$, then
coordinates $(q^\zk,p_\zl,\dot q^\zm,\dot p_\zn)$ of elements of $D$ satisfy the
equations
        $$\aligned
            \zF_A(q^\zk,p_\zl) &= 0 \\
            \dot q^\zk&= \partial^\zk \overline H + v^A\partial^\zk\zF_A \\
            \dot p_\zk&= -\partial_\zk \overline H + v^A\partial_\zk\zF_A
        \endaligned
                                                                                \tag \label{Fxx361}$$
    derived from the variational principle
        $$\aligned
            \zF_A(q^\zk,p_\zl) &= 0 \\
            \dot p_\zm\zd q^\zm - \dot q^\zm\zd p_\zm &= \zd \overline
H(q^\zk,p_\zl) = - \partial_\zm \overline H(q^\zk,p_\zl) \zd q^\zm - \partial^\zm
\overline H(q^\zk,p_\zl) \zd p_\zk
        \endaligned
                                                                                \tag \label{Fxx362}$$
    with variations $(\zd q^\zk,\zd p_\zl)$ satisfying
        $$\partial_\zk \zF_A \zd q^\zk + \partial^\zk \zF_A \zd p_\zk = 0.
                                                                                \tag \label{Fxx363}$$
    Lagrange multipliers $(v^A \in \R^k)$ appear in these equations.  The same
equations are derived from the variational principle
        $$\dot p_\zm\zd q^\zm - \dot q^\zm\zd p_\zm = \zd\widetilde H(q^\zk,p_\zl,v^A) =
- \partial_\zm H(q^\zk,p_\zl) \zd q^\zm - \partial^\zm H(q^\zk,p_\zl) \zd p_\zk +
\zF_A(q^\zk,p_\zl)\zd v^A
                                                                                \tag \label{Fxx364}$$
    corresponding to the Morse family
        $$\widetilde H(q^\zk,p_\zl,v^A) = \overline H(q^\zk,p_\zl) -
v^A\zF_A(q^\zk,p_\zl)
                                                                                \tag \label{Fxx365}$$
    of functions of the variables $(v^A \in \R^k)$.  The set $D$ is a Lagrangian
submanifold since it is generated by a Morse family.  At each point $a \in C$ the
set $D_a = D \cap \sT_a\sT^\*Q$ is an affine subspace of the vector space
$\sT_a\sT^\*Q$.

    In Dirac's construction the dynamics is described by vector fields
        $$X = \zb_{(\sT^\*Q,\zw_Q)}^{-1} \circ (v^A \xd\zF_A - \xd\overline H) \colon
\sT^\*Q \rightarrow \sT\sT^\*Q
                                                                                \tag \label{Fxx366}$$
    with arbitrary functions $v^A(q^\zk,p_\zl)$.  These fields are restricted to the
constraint set $C$.

        \claim \c{e}{Example}{}{\rm                                                     \label{Cxx10}
    The dynamics of the charged particle of Example \Ref{Cxx5} is a Dirac system.
The Hamiltonian is the function
        $$\overline H(q,q^i,p,p_j) = \frac{1}{2m} g^{ij}(p_i - eA_i)(p_j - eA_j) + e\zf
                                                                                \tag \label{Fxx367}$$
    and the constraint is the set characterized by
        $$\zF(q,q^i,p,p_j) = p - e = 0.
                                                                                \tag \label{Fxx368}$$
    The function
        $$\widetilde H(q,q^i,p,p_j,v) = \frac{1}{2m} g^{ij}(p_i - eA_i)(p_j - eA_j)
+ e\zf - v(p - e)
                                                                                \tag \label{Fxx369}$$
    is a Morse family of functions of $v \in \R$.  Equations
        $$\aligned
            p&= e \\
            \dot q &= v \\
            \dot q^i&= \frac{1}{m} g^{ij}(p_j - eA_j) \\
            \dot p &= 0 \\
            \dot p_i&= \frac{1}{2m} \partial_i g_{jk}g^{jl}g^{km}(p_l - eA_l)(p_m -
eA_m) - e \partial_i\zf + \frac{e}{m}  \partial_i A_j g^{jl}(p_l - eA_l).
        \endaligned
                                                                                \tag \label{Fxx370}$$
    obtained with this Morse family are equivalent to equations \Ref{Fxx313}.
        }{\hfill    $\blacktriangle$}\endclaim

    The dynamics of a non relativistic charged particle in the above example is the
only Dirac system known to us.  Hamiltonian formulations of relativistic dynamics
require a higher level of complexity.  Differential equations generated by
Lagrangians and by Lagrangian Morse families are Lagrangian submanifolds of
$(\sT\sT^\*Q,\xd_T\zw_Q)$.  The same is true of Dirac systems.  We define a {\it
generalized Dirac system} as a differential equation $D \subset \sT\sT^\*Q$, which
is a Lagrangian submanifold of $(\sT\sT^\*Q,\xd_T\zw_Q)$.  Existence of Morse
families for open subsets of Lagrangian submanifolds is guaranteed by H\"ormander's
theorem.  Known generalized Dirac systems are globally generated by Hamiltonian
Morse families.

        \claim \c{e}{Example}{}{\rm                                                     \label{Cxx11}
    Lagrange equations \Ref{Fxx318} of Example \Ref{Cxx6} have an equivalent form
        $$\aligned
            g^{\zk\zl}(p_\zk - eA_\zk)(p_\zl - eA_\zl) &= m^2 \\
            \dot q^\zk &= \frac{v}{m} g^{\zk\zl}(p_\zl - eA_\zl) \\
            \dot p_\zk &= -\frac{v}{2m} \partial_\zk g^{\zm\zn}(p_\zm -
eA_\zm)(p_\zn - eA_\zn) + \frac{ve}{m} g^{\zm\zn} \partial_\zk A_\zm(p_\zn - eA_\zn)
        \endaligned
                                                                                \tag \label{Fxx371}$$
    with arbitrary $v > 0$.  These equations are obtained from the Morse family
        $$H(q^\zk,p_\zl,v) = v\left(\sqrt{g^{\zk\zl}(p_\zk - eA_\zk)(p_\zl -
eA_\zl)} - m\right)
                                                                                \tag \label{Fxx372}$$
    of functions of the variable $v > 0$.
        }{\hfill    $\blacktriangle$}\endclaim

        \claim \c{e}{Example}{}{\rm                                                     \label{Cxx12}
    The gauge independent dynamics in Example \Ref{Cxx7} is a generalized Dirac
system.  Equations
        $$\aligned
            g^{\zk\zl}(p_\zk - eA_\zk)(p_\zl - eA_\zl) &= m^2 \\
            p &= e \\
            \dot q &= v^2 \\
            \dot q^\zk &= \frac{v^1}{m} g^{\zk\zl}(p_\zl - eA_\zl) \\
            \dot p &= 0 \\
            \dot p_\zk &= -\frac{v^1}{2m} \partial_\zk g^{\zm\zn}(p_\zm -
eA_\zm)(p_\zn - eA_\zn) + \frac{v^1e}{m} g^{\zm\zn} \partial_\zk A_\zm(p_\zn - eA_\zn)
        \endaligned
                                                                                \tag \label{Fxx373}$$
    with $v^1 >0$ and arbitrary $v^2$ are equivalent to equations \Ref{Fxx328}.
These equations are generated by the Morse family
        $$H(q^\zk,p_\zl,v^1,v^2) = v^1\left(\sqrt{g^{\zk\zl}(p_\zk - eA_\zk)(p_\zl -
eA_\zl)} - m\right) + v^2(p - e)
                                                                                \tag \label{Fxx374}$$
    of functions of the variables $v^1 > 0$ and $v^2$.
        }{\hfill    $\blacktriangle$}\endclaim

        \claim \c{e}{Example}{}{\rm                                                     \label{Cxx13}
    The generalized Dirac system of Example \Ref{Cxx8} is described by equations
        $$\aligned
            g^{\zk\zl}p_\zk p_\zl &= 0 \\
            \dot q^\zk &= v g^{\zk\zl}p_\zl \\
            \dot p_\zk &= -\frac{v}{2} \partial_\zk g^{\zm\zn}p_\zm p_\zn
        \endaligned
                                                                                \tag \label{Fxx375}$$
    derived from the Morse family
        $$H(q^\zk,p_\zl,v) = \frac{v}{2} g^{\zk\zl}p_\zk p_\zl
                                                                                \tag \label{Fxx376}$$
    with $v > 0$.
        }{\hfill    $\blacktriangle$}\endclaim

    \vskip3mm
        \leftline{\black 10. The Legendre transformation.}
    \vskip1mm

    A Lagrangian $L(q^\zk,\dot q^\zl)$ is said to be {\it hyperregular} if the {\it
Legendre mapping}
        $$\zl \colon \sT Q \rightarrow \sT^\*Q
                                                                                \tag \label{Fxx377}$$
    defined by
        $$(q^\zk,p_\zl) \circ \zl = (q^\zk,\partial_{\dot\zl} L(q^\zr,\dot q^\zs))
                                                                                \tag \label{Fxx378}$$
    is a diffeomorphism.  Let the mapping
        $$\zq \colon \sT^\*Q \rightarrow \sT Q
                                                                                \tag \label{Fxx379}$$
    represented by
        $$(q^\zk,\dot q^\zl) \circ \zq = (q^\zk,\zq^\zl(q^\zr,p_\zs))
                                                                                \tag \label{Fxx380}$$
    be the inverse diffeomorphism.  Relations
        $$\partial_{\dot\zk}L(q^\zm,\zq^\zn(q^\zr,p_\zs)) = p_\zk
                                                                                \tag \label{Fxx381}$$
        $$\zq^\zk(q^\zm,\partial{\dot\zn}L(q^\zr,\dot q^\zs)) = \dot q^\zk
                                                                                \tag \label{Fxx382}$$
    hold.  Using the diffeomorphism $\zq$ to eliminate the velocities $(\dot q^\zk)$
from the {\it energy function}
        $$E(q^\zk,p_\zl,\dot q^\zm) = p_\zl\dot q^\zl - L(q^\zk,\dot q^\zm)
                                                                                \tag \label{Fxx383}$$
    we obtain the Hamiltonian
        $$H(q^\zm,p_\zn) = p_\zl\zq^\zl(q^\zm,p_\zn) - L(q^\zm,\zq^\zn(q^\zr,p_\zs)).
                                                                                \tag \label{Fxx384}$$
    The energy function is defined on $\sT^\*Q \underset{(\zp_Q,\zt_Q)}\to\times \sT
Q$.  The passage from the Lagrangian to the above Hamiltonian is the {\it Legendre
transformation} for a hyperregular Lagrangian.

    Let $H$ be the Hamiltonian \Ref{Fxx384} obtained from a hyperregular Lagrangian
$L$.  For the mappings
        $$\xd H \colon \sT^\*Q \rightarrow \sT^\*\sT^\*Q,
                                                                                \tag \label{Fxx385}$$
        $$\zb^{-1}_{(\sT^\*Q,\zw_Q)} \colon \sT^\*\sT^\*Q \rightarrow \sT\sT^\*Q,
                                                                                \tag \label{Fxx386}$$
    and
        $$\zb^{-1}_{(\sT^\*Q,\zw_Q)} \circ (-\xd H) \colon \sT^\*Q \rightarrow \sT\sT^\*Q
                                                                                \tag \label{Fxx387}$$
    we have
        $$(q^\zk,p_\zl,u_\zm,v^\zn) \circ \xd H = (q^\zk,p_\zl,\partial_\zm
H(q^\zr,p_\zs),\partial^\zn H(q^\zr,p_\zs)),
                                                                                \tag \label{Fxx388}$$
        $$(q^\zk,p_\zl,\dot q^\zm,\dot p_\zn) \circ \zb^{-1}_{(\sT^\*Q,\zw_Q)} =
(q^\zk,p_\zl,-v^\zm,u_\zn),
                                                                                \tag \label{Fxx389}$$
    and
        $$(q^\zk,p_\zl,\dot q^\zm,\dot p_\zn) \circ \zb^{-1}_{(\sT^\*Q,\zw_Q)} \circ
(-\xd H) = (q^\zk,p_\zl,\partial^\zm H(q^\zr,p_\zs),-\partial_\zn H(q^\zr,p_\zs)).
                                                                                \tag \label{Fxx390}$$
    On the other hand, we have
        $$(q^\zk,\dot q^\zl,a_\zm,b_\zn) \circ \xd L = (q^\zk,\dot
q^\zl,\partial_\zm L(q^\zr,\dot q^\zs),\partial_{\dot\zn}L(q^\zr,\dot q^\zs)),
                                                                                \tag \label{Fxx391}$$
        $$(q^\zk,p_\zl,\dot q^\zm,\dot p_\zn) \circ \za^{-1}_Q = (q^\zk,b_\zl,\dot
q^\zm,a_\zn),
                                                                                \tag \label{Fxx392}$$
        $$(q^\zk,p_\zl,\dot q^\zm,\dot p_\zn) \circ \za^{-1}_Q \circ \xd L =
(q^\zk,\partial_{\dot\zl}L(q^\zr,\dot q^\zs),\dot q^\zm,\partial_\zn L(q^\zr,\dot
q^\zs)),
                                                                                \tag \label{Fxx393}$$
    and
        $$(q^\zk,p_\zl,\dot q^\zm,\dot p_\zn) \circ \za^{-1}_Q \circ \xd L \circ \zq
= (q^\zk,\partial_{\dot\zl}L(q^\zr,\zq^\zs(q^\zw,p_\zt)),\zq^\zm(q^\zw,p_\zt),\partial_\zn
L(q^\zr,\zq^\zs(q^\zw,p_\zt)))
                                                                                \tag \label{Fxx394}$$
    for the mappings
        $$\xd L \colon \sT Q \rightarrow \sT^\*\sT Q,
                                                                                \tag \label{Fxx395}$$
        $$\za^{-1}_Q \colon \sT^\*\sT Q \rightarrow \sT\sT^\*Q,
                                                                                \tag \label{Fxx396}$$
        $$\za^{-1}_Q \circ \xd L \colon \sT Q \rightarrow \sT\sT^\*Q,
                                                                                \tag \label{Fxx397}$$
    and
        $$\za^{-1}_Q \circ \xd L \circ \zq \colon \sT^\*Q \rightarrow \sT\sT^\*Q.
                                                                                \tag \label{Fxx398}$$
    From
        $$\align
    \partial_\zk H(q^\zm,p_\zn)&= p_\zm\partial_\zk \zq^\zm(q^\zr,p_\zs) -
\partial_\zk L(q^\zm,\zq^\zn(q^\zr,p_\zs)) -
\partial_{\dot\zr}L(q^\zm,\zq^\zn(q^\zw,p_\zt)) \partial_\zk \zq^\zn(q^\zr,p_\zs) \\
            &= -\partial_\zk L(q^\zm,\zq^\zn(q^\zr,p_\zs))
                                                                                \tag \label{Fxx399}\endalign$$
    and
        $$\align
    \partial^\zk H(q^\zm,p_\zn)&= \zq^\zk(q^\zr,p_\zs) + p_\zm \partial^\zk \zq^\zm(q^\zr,p_\zs)
- \partial_{\dot\zr}L(q^\zm,\zq^\zn(q^\zw,p_\zt)) \partial^\zk \zq^\zn(q^\zr,p_\zs) \\
            &= \zq^\zk(q^\zr,p_\zs)
                                                                                \tag \label{Fxx400}\endalign$$
    it follows that
        $$\za^{-1}_Q \circ \xd L \circ \zq = \zb^{-1}_{(\sT^\*Q,\zw_Q)} \circ (-\xd H)
                                                                                \tag \label{Fxx401}$$
    We see that the Hamiltonian and the Lagrangian generate the same dynamics
        $$D = \im(\zb^{-1}_{(\sT^\*Q,\zw_Q)} \circ (-\xd H)) = \im(\za^{-1}_Q \circ \xd L).
                                                                                \tag \label{Fxx402}$$

        \claim \c{e}{Example}{}{\rm                                                     \label{Cxx14}
    The Lagrangian
        $$L(q^i,\dot q^j) = \frac{m}{2} g_{ij}\dot q^i\dot q^j - e\zf + eA_i \dot q^i
                                                                                \tag \label{Fxx403}$$
    of Example \Ref{Cxx4} is hyperregular.  The Legendre mapping and its inverse are
the mappings
        $$(q^i,p_j) \circ \zl = (q^i,mg_{jk}\dot q^k + eA_j)
                                                                                \tag \label{Fxx404}$$
    and
        $$(q^i,\dot q^j) \circ \zq = \left(q^i,\frac{1}{m} g^{ij}(p_j - eA_j)\right).
                                                                                \tag \label{Fxx405}$$
    From the energy function
        $$E(q^i,p_j,\dot q^k) = p_i\dot q^i - \frac{m}{2} g_{ij}\dot q^i\dot q^j +
e\zf - eA_i \dot q^i
                                                                                \tag \label{Fxx406}$$
    we derive the Hamiltonian
        $$H(q^i,p_j) = \frac{1}{2m} g^{ij}(p_i - eA_i)(p_j - eA_j) + e\zf.
                                                                                \tag \label{Fxx407}$$
        }{\hfill    $\blacktriangle$}\endclaim

    If a Lagrangian is not hyperregular, then the Legendre mapping is not
invertible.  It may happen that the image of the Legendre mapping $\zl$ is a
submanifold $C \subset \sT^\*Q$.  Let $G \subset \sT^\*Q
\underset{(\zp_Q,\zt_Q)}\to\times \sT Q$ be the graph of $\zl$ (intersected with
$\sT^\*Q \underset{(\zp_Q,\zt_Q)}\to\times \sT Q \subset \sT^\*Q \times \sT Q$).
The energy function \Ref{Fxx383} restricted to the graph $G$ does not depend on
$\dot q^\zk$ since
        $$\partial_{\dot\zk} E = p_\zk - \partial_{\dot\zk} L.
                                                                                \tag \label{Fxx408}$$
    If fibres of the projection $pr_1 \colon \sT^\*Q
\underset{(\zp_Q,\zt_Q)}\to\times \sT Q \rightarrow \sT^\*Q$ are connected, then a
function
        $$H(q^\zk,p_\zl) = E(q^\zk,p_\zl,\dot q^\zm)
                                                                                \tag \label{Fxx409}$$
    can be defined on $C$ by substituting in $E$ any values of the velocities $(\dot
q^\zm)$ such that $p_\zk - \partial_{\dot\zk} L = 0$.  This is the construction of the
constrained Hamiltonian used by Dirac.  This is also the construction of the {\it
generalized Legendre transformation} introduced by Cendra, Holm, Hoyle and Marsden.
The Dirac system generated by the constrained Hamiltonian is sometimes equal to the
Lagrangian system generated by the original Lagrangian.

        \claim \c{e}{Example}{}{\rm                                                     \label{Cxx15}
    Applying the Cendra-Holm-Hoyle-Marsden generalized Legendre transformation to
the Lagrangian
        $$L(q,q^i,\dot q,\dot q^j) = \frac{m}{2} g_{ij}\dot q^i\dot q^j - e\zf +
eA_i \dot q^i + e\dot q.
                                                                                \tag \label{Fxx410}$$
    of Example \Ref{Cxx5} we obtain the constraint $C$ described by
        $$p = e
                                                                                \tag \label{Fxx411}$$
     and the Hamiltonian
        $$H(q,q^i,p_j) = \frac{1}{2m} g^{ij}(p_i - eA_i)(p_j - eA_j) + e\zf
                                                                                \tag \label{Fxx412}$$
    defined on the constraint $C$ with coordinates $(q,q^i,p_j)$.  This Hamiltonian
generates the equations \Ref{Fxx370} of Example \Ref{Cxx10} equivalent to the
Lagrange equations \Ref{Fxx313} of Example \Ref{Cxx5}.  The
Cendra-Holm-Hoyle-Marsden version of the Legendre transformation gives correct
results for this example of a mechanical system.
        }{\hfill    $\blacktriangle$}\endclaim

    The mechanical system in the above example is the only mechanical system known
to us for which this version of the Legendre transformation functions correctly.

        \claim \c{e}{Example}{}{\rm                                                     \label{Cxx16}
    The Lagrangian
        $$L(q^\zk,\dot q^\zl) = m \sqrt{g_{\zk\zl}\dot q^\zk
\dot q^\zl} + eA_\zk\dot q^\zk
                                                                                \tag \label{Fxx413}$$
    of Example \Ref{Cxx6} is singular.  The image of the Legendre mapping
        $$(q^\zk,p_\zl) \circ \zl = (q^\zk,mg_{\zl\zm}\frac{\dot q^\zm}{\|\dot q\|}
+ e A_\zl)
                                                                                \tag \label{Fxx414}$$
    is the constraint set C described by
        $$g^{\zk\zl}(p_\zk - eA_\zk)(p_\zl - eA_\zl) = m^2.
                                                                                \tag \label{Fxx415}$$
    The energy function
        $$E(q^\zk,p_\zl,\dot q^\zm) = p_\zl\dot q^\zl - m \sqrt{g_{\zk\zl}\dot q^\zk
\dot q^\zl} - eA_\zk\dot q^\zk
                                                                                \tag \label{Fxx416}$$
    vanishes on the graph $G$ of the Legendre mapping and the Dirac Hamiltonian is
zero.  Differential equations derived from this constrained Hamiltonian are the
equations
        $$\aligned
            g^{\zk\zl}(p_\zk - eA_\zk)(p_\zl - eA_\zl) &= m^2 \\
            \dot q^\zk &= \frac{v}{m} g^{\zk\zl}(p_\zl - eA_\zl) \\
            \dot p_\zk &= -\frac{v}{2m} \partial_\zk g^{\zm\zn}(p_\zm -
eA_\zm)(p_\zn - eA_\zn) + \frac{ve}{m} g^{\zm\zn} \partial_\zk A_\zm(p_\zn - eA_\zn)
        \endaligned
                                                                                \tag \label{Fxx417}$$
    with arbitrary values of the Lagrange multiplier $v$.  The Dirac system
represented by these equations is the union $D_+ \cup D_0 \cup D_-$ of three sets
corresponding to $v > 0$, $v = 0$, and $v < 0$ respectively.  The set $D_+$ is the
generalized Dirac system $D$ of Example \Ref{Cxx11} equivalent to the Lagrangian
system of Example \Ref{Cxx6}.  The set $D_0$ described by
        $$\aligned
            g^{\zk\zl}(p_\zk - eA_\zk)(p_\zl - eA_\zl) &= m^2 \\
            \dot q^\zk &= 0 \\
            \dot p_\zk &= 0
        \endaligned
                                                                                \tag \label{Fxx418}$$
    must be excluded since the velocities $\dot q\zk$ evaluated on vectors tangent
to world lines are never zero.  The set $D_-$ is a generalized Dirac system obtained
from the Lagrangian
        $$L_- (q^\zk,\dot q^\zl) = - m \sqrt{g_{\zk\zl}\dot q^\zk
\dot q^\zl} + eA_\zk\dot q^\zk.
                                                                                \tag \label{Fxx419}$$
    Setting $v = 1$ in equations \Ref{Fxx417} we obtain equations
        $$\aligned
            p_\zk &= mg_{\zk\zl}\dot q^\zl + e A_\zk \\
        m(\ddot q^\zk + \zG_{\zl\zm}^{\,\zk}\dot q^\zl \dot q^\zm) &= - eg^{\zk\zl}
F_{\zl\zm}\dot q^\zm
        \endaligned
                                                                                \tag \label{Fxx420}$$
    correctly describing the dynamics of charged particles.  Solutions of these
equations are curves using proper time as the parameter.  With $y = -1$ equations
\Ref{Fxx417} result in the equation
        $$p_\zk = - mg_{\zk\zl}\dot q^\zl + e A_\zk
                                                                                \tag \label{Fxx421}$$
    and the second order system
        $$m(\ddot q^\zk + \zG_{\zl\zm}^{\,\zk}\dot q^\zl \dot q^\zm) = eg^{\zk\zl}
F_{\zl\zm}\dot q^\zm.
                                                                                \tag \label{Fxx422}$$
    Solutions of the second order equations are world lines of particles with mass
$m$ and charge $-e$ or particles with mass $-m$ and charge $e$.  The equation
\Ref{Fxx421} suggests that we are dealing with particles with negative mass $-m$.
The principle that world lines of particles with positive energy should be oriented
towards the future and that world lines of particles with negative energy
(antiparticles) should be oriented towards the past (Stueckelberg [11], Feynman [3])
is violated.  We conclude that the Cendra-Holm-Hoyle-Marsden version of the Legendre
transformation is too fast to provide correct results for this important example of
a mechanical system.
        }{\hfill    $\blacktriangle$}\endclaim

    The Dirac system generated by a constrained Hamiltonian
        $$H \colon C \rightarrow \R
                                                                                \tag \label{Fxx423}$$
    is characterized by the variational relation
        $$\dot p_\zk \zd q^\zk - \dot q^\zk \zd p_\zk = -\partial_zk H \zd q^\zk -
\partial^\zk H \zd p_\zk
                                                                                \tag \label{Fxx424}$$
    on $C$.  A constrained Hamiltonian derived from a singular Lagrangian can be
considered a function on the graph $G$ of the Legendre mapping defined by
        $$H(q^\zk,p_\zl) = E(q^\zk,p_\zl,\dot q^\zm) = p_\zl\dot q^\zl -
L(q^\zk,\dot q^\zm)
                                                                                \tag \label{Fxx425}$$
    The variational relation
        $$\dot p_\zk \zd q^\zk - \dot q^\zk \zd p_\zk = -\partial_zk E \zd q^\zk -
\partial^\zk E \zd p_\zk - \partial_{\dot\zk} E \zd \dot q^\zk
                                                                                \tag \label{Fxx426}$$
    on $G$ is equivalent to the relation \Ref{Fxx424} for the Hamiltonian
\Ref{Fxx425}.  The variations $(\zd q^\zk,\zd p_\zl,\zd \dot q^\zm)$ are components of
a vector tangent to $G$.  Hence
        $$\zd p_\zk = \partial_\zm\partial_{\dot\zk}\zd q^\zm +
\partial_{\dot\zm}\partial_{\dot\zk}\zd \dot q^\zm.
                                                                                \tag \label{Fxx427}$$
    If Lagrange equations
        $$\aligned
            \dot p_\zm&= \partial_\zm L(q^\zk,\dot q^\zl) \\
            p_\zn&= \partial_{\dot\zn} L(q^\zk,\dot q^\zl)
        \endaligned
                                                                                \tag \label{Fxx428}$$
    are satisfied, then
        $$\dot p_\zk \zd q^\zk - \dot q^\zk \zd p_\zk = \partial_\zk L \zd q^\zk -
\dot q^\zk \zd p_\zk
                                                                                \tag \label{Fxx429}$$
    and
        $$-\partial_zk E \zd q^\zk - \partial^\zk E \zd p_\zk - \partial_{\dot\zk} E
\zd \dot q^\zk = \partial_\zk L \zd q^\zk - \dot q^\zk \zd p_\zk - (p_\zk -
\partial_{\dot\zk} L)\zd \dot q^\zk = \partial_\zk L \zd q^\zk - \dot q^\zk \zd
p_\zk.
                                                                                \tag \label{Fxx430}$$
    This seems to imply that the Dirac system generated by the Hamiltonian
\Ref{Fxx425} is equivalent to the Lagrangian system \Ref{Fxx428}.  This conclusion is
not correct.  It is true that the variational relation \Ref{Fxx424} follows from
the Lagrange equations.  The converse is not true since the same constrained
Hamiltonian may be in the relation \Ref{Fxx425} with different energy functions
constructed from different Lagrangians.  The Lagrangians $L_+ = L$ and $L_-$ of
Example \Ref{Cxx16} generate different Lagrangian systems $D_+$ and $D_-$ but lead to
the same constrained Hamiltonian.

    We observe that if $E(q^\zk,p_\zl,\dot q^\zm)$ is the energy function associated
with a Lagrangian $L(q^\zk,\dot q^\zl)$, then $E(q^\zk,p_\zl,v^\zm)$ is a Morse
family of the variables $(v^\zm)$.  The rank of the matrix
    $$\pmatrix
        \dfrac{\partial^2 E}{\partial v^\zk \partial v^\zl} & \dfrac{\partial^2
E}{\partial\dot v^\zk \partial q^\zm} & \dfrac{\partial^2 E}{\partial v^\zk \partial p_\zn}
        \endpmatrix
        = \pmatrix
        \dfrac{-\partial^2 L}{\partial v^\zk \partial v^\zl} & \dfrac{-\partial^2
L}{\partial\dot v^\zk \partial q^\zm} & \zd_\zk^\zn
        \endpmatrix
                                                                                \tag \label{Fxx431}$$
    is maximal.  No requirements are imposed on the Lagrangian.  The generalized
Dirac system generated by the Morse family $E(q^\zk,p_\zl,v^\zm)$ is obtained from
the variational relation
        $$\dot p_\zk \zd q^\zk - \dot q^\zk \zd p_\zk = -\partial_zk E \zd q^\zk -
\partial^\zk E \zd p_\zk + \partial_{\dot\zk} E \zd v^\zk
                                                                                \tag \label{Fxx432}$$
    with arbitrary variations $(\zd q^\zk,\zd p_\zl,\zd v^\zm)$.  With
        $$E(q^\zk,p_\zl,v^\zm) = p_\zl v^\zl - L(q^\zk,v^\zm)
                                                                                \tag \label{Fxx433}$$
    the relation takes the form
        $$\dot p_\zk \zd q^\zk - \dot q^\zk \zd p_\zk = \partial_\zk L(q^\zk,v^\zm)
\zd q^\zk - v^\zk \zd p_\zk - (p_\zk - \partial_{\dot\zk} L(q^\zk,v^\zm))\zd v^\zk.
                                                                                \tag \label{Fxx434}$$
    Equations
        $$\aligned
            \dot p_\zm &= \partial_\zm L(q^\zk,v^\zl) \\
            p_\zn &= \partial_{\dot\zn} L(q^\zk,v^\zl) \\
            \dot q^\zl &= v^\zl
        \endaligned
                                                                                \tag \label{Fxx435}$$
    obtained from this relation are equivalent to the Lagrange equations.

    The passage from the Lagrangian $L(q^\zk,\dot q^\zl)$ to the global Hamiltonian
Morse family $E(q^\zk,p_\zl,v^\zm)$ is the {\it slow and careful Legendre
transformation} required to provide a correct Hamiltonian formulation of any
Lagrangian system.  The Morse family generating a Lagrangian submanifold is not
unique.  Modifications resulting in a reduction of the number of variables are
usually possible.

        \claim \c{e}{Example}{}{\rm                                                     \label{Cxx17}
    The Hamiltonian Morse family
        $$E(q,q^i,p,p_j,v,v^k) = pv + p_i v^i - \frac{m}{2} g_{ij}v^i v^j + e\zf -
eA_i v^i - ev
                                                                                \tag \label{Fxx436}$$
    for the charged particle of Example \Ref{Cxx5} reduces to the Hamiltonian Morse
family
        $$H(q,q^i,p,p_j,v) = \frac{1}{2m} g^{ij}(p_i - eA_i)(p_j - eA_j)
+ e\zf - v(p - e)
                                                                                \tag \label{Fxx437}$$
    of Example \Ref{Cxx10}.  The reduction is obtained by using in \Ref{Fxx436} the
equality
        $$v^k = \frac{1}{m} g^{kj}(p_j - eA_j)
                                                                                \tag \label{Fxx438}$$
    obtained from
        $$\frac{\partial E}{\partial v^i} = p_i - mg_{ij}v^j - eA_i = 0.
                                                                                \tag \label{Fxx439}$$
    }{\hfill    $\blacktriangle$}\endclaim

        \claim \c{e}{Example}{}{\rm                                                     \label{Cxx18}
        $$E(q^\zk,p_\zl,v^\zm) = p_\zk v^\zk - m \sqrt{g_{\zk\zl}v^\zk v^\zl} -
eA_\zk v^\zk
                                                                                \tag \label{Fxx440}$$
    for the charged particle of Example \Ref{Cxx6} reduces to the Hamiltonian Morse
family
        $$H(q^\zk,p_\zl,v) = v\left(\sqrt{g^{\zk\zl}(p_\zk - eA_\zk)(p_\zl -
eA_\zl)} - m\right)
                                                                                \tag \label{Fxx441}$$
    of a single variable $v > 0$ introduced in Example \Ref{Cxx11}.  Variables
$(q^\zk,p_\zl,v^\zm)$ are coordinates in the manifold $\sT^\*Q
\underset{(\zp_Q,\zt_Q)}\to\times \sT Q$.  Only the open subset described by the
inequality $g_{\zk\zl}v^\zk v^\zl > 0$ is considered.  Variables $(q^\zk,p_\zl,v)$
are coordinates in $\sT^\*Q \times \R_+$.  The function $E(q^\zk,p_\zl,v^\zm)$ is a
Morse family of functions on fibres of the fibration
        $$\zz \colon \sT^\*Q \underset{(\zp_Q,\zt_Q)}\to\times \sT Q \rightarrow
\sT^\*Q
                                                                                \tag \label{Fxx442}$$
     characterized by
        $$(q^\zk,p_\zl) \circ \zz = (q^\zk,p_\zl).
                                                                                \tag \label{Fxx443}$$
    This fibration can be interpreted as the fibration
        $$\zz' \colon \sT^\*Q \underset{(\zp_Q,\zt_Q)}\to\times \sT Q \rightarrow
\sT^\*Q \times \R_+
                                                                                \tag \label{Fxx444}$$
     characterized by
        $$(q^\zk,p_\zl,v) \circ \zz' = \left(q^\zk,p_\zl,\sqrt{g_{\zm\zn}v^\zm
v^\zn}\right)
                                                                                \tag \label{Fxx445}$$
    followed by the projection
        $$pr_{\sT^\*Q} \colon \sT^\*Q \times \R_+ \rightarrow \sT^\*Q.
                                                                                \tag \label{Fxx446}$$
    Fibres of $\zz'$ are hyperboloids $g_{\zm\zn}v^\zm v^\zn = v^2$.  From
        $$\zd E(q^\zk,p_\zl,v^\zm) = \frac{\partial}{\partial v^\zn} \zd v^\zn = 0
                                                                                \tag \label{Fxx447}$$
    with variations $\zd v^\zn$ satisfying
        $$\zd(g_{\zm\zn}v^\zm v^\zn) = 2g_{\zm\zn}v^\zm \zd v^\zn = 0
                                                                                \tag \label{Fxx448}$$
    we obtain two critical points
        $$v^\zk = \frac{\pm v}{\sqrt{g_{\zm\zn}v^\zm v^\zn}} g^{\zk\zl}(p_\zl -
eA_\zl)
                                                                                \tag \label{Fxx449}$$
    in a fibre over the point with coordinates $(q^\zk,p_\zl,v)$.  Evaluating the
function $E$ at these points results in two Morse families
        $$H\pm(q^\zk,p_\zl,v) = \pm v\left(\sqrt{g^{\zk\zl}(p_\zk - eA_\zk)(p_\zl -
eA_\zl)} \mp m\right)
                                                                                \tag \label{Fxx450}$$
    of functions of $v$.  One of these is the reduced Morse family \Ref{Fxx441}.  The
other generates an empty set since it has no critical points.  The family
\Ref{Fxx441} depends linearly on the variable $v$ restricted to positive values.  No
further reduction is possible.
    }{\hfill    $\blacktriangle$}\endclaim

        \claim \c{e}{Example}{}{\rm                                                     \label{Cxx19}
    The dynamics of the charged particle of Example \Ref{Cxx7} is generated by the
Hamiltonian Morse family
        $$E(q,q^\zk,p,p_\zl,v,v^\zm) = pv + p_\zk v^\zk - m \sqrt{g_{\zk\zl}v^\zk v^\zl} -
eA_\zk v^\zk - ev
                                                                                \tag \label{Fxx451}$$
    or by the Hamiltonian Morse family
        $$H(q^\zk,p_\zl,v^1,v^2) = v^1\left(\sqrt{g^{\zk\zl}(p_\zk - eA_\zk)(p_\zl -
eA_\zl)} - m\right) + v^2(p - e)
                                                                                \tag \label{Fxx452}$$
    of functions of two variables $v^1 > 0$ and $v^2$ as in Example \Ref{Cxx12}.
    }{\hfill    $\blacktriangle$}\endclaim

    The slow Legendre transformation can be extended to Lagrangian Morse families.
If
        $$L(q^\zk,\dot q^\zl,y^A)
                                                                                \tag \label{Fxx453}$$
    is a Morse family, then the $k \times (k + 2m)$ matrix
    \vskip2mm
        $$\pmatrix
    \dfrac{\partial^2 L}{\partial y^A \partial y^B} & \dfrac{\partial^2 L}{\partial
y^A \partial v^\zl} & \dfrac{\partial^2 L}{\partial y^A \partial p_\zm}
        \endpmatrix
                                                                                \tag \label{Fxx454}$$
    \vskip2mm
    \noindent is of maximal rank $k$.  It follows that the function
        $$E(q^\zk,p_\zl,y^A,v^\zm) = p_\zk v^\zk - L(q^\zk,v^\zl,y^A)
                                                                                \tag \label{Fxx455}$$
    is a Morse family of functions of the variables $(y^A,v^\zm)$ since the $(k + m)
\times (k + 3m)$ matrix
    \vskip2mm
        $$\pmatrix
    \dfrac{\partial^2 E}{\partial y^A \partial y^B} & \dfrac{\partial^2 E}{\partial
y^A \partial v^\zl} & \dfrac{\partial^2 E}{\partial y^A \partial q^\zm} & \dfrac{\partial^2 E}{\partial y^A \partial p_\zn} \\
    \dfrac{\partial^2 E}{\partial v^\zk \partial y^B} & \dfrac{\partial^2 E}{\partial
v^\zk \partial v^\zl} & \dfrac{\partial^2 E}{\partial v^\zk \partial q^\zm} & \dfrac{\partial^2 E}{\partial v^\zk \partial p_\zn}
        \endpmatrix
        = \pmatrix
    \dfrac{-\partial^2 L}{\partial y^A \partial y^B} & \dfrac{-\partial^2 L}{\partial
y^A \partial v^\zl} & \dfrac{-\partial^2 L}{\partial y^A \partial q^\zm} & 0 \\
    \dfrac{-\partial^2 L}{\partial v^\zk \partial y^B} & \dfrac{-\partial^2 L}{\partial
v^\zk \partial v^\zl} & \dfrac{-\partial^2 L}{\partial v^\zk \partial q^\zm} & \zd_\zk^\zn
        \endpmatrix
                                                                                \tag \label{Fxx456}$$
    \vskip2mm
    \noindent is of rank $k + m$.

        \claim \c{e}{Example}{}{\rm                                                     \label{Cxx20}
    The Hamiltonian Morse family
        $$E(q^\zk,p_\zl,y,v^\zm) = p_\zk v^\zk - \frac{1}{2y} g_{\zk\zl}v^\zk v^\zl
                                                                                \tag \label{Fxx457}$$
    generates the generalized Dirac system of Example \Ref{Cxx8}.  The equations
        $$\partial_{\dot\zm} E(q^\zk,p_\zl,y,v^\zm) = p_\zm - \frac{1}{y} g_{\zm\zl}
v^\zl
                                                                                \tag \label{Fxx458}$$
    permits the elimination of $v^\zm$ from $E(q^\zk,p_\zl,y,v^\zm)$.  The result is
the Hamilton Morse family
        $$H(q^\zk,p_\zl,y) = \frac{y}{2} g^{\zk\zl}p_\zk p_\zl
                                                                                \tag \label{Fxx459}$$
    with $y > 0$.  It is the Morse family of Example \Ref{Cxx13} with $v$ replaced by $y$.
    }{\hfill    $\blacktriangle$}\endclaim

    \vskip3mm
        \leftline{\black 11. Integrability.}
    \vskip1mm

    In Section 4 we have established integrability criteria for a class of
differential equations which can be specified in terms of vector fields restricted
to a submanifold.  The dynamics of relativistic mechanical systems presented in our
examples all admit the Dirac-style formulations in terms of Hamiltonian vector
fields.  The integrability criterion and the first integrability algorithm
formulated in Section 4 can be adapted to this situation.

    Let $Q$ be the configuration manifold of a mechanical system and let the
dynamics of the system be represented by the union
        $$D = \bigcup_{\za \in \sA}\{\im(X_\za|C)\}
                                                                                \tag \label{Fxx460}$$
    of a family of Hamiltonian vector fields
        $$X_\za \colon \sT^\*Q \rightarrow \sT\sT^\* Q
                                                                                \tag \label{Fxx461}$$
    generated by a family of Hamiltonians
        $$H_\za \colon \sT^\*Q \rightarrow \R
                                                                                \tag \label{Fxx462}$$
    and restricted to a submanifold $C \subset \sT^\*Q$ specified as
        $$C = \left\{p \in \sT^\*Q ;\; \all{A}\zF_A(p) = 0 \right\},
                                                                                \tag \label{Fxx463}$$
    where $\zF_A$ are independent functions on $\sT^\*Q$ called {\it primary
constraints}.  The condition $D \subset \sT C$ means that at points of $C$ the vector
fields $X_\za$ are tangent to $C$ or that
        $$\langle \xd \zF_A, X_\za \rangle|C = 0
                                                                                \tag \label{Fxx464}$$
    for each $\za \in \sA$ and each function $\zF_A$.  In view of
        $$X_\za = \zb_{(\sT^\*Q,\zw_Q)}^{-1} \circ (- \xd H_\za)
                                                                                \tag \label{Fxx465}$$
    the integrability criterion for the dynamics $D$ assumes the form
        $$\{H_\za,\zF_A\}|C = 0
                                                                                \tag \label{Fxx466}$$
    for each $\za \in \sA$ and each function $\zF_A$.

        \claim \c{e}{Example}{}{\rm                                                     \label{Cxx21}
    The dynamics of the system in Example \Ref{Cxx9} is integrable since it is the
image of a Hamiltonian vector field.
        }{\hfill    $\blacktriangle$}\endclaim

        \claim \c{e}{Example}{}{\rm                                                     \label{Cxx22}
    The dynamics in Example \Ref{Cxx10} is a Dirac system.  It is the union of images
of the family of Hamiltonian vector fields generated by the family
        $$H_\za(q,q^i,p,p_j) = \frac{1}{2m} g^{ij}(p_i - eA_i)(p_j - eA_j)
+ e\zf - \za(p - e)
                                                                                \tag \label{Fxx467}$$
    restricted to a constraint set $C$.  The parameter $\za$ is an arbitrary function
on $\sT^\*Q$.  There is one primary constraint.  It is the function
        $$\zF(q,q^i,p,p_j) = p - e.
                                                                                \tag \label{Fxx468}$$
    The system is integrable since
        $$\{H_\za,\zF\}|C = 0.
                                                                                \tag \label{Fxx469}$$
        }{\hfill    $\blacktriangle$}\endclaim

        \claim \c{e}{Example}{}{\rm                                                     \label{Cxx23}
    The dynamics in Example \Ref{Cxx11} is a generalized Dirac system.  It is
described by the family of Hamiltonian vector fields generated by the family
        $$H_\za(q^\zk,p_\zl) = \za\left(\sqrt{g^{\zk\zl}(p_\zk - eA_\zk)(p_\zl -
eA_\zl)} - m\right)
                                                                                \tag \label{Fxx470}$$
    of Hamiltonians parametrized by a function $\za > 0$ on $\sT^\*Q$.
There is one primary constraint
        $$\zF(q^\zk,p_\zl) = \sqrt{g^{\zk\zl}(p_\zk - eA_\zk)(p_\zl - eA_\zl)} - m.
                                                                                \tag \label{Fxx471}$$
    The integrability criterion is satisfied.

        }{\hfill    $\blacktriangle$}\endclaim

        \claim \c{e}{Example}{}{\rm                                                     \label{Cxx24}
    The generalized Dirac system of Example \Ref{Cxx12} is described by Hamiltonian
vector fields generated by the family
        $$H_{(\za^1,\za^2)}(q^\zk,p_\zl) = \za^1\left(\sqrt{g^{\zk\zl}(p_\zk - eA_\zk)(p_\zl -
eA_\zl)} - m\right) + \za^2(p - e)
                                                                                \tag \label{Fxx472}$$
    parametrized by functions $\za^1 > 0$ and $\za^2$ on $\sT^\*Q$.  There are two
primary constraints:
        $$\zF_1(q,q^\zk,p,p_\zl) = \sqrt{g^{\zk\zl}(p_\zk - eA_\zk)(p_\zl - eA_\zl)} - m
                                                                                \tag \label{Fxx473}$$
    and
        $$\zF_2(q,q^\zk,p,p_\zl) = p - e.
                                                                                \tag \label{Fxx474}$$
    Integrability criteria are again satisfied.
        }{\hfill    $\blacktriangle$}\endclaim

        \claim \c{e}{Example}{}{\rm                                                     \label{Cxx25}
    For the generalized Dirac system in Example \Ref{Cxx13} we use the family of
Hamiltonians
        $$H_\za(q^\zk,p_\zl) = \frac{\za}{2} g^{\zk\zl}p_\zk p_\zl
                                                                                \tag \label{Fxx475}$$
    depending on a function $\za > 0$.  There is one primary constraint
        $$\zF(q^\zk,p_\zl) = g^{\zk\zl}p_\zk p_\zl.
                                                                                \tag \label{Fxx476}$$
    The system is integrable.
        }{\hfill    $\blacktriangle$}\endclaim

        \claim \c{e}{Example}{}{\rm                                                     \label{Cxx26}
    Let $Q$ be the affine space-time of special relativity with Cartesian
coordinates $(q^\zk)$ and a constant Minkowski metric tensor $(g_{\zk\zl})$.  We
analyse the dynamics of two interacting relativistic particles [15] [16].  The
configuration space is the product $Q \times Q$ with coordinates
$(q^\zk_1,q^\zl_2)$.  Coordinates $(q^\zk_1,q^\zl_2,\dot q^\zm_1,\dot q^\zl_2)$,
$(q^\zk_1,q^\zl_2,p_\zm^1,p_\zn^2)$, and $(q^\zk_1,q^\zl_2,p_\zm^1,p_\zn^2,\dot
q^\zr_1,\dot q^\zs_2,\dot p_\zt^1,\dot p_\zw^2)$ will be used in $\sT(Q \times Q)$,
$\sT^\*(Q \times Q)$, and $\sT\sT^\*(Q \times Q)$ respectively.  Masses of the
particles are denoted by $m_1$ and $m_2$.  The interaction potential is a function
$V$ of a real positive argument.  Relations
        $$\aligned
            g_{\zk\zl}\dot q^\zk_1 \dot q^\zl_1 &> 0 \\
            g_{\zk\zl}\dot q^\zk_2 \dot q^\zl_2 &> 0 \\
            g^{\zk\zl} p_\zk^1 p_\zl^1 &> 0 \\
            g^{\zk\zl} p_\zk^2 p_\zl^2 &> 0 \\
            g_{\zk\zl}(q^\zk_2 - q^\zk_1)(q^\zl_2 - q^\zl_1) &< 0
        \endaligned
                                                                                \tag \label{Fxx477}$$
    are assumed to be satisfied.  Abbreviations
        $$\aligned
            \|\dot q_1\| &= \sqrt{g_{\zk\zl}\dot q^\zk_1 \dot q^\zl_1} \\
            \|\dot q_2\| &= \sqrt{g_{\zk\zl}\dot q^\zk_2 \dot q^\zl_2} \\
            \|\dot p^1\| &= \sqrt{g^{\zk\zl} p_\zk^1 p_\zl^1} \\
            \|\dot p^2\| &= \sqrt{g^{\zk\zl} p_\zk^2 p_\zl^2} \\
            \|q_2 - q_1\| &= \sqrt{-g_{\zk\zl}(q^\zk_2 - q^\zk_1)(q^\zl_2 - q^\zl_1)}\\
            \overline{m}_1 &= \sqrt{m_1{}^2 + V(\|x_2 - x_1\|)} \\
            \overline{m}_2 &= \sqrt{m_2{}^2 + V(\|x_2 - x_1\|)}
        \endaligned
                                                                                \tag \label{Fxx478}$$
    will be used.

    The dynamics of the particles is a generalized Dirac system described by
Lagrange equations
        $$\aligned
            \dot p_\zk^1 &= \frac{\xD V(\|q_2 - q_1\|)}{2\|q_2 - q_1\|}\left(\frac{\|\dot
q_1\|}{\overline{m}_1} + \frac{\|\dot
q_2\|}{\overline{m}_2}\right)g_{\zk\zl}(q^\zl_2 - q^\zl_1) \\
            \dot p_\zk^2 &= \frac{\xD V(\|q_2 - q_1\|)}{2\|q_2 - q_1\|}\left(\frac{\|\dot
q_1\|}{\overline{m}_1} + \frac{\|\dot
q_2\|}{\overline{m}_2}\right)g_{\zk\zl}(q^\zl_1 - q^\zl_2) \\
            p_\zk^1 &= \frac{\overline{m}_1}{\|\dot q_1\|} g_{\zk\zl} \dot q^\zl_1 \\
            p_\zk^2 &= \frac{\overline{m}_2}{\|\dot q_2\|} g_{\zk\zl} \dot q^\zl_2 \\
        \endaligned
                                                                                \tag \label{Fxx479}$$
    derived from the Lagrangian
        $$L(q^\zk_1,q^\zl_2,\dot q^\zm_1,\dot q^\zn_2) = \overline{m}_1\,\|\dot q_1\| +
\overline{m}_2\,\|\dot q_2\|.
                                                                                \tag \label{Fxx480}$$

    The Hamiltonian is the Morse family
        $$H(q^\zk_1,q^\zl_2,p_\zm^1,p_\zn^2,v^1,v^2) = v^1 \left(\|p_1\| -
\overline{m}_1\right) + v^2 \left(\|p_2\| - \overline{m}_2\right)
                                                                                \tag \label{Fxx481}$$
    of functions of the variables $v^1 > 0$ and $v^2 > 0$.  Hamilton equations
        $$\aligned
            \dot p_\zk^1 &= \frac{\xD V(\|q_2 - q_1\|)}{2\|q_2 -
q_1\|}\left(\frac{v^1}{\overline{m}_1} +
\frac{v^2}{\overline{m}_2}\right)g_{\zk\zl}(q^\zl_2 - q^\zl_1) \\
            \dot p_\zk^2 &= \frac{\xD V(\|q_2 - q_1\|)}{2\|q_2 -
q_1\|}\left(\frac{v^1 }{\overline{m}_1} +
\frac{v^2}{\overline{m}_2}\right)g_{\zk\zl}(q^\zl_1 - q^\zl_2) \\
            \dot q^\zk_1 &= \frac{v^1}{\|p^1\|} g^{\zk\zl} p_\zl^1 \\
            \dot q^\zk_2 &= \frac{v^2}{\|p^2\|} g^{\zk\zl} p_\zl^2 \\
            \|p^1\| &= \overline{m}_1 \\
            \|p^2\| &= \overline{m}_2
        \endaligned
                                                                                \tag \label{Fxx482}$$
    are equivalent to the Lagrange equations \Ref{Fxx479}.

    The dynamics is the union of the family of Hamiltonian vector fields
        $$X_{(\za^1,\za^2)} \colon \sT^\*(Q \times Q) \rightarrow \sT\sT^\*(Q \times Q)
                                                                                \tag \label{Fxx483}$$
    generated by the family of Hamiltonians
        $$H_{(\za^1,\za^2)}(q^\zk_1,q^\zl_2,p_\zm^1,p_\zn^2) = \za^1 \left(\|p_1\| -
\overline{m}_1\right) + \za^2 \left(\|p_2\| - \overline{m}_2\right)
                                                                                \tag \label{Fxx484}$$
    and restricted to a submanifold $C \subset \sT^\*(Q \times Q)$ described by two
primary constraints:
        $$\zF_1(q^\zk_1,q^\zl_2,p_\zm^1,p_\zn^2) = \|p_1\| - \overline{m}_1
                                                                                \tag \label{Fxx485}$$
    and
        $$\zF_2(q^\zk_1,q^\zl_2,p_\zm^1,p_\zn^2) = \|p^2\| - \overline{m}_2.
                                                                                \tag \label{Fxx486}$$
    The parameters $(\za^1,\za^2)$ are positive functions on $\sT^\*(Q \times Q)$.
The Poisson brackets
        $$\{H_{(\za^1,\za^2)},\zF_1\}|C = \za^2 \frac{\xD V(\|q_2 -
q_1\|)}{2\overline{m}_1\overline{m}_2\|q_2 - q_1\|}(p_\zk^1 + p_\zk^2)(q^\zk_1 -
q^\zk_2)
                                                                                \tag \label{Fxx487}$$
    and
        $$\{H_{(\za^1,\za^2)},\zF_2\}|C = \za^1 \frac{\xD V(\|q_2 -
q_1\|)}{2\overline{m}_1\overline{m}_2\|q_2 - q_1\|}(p_\zk^1 + p_\zk^2)(q^\zk_2 -
q^\zk_1)
                                                                                \tag \label{Fxx488}$$
    indicate that the dynamics is not integrable unless the potential $V$ is
constant.
        }{\hfill    $\blacktriangle$}\endclaim

    The generalized Dirac system in Example \Ref{Cxx26} is the only case of a non
integrable dynamics known to us.  We will extract the integrable part of this system
by applying different versions of the extraction algorithm.

        \claim \c{e}{Example}{}{\rm                                                     \label{Cxx27}
    We apply the first algorithm of Section 4 to Hamilton equations \Ref{Fxx482} and
constraint set $C$ described by
        $$\aligned
            \|p^1\| &= \overline{m}_1 \\
            \|p^2\| &= \overline{m}_2
        \endaligned
                                                                                \tag \label{Fxx489}$$
    We obtain equations
        $$\aligned
            \|p^1\| &= \overline{m}_1 \\
            \|p^2\| &= \overline{m}_2 \\
            g^{\zk\zl}p_\zk^1\dot p_\zl^1 &= \frac{\xD V(\|q_2 - q_1\|)}{2\|q_2 -
q_1\|}g_{\zk\zl}(q^\zk_2 - q^\zk_1)(\dot q^\zl_2 - \dot q^\zl_1) \\
            g^{\zk\zl}p_\zk^2\dot p_\zl^2 &= \frac{\xD V(\|q_2 - q_1\|)}{2\|q_2 -
q_1\|}g_{\zk\zl}(q^\zk_2 - q^\zk_1)(\dot q^\zl_2 - \dot q^\zl_1)
        \endaligned
                                                                                \tag \label{Fxx490}$$
    for $\sT C$ and equations
        $$\aligned
            \dot p_\zk^1 &= \frac{\xD V(\|q_2 - q_1\|)}{2\|q_2 -
q_1\|}\left(\frac{v^1}{\overline{m}_1} +
\frac{v^2}{\overline{m}_2}\right)g_{\zk\zl}(q^\zl_2 - q^\zl_1) \\
            \dot p_\zk^2 &= \frac{\xD V(\|q_2 - q_1\|)}{2\|q_2 -
q_1\|}\left(\frac{v^1 }{\overline{m}_1} +
\frac{v^2}{\overline{m}_2}\right)g_{\zk\zl}(q^\zl_1 - q^\zl_2) \\
            \dot q^\zk_1 &= \frac{v^1}{\|p^1\|} g^{\zk\zl} p_\zl^1 \\
            \dot q^\zk_2 &= \frac{v^2}{\|p^2\|} g^{\zk\zl} p_\zl^2 \\
            \|p^1\| &= \overline{m}_1 \\
            \|p^2\| &= \overline{m}_2 \\
            0 &= (p_\zk^1 + p_\zk^2)(q^\zk_2 - q^\zk_1) \\
            v^1 &> 0 \\
            v^2 &> 0
        \endaligned
                                                                                \tag \label{Fxx491}$$
    for the intersection $\overline D^1 = D \cap \sT C$ if the potential $V$ is not
constant.  For the new constraint set $\overline{C}^1 = \zt_Q(\overline{D}^1)$ we
have equations
        $$\aligned
            \|p^1\| &= \overline{m}_1 \\
            \|p^2\| &= \overline{m}_2 \\
            0 &= (p_\zk^1 + p_\zk^2)(q^\zk_2 - q^\zk_1)
        \endaligned
                                                                                \tag \label{Fxx492}$$
    and equations
        $$\aligned
            \|p^1\| &= \overline{m}_1 \\
            \|p^2\| &= \overline{m}_2 \\
            0 &= (p_\zk^1 + p_\zk^2)(q^\zk_2 - q^\zk_1) \\
            g^{\zk\zl}p_\zk^1\dot p_\zl^1 &= \frac{\xD V(\|q_2 - q_1\|)}{2\|q_2 -
q_1\|}g_{\zk\zl}(q^\zk_2 - q^\zk_1)(\dot q^\zl_2 - \dot q^\zl_1) \\
            g^{\zk\zl}p_\zk^2\dot p_\zl^2 &= \frac{\xD V(\|q_2 - q_1\|)}{2\|q_2 -
q_1\|}g_{\zk\zl}(q^\zk_2 - q^\zk_1)(\dot q^\zl_2 - \dot q^\zl_1) \\
            0 &= (\dot p_\zk^1 + \dot p_\zk^2)(q^\zk_2 - q^\zk_1) + (p_\zk^1 +
p_\zk^2)(\dot q^\zk_2 - \dot q^\zk_1)
        \endaligned
                                                                                \tag \label{Fxx493}$$
    for $\sT \overline C^1$.  The equations
        $$\aligned
            \dot p_\zk^1 &= \frac{\xD V(\|q_2 - q_1\|)}{2\|q_2 -
q_1\|}\left(\frac{v^1}{\overline{m}_1} +
\frac{v^2}{\overline{m}_2}\right)g_{\zk\zl}(q^\zl_2 - q^\zl_1) \\
            \dot p_\zk^2 &= \frac{\xD V(\|q_2 - q_1\|)}{2\|q_2 -
q_1\|}\left(\frac{v^1 }{\overline{m}_1} +
\frac{v^2}{\overline{m}_2}\right)g_{\zk\zl}(q^\zl_1 - q^\zl_2) \\
            \dot q^\zk_1 &= \frac{v^1}{\|p^1\|} g^{\zk\zl} p_\zl^1 \\
            \dot q^\zk_2 &= \frac{v^2}{\|p^2\|} g^{\zk\zl} p_\zl^2 \\
            \|p^1\| &= \overline{m}_1 \\
            \|p^2\| &= \overline{m}_2 \\
            p_\zk^1 + p_\zk^2 &\neq 0 \\
            0 &= (p_\zk^1 + p_\zk^2)(q^\zk_2 - q^\zk_1) \\
            v^1 &> 0 \\
            v^2 &> 0 \\
            \frac{v^1}{\overline m_1} g^{\zk\zl}(p_\zk^1 + p_\zk^2)p_\zl^1 &=
\frac{v^2}{\overline m_2} g^{\zk\zl}(p_\zk^1 + p_\zk^2)p_\zl^2
        \endaligned
                                                                                \tag \label{Fxx494}$$
    for $\overline D^2 = D \cap \sT \overline{C}^1$ of $D$ are the last step in the
algorithm.  We have excluded from $\overline D^2$ the case $p_\zk^1 + p_\zk^2 = 0$.
The system $\overline D = \overline D^2$ will be shown to be integrable.
        }{\hfill    $\blacktriangle$}\endclaim

        \claim \c{e}{Example}{}{\rm                                                     \label{Cxx28}
    We apply the second algorithm of Section 4 to the Lagrange equations
\Ref{Fxx479}.  The prolongation of the Lagrange equations is the system of second
order equations
        $$\aligned
        \dot p_\zk^1 &= \frac{\xD V(\|q_2 - q_1\|)}{2\|q_2 - q_1\|}\left(\frac{\|\dot
q_1\|}{\overline{m}_1} + \frac{\|\dot
q_2\|}{\overline{m}_2}\right)g_{\zk\zl}(q^\zl_2 - q^\zl_1) \\
        \dot p_\zk^2 &= \frac{\xD V(\|q_2 - q_1\|)}{2\|q_2 - q_1\|}\left(\frac{\|\dot
q_1\|}{\overline{m}_1} + \frac{\|\dot
q_2\|}{\overline{m}_2}\right)g_{\zk\zl}(q^\zl_1 - q^\zl_2) \\
        p_\zk^1 &= \frac{\overline{m}_1}{\|\dot q_1\|} g_{\zk\zl} \dot q^\zl_1 \\
        p_\zk^2 &= \frac{\overline{m}_2}{\|\dot q_2\|} g_{\zk\zl} \dot q^\zl_2 \\
        \ddot p_\zk^1 &= f^1(q^\zk_1,q^\zl_2,\dot q^\zm_1,\dot q^\zl_2,\ddot
q^\zr_1,\ddot q^\zs_2) \\
        \ddot p_\zk^2 &= f^2(q^\zk_1,q^\zl_2,\dot q^\zm_1,\dot q^\zl_2,\ddot
q^\zr_1,\ddot q^\zs_2) \\
        \dot p_\zk^1 &= - \frac{\xD V(\|q_2 - q_1\|)}{2\overline{m}_1\|q_2 -
q_1\|} g_{\zm\zn}(q^\zm_2 - q^\zm_1)(\dot q^\zn_2 - \dot
q^\zn_1)g_{\zk\zl}\frac{\dot q^\zl_1}{\|\dot q_1\|} + \overline{m}_1
g_{\zk\zl}\left(\zd^\zl_\zm - g_{\zm\zn}\frac{\dot q^\zl_1 \dot q^\zn_1}{\|\dot
q_1\|}\right)\frac{\ddot q^\zm_1}{\|\dot q_1\|}  \\
        \dot p_\zk^2 &= - \frac{\xD V(\|q_2 - q_1\|)}{2\overline{m}_2\|q_2 -
q_1\|} g_{\zm\zn}(q^\zm_2 - q^\zm_1)(\dot q^\zn_2 - \dot
q^\zn_1)g_{\zk\zl}\frac{\dot q^\zl_2}{\|\dot q_2\|} + \overline{m}_1
g_{\zk\zl}\left(\zd^\zl_\zm - g_{\zm\zn}\frac{\dot q^\zl_2 \dot q^\zn_2}{\|\dot
q_2\|}\right)\frac{\ddot q^\zm_2}{\|\dot q_2\|}
        \endaligned
                                                                                \tag \label{Fxx495}$$
    The exact form of the functions $f^1(q^\zk_1,q^\zl_2,\dot q^\zm_1,\dot
q^\zl_2,\ddot q^\zr_1,\ddot q^\zs_2)$ and $f^2(q^\zk_1,q^\zl_2,\dot q^\zm_1,\dot
q^\zl_2,\ddot q^\zr_1,\ddot q^\zs_2)$ is of no interest since equations for $\ddot
p_\zk^1$ and $\ddot p_\zk^2$ can not impose restrictions on first derivatives.
Projection in $\sT(Q \times Q)$ eliminates these equations and also the components of
the last two equations orthogonal to $(\dot q^\zk_1)$ and $(\dot q^\zk_2)$
respectively.  The resulting first order equations
        $$\aligned
            \dot p_\zk^1 &= \frac{\xD V(\|q_2 - q_1\|)}{2\|q_2 - q_1\|}\left(\frac{\|\dot
q_1\|}{\overline{m}_1} + \frac{\|\dot
q_2\|}{\overline{m}_2}\right)g_{\zk\zl}(q^\zl_2 - q^\zl_1) \\
            \dot p_\zk^2 &= \frac{\xD V(\|q_2 - q_1\|)}{2\|q_2 - q_1\|}\left(\frac{\|\dot
q_1\|}{\overline{m}_1} + \frac{\|\dot
q_2\|}{\overline{m}_2}\right)g_{\zk\zl}(q^\zl_1 - q^\zl_2) \\
            p_\zk^1 &= \frac{\overline{m}_1}{\|\dot q_1\|} g_{\zk\zl} \dot q^\zl_1 \\
            p_\zk^2 &= \frac{\overline{m}_2}{\|\dot q_2\|} g_{\zk\zl} \dot q^\zl_2 \\
            \dot p_\zk^1 \dot q^\zk_1 &= -\frac{\xD V(\|q_2 - q_1\|)}{2\|q_2 - q_1\|}\frac{\|\dot
q_1\|}{\overline{m}_1}g_{\zk\zl}(q^\zl_2 - q^\zl_1)\dot q^\zk_1 \\
            \dot p_\zk^2 \dot q^\zk_2 &= -\frac{\xD V(\|q_2 - q_1\|)}{2\|q_2 - q_1\|}\frac{\|\dot
q_2\|}{\overline{m}_2}g_{\zk\zl}(q^\zl_1 - q^\zl_2)\dot q^\zk_2
        \endaligned
                                                                                \tag \label{Fxx496}$$
    are equivalent to the simplified equations
        $$\aligned
            \dot p_\zk^1 &= \frac{\xD V(\|q_2 - q_1\|)}{2\|q_2 - q_1\|}\left(\frac{\|\dot
q_1\|}{\overline{m}_1} + \frac{\|\dot
q_2\|}{\overline{m}_2}\right)g_{\zk\zl}(q^\zl_2 - q^\zl_1) \\
            \dot p_\zk^2 &= \frac{\xD V(\|q_2 - q_1\|)}{2\|q_2 - q_1\|}\left(\frac{\|\dot
q_1\|}{\overline{m}_1} + \frac{\|\dot
q_2\|}{\overline{m}_2}\right)g_{\zk\zl}(q^\zl_1 - q^\zl_2) \\
            p_\zk^1 &= \frac{\overline{m}_1}{\|\dot q_1\|} g_{\zk\zl} \dot q^\zl_1 \\
            p_\zk^2 &= \frac{\overline{m}_2}{\|\dot q_2\|} g_{\zk\zl} \dot q^\zl_2 \\
            0 &= (p_\zk^1 + p_\zk^2)(q^\zk_2 - q^\zk_1)
        \endaligned
                                                                                \tag \label{Fxx497}$$

    These equations are not yet integrable.  They were falsely declared the
integrable part of the dynamics in [8].  The prolongation of equations \Ref{Fxx497}
projected in $\sT\sT^\*(Q \times Q)$ results in equations
        $$\aligned
            \dot p_\zk^1 &= \frac{\xD V(\|q_2 - q_1\|)}{2\|q_2 - q_1\|}\left(\frac{\|\dot
q_1\|}{\overline{m}_1} + \frac{\|\dot
q_2\|}{\overline{m}_2}\right)g_{\zk\zl}(q^\zl_2 - q^\zl_1) \\
            \dot p_\zk^2 &= \frac{\xD V(\|q_2 - q_1\|)}{2\|q_2 - q_1\|}\left(\frac{\|\dot
q_1\|}{\overline{m}_1} + \frac{\|\dot
q_2\|}{\overline{m}_2}\right)g_{\zk\zl}(q^\zl_1 - q^\zl_2) \\
            p_\zk^1 &= \frac{\overline{m}_1}{\|\dot q_1\|} g_{\zk\zl} \dot q^\zl_1 \\
            p_\zk^2 &= \frac{\overline{m}_2}{\|\dot q_2\|} g_{\zk\zl} \dot q^\zl_2 \\
            0 &= (p_\zk^1 + p_\zk^2)(q^\zk_2 - q^\zk_1) \\
            0 &= (p_\zk^1 + p_\zk^2)(\dot q^\zk_2 - \dot q^\zk_1)
        \endaligned
                                                                                \tag \label{Fxx498}$$
    From
        $$(p_\zk^1 + p_\zk^2)(\dot q^\zk_2 - \dot q^\zk_1) = \frac{\|\dot
q_2\|}{\overline{m}_2} g^{\zk\zl}(p_\zk^1 + p_\zk^2)p_\zl^2 - \frac{\|\dot
q_1\|}{\overline{m}_1} g^{\zk\zl}(p_\zk^1 + p_\zk^2)p_\zl^1
                                                                                \tag \label{Fxx499}$$
    we see that the last equation in \Ref{Fxx498} imposes a synchronization relation
between parametrizations of the world lines of the particles unless $p_\zk^1 +
p_\zk^2 = 0$.  The case $p_\zk^1 + p_\zk^2 = 0$ seems to be too restrictive to be of
interest.  We will exclude this case.  The integrability algorithm applied to
equations \Ref{Fxx498} produces no further restrictions.
        }{\hfill    $\blacktriangle$}\endclaim

        \claim \c{e}{Example}{}{\rm                                                     \label{Cxx29}
    Although the dynamics of two relativistic particles is not a Dirac system it
admits the Dirac-style representation in terms of constrained Hamiltonian vector
fields.  This representation can be used to simplify the algorithm in Example
\Ref{Cxx27} and to prove the integrability of the resulting equations.  The
simplified algorithm is an adaptation of Dirac's original algorithm [1].  Since the
Poisson brackets \Ref{Fxx487} and \Ref{Fxx488} do not vanish on $C$ we restrict the set
$C$ by adding the {\it secondary constraint}
        $$\zC(q^\zk_1,q^\zl_2,p_\zm^1,p_\zn^2) = (p_\zk^1 + p_\zk^2)(q^\zk_2 -
q^\zk_1).
                                                                                \tag \label{Fxx500}$$
    The resulting system $\overline D^1$ is now the union of images of Hamiltonian
vector fields generated by the family \Ref{Fxx484} restricted to the constraint set
$\overline{C}^1 \subset \sT^\*(Q \times Q)$ satisfying the equations
        $$\zF_1 = 0,\; \zF_2 = 0,\; \zC = 0.
                                                                                \tag \label{Fxx501}$$
    The system is not yet integrable.  In the second step we require the vanishing of
the Poisson bracket
        $$\{H_{(\za^1,\za^2)},\zC\} = \frac{\za^1}{\|p^1\|}
g^{\zk\zl}(p_\zk^1 + p_\zk^2)p_\zl^1 - \frac{\za^2}{\|p^2\|}
g^{\zk\zl}(p_\zk^1 + p_\zk^2)p_\zl^2
                                                                                \tag \label{Fxx502}$$
    on the new constraint set.  We will exclude the case $p_\zk^1 + p_\zk^2 = 0$ and
impose the condition
        $$\frac{\za^1}{\overline m_1} g^{\zk\zl}(p_\zk^1 + p_\zk^2)p_\zl^1 -
\frac{\za^2}{\overline m_2} g^{\zk\zl}(p_\zk^1 + p_\zk^2)p_\zl^2 = 0
                                                                                \tag \label{Fxx503}$$
    on the functions $(\za^1,\za^2)$.  The resulting system $\overline D^2$ is the
union of images of constrained Hamiltonian vector fields generated by the family
\Ref{Fxx484} with the positive functions $(\za^1,\za^2)$ satisfying the condition
\Ref{Fxx503}.  The vector fields are restricted to the constraint set
$\overline{C}^1$.  The system is exactly the system described by equations
\Ref{Fxx494} and \Ref{Fxx498}.  The system is integrable since
        $$\{H_{(\za^1,\za^2)},\zF_1\}|\overline{C}^1 = 0,
                                                                                \tag \label{Fxx504}$$
        $$\{H_{(\za^1,\za^2)},\zF_2\}|\overline{C}^1 = 0,
                                                                                \tag \label{Fxx505}$$
    and
        $$\{H_{(\za^1,\za^2)},\zC\}|\overline{C}^1 = 0.
                                                                                \tag \label{Fxx506}$$
        }{\hfill    $\blacktriangle$}\endclaim

    \vskip3mm
        \leftline{\black References}
    \vskip1mm
    \noindent [1] P. A. M. Dirac, {\it Generalized Hamiltonian
Dynamics}, Canad. J. Math. {\bf 2} (1950), 129--148.
    \noindent \phantom{[1] P. A. M. Dirac,}
{\it Lectures in Quantum Mechanics}, Belfer Graduate School
of Science, Yeshiva University, New York 1967.
    \vskip3mm
    \noindent \phantom{[1]} J. L. Anderson, P. G. Bergmann, {\it Constraints in
Covariant Field Theories}, Phys. Rev., {\bf 83} (1951), 1018--1026.
    \vskip3mm
    \noindent [2] H. Cendra, D. D. Holm, M. J. W. Hoyle, J. E. Marsden, {\it The
Maxwell-Vlasov equations in Euler-Poincar\'e form}, J. Math. Phys. {\bf 39} (1998),
3138--3157.
    \vskip3mm
    \noindent [3] R. P. Feynman, {\it The Theory of Positrons}, Phys. Rev. {\bf 76}
(1949), 749--759.
    \vskip3mm
    \noindent [4] L. H\"ormander, {\it Fourier integral operators}, Acta Mathematica
{\bf 127} (1971), 79--183.
    \vskip3mm
    \noindent [5] Th. Kaluza, {\it Zum Unit\"atsproblem der Physik}, Berl. Berichte
(1921), 966.
    \vskip3mm
    \noindent [6] R. Kerner, {\it Generalization of the Kaluza-Klein theory for an
arbitrary non-abelian gauge group}, Ann. Inst. Henri Poincar\'e {\bf IX} (1968),
143--152.
    \vskip3mm
    \noindent [7] P. Libermann, M.-C. Marle, {\it Symplectic Geometry and Analytical
Mechanics}, D. Reidel Publishing Company, Dordrecht, 1987.
    \vskip3mm
    \noindent [8] G. Marmo, G. Mendela, W. M. Tulczyjew,  {\it Constrained
Hamiltonian systems as implicit differential equations}, J. Phys. A: Gen. {\bf 30}
(1997) 277--293,
    \vskip3mm
    \noindent [9] J. E. Marsden, T. S. Ratiu, {\it Introduction to Mechanics and
Symmetry}, Texts in Applied Mathematics, vol. 17, Springer-Verlag, New York Berlin
Heidelberg London Paris Tokyo Hong Kong Barcelona Budapest, 1994.
    \vskip3mm
    \noindent [10] Maria Rosa Menzio, W. M. Tulczyjew, {\it Infinitesimal symplectic
relations and generalized Hamiltonian dynamics}, Ann. Inst. H. Poincar\'e {\bf 28}
(1978), 349--367.
    \vskip3mm
    \noindent [11] E. C. C. Stueckelberg, {\it La m\'ecanique du point mat\'eriel en
th\'eorie de relativit\'e et en th\'eorie des quanta}, Helv. Phys. Acta {\bf 15},
(1942).
    \vskip3mm
    \noindent [12] W. M. Tulczyjew, {\it Hamiltonian Systems, Lagrangian Systems and the
Legendre Transformation}, Symposia Mathematica {\bf 16} (1974), 247-258.
    \vskip3mm
    \noindent \phantom{[12] W. M. Tulczyjew,} {\it The
Legendre Transformation}, Ann. Inst. H. Poincar\'e {\bf 27} (1977), 101--114.
    \vskip3mm
    \noindent [13] W. M. Tulczyjew, P. Urbanski,  {\it Homogeneous Lagrangian
systems}, in {\it Gravitation, Electromagnetism and Geometric Structures}, Pitagora
Editrice, 1996
    \vskip3mm
    \noindent [14] R. Utiyama, {\it Invariant theory of interactions}, Phys. Rev.
{\bf 101} (1956), 1597.
    \vskip3mm
    \noindent [15] {\it The Theory of Action-at-a-Distance in Relativistic Particle Mechanics},
ed. by E. H. Kerner, Gordon and Breach, New York, NY, 1972.
    \vskip3mm
    \noindent [16] {\it Relativistic Action-at-a-Distance: Classical and Quantum
Aspects}, ed. by J. Llosa, Proc. of the Barcelona Workshop, Lecture Notes in
Physics, vol. 162, Springer-Verlag, Berlin, 1982.

\enddocument